\documentclass[11pt,a4paper]{article}
\usepackage{jheppub}
\usepackage{amsmath}
\allowdisplaybreaks
\usepackage{hyperref}
\usepackage[T1]{fontenc}
\usepackage{fix-cm}
\usepackage{lmodern}
\usepackage{mathtools}
\usepackage{mathrsfs}
\usepackage{empheq}
\usepackage{physics}
\usepackage{bbm}
\usepackage{bm}
\usepackage{hhline}
\usepackage{tensor}
\usepackage{tikz}
\usepackage[dvipsnames, svgnames]{xcolor}
\definecolor{customOrange}{HTML}{E69F00}
\definecolor{skyBlue}{HTML}{56B4E9}
\definecolor{bluishGreen}{HTML}{009E73}
\definecolor{reddishPurple}{HTML}{CC79A7}
\definecolor{darkBlue}{HTML}{0072B2}
\definecolor{vermillion}{HTML}{D55E00}
\definecolor{charcoal}{HTML}{333333}
\definecolor{yellow}{HTML}{F0E442}
\definecolor{lightGray}{HTML}{CCCCCC}
\definecolor{purple}{HTML}{7570B3}
\definecolor{lightGreen}{HTML}{A6D854}
\definecolor{terracotta}{HTML}{E7298A}
\usepackage{pifont}
\usepackage{latexsym}
\usepackage{slashed}
\usepackage{multirow}
\usepackage{float}
\usepackage{afterpage}
\usepackage{booktabs}
\usepackage{tabularx}
\newcolumntype{C}{>{\centering\arraybackslash}X}
\renewcommand\arraystretch{1.5}
\usepackage{ytableau}
\usepackage{adjustbox}
\makeatletter
\newcommand*{\rom}[1]{\expandafter\@slowromancap\romannumeral #1@}
\makeatother
\newcommand{\be}{\begin{equation}} \newcommand{\ee}{\end{equation}}
\newcommand{\bea}{\begin{equation} \begin{aligned}} 
\newcommand{\eea}{\end{aligned} \end{equation}}

\newcommand{\cA}{\mathcal{A}}
\newcommand{\cB}{\mathcal{B}}
\newcommand{\cN}{\mathcal{N}}

\newcommand{\bC}{\mathbb{C}}
\newcommand{\bQ}{\mathbb{Q}}
\newcommand{\bZ}{\mathbb{Z}}

\newcommand{\bH}{\mathbb{H}}

\newcommand{\bN}{\mathbb{N}}

\newcommand{\bP}{\mathbb{P}}
\newcommand{\bR}{\mathbb{R}}

\newcommand{\SU}{{\rm SU}}
\newcommand{\U}{{\rm U}}

\newcommand{\cF}{\mathcal F}
\newcommand{\cG}{\mathcal{G}}
\newcommand{\cH}{\mathcal H}
\newcommand{\cI}{\mathcal I}

\newcommand{\cL}{\mathcal L}
\newcommand{\cO}{\mathcal O}

\newcommand{\Li}{\operatorname{Li}}

\newcommand\fc{\mathfrak{c}}
\newcommand\fg{\mathfrak{g}}

\newcommand{\fn}{\mathfrak n}
\newcommand{\fl}{\mathfrak l}
\newcommand{\fm}{\mathfrak m}
\newcommand{\fq}{\mathfrak q}
\newcommand{\fr}{\mathfrak r}
\newcommand{\tDelta}{\widetilde\Delta}
\newcommand{\tfn}{\widetilde{\mathfrak n}}
\newcommand{\tq}{\tilde q}
\newcommand{\wtd}[1]{\widetilde{#1}}
\newcommand{\ii}{\mathrm i}

\newcommand{\cV}{\mathcal{V}}
\newcommand{\Q}{\mathbb{Q}}

\newcommand{\cQ}{\mathcal Q}

\newcommand{\R}{\mathbb{R}}
\newcommand{\Z}{\mathbb{Z}}
\newcommand{\fracpart}[1]{\left\{#1\right\}}
\newcommand{\floor}[1]{\left\lfloor #1\right\rfloor}
\newcommand{\den}{\operatorname{den}}

\newcommand{\dist}{\operatorname{dist}}
\DeclareMathOperator{\re}{\mathbb{R}e}
\DeclareMathOperator{\im}{\mathbb{I}m}
\DeclareMathOperator{\sgn}{sgn}
\DeclareMathOperator{\Ai}{Ai}
\def\PSL{\mathrm{PSL}}
\newcommand{\wt}{\widetilde}
\def\Sp{\mathrm{Sp}}

\def\SU{\mathrm{SU}}
\newcommand{\su}{\mathfrak{su}}
\DeclareMathOperator{\diag}{diag}

\newtheorem{property}{Property}
\newtheorem{definition}{Definition}
\newtheorem{conjecture}{Conjecture}
\newtheorem{corollary}{Corollary}
\newtheorem{lemma}{Lemma}
\newtheorem{proof}{Proof}
\newtheorem{theorem}{Theorem}

\newtheorem{proposition}{Proposition}
\title{ Mapping AdS$_4$ quantum gravity via the Bethe Ansatz formula }

\author[a,b]{Cheng Peng}
\author[a]{, Xuao Zhang}
\author[a]{, and Ziruo Zhang}

\affiliation[a]{University of Chinese Academy of Sciences, Beijing 100190, China}
\affiliation[b]{Peng Huanwu Center for Fundamental Theory, Hefei, Anhui 230026, China}

\emailAdd{pengcheng@ucas.ac.cn}
\emailAdd{zhangxuao@ucas.ac.cn}
\emailAdd{zhangziruo@ucas.ac.cn}

\abstract{The topologically twisted index of 3d ADHM theory can be written as a sum over solutions to the Bethe Ansatz Equations (BAEs) of the XXZ-type model constructed from quantum toroidal $\fg\fl_1$. Via a novel numerical algorithm, we solve for all $p(N)$ physical solutions at $N\leq 21$, where $p(N)$ denotes the number of partitions of $N$. Surveying these solutions, we find that in addition to the known ``1-cut'' solution dual to dyonic black holes, where the gauge holonomies are distributed along a single curve in the complex plane, there are ``dressed'' 1-cut solutions dual to combined configurations of the black hole and M2-branes, $m$-cut solutions dual to $\bZ_m$ orbifolds of the black hole, a discrete family of solutions interpolating between the 1-cut and 3-cut solutions whose bulk dual is not yet identified, together with other exotic solutions, at each value of $N$. We show that the $1,2,3$-cut solutions and interpolating solutions all have a well-defined large $N$ limit, and analytically compute their leading order distribution of gauge holonomies. Whenever possible, we fit the contributions of the solutions to the index to all orders in the $1/N$ expansion with high-precision numerics, and extract the leading non-perturbative correction. With exceptions, these are matched with the results of equivariant localization in the bulk, and the action of wrapped M2-branes respectively.}

\begin{document}

\begin{flushright}
USTC-ICTS/PCFT-26-75
\end{flushright}

	\maketitle
	%%%%%%%%%%%%%%%%%%%%%%%%

	%%%%%%%%%%%%%%%%%%%%%%%%
	\section{Introduction}
	\label{sec:intro}
	%%%%%%%%%%%%%%%%%%%%%%%%
	
	The AdS/CFT correspondence \cite{Maldacena:1997re,Gubser:1998bc,Witten:1998qj} in its strongest form is an exact duality between quantum gravity on asymptotically anti-de-Sitter (AdS) spacetimes and a conformal field theory (CFT) living on the boundary of AdS. This opens up an important avenue towards further understanding quantum gravity on asymptotically AdS spacetimes since we can define it as the dual CFT, which is often better understood and computationally more tractable. This has proven to be especially fruitful in the context of supersymmetric observables such as supersymmetric partition functions or indices, since these can be computed exactly via supersymmetric localization \cite{Pestun:2007rz,Pestun:2016zxk}. In particular, the entropy of many supersymmetric (BPS) asymptotically AdS black holes in various dimensions have been successfully accounted for by the counting of BPS states in the dual CFT~\cite{Benini:2015eyy,Choi:2018hmj,Benini:2018ywd,Cabo-Bizet:2018ehj}.

    A tool that has been useful in some of these computations is the Bethe ansatz formula. For example, the 4d superconformal index (SCI) \cite{Romelsberger:2005eg, Kinney:2005ej} and 3d topologically twisted index (TTI) \cite{Benini:2015noa, Benini:2016hjo} can both be written as a sum over residues, where the locations of the poles are at solutions to a set of Bethe ansatz equations (BAEs) \cite{Closset:2017zgf,Benini:2018mlo}, so named because their form resembles the Bethe ansatz equations determining the energy eigenstates of an integrable system. For some field theories, one can show that this resemblance is not merely
    cosmetic; one can identify the specific integrable systems whose Bethe ansatz equations are precisely the equations obtained from the supersymmetric indices. In such cases, this coincidence is also called the Bethe/gauge correspondence \cite{Nekrasov:2009ui, Nekrasov:2009rc}. The 3d ADHM theory \cite{Hosomichi:2014rqa, Crew:2020psc} that we will focus on in this paper is exactly such an example where the equations coincide, and the corresponding integrable system is an XXZ-type model constructed from quantum toroidal $\fg\fl_1$ \cite{Koroteev:2020mxs, Crew:2024fom}. 
    
    In the computations that established the microscopic interpretation of the entropy, a particular Bethe vacuum (i.e. solution to the BAEs) was singled out, and its leading large $N$ contribution to an index was found to reproduce the entropy of the corresponding BPS AdS black hole after the appropriate Legendre transform or extremization with respect to the chemical potentials. In three dimensions, this was demonstrated for dyonically charged AdS$_4$ black holes using the 3d TTI \cite{Benini:2015eyy,Benini:2016rke,Hosseini:2016tor,Azzurli:2017kxo,Hosseini:2017fjo,Benini:2017oxt}, while analogous results were obtained in four dimensions for supersymmetric AdS$_5$ black holes using the 4d SCI \cite{Benini:2018ywd, GonzalezLezcano:2019nca, Lanir:2019abx, Benini:2020gjh}.
    
    This perspective raises two natural questions. First, can the subleading terms in the large $N$ expansion of the contribution of a Bethe vacuum be given a gravitational interpretation? For the 3d TTI of M2-brane theories, the contribution of the Bethe vacuum corresponding to the black hole solution can be determined to all perturbative orders in $N$ \cite{Bobev:2022jte,Bobev:2022eus}, and recent work using equivariant localization in M-theory has reproduced the perturbative Airy structure directly from the 11d quantum effective action \cite{Bobev:2026gir,BenettiGenolini:2026cyc}. In parallel, \cite{Aharony:2021zkr} studied the corresponding subleading and non-perturbative corrections in the 4d SCI. In particular, the exponentially small corrections in $N$ were matched with the classical actions of supersymmetric D3-branes wrapped on suitable cycles of the Euclidean black hole geometry.
    
    Second, and more importantly, what is the gravitational interpretation, if any, of the other Bethe vacua appearing in the Bethe ansatz formula? This question has been investigated in four dimensions in the previously mentioned work \cite{Aharony:2021zkr}. Apart from the BAE solution whose leading contribution reproduces the black hole entropy, the authors found that the other Hong-Liu solutions of \cite{Hong:2018viz, ArabiArdehali:2019orz} correspond to orbifolds of Euclidean black holes. 
    
    In this work, we pursue these questions further in the context of the 3d ADHM theory with one flavour, whose BAEs are those of the XXZ system constructed from quantum toroidal $\fg\fl_1$ \cite{Feigin:2015raa}. The theory is holographically dual to M-theory on AdS$_4\times S^7$ \cite{Bashkirov:2010kz,Hatsuda:2014vsa}. This setup is particularly suited to tackle these questions since we can leverage results from the mathematical literature. In particular, the BAE solutions that are physical from the perspective of the integrable system are completely classified, and are in one-to-one correspondence with the partitions of $N$ \cite{Feigin:2015raa}. Their total count is given by $p(N)$, the number of partitions of $N$. Assuming that the Bethe/gauge correspondence implies that only physical BAE solutions in the integrable system are physical in the gauge theory, we only need to consider these $p(N)$ solutions. Moreover, when the FI parameter $\zeta$ is zero, the BAE solutions are analytically known and take a very simple form \cite{Koroteev:2017nab, Crew:2024fom}. This enables us to develop a numerical algorithm to find all $p(N)$ solutions to the BAEs by slowly tuning $\zeta$ from $0$ to the required value, building upon the work of \cite{Crew:2024fom}. We introduce the setup and our numerical method in Section \ref{sec:TTI}.

    By examining the complete set of $p(N)$ solutions for $N\leq 21$, we found a variety of solutions with a well-defined large $N$ limit, whose contributions to the logarithm of the index scale as $N^{3/2}\sim 1/G_N$. Apart from the solution in \cite{Hosseini:2016ume} corresponding to dyonic black holes, where the gauge holonomies are distributed along a single cut in the complex plane, we also find new ``dressed'' one-cut solutions where a few gauge holonomies are outside of the cut, as well as 2-cut and 3-cut solutions where the holonomies are spread equally between 2 and 3 cuts respectively. More exotically, for a specific region of flavour chemical potentials, there also exist a discrete family of solutions which interpolate between the 1-cut and 3-cut solutions. We call these the ``interpolating solutions''. We will focus our subsequent analyses on the solutions mentioned above, even though many more solutions were found to exist, including a family of solutions interpolating between the 1-cut and 2-cut solutions, and general $m$-cut solutions for $m>3$. We emphasize that the different classes of solutions all exist at each $N$ we examined and our approach can potentially find \emph{all} of the Bethe vacua with a well-defined large $N$ limit, provided that the solution finding algorithm is performed at large enough $N$. In the following, we list the main results that we have obtained regarding the new Bethe solutions.

    \paragraph{Main results}
    \begin{itemize}

    \item We begin in Section \ref{sec:oneCut} by revisiting the one-cut solution in \cite{Hosseini:2016ume} corresponding to dyonic black holes, where the gauge holonomies are distributed along a single cut in the complex plane. This will introduce the techniques and methods that will be used repeatedly to analyze the new solutions. Firstly, we determine the region of chemical potentials where the one-cut solutions exist, and review how the distribution of gauge holonomies in the large $N$ limit can be computed. Let $\cI_\lambda$ be the contribution to the index of a Bethe root $\lambda$, while $F_\lambda = -\log\cI_\lambda$ is the corresponding free energy. The large $N$ distribution of holonomies determines $F_\lambda$ to leading order in large $N$. By constructing approximate initial solutions using the analytical distribution of the gauge holonomies, we can extend the solution to large values of $N\sim 300$. Fitting $F_\lambda$ at various $N$ to an appropriate ansatz as in \cite{Bobev:2022jte} then allows $F_\lambda$ to be determined to all perturbative orders in $1/N$, while extracting the leading non-perturbative correction. The perturbative contributions can be reproduced via equivariant localization in M-theory \cite{BenettiGenolini:2026cyc}, while the leading non-perturbative correction corresponds to a probe M2-brane wrapped on $S^3\subset S^7$. 

    \item In Section \ref{sec:dressed}, we repeat the same analysis for the dressed 1-cut solutions. The leading order distribution of gauge holonomies is identical to that of the 1-cut solution, and therefore the leading $\cO(N^{3/2})$ term in $F_\lambda$ is also the same for the 1-cut and dressed 1-cut solutions. The free energy starts to differ from that of the 1-cut solution at $\cO(N^{1/2})$, where the correction can be matched with the classical action of a M2-brane wrapping $\Sigma_\fg \times S^1$, with $\Sigma_\fg$ being the horizon and $S^1\subset S^7$. Loosely speaking, this wrapping is the Euclidean analogue of Lorentzian dual giant gravitons \cite{Grisaru:2000zn,Basu:2006id}. Their counterparts in the uplift of black holes in AdS$_5\times S^5$ have already been constructed in \cite{Aharony:2021zkr}. Thus we propose that the dressed 1-cut solutions correspond to bound states of the black hole and Euclidean M2-branes in a dual giant graviton-like configuration.

    \item We then move on to the 2-cut and 3-cut solutions in Section \ref{sec:multi-cut}. In each case, we can analytically compute the distribution of gauge holonomies in the large $N$ limit, and find $F_\lambda$ at leading order in large $N$. The leading $F_\lambda$ of the 2-cut and 3-cut solutions coincide with the on-shell actions of the $\bZ_2$ and $\bZ_3$ orbifolds of the Euclidean black saddles in \cite{Bobev:2020pjk}, which are therefore established as their bulk duals. More generally, we expect a $m$-cut solution to be dual to the $\bZ_m$ orbifold of a black saddle. Because the gauge holonomies are distributed across $m$ cuts and their contribution to the index matches that of the $\bZ_m$ orbifolds, these multi-cut solutions are directly analogous to the Hong-Liu solutions in \cite{ArabiArdehali:2019orz, Aharony:2021zkr}. 
    
    By finding the numerical solutions at a range of large $N$, we can also compute their free energies $F_\lambda$ to all perturbative orders in $1/N$, while extracting the leading non-perturbative correction. Exceptions apply when the orbifold action has a 7-dimensional fixed locus $\Sigma_\fg\times S^5$ (where $S^5 \subset S^7$). In this case, the fit using the usual ansatz fails because one must include the free energy of the 7d super-Yang-Mills (SYM) theory living on the fixed locus \cite{Anderson:2006pb}. Since we do not know what is an appropriate ansatz for the free energy as a function of the gauge coupling $g_7^2\sim N^{-1/2}$, we were not able to compute the full perturbative corrections in these cases. Even so, the presence of the 7d SYM theory can be confirmed by the fact that the coefficients of $\log N$ in $F_\lambda$ can only be reproduced if the zero modes of both 11d supergravity and 7d SYM are taken into account. If the orbifold action has no fixed points, the perturbative expansions of $F_\lambda$ can be reproduced via equivariant localization in M-theory following \cite{BenettiGenolini:2026cyc}. The exponent of the leading non-perturbative corrections can be matched with the classical M2-brane action for an M2-brane wrapping $S^3\subset S^7$ in the uplifted orbifold geometry.

    \item In Section \ref{sec:multi-cut:mixing}, we analyse the interpolating solutions. As before, we analytically compute the large $N$ distribution of gauge holonomies, and $F_\lambda$ at leading order in large $N$. However, we were unable to identify the leading term in $F_\lambda$ with any known on-shell action, and it would be very interesting to find the bulk duals of these interpolating solutions. Via the same methods as before, we numerically compute $F_\lambda$ to all perturbative orders in $1/N$ and extract the leading non-perturbative correction. The ansatz for $F_\lambda$ takes a suggestive form, hinting at the structure of the bulk dual.  

    \item The finding of new solutions to the BAEs also advances our understanding of the 3d superconformal index (SCI), and asymptotically AdS quantum gravity under the corresponding boundary conditions of \cite{Cassani:2019mms, Bobev:2019zmz, Benini:2019dyp}, which are different from those of the TTI. This is because the SCI in the Cardy limit can also be written as a sum over Bethe vacua \cite{Choi:2019zpz,Bobev:2022wem,Bobev:2024mqw}. The Cardy limit here refers to $\omega\rightarrow i0^+$, where $\omega$ is a chemical potential conjugate to spatial rotations. In Section \ref{sec:SCI}, we compute the leading $\cO(\omega^{-1})$ and first subleading $\cO(1)$ contributions to the SCI coming from the $1,3$-cut solutions and interpolating solutions. For the 1-cut and 3-cut solutions, one finds that the leading contributions correspond to the on-shell actions of AdS$_4$ Kerr-Newman black holes \cite{Kostelecky:1995ei,Cassani:2019mms} and their $\bZ_3$ orbifolds. For the interpolating solutions, as was the case in the TTI, we were unable to match their leading contribution to a known gravitational on-shell action. Notably, a one-to-one correspondence seems to exist between Bethe vacua and gravitational solutions for both boundary conditions: one specified by the TTI and the other by the SCI. It would also be interesting to construct the geometry that is dual to the interpolating solutions under the SCI boundary conditions.

    \item We also speculate about the possible bulk duals of the Bethe vacua which do not have a well-defined large $N$ limit in Section \ref{sec:gravitons}. Notably, there is a topological constraint on the bulk geometries when there are nonzero flavour fluxes.
    \end{itemize}
     
    We conclude in Section \ref{sec:discussion} with future directions. 
    
    %%%%%%%%%%%%%%%%%%%%%%%%
	\section{Topologically twisted index}
	\label{sec:TTI}
	%%%%%%%%%%%%%%%%%%%%%%%%
	
	%%%%%%%%%%%%%%%%%%%%%%%%
	\subsection{Review of the ADHM theory and its index}
	\label{subsec:TTI:ADHM}
	%%%%%%%%%%%%%%%%%%%%%%%%
	
	The ADHM theory with \(N_f\) flavours \cite{Mezei:2013gqa,Grassi:2014vwa,Hosseini:2016ume,Bobev:2022wem,Bobev:2023lkx} is a \(3\)d \(\cN=4\) theory with gauge group \(\U(N)\), one adjoint hypermultiplet, \(N_f\) fundamental hypermultiplets, and zero Chern-Simons level. In terms of \(\cN = 2\) multiplets, the \(\cN = 4\) vector multiplet decomposes into a \(\cN=2\) vector multiplet and an adjoint chiral multiplet \( C\), the adjoint hypermultiplet splits into two adjoint chiral multiplets \((A,B)\), while each fundamental hypermultiplet consists of a fundamental chiral multiplet \(I\) and an anti-fundamental chiral multiplet \(J\). 

    The theory arises as the low-energy worldvolume theory of \(N\) coincident M2-branes probing a \(\bC^2\times(\bC^2/\bZ_{N_f})\) singularity, where the \(Z_{N_f}\) action has fixed points \cite{Benini:2009qs,Bashkirov:2010kz,Mezei:2013gqa}. At large \(N\), it is holographically dual to M-theory on \(AdS_4\times S^7/\bZ_{N_f}\). Unless otherwise stated, we will only consider the theory with \(N_f = 1\) in the rest of this paper. In this case, the ADHM theory is dual in the infrared to the \(k=1\) ABJM theory \cite{Aharony:2008ug,Kapustin:2010xq}. This duality follows from reducing the same configuration of \(N\) M2-branes in flat space along different choices of the M-theory circle\footnote{When \(N_f>1\), although both the ADHM theory and ABJM theory with \(k=N_f\) arise from M2-branes probing \(\bC^4/\bZ_k\), the \(\bZ_k\) actions of the two theories are distinct, and there is no duality between them. In particular, the \(\bZ_k\) action of ABJM theory has no fixed points.} \cite{Aganagic:2009zk,Benini:2009qs,Boido:2023ojv}. Under 3d mirror symmetry, the \(N_f=1\) ADHM theory is self dual\footnote{The type IIB Hanany-Witten brane set-up that engineers the ADHM theory is self-dual under S-duality.}.

	The \(\cN=4\) R-symmetry is \(\SU(2)_H\times \SU(2)_C\). With respect to a chosen \( \cN=2\) subalgebra, \(\SU(2)_H\) is a flavour symmetry with Cartan generator \(F_H\), while the Cartan generator \(R_C\) of \(\SU(2)_C\) generates a \(\U(1)\) R-symmetry. There is also an additional R-symmetry \(R_H\) assigning unit charge to scalars in the adjoint and fundamental hypermultiplets. The superconformal R-charge combination is $R_{\text{sc}}\equiv\frac{1}{2}(R_H+R_C)$ \cite{Kapustin:2010xq}. Lastly, there is a topological flavour symmetry
	\(F_C\) acting on monopole operators, whose current is \(J_C=*\Tr F\). Following the conventions of \cite{Crew:2024fom}, we take 
	\begin{equation}
		F_1=\frac{R_H-R_C+F_H}{2},\qquad
		F_2=\frac{R_H-R_C-F_H}{2},\qquad
		F_C\,,
		\label{eq:CZZflavorbasis}        
	\end{equation}
	as the basis of flavour charges. Concretely, the scalar components of the \(\cN=2\) chiral multiplets have the charges listed in Table \ref{tab: charges}.
	\begin{table}[h]
    \centering
    \small
    \begin{equation*}
    \arraycolsep=5mm
    \begin{array}{c | c | c c | c | c c | c c}
        \toprule %\rule[-0.6em]{0pt}{1.9em}
            & G & R_H & R_C & R_{\text{sc}} & \frac{R_H-R_C}{2} & F_H & F_1 & F_2 \\
        \midrule
         I  & \square & 1 & 0 & \frac{1}{2} & \frac{1}{2} & 0 & \frac{1}{2} & \frac{1}{2} \\[.2em]
        %\hline
         J  & \overline{\square} & 1 & 0 & \frac{1}{2} & \frac{1}{2} & 0 & \frac{1}{2} & \frac{1}{2} \\[.2em]
        %\hline
         A  &  \text{adj} & 1 & 0 & \frac{1}{2} & \frac{1}{2} & 1 & 1 & 0 \\[.2em]
        %\hline
         B  &  \text{adj} & 1 & 0 & \frac{1}{2} & \frac{1}{2} & -1 & 0 & 1 \\[.2em]
        %\hline
         C  &  \text{adj} & 0 & 2 & 1 & -1 & 0 & -1 & -1 \\
        \bottomrule
    \end{array}
    \end{equation*}
    \normalsize
    \vspace{-5mm}
    \caption{\small Charges of scalar components of $\cN=2$ chiral multiplets in ADHM with one flavour.
    \label{tab: charges}}
    \end{table}
    
	The superpotential 
    \begin{equation}
    	W=\Tr\!\left(C[A,B]+JCI\right)
    	\label{eq:CZZsuperpotential}
    \end{equation}
	has charge \(2\) under both
	\(R_H\) and \(R_C\), and it is neutral under the flavour symmetries.
	
	%%%%%%%%%%%%%%%%%%%%%%%%
	\subsubsection{The index and its Bethe ansatz formula}
	\label{subsubsec:TTI:ADHM:TTI-BA}
	%%%%%%%%%%%%%%%%%%%%%%%%
	
	Our central object of study is the topologically twisted index \cite{Benini:2015noa,Benini:2016hjo,Closset:2016arn}, which is defined for any 3d \(\cN=2\) theory with an R-symmetry. Since the ADHM theory has \(\cN=4\) supersymmetry, there are two \(\cN=2\) subalgebras that can be used to define the index, corresponding to the R symmetries \(R_H\) and \(R_C\). Without loss of generality\footnote{One can check that if the topological twist is performed with respect a generic R symmetry \(R_C-\fr_1F_1-\fr_2F_2\), for integers \(\fr_{1,2}\), the formula for the index is unchanged if we simultaneously redefine the flavour fluxes as \(\fn_{1,2}\rightarrow \fn_{1,2} + \fr_{1,2}(\fg-1)\).}, we shall choose the subalgebra with \(R_C\) in the following. The index can be defined as\footnote{The index can be further refined by a fugacity for the angular momentum when \(\Sigma_\fg=S^2\), but we do not consider it here for simplicity.} 
	\begin{equation}
		\cI_{\fg}(\Delta_1,\Delta_2,\Delta_C;\fn_1,\fn_2,\fn_C)
		=
		\Tr_{\mathcal H^B_{\Sigma_\fg}}
		\left[(-1)^F t_1^{F_1}t_2^{F_2}\zeta^{F_C}\right],
		\label{eq:CZZtrace}
	\end{equation}
    where \(\cH^B_{\Sigma_\fg}\) denotes the Hilbert space on a Riemann surface \(\Sigma_\fg\) of genus \(\fg\) in the presence of a topological twist with respect to \(R_C\), and \(F\) is the fermion number. The fugacities and chemical potentials corresponding to \(F_1\), \(F_2\), and \(F_C\) are 
    \begin{equation}
		t_1 = e^{i\Delta_1},\qquad t_2  = e^{i\Delta_2},\qquad \zeta = e^{i\Delta_C}\,.
	\end{equation}
    To make subsequent formulae more compact, it will also be convenient to introduce the auxiliary variables
	\begin{equation}
		t_3=(t_1t_2)^{-1}=e^{i\Delta_3}
		\label{eq:t3def}
	\end{equation}
    The index can be computed as a path integral on \(\Sigma_\fg\times S^1\), where the background gauge field for \(R_C\) has \(2(\fg-1)\) units of flux. We shall also include background fluxes \(\fn_{1,2}\), \(\fn_C\) for the flavour symmetries \(F_{1,2}\) and \(F_C\). Since the fundamental chirals \(I,J\) have half-integer charges under \(F_{1,2}\), Dirac quantization gives
	\begin{equation}
		\fn_1,\fn_2,\fn_C\in \bZ,\qquad
		\fn_1+\fn_2\in 2\bZ.
		\label{eq:CZZDirac}
	\end{equation}
	Similar to \(\Delta_3\), it is also convenient to introduce
    \begin{equation}
        \fn_3\equiv 2(\fg-1)-\fn_1-\fn_2\,.
    \end{equation}
    By supersymmetric localization, the path integral for the index reduces to a \(N\)-dimensional contour integral over the gauge fugacities \(s_a = e^{iu_a}\,, a=1,\ldots,N\):
    \begin{align}
		\cI_{\fg}
		&=
		\frac{(-1)^N}{N!}
		\sum_{\fm\in\bZ^N}
		\int_{\rm JK}
		\prod_{a=1}^N
		\frac{ds_a}{2\pi i s_a}\det\left(\frac{\partial W_a}{\partial u_b}\right)^{\fg}\,
		(-\zeta^{-1})^{\sum_a\fm_a}
		\left(\prod_a s_a\right)^{\fn_C}
		\notag\\
		&\quad\times\prod_{\substack{a,b=1 \\ a\ne b}}^N
		(1-s_{ab})^{1-\fg}
		\prod_{I=1}^3\prod_{a,b=1}^N
		\left(\frac{s_{ab}^{1/2}t_I^{1/2}}{1-s_{ab}t_I}\right)^{
			\fm_a-\fm_b+\fn_I+1-\fg}
		\notag\\
		&\quad\times
		\prod_{a=1}^N
		\left(\frac{s_a^{1/2}t_3^{-1/4}}{1-s_a t_3^{-1/2}}\right)^{
			\fm_a-\frac{\fn_3}{2}}
		\left(\frac{s_a^{-1/2}t_3^{-1/4}}{1-s_a^{-1} t_3^{-1/2}}\right)^{
			-\fm_a-\frac{\fn_3}{2}}.
		\label{eq:CZZindex}
	\end{align}
    The \(\fm_a\) are integer valued gauge fluxes on \(\Sigma_\fg\) and JK denotes the Jeffrey-Kirwan residue \cite{Benini:2015noa,Benini:2016hjo}. We have also defined \(s_{ab}\equiv s_a/s_b\), and 
    \be
    e^{iW_a(u)}\equiv(-\zeta)^{-1}t_1^{-1/2}t_2^{-1/2}
		\frac{s_a-t_1^{1/2}t_2^{1/2}}
		{s_a-t_1^{-1/2}t_2^{-1/2}}
		\prod_{I=1}^3
		\prod_{\substack{b=1\\b\ne a}}^N
		\frac{s_b-t_I^{-1}s_a}{s_b-t_Is_a}\,.
    \ee
    Formally performing the sum over the gauge fluxes, we obtain the Bethe ansatz formula for the index
    \begin{equation}\label{eq: index BA formula}
    \begin{aligned}             &\cI_{\fg}=\sum_{\lambda\in\text{BAE}}\cI_{\fg,\lambda}=\sum_{\lambda\in\text{BAE}}\det\left(\frac{\partial W_a}{\partial u_b}\right)^{\fg-1}\left(\prod_a s_a\right)^{\fn_C}\prod_{\substack{a,b=1 \\ a\ne b}}^N
		(1-s_{ab})^{1-\fg}\\
        &\quad \times \prod_{I=1}^3\prod_{a,b=1}^N
		\left(\frac{s_{ab}^{1/2}t_I^{1/2}}{1-s_{ab}t_I}\right)^{\fn_I+1-\fg}\prod_{a=1}^N
		\left(\frac{s_a^{1/2}t_3^{-1/4}}{1-s_a t_3^{-1/2}}\right)^{-\frac{\fn_3}{2}}
		\left(\frac{s_a^{-1/2}t_3^{-1/4}}{1-s_a^{-1} t_3^{-1/2}}\right)^{-\frac{\fn_3}{2}}\,,
    \end{aligned}
    \end{equation}
	where the sum ranges over solutions to the Bethe ansatz equations (BAEs)
	\begin{equation}
		e^{iW_a(u)}
		=1,\qquad a=1,\ldots,N\,.
		\label{eq:CZZBAE}
	\end{equation}
	We shall assume in the following that $(\Delta_1,\Delta_2,\Delta_C)$ are real. This implies that solutions to the BAEs are either self-conjugate, $s^*=s$, or come in complex conjugate pairs $(s,s^*)$. The contribution of each solution labeled by \(\lambda\) is \(\cI_{\fg,\lambda}\), and let 
    \be F_{\fg,\lambda}\equiv-\log\cI_{\fg,\lambda}
\label{eq:F=logI}\ee 
be the corresponding free energy. We will also refer to solutions as Bethe roots. Note that for genus $\fg\neq 1$, the free energy is related to that of $\fg = 0$ in a simple way: 
	\begin{equation} \label{eq:general-g}
			F_{\fg,\lambda}(\Delta; \fn) = (1-\fg) F_{0,\lambda}\left(\Delta; \frac{\fn}{1-\fg}\right)\,.
	\end{equation}
	Here, \(\Delta\) collectively stands for \(\Delta_{1,2}\), \(\Delta_C\) and \(\fn\) collectively stands for \(\fn_{1,2}\), \(\fn_C\). This is straightforward to show using the form of the summand in \eqref{eq: index BA formula}. Thus we can recover the free energy for $\fg\neq 1$ from the $\fg = 0$ result. As is apparent from \eqref{eq: index BA formula}, $\fg = 1$ is a special case and will not be considered in this paper. In this case, a $R$-symmetry flux is not required to preserve supersymmetry. The topologically twisted index reduces to the ordinary 3d Witten index and returns the number of Bethe vacua when all the flavour fluxes are turned off \cite{Benini:2016hjo,Closset:2016arn, Closset:2023vos}.  
    
    The BAEs coincide with the Bethe equations of the integrable XXZ type system associated with quantum toroidal \(\fg\fl_1\) and its graded Fock module \(\cF(t_1^{\frac{1}{2}}t_2^{\frac{1}{2}})\) at degree $N$ \cite{Koroteev:2015dja, Crew:2024fom}\footnote{The Fock module \(\cF(u)\) and its lowest weight depends on a nonzero complex number \(u\), which we identify with \(t_1^{\frac{1}{2}}t_2^{\frac{1}{2}}\). The parameters \(q_{1,2,3}\) of quantum toroidal \(\fg\fl_1\) in the conventions of \cite{Feigin:2015raa} are identified as \(q_2=t_1^{-1}t_2^{-1}\), \(q_1=t_2\), \(q_3=t_1\). The parameter \(p\) of the transfer matrix is identified as \(p=-\zeta\). }. The dimension of the $N^\text{th}$ component is \(p(N)\), the number of partitions of \(N\) \cite{Feigin:2015raa}. If the Bethe ansatz were to be complete, meaning that the Bethe vectors form a basis of the given module, the number of solutions to the BAEs must be \(p(N)\). 

    \subsection{Numerically solving the Bethe ansatz equations}
	\label{subsec:TTI:solving BAE}
    
    We start by clearing the denominators of \eqref{eq:CZZBAE}, rewriting it as
    \begin{equation}\label{eq:BAEexp}
    B_a = (s_a - t_1^{\frac{1}{2}}t_2^{\frac{1}{2}} ) \prod_{b\ne a} \prod_{I=1}^3 (s_b - t_I^{-1}s_a) +  \zeta t_1^{\frac{1}{2}}t_2^{\frac{1}{2}} (s_a - t_1^{-\frac{1}{2}}t_2^{-\frac{1}{2}} ) \prod_{b\ne a} \prod_{I=1}^3 (s_b - t_I s_a)=0\,.
    \end{equation}
    If \(\zeta=0\), a complete set of \(p(N)\) solutions to the above is \cite{Koroteev:2015dja,Crew:2024fom}
    \begin{equation}\label{eq: classical BAE sols}
        s_{\lambda,a} = t_1^{i_a-\frac{1}{2}}t_2^{j_a-\frac{1}{2}}\,.
    \end{equation}
    Here, \(\lambda\) is a partition of \(N\), with an associated Young diagram. The index \(a=1,\ldots,N\) runs over the boxes in the Young diagram, where \((i_a, j_a\) are the row and column numbers of the box \(a\), labeled from top-left to bottom-right in a Young diagram displayed in the ``English'' convention. The strategy is then to introduce a one-parameter family of equations depending on the parameter \(v\in[0,1]\):
    \begin{equation}\label{eq: Bethe denom cleared}
    B_a(s,v)\equiv\Big(s_a-t_1^\frac{1}{2}t_2^\frac{1}{2}\Big)\prod_{I=1}^3\prod_{\substack{b=1\\b\neq a}}^N(s_b-t_I^{-1}s_a)+v\,\zeta\, t_1^{\frac{1}{2}}t_2^{\frac{1}{2}}\Big(s_a-t_1^{-\frac{1}{2}}t_2^{-\frac{1}{2}}\Big)\prod_{I=1}^3\prod_{\substack{b=1\\b\neq a}}^N(s_b-t_Is_a)=0\,.
    \end{equation}
    At $v=0$, these are the equations at $\zeta = 0$ for which we know that the solutions are \eqref{eq: classical BAE sols}. At $v = 1$, we recover the original equations \eqref{eq:BAEexp} for generic $\zeta$. We can then solve the differential equation 
    \begin{equation}
        \frac{dB_a(s(v),v)}{dv}=0\,,\quad s_a(v=0)=s_{\lambda,a}\,,
    \end{equation}
    numerically tracking the solutions from \(s_{\lambda,a}\) at \(v=0\) to the desired set of \(p(N)\) solutions at \(v=1\). Practically, this is done using a standard predictor-corrector algorithm. Namely, given a solution \(s(v)\) satisfying \(B_a(s(v),v)=0\), we predict the solution at \(v+\delta v\) to be approximately
    \begin{equation}\label{eq: predictor}
        s_a(v+\delta v)\approx s_a(v)-J^{-1}_{ab}\frac{\partial B_b(s(v),v)}{\partial v}\,,\quad J_{ab}\equiv\frac{\partial B_a(s(v),v)}{\partial s_b}\,.
    \end{equation}
    This prediction is then corrected to be closer to the actual value of \(s_a(v+\delta v)\) by applying succesive Newton iterations, before repeating the same steps with another increment \(v\rightarrow v+\delta v\). 
    
    Examining the off-shell Bethe vectors in \cite{Feigin:2015raa}, we notice that in order for them to be nonzero and finite when evaluated on solutions to the BAEs, the solutions must satisfy certain conditions. Firstly, the gauge fugacities must be distinct from one another, i.e.
    \begin{equation}\label{eq: noncoincident s}
        s_a \neq s_b\qquad \text{for}\quad a\neq b\,.
    \end{equation}
    This enabled us to cancel the overall factor of \(N!\) in going from \eqref{eq: index BA formula} to \eqref{eq:CZZBAE}, since \(\lambda\) labels solutions up to permutations. In addition, there must not exist 3 fugacities, say \(s_{a,b,c}\) with \(a\neq b\neq c\) satisfying the wheel relations
    \begin{equation}\label{eq: wheel combi}
    (s_a,s_b,s_c) = (x, t_1x, t_1t_2 x)\quad \text{or}\quad (x, t_2x, t_1t_2 x)\,,\quad \exists x\in\bC\,.
    \end{equation}
    At every step in \(v\) during the predictor-corrector algorithm, one must ensure that these conditions are satisfied, so that the same holds for the desired solution at the endpoint \(v=1\). If the predictor-corrector step lands on a solution that violates these conditions, it is repeated using a smaller step size \(\delta v\) until an acceptable solution is found.  

    Another subtlety occurs for the partitions \(\lambda\) whose Young diagram contains the box at position \((i,j)=(2,2)\). In other words, the side length of the Durfee square of the partition, also known as the rank, is greater than \(1\). Such partitions occur with increasing frequency as \(N\) becomes large. In this case, the initial prediction step from \(v=0\) to \(v=\delta v\) fails because the Jacobian \(J\) in \eqref{eq: predictor} is singular when evaluated on \(s_{\lambda}\) (See Appendix C.1 of \cite{Crew:2024fom} for a proof). We solve this by examining the structure of the higher derivative equations \(\frac{d^nB_a}{dv^n}=0\). Let \(\delta s_a\equiv s_a(\delta v)-s_a(0)\sim \delta v^{D_a}\). Remarkably, by assuming that \(D_a\) decreases from top-left to bottom-right as \(a\) ranges over the boxes of the Young diagram, the higher derivative equations completely fix the leading \(\cO(\delta v^{D_a})\) term of \(\delta s_a\). The detailed analysis can be found in Appendix \ref{App:tracking}. In this way, we can find the prediction for $s_a(\delta v)$ in the initial step, after which tracking to $v=1$ proceeds as per usual. 
    
    This numerical algorithm is relatively efficient, and we are able to compute the complete set of BAE solutions up to $N=21$, where there are $p(21)=792$ of them. If we assume that the only solutions which are physical from the point of view of the CFT are those which are also physical from the point of view of the integral system (quantum toroidal $\fg\fl_1$), it is worth emphasizing that our algorithm generates all $p(N)$ of such solutions. With the further conjecture that every Bethe root with the appropriate large $N$ scaling $F_{\fg,\lambda}\sim N^\frac{3}{2}$ corresponds to a classical solution of M-theory on AdS$_4\times S^7$, our approach has the potential of finding all such solutions with the boundary conditions specified by the topologically twisted index.  
	
	%%%%%%%%%%%%%%%%%%%%%%%%
	\subsection{Fundamental region of the chemical potentials}
	\label{subsec:TTI:FundamentalRegions}
	%%%%%%%%%%%%%%%%%%%%%%%%
	
	The topologically twisted index is invariant up to a sign under various transformations of the flavour chemical potentials and fluxes, which allows us to restrict our attention to a fundamental domain in \(\Delta_1, \Delta_2, \Delta_C\). The sign is harmless and expected since its origin is the parity anomaly \cite{Closset:2019hyt}. The invariance of the index can be shown by substituting the transformed values of the chemical potentials and fluxes into the formula \eqref{eq:CZZindex}. In some cases, one must also perform a change of variables in the gauge fugacities and gauge fluxes, which are dummy variables that are integrated and summed over. For a transformation \(T\), let \(T(\Delta)\) and \(T(s)\) denote its action on the flavour chemical potentials and gauge fugacities. From the invariance of the index, it follows that \(B_a(\Delta; s) = 0\) implies \(B_a(T(\Delta); T(s)) = 0\), namely \(T\) is a symmetry of the Bethe equations. Among these symmetries, some of them do not act on the Bethe vacua, i.e. \(T(s)=s\), and we shall call them vacuum-preserving symmetries. The remaining symmetries do act on the Bethe vacua, and we shall call them joint symmetries. 

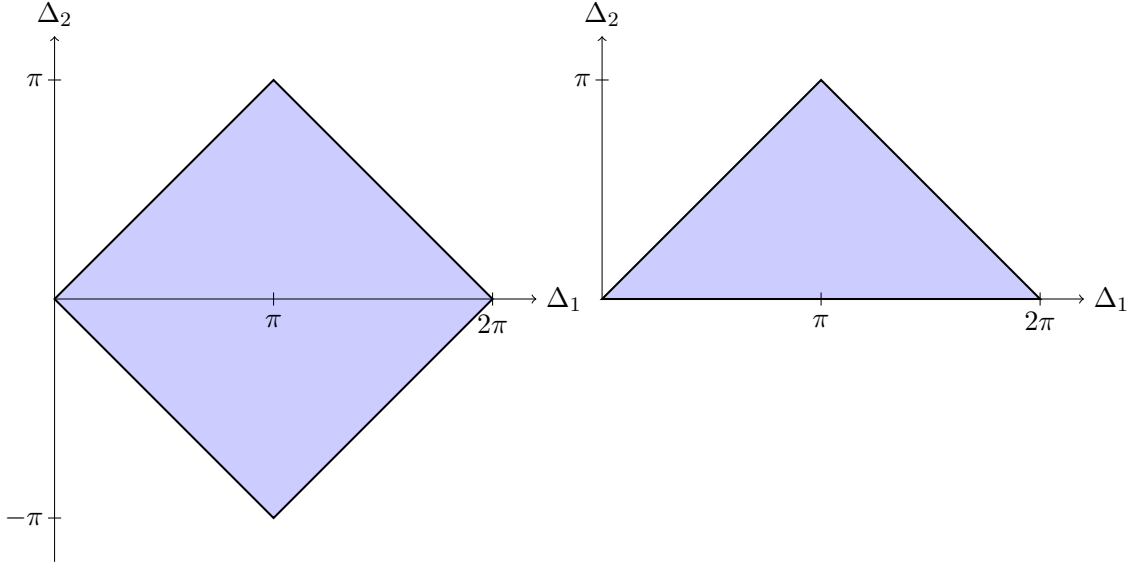
\begin{figure}[!h]
		\centering 
        \begin{adjustbox}{width=\textwidth,keepaspectratio}
        \begin{tikzpicture}[scale=3]
        
        %-------------------------------------------------
        % Left panel
        %-------------------------------------------------
        \begin{scope}

        % shaded region
        \fill[blue!20]
          (0,0)
          -- (1,1)
          -- (2,0)
          -- (1,-1)
          -- cycle;
        
        % axes
        \draw[->] (0,0) -- (2.2,0) node[right] {$\Delta_1$};
        \draw[->] (0,-1.2) -- (0,1.2) node[above] {$\Delta_2$};
        
        % boundary
        \draw[thick]
          (0,0) -- (1,1) -- (2,0) -- (1,-1) -- cycle;
        
        % ticks
        \draw (1,-0.03)--(1,0.03) node[below=5pt] {$\pi$};
        \draw (2,-0.03)--(2,0.03) node[below=5pt] {$2\pi$};
        
        \draw (-0.03,1)--(0.03,1) node[left=3pt] {$\pi$};
        \draw (-0.03,-1)--(0.03,-1) node[left=3pt] {$-\pi$};
        
        \end{scope}
        
        %-------------------------------------------------
        % Right panel
        %-------------------------------------------------
        \begin{scope}[xshift=2.5cm]
        
        % axes
        \draw[->] (0,0) -- (2.2,0) node[right] {$\Delta_1$};
        \draw[->] (0,0) -- (0,1.2) node[above] {$\Delta_2$};
        
        % shaded region
        \fill[blue!20]
          (0,0)
          -- (1,1)
          -- (2,0)
          -- cycle;
        
        % boundary
        \draw[thick]
          (0,0) -- (1,1) -- (2,0) -- cycle;
        
        % ticks
        \draw (1,-0.03)--(1,0.03) node[below=5pt] {$\pi$};
        \draw (2,-0.03)--(2,0.03) node[below=4pt] {$2\pi$};
        
        \draw (-0.03,1)--(0.03,1) node[left=3pt] {$\pi$};
        
        \end{scope}
        
        \end{tikzpicture}
        \end{adjustbox}
		\caption{\rm The left panel shows the region $0<\Delta_1\pm\Delta_2<2\pi$ of inequivalent chemical potentials after using the four vacua-preserving symmetries. After further applying joint symmetries $1$ and $2$, the region is reduced to the fundamental domain in the right panel, with $0<\Delta_C<\pi$. }
		\label{fig:fundamentalDomain}
	\end{figure}

    Firstly, only integer powers of $\zeta$ appear in the index and BAEs. Therefore we can take the domain of $\Delta_C$ to be $(-\pi,\pi)$. The inequivalent vacuum-preserving symmetries that we found are the following: 
	\begin{itemize}
		\item background gauge transformations with respect to \(F_{1,2}\):
        \be
        \Delta_{1,2} \mapsto \Delta_{1,2} + 4\pi \mathbb{Z}\,.
        \ee
        Here, $\Delta_1$ and $\Delta_2$ can be independently shifted by integer multiples of $4\pi$.
		
		\item background gauge transformations with respect to $\frac{F_1+F_2}{2}=\frac{R_H-R_C}{2}$:
        \be
        (\Delta_1, \Delta_2) \mapsto (\Delta_1, \Delta_2) + 2\pi \mathbb{Z}\,.
        \ee
        Both $\Delta_1$ and $\Delta_2$ are simultaneously shifted by the same integer multiple of $2\pi$.

        \renewcommand{\arraystretch}{0.7}
		\item $i\sigma_{1,H}=\begin{pmatrix}0 & i \\ i & 0\end{pmatrix}\in \SU(2)_H$, which permutes the adjoint chirals $A\leftrightarrow B:\;$ 
        \be
        (\Delta_1, \Delta_2) \mapsto (\Delta_2, \Delta_1).
        \ee
        The index is invariant if the corresponding fluxes are also swapped:
        \begin{equation}
            (\fn_1,\fn_2)\mapsto(\fn_2,\fn_1).
        \end{equation}

        \item Parity $P:\;$
        \be
        (\Delta_1,\Delta_2,\Delta_C)\mapsto -(\Delta_1,\Delta_2,\Delta_C)\,,\quad \fm_a \mapsto -\fm_a
        \ee
	\end{itemize}
    Using these four symmetries, the inequivalent region in $(\Delta_1,\Delta_2)$ can be reduced to $0<\Delta_1\pm\Delta_2<2\pi$, displayed in the left panel of Figure \ref{fig:fundamentalDomain}. The range of $\Delta_C$ at this point is still $\Delta_C\in (-\pi,\pi)$.

	Next, we consider the joint symmetries which transform both the parameters (flavour chemical potentials and fluxes), and the Bethe vacua $s$. They are:
	\begin{itemize}
		\item joint symmetry 1: 
		\begin{equation}\label{eq: joint symm 1}
			\Delta_1 \mapsto 2\pi - \Delta_1, \quad \Delta_2 \mapsto - \Delta_2, \quad \Delta_C \mapsto \pi - \Delta_C,\quad \fm_a\mapsto -\fm_a,\quad  s_a \mapsto - s_a.
		\end{equation}
        Note that the index gains a phase $(-1)^{N\left( 1 - \fg + \frac{\fn_1 - \fn_2}{2} + \fn_C \right)}$ and thus might be different up to a sign. The sign itself is harmless as the contributions $\cI_{\fg,\lambda}$ of all the Bethe roots gain the same sign at once.
        
        \item joint symmetry 2:
        \be
        (\Delta_1, \Delta_2, \fn_C) \mapsto - (\Delta_1, \Delta_2, \fn_C)\,,\quad s_a\mapsto s_a^{-1}\,. 
        \ee
        Because of the vacuum-preserving parity symmetry $P$, this transformation is equivalent to $(\Delta_C, \fn_C) \mapsto - (\Delta_C, \fn_C)$, $s_a\mapsto s_a^{-1}$.
	\end{itemize}

	Joint symmetry $1$ maps the lower half ($\Delta_2 < 0$) of the left panel in Figure \ref{fig:fundamentalDomain} to the upper half ($\Delta_2 > 0$), while joint symmetry $2$ allows us to restrict to $\Delta_C \in (0,\pi)$. This leaves us with the fundamental domain displayed in the right panel of Figure \ref{fig:fundamentalDomain}, which is
    \be\label{eq: fund dom}
    0<\Delta_2<\pi\,,\quad \Delta_2<\Delta_1<2\pi-\Delta_2\,,\quad 0<\Delta_C<\pi\,.
    \ee
    We set 
    \be
    \Delta_3 = 2\pi -\Delta_1-\Delta_2\,,\quad 0<\Delta_3<2\pi\,,
    \ee
    where the range for $\Delta_3$ follows from \eqref{eq: fund dom}.

	The symmetries introduced thus far can all be proven by substituting them directly into the index formula in \eqref{eq:CZZindex}. There is however one additional joint symmetry that we were unable to prove from the index formula through direct substitution, but it follows from the duality between ADHM and ABJM theory at $N_f=k=1$ and it can also be verified numerically \cite{Bobev:2022eus, Bobev:2023lkx}. By matching the flavour symmetries of the two dual theories, one can check that the redundant chemical potentials $\hat{\Delta}_I$ and fluxes $\hat{\fn}_I$, $I=1,\ldots 4$ used in the topologically twisted index of ABJM theory \cite{Benini:2015eyy} should be identified in terms of our chemical potentials and fluxes as \cite{Bobev:2023lkx}
    \begin{equation}\label{eq:CZZnotation}
		\hat{\Delta}_I
		=
		\left(\Delta_1,\Delta_2,
		\frac{\Delta_3}{2}+\Delta_C,
		\frac{\Delta_3}{2}-\Delta_C\right),\quad \hat{\fn}_I
		=
		\left(\fn_1,\fn_2,
		\frac{\fn_3}{2}-\fn_C,
		\frac{\fn_3}{2}+\fn_C\right)\,,\quad I = 1,2,3,4\,.
	\end{equation}
    The constraints $\sum_{I=1}^4\hat{\Delta}_I=2\pi$, $\sum_{I=1}^4\hat{\fn}_I=2(\fg-1)$ then coincide with our constraints $\sum_{I=1}^3\Delta_I=2\pi$, $\sum_{I=1}^3\fn_I=2(\fg-1)$. These hatted variables are also natural from the point of view of the bulk geometry, where they are conjugate to the toric action on $S^7$. The index of ABJM theory is invariant under the involution
    \be
    (\hat{\Delta}_1, \hat{\Delta}_2, \hat{\fn}_1, \hat{\fn}_2) \ \mapsto\ (\hat{\Delta}_3, \hat{\Delta}_4, \hat{\fn}_3, \hat{\fn}_4)\,.
    \ee
    Since the indices of dual theories should be equal up to parameter redefinitions such as \eqref{eq:CZZnotation}, it follows that the index of ADHM theory has the same symmetry. This gives  

	\begin{itemize}
		\item joint symmetry 3:
		\bea\label{eq:Transformation3}
			(\Delta_1, \Delta_2, \Delta_C) &\mapsto \left( \frac{\Delta_3}{2} + \Delta_C, \frac{\Delta_3}{2} - \Delta_C, \frac{\Delta_1 - \Delta_2}{2}  \right)\,, \\
        (\fn_1, \fn_2, \fn_C) &\mapsto \Big( \frac{\fn_3}{2} + \fn_C, \frac{\fn_3}{2} - \fn_C, \frac{\fn_1 - \fn_2}{2} \Big)\,.
		\eea
	\end{itemize}
	
	Numerically, we found a one-to-one correspondence between the Bethe vacua at any two sets of parameter values related by \eqref{eq:Transformation3}, where the contributions $\cI_{\fg,\lambda}$ of the vacua related by the correspondence are equal up to a harmless sign $(-1)^N$ that is independent of $\lambda$. Since the index $\cI_\fg = \sum_\lambda \cI_{\fg,\lambda}$ is the sum of these contributions, this numerically verifies that the index is invariant up to an overall sign under joint symmetry 3. This is stronger than what was tested in \cite{Bobev:2022eus}, which is a computation in the large $N$ limit. In addition, the vacua related by the correspondence have roughly the same shape. For example, a one-cut solution is mapped to a one-cut solution, a one-pair dressed solution is mapped to a one-pair dressed solution, and self-conjugate solutions are always mapped to self-conjugate solutions. Because the Bethe vacua transform in a very non-trivial way under joint symmetry 3, unlike the vacuum-preserving symmetries and joint symmetries 1 and 2, we will not use it to reduce the fundamental domain further, but it is still useful as a check for our numerics.

	%%%%%%%%%%%%%%%%%%%%%%%%
	\section{The one-cut solution and dual black hole}
	\label{sec:oneCut}
	%%%%%%%%%%%%%%%%%%%%%%%%
    
\begin{figure}[!h]
		\centering 
        \includegraphics[width=\textwidth]{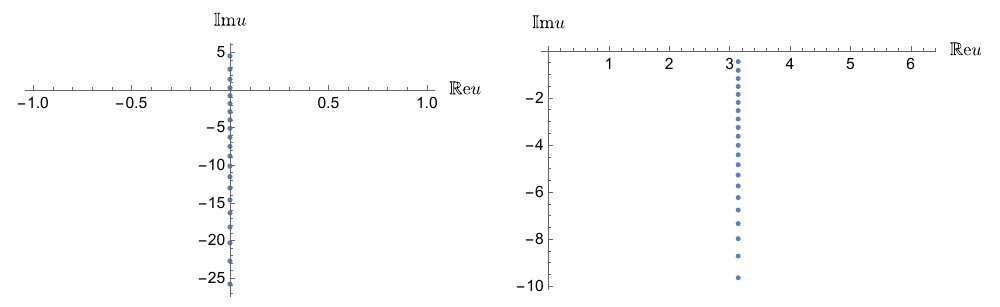}
		\caption{\rm Examples of the two classes of one-cut solutions. The left panel plots the positions of $u_a$ for a solution in the first class with $\re u_a = 0$, where we have chosen $N=21$, $\Delta_C=\frac{\pi}{9}$, $\Delta_1 = \frac{\pi^2}{4}$, and $\Delta_2 = e$. The right panel plots a solution in the second class with $\re u_a = \pi$, where $N=21$, $\Delta_C=\frac{\pi}{9}$, $\Delta_1=\frac{17\pi}{9}$, and $\Delta_2=-\frac{\pi}{18}$.}
		\label{fig: bhsol eg}
	\end{figure}

    The first class of Bethe roots with the scaling $F_{\fg,\lambda}\sim N^\frac{3}{2}$ that we will consider are what we call one-cut solutions, so named because their distribution of $u_a=-i\log s_a$ is supported on a single cut or curve. We are using this terminology in the same context as the matrix model literature \cite{Brezin:1977sv,Ambjorn:1993vz}. There are two classes of one-cut solutions: the first has $\re u_a = 0$ while the second has $\re u_a = \pi$, as shown in Figure \ref{fig: bhsol eg}.  

\begin{figure}[h!]
        \centering        
        \begin{tikzpicture}[scale=1.2]
          
          % Vertices of the blue diamond
          \coordinate (A)  at (0,0);
          \coordinate (B)  at ({3.1415926535},{3.1415926535});
          \coordinate (C)  at ({2*3.1415926535},0);
          \coordinate (D)  at ({3.1415926535},{-3.1415926535});
          % Vertices of the orange triangle
          \coordinate (E)  at ({16*3.1415926535/9},0);
          \coordinate (F)  at ({8*3.1415926535/9},{8*3.1415926535/9});
          % Vertex of the green triangle
          \coordinate (G)  at ({17*3.1415926535/9},{-3.1415926535/9});
          
          % Fill regions with transparency
          \fill[blue,   opacity=0.3] (A) -- (B) -- (C) -- (D) -- cycle;
          \fill[orange, opacity=0.5] (A) -- (E) -- (F) -- cycle;
          \fill[green,  opacity=0.5] (E) -- (C) -- (G) -- cycle;
          
          % Draw region boundaries
          \draw[blue,   thick] (A) -- (B) -- (C) -- (D) -- cycle;
          \draw[orange, thick] (A) -- (E) -- (F) -- cycle;
          \draw[green,  thick] (E) -- (C) -- (G) -- cycle;
          
          % Coordinate axes
          \draw[->] (-0.5,0) -- ({2*3.1415926535+0.5},0) node[right] {$\Delta_1$};
          \draw[->] (0,{-3.1415926535-0.5}) -- (0,{3.1415926535+0.5}) node[above] {$\Delta_2$};
          
          % Axis labels
          \node[below left]  at (0,0)       {$0$};
          \node[below]       at ({3.1415926535},0)   {$\pi$};
          \node[below]       at ({2*3.1415926535},0) {$2\pi$};
          \node[left]        at (0,{3.1415926535})   {$\pi$};
          \node[left]        at (0,{-3.1415926535})  {$-\pi$};
    
          % Label inside orange triangle (centroid)
          \node[inner sep=1pt] at ({8*3.1415926535/9},{8*3.1415926535/27}) {BH$_+$};
          
          % Label outside green triangle with arrow
          \node[below] (BHminus) at ({17*3.1415926535/9},-0.7) {BH$_-$};
          \draw[->, thick] (BHminus) -- ({17*3.1415926535/9},{-3.1415926535/27});
        \end{tikzpicture}
        \caption{\small Plots of the region $0<\Delta_1\pm\Delta_2<2\pi$ in blue, the region BH$_+$ in orange, and the region BH$_-$ in green. We take $\Delta_C=\pi/9$ as an example. }
        \label{fig: BH regions}
    \end{figure}
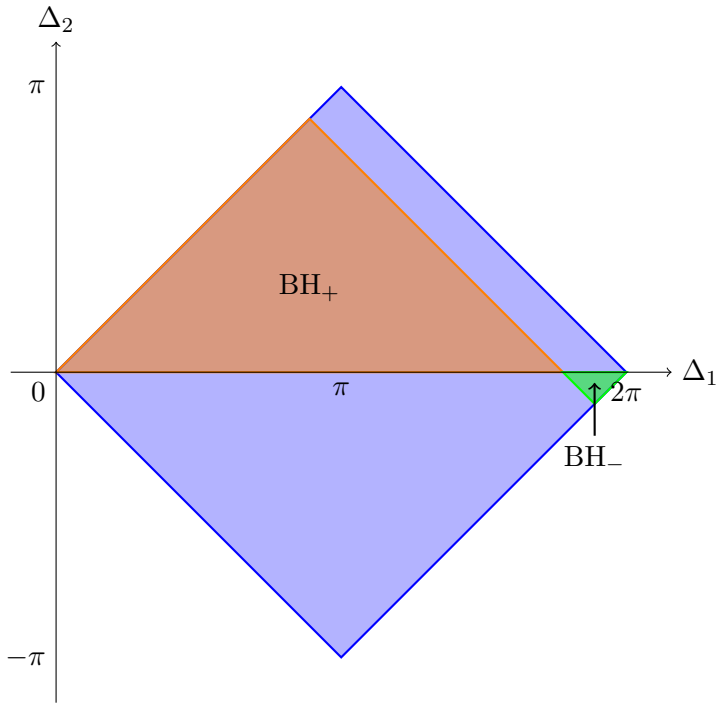

    Restricting to the region of chemical potentials $0<\Delta_1\pm\Delta_2<2\pi$ shown in the left panel of Figure \ref{fig:fundamentalDomain} and $0<\Delta_C<\pi$, we find numerically that the one-cut solutions with $\re u_a = 0$ and $\re u_a = \pi$ only exist in the subregions BH$_+$ and BH$_-$ respectively, where 
    \begin{gather}
    \text{BH}_+: \quad 0<\Delta_2<\pi-\Delta_C\,,\quad \Delta_2<\Delta_1<2\pi-2\Delta_C-\Delta_2\,, \notag\\
    \text{BH}_-: \quad -\Delta_C <\Delta_2 <0\,,\quad 2\pi-2\Delta_C-\Delta_2 <\Delta_1<2\pi+\Delta_2\,.\label{eq: BH reg exists}
    \end{gather}
    As a visual aid, we have plotted these regions in Figure \ref{fig: BH regions}. 

    The plan for the rest of Section \ref{sec:oneCut} is as follows. To begin with, we will study the one-cut solutions both analytically and numerically on the field theory side. Then, we will independently account for various contributions to their free energy on the gravity side. This constitutes an extremely stringent test of the AdS/CFT correspondence.

    Firstly, in Section \ref{subsec:oneCut:largeN}, we analytically examine the Bethe equations at large $N$ under a one-cut ansatz for $u_a$, and show that the positivity of the distribution function $\rho(t)$ of $t_a=\im u_a/N^\frac{1}{2}$ precisely implies that the one-cut solutions only exist in regions BH$_\pm$. This is mostly a repeat of \cite{Hosseini:2016tor,Hosseini:2016ume}, but the class of solutions in region BH$_-$ with $\re u_a = \pi$ is new, as is the analysis of the regions of existence. Having solved for $\rho(t)$ analytically at leading order in $N$, we can then compute the free energy of the one-cut solutions to leading order in $N$. While this is the limit of the analytical approach for now, one can go much further with a numerical analysis. Next, in Section \ref{subsec:oneCut:numerics}, by fitting an ansatz to numerical data, we compute the free energy of the solution in BH$_+$ to all perturbative orders in $1/N$ and at leading non-perturbative order. Our result generalizes the analysis in \cite{Bobev:2022jte,Bobev:2022eus,Bobev:2023lkx} by providing the non-perturbative correction at generic chemical potentials away from the so-called ``superconformal'' point. 
    
    On the other hand, almost all of the aforementioned contributions to the free energy can be obtained via a bulk computation in M-theory on AdS$_4\times S^7$. In Section \ref{subsec:oneCut:bulk}, we show that the Bekenstein-Hawking entropy of a class of magnetically charged black holes exactly matches the Legendre transform of the leading free energy of the one-cut solutions in BH$_\pm$, where BH$_+$ and BH$_-$ correspond to two different branches of black hole solutions. This match for the solutions in BH$_+$ was already completed in \cite{Benini:2015eyy}, but we note here that there is a second branch of solutions in BH$_-$. Then, in Section \ref{subsec:equiv loc BH}, we show that the perturbative $1/N$ corrections in the free energy can be reproduced via equivariant localization in supergravity. Finally, in Section \ref{subsec:non-pert M2}, we match the leading exponent in the non-perturbative corrections with the action of M2-branes wrapping $S^3\subset S^7$.   
        
    %%%%%%%%%%%%%%%%%%%%%%%%
	\subsection{The analytic large $N$ solution}
	\label{subsec:oneCut:largeN}
	%%%%%%%%%%%%%%%%%%%%%%%%

    We begin by reviewing how the one-cut solution can be described by a density $\rho$ to leading order in $N$ \cite{Hosseini:2016tor,Hosseini:2016ume}. Taking the logarithm of \eqref{eq:CZZBAE}, the Bethe equations can be written as
    \bea\label{eq: W'}
    &-i\Delta_C+iu_a+\log\left(1-e^{i\left(-u_a+\frac{\Delta_1+\Delta_2}{2}\right)}\right)-\log\left(1-e^{i\left(u_a+\frac{\Delta_1+\Delta_2}{2}\right)}\right)\\
    &+\sum_{I=1}^3\sum_{\substack{b=1 \\ b\neq a}}^N\left[i\pi + i(u_a-u_b)+\log\left(1-e^{i(u_b-u_a+\Delta_I)}\right)-\log\left(1-e^{i(u_a-u_b+\Delta_I)}\right)\right]=2\pi in_a\,,
    \eea
    where the $n_a\in\bZ$ are integer ambiguities that can be freely chosen. Motivated by the form of the numerical one-cut solution in Figure \ref{fig: bhsol eg}, a natural large $N$ ansatz is
    \be\label{eq: BH ansatz}
    u_a = iN^\frac{1}{2}t_a + f(t_a)\,,\quad a=1,\ldots,N\,,
    \ee
    where $f(t_a) \in (-\pi,\pi]$. We assume that $t_a\neq t_b$ for $a\neq b$. Using the Weyl group symmetry $S_N$ permuting $u_a$, we choose to order them such that $t_a > t_b$ for $a>b$. We then parametrize the variables $u$ by the continuous variable $t$ instead of the discrete index $a$. The two are related via the density $\rho$, which is defined as
    \be
    \rho(t) \equiv \frac{1}{N}\frac{da}{dt} = \frac{1}{N}\sum_{a=1}^N \delta(t - t_a)\,.
    \ee
    By construction, $\rho \ge 0$ and its integral is normalized to $1$:
    \be
    \int dt \rho(t) = 1\,.
    \ee
    We assume that in the large $N$ limit, $\rho$ tends to a piecewise smooth function of $t$. Plugging this ansatz into the Bethe equations \eqref{eq: W'}, one finds that their large $N$ limit boils down to 
    \be\label{eq: large N W'}
    \Delta_C+\sgn t\frac{\Delta_3}{2} + 2\pi\sgn t\left\lfloor\frac{\frac{\Delta_1+\Delta_2}{2}-\sgn(t)f(t)}{2\pi}\right\rfloor + 2\rho'(t)G(\Delta)= 0\,,
    \ee
    where we have defined
    \bea\label{eq: bracket gp G def}
    [\Delta_I]\equiv &\Delta_I-2\pi\left\lfloor\frac{\Delta_I}{2\pi}\right\rfloor\,,\quad g_+(\Delta)\equiv \frac{1}{6}\Delta^3-\frac{\pi}{2}\Delta^2+\frac{\pi^2}{3}\Delta\,,\quad G(\Delta)\equiv\sum_{I=1}^3g_+([\Delta_I])\,.
    \eea
    The details of this computation can be found in Appendix \ref{App: large N Bethe 1cut}. This fixes
    \be\label{eq: BH density}
    \rho(t) = \frac{\mu - \left(\frac{\Delta_3}{2}+2\pi\tilde{n}+\sgn (t)\Delta_C\right)|t|}{2G}\,,\quad \tilde{n}\equiv \left\lfloor\frac{\frac{\Delta_1+\Delta_2}{2}-\sgn(t)f(t)}{2\pi}\right\rfloor\,,
    \ee
    where $\mu$ is a constant to be determined. The Bethe equations at leading order in $N$ do not place any constraints on $f(t)$, but we observe numerically that the solutions have either $f(t) = 0$ or $f(t) = \pi$. 
    
    Let us consider $f(t) = 0$ first. In the fundamental domain \eqref{eq: fund dom}, one can check that $2G=\Delta_1\Delta_2\Delta_3>0$ and $\tilde{n}=0$. We assume that $\rho(t)$ has compact support between two zeros, which is also observed numerically. Given that $\rho(t)$ is piecewise linear in $t$ and positive, we must have $\rho'(t)<0$ for $t>0$ and $\rho'(t)>0$ for $t<0$, implying that $\frac{\Delta_3}{2}\pm\Delta_C>0$. This is equivalent to the definition of region BH$_+$ in \eqref{eq: BH reg exists}. The density $\rho$ is supported on $t\in\left(-\frac{\mu}{\frac{\Delta_3}{2}-\Delta_C},\frac{\mu}{\frac{\Delta_3}{2}+\Delta_C}\right)$, up to the points where $\rho = 0$. Normalizing $\rho$ then fixes $\mu$, and the solution is
    \bea\label{eq: BH 0 density}
    &\rho(t) = \frac{\mu-(\frac{\Delta_3}{2}+\sgn(t)\Delta_C)|t|}{\Delta_1\Delta_2\Delta_3}\quad t\in\left(-\frac{\mu}{\frac{\Delta_3}{2}-\Delta_C},\frac{\mu}{\frac{\Delta_3}{2}+\Delta_C}\right)\\
    &\mu=\sqrt{\frac{1}{2}\Delta_1\Delta_2(\Delta_3+2\Delta_C)(\Delta_3-2\Delta_C)}\,.
    \eea
    Due to the inequalities satisfied by BH$_+$ in \eqref{eq: BH reg exists}, the density can also be written as
    \bea\label{eq: BH+ rho uniform}
    &{\rho(t) = \begin{cases}\frac{\sqrt{2[\hat{\Delta}_1][\hat{\Delta}_2][\hat{\Delta}_3][\hat{\Delta}_4]}-[\hat{\Delta}_3]t}{[\hat{\Delta}_1][\hat{\Delta}_2]([\hat{\Delta}_3]+[\hat{\Delta}_4])}\,, & t>0 \\[0.2cm] \frac{\sqrt{2[\hat{\Delta}_1][\hat{\Delta}_2][\hat{\Delta}_3][\hat{\Delta}_4]}+[\hat{\Delta}_4]t}{[\hat{\Delta}_1][\hat{\Delta}_2]([\hat{\Delta}_3]+[\hat{\Delta}_4])}\,, & t<0\end{cases}}\\[0.2cm]
    &t\in\left(-\frac{\sqrt{2[\hat{\Delta}_1][\hat{\Delta}_2][\hat{\Delta}_3][\hat{\Delta}_4]}}{[\hat{\Delta}_4]}\,,\frac{\sqrt{2[\hat{\Delta}_1][\hat{\Delta}_2][\hat{\Delta}_3][\hat{\Delta}_4]}}{[\hat{\Delta}_3]}\right)\,,
    \eea
    where $[\cdot]$ is defined in \eqref{eq: bracket gp G def}, and $\hat{\Delta}_I$ are defined in \eqref{eq:CZZnotation}.
    
    If $f(t) = \pi$, we have $\tilde{n}=-\Theta(t)$, where $\Theta$ is the Heaviside theta function. In the fundamental domain, $2G=\Delta_1\Delta_2\Delta_3>0$ like before, and we would need $\frac{\Delta_3}{2}+\Delta_C-2\pi >0$ and $\frac{\Delta_3}{2}-\Delta_C >0$ for $\rho$ to have compact support between two zeros. But $\frac{\Delta_3}{2}+\Delta_C-2\pi >0$ cannot be satisfied in the fundamental domain, so there are no solutions with $f(t)=\pi$ there. If we broaden our search to the region $-\pi<\Delta_2<0$, $-\Delta_2 < \Delta_1 <\Delta_2 +2\pi$, which is the bottom half with $\Delta_2 < 0$ in the left panel of Figure \ref{fig:fundamentalDomain}, one can check that $2G = (\Delta_1-2\pi)\Delta_2(\Delta_3-2\pi) < 0$, and our assumptions on $\rho$ instead require $\frac{\Delta_3}{2}+\Delta_C-2\pi <0$ and $\frac{\Delta_3}{2}-\Delta_C <0$. The region which satisfies this is precisely region BH$_-$ in \eqref{eq: BH reg exists}. As before, normalizing $\rho$ fixes $\mu$ and the solution is given by
    \bea\label{eq: BH pi density}
    &\rho(t) = \begin{cases}\frac{-\mu-\left(\frac{\Delta_3}{2}+\Delta_C-2\pi\right)t}{(\Delta_1-2\pi)\Delta_2(\Delta_3-2\pi)}\,, & t>0 \\ \frac{-\mu +\left(\frac{\Delta_3}{2}-\Delta_C\right)t}{(\Delta_1-2\pi)\Delta_2(\Delta_3-2\pi)}\,, &t<0\end{cases}\quad t\in\left(\frac{\mu}{\frac{\Delta_3}{2}-\Delta_C},-\frac{\mu}{\frac{\Delta_3}{2}+\Delta_C-2\pi}\right)\\
    &\mu=\sqrt{2(\Delta_1-2\pi)\Delta_2\left(\frac{\Delta_3}{2}+\Delta_C-2\pi\right)\left(\frac{\Delta_3}{2}-\Delta_C\right)}\,,
    \eea
    with $f(t) = \pi$. Note that if we apply joint symmetry 1 defined in \eqref{eq: joint symm 1}, the region BH$_+$ is mapped to BH$_-$, and \eqref{eq: BH 0 density} is mapped to \eqref{eq: BH pi density}. Due to the restrictions on BH$_-$ in \eqref{eq: BH reg exists}, the density can also be written as
    \bea\label{eq: BH- rho uniform}
    &{\rho(t)=\begin{cases}\frac{-\sqrt{2[\hat{\Delta}_1]'[\hat{\Delta}_2]'[\hat{\Delta}_3]'[\hat{\Delta}_4]'}-[\hat{\Delta}_3]'t}{[\hat{\Delta}_1]'[\hat{\Delta}_2]'([\hat{\Delta}_3]'+[\hat{\Delta}_4]')}\,, & t>0 \\[0.2cm] 
    \frac{-\sqrt{2[\hat{\Delta}_1]'[\hat{\Delta}_2]'[\hat{\Delta}_3]'[\hat{\Delta}_4]'}+[\hat{\Delta}_4]'t}{[\hat{\Delta}_1]'[\hat{\Delta}_2]'([\hat{\Delta}_3]'+[\hat{\Delta}_4]')}\,, & t<0\end{cases}}\\[0.2cm]    &t\in\left(\frac{\sqrt{2[\hat{\Delta}_1]'[\hat{\Delta}_2]'[\hat{\Delta}_3]'[\hat{\Delta}_4]'}}{[\hat{\Delta}_4]'}\,,-\frac{\sqrt{2[\hat{\Delta}_1]'[\hat{\Delta}_2]'[\hat{\Delta}_3]'[\hat{\Delta}_4]'}}{[\hat{\Delta}_3]'}\right)\,,
    \eea
    where we defined
    \be\label{eq: bracket prime def}
    [\Delta]'\equiv \Delta-2\pi\left\lceil\frac{\Delta}{2\pi}\right\rceil\in(-2\pi,0]\,.
    \ee 

    \subsubsection{Leading contribution to the index}

    Neglecting the contribution of the Jacobian $\det\frac{\partial W_a}{\partial u_b}$, which one can show to be subleading in $N$ for the Bethe solutions considered in this paper, each solution $\lambda = \{s_a\}$ to the Bethe equations contributes
    \bea\label{eq: log Z}
    -F_{\fg,\lambda}  \simeq &\, i\frac{N^2}{2}\sum_{I=1}^3(\fn_I+1-\fg)\Delta_I+iN\frac{\fn_3}{4}\Delta_3 +i\fn_C\sum_{a=1}^N u_a \\
    &+ (1-\fg)\sum_{\substack{a,b=1 \\ a \neq b}}^N\log\left(1-e^{i(u_a-u_b)}\right)-\sum_{I=1}^3(\fn_I+1-\fg)\sum_{a,b=1}^N\log\left(1-e^{i(u_a-u_b+\Delta_I)}\right)\\
    &+\frac{\fn_3}{2}\sum_{a=1}^N\left[\log\left(1-e^{i\left(u_a+\frac{\Delta_1+\Delta_2}{2}\right)}\right) + \log\left(1-e^{i\left(-u_a+\frac{\Delta_1+\Delta_2}{2}\right)}\right)\right]\,.
    \eea
    Substituting the one-cut ansatz \eqref{eq: BH ansatz} into \eqref{eq: log Z} gives
    \begin{gather}
    -\frac{F_{\fg,\text{BH}}}{N^\frac{3}{2}}=\left[\sum_{I=1}^3(\fn_I+1-\fg)g_+'([\Delta_I])-\frac{\pi^2}{3}(1-\fg)\right]\int dt\rho(t)^2-\fn_C\int dt\rho(t)t+\frac{\fn_3}{2}\int dt\rho(t)|t|\notag\\
    +\cO\left(N^{-\frac{1}{2}}\right)\label{eq: one-cut F}.
    \end{gather}
    If we further substitute the one-cut solution in region BH$_+$, given by \eqref{eq: BH+ rho uniform}, we obtain that its contribution is 
    \bea\label{eq: logZ BH}
    &-\frac{F_{\fg,\text{BH}_+}}{N^\frac{3}{2}}=\frac{1}{3}\sqrt{2[\hat{\Delta}_1][\hat{\Delta}_2][\hat{\Delta}_3][\hat{\Delta}_4] }\sum_{I=1}^4\frac{\hat{\fn}_I}{[\hat{\Delta}_I]}+\cO\left(N^{-\frac{1}{2}}\right)\,,\\
    &\hspace{2cm}\sum_{I=1}^4[\hat{\Delta}_I]=\sum_{I=1}^4\hat{\Delta}_I = 2\pi\,. 
    \eea
    Here, $\hat{\Delta}_I$ and $\hat{\fn}_I$ are the variables defined in \eqref{eq:CZZnotation}. On the other hand, substituting the other one-cut solution \eqref{eq: BH- rho uniform} in region BH$_-$ gives
    \bea\label{eq: logZ BH branch 2}
    &-\frac{F_{\fg,\text{BH}_-}}{N^\frac{3}{2}}=-\frac{1}{3}\sqrt{2[\hat{\Delta}_1]'[\hat{\Delta}_2]'[\hat{\Delta}_3]'[\hat{\Delta}_4]' }\sum_{I=1}^4\frac{\hat{\fn}_I}{[\hat{\Delta}_I]'}+\cO\left(N^{-\frac{1}{2}}\right)\,,\\ 
    &\hspace{2cm}\sum_{I=1}^4[\hat{\Delta}_I]'=\sum_{I=1}^4\hat{\Delta}_I-4\pi=-2\pi\,.
    \eea
    As we shall show in Section \ref{subsec:oneCut:bulk}, the Legendre transforms of \eqref{eq: logZ BH} and \eqref{eq: logZ BH branch 2} produce the entropy of black holes in two distinct branches.
    
	%%%%%%%%%%%%%%%%%%%%%%%%
	\subsection{Finite $N$ numerics}
	\label{subsec:oneCut:numerics}
	%%%%%%%%%%%%%%%%%%%%%%%%

    Using the numerical data, we can determine $F_{\fg,\text{BH}_+}$ to all perturbative orders in $N$, and extract the leading non-perturbative contribution. This follows the same method as the precision numerics conducted in \cite{Bobev:2022jte,Bobev:2022eus,Bobev:2023lkx}, which we briefly summarize. Starting from a seed solution generated using the large-$N$ density $\rho(t)$ in \eqref{eq: BH density}, one can use Newton's method to get the numerical one-cut solution between $N = 100$ and $350$ with spacing $10$, up to very high precision. Using the values of $F_{\fg,\text{BH}_+}$ obtained from evaluating \eqref{eq: index BA formula} on these solutions, the large-$N$ expansion coefficients are obtained numerically using {\tt LinearModelFit[]} in Mathematica, and a closed form expression can be extrapolated. Since the free energy at generic genus $\fg\neq 1$ is related to that at $\fg=0$ via \eqref{eq:general-g}, we will focus our attention on $\fg = 0$, leaving $\fg = 1$ for future analyses. Explicitly, in terms of the chemical potentials in \eqref{eq:CZZnotation}, the free energy at $\fg=0$ is
	\begin{equation} \label{eq:TTI-BH}
		\begin{aligned}
			F_{0,\text{BH}_+}\left(N, \fc_I^{\rm (BH)}, \delta^{\rm (BH)}\right) &= - \frac{\sqrt{2\hat{\Delta}_1\hat{\Delta}_2\hat{\Delta}_3\hat{\Delta}_4}}{3} \sum_{I=1}^4 \frac{\hat{\fn}_I}{\hat{\Delta}_I} \Big( (N + \delta^{\rm (BH)})^{\frac{3}{2}} + \fc_I^{\rm (BH)} (N + \delta^{\rm (BH)})^{\frac{1}{2}} \Big)\\
			&\qquad  + \frac{1}{2} \log (N + \delta^{\rm (BH)})  + \hat{f}_0 + \hat{f}_{\rm np} .
		\end{aligned}
	\end{equation}
	As shown in \cite{Bobev:2023lkx}, the coefficients $\fc_I^{\rm (BH)}$ and $\delta^{\rm (BH)}$ admit the following closed-form expressions:
	\begin{equation} \label{eq:coeExactBH}
		\fc_I^{\rm (BH)}(\hat{\Delta}) = - \frac{  \prod_{J\ne I} (\hat{\Delta}_I + \hat{\Delta}_J ) }{ 8 \hat{\Delta}_1 \hat{\Delta}_2 \hat{\Delta}_3 \hat{\Delta}_4 } \sum_{J\ne I} \hat{\Delta}_J,\qquad \delta^{\rm (BH)} (\hat{\Delta}) = - \frac{1}{24} + \frac{1}{12} \sum_I \frac{\pi}{\hat{\Delta}_I}.
	\end{equation}
	The comparison between our notation and the convention of \cite{Bobev:2022jte, Bobev:2022eus, Bobev:2023lkx} is shown in Appendix \ref{Appsubsec:TTI:ConventionComparison}. As Section \ref{subsec:equiv loc BH} will soon show, equivariant localization in the bulk produces both the $1/N$ expansion induced by $\fc_I^{\rm (BH)}$ and $\delta^{\rm (BH)}$, as well as the $\log N$ term in \eqref{eq:TTI-BH}. The constant $\hat{f}_0$ and non-perturbative contribution $\hat{f}_{\rm np}$ can be decomposed proportionally to the four-variable fluxes as\footnote{Note that although the result for $\hat{f}_0$ in \cite{Bobev:2022jte, Bobev:2022eus, Hong:2026zul} is not written as $\sum_{I=1}^4\hat{\fn}_I\hat{f}_{0,I}$, it can be put into this form using the constraint $\sum_{I=1}^4\hat{\fn}_I=2(\fg-1)$.}
	\begin{equation}\label{eq: f0 fnp flux decomp}
		 \hat{f}_0 + \hat{f}_{\rm np} = \sum_{I=1}^4 \hat{\fn}_I \left( \hat{f}_{0,I} + \hat{f}_{{\rm np}, I} \right).
	\end{equation}
	On the field theory side, the constant term $\hat{f}_0$ has been studied in great detail in \cite{Bobev:2022jte, Bobev:2022eus}, and a closed-form formula is recently found in \cite{Hong:2026zul} for the ``superconformal'' point $\hat{\Delta}_I=\frac{\pi}{2}$. More recently, an analytic expression is obtained using equivariant localisation \cite{BenettiGenolini:2026kjy} with some assumptions, which is consistent with the known cases.
    
    In the following, we shall focus our attention on the non-perturbative corrections $\hat{f}_{\rm np}$. In general, we expect $\hat{f}_{\rm np}$ to take the form: 
	\begin{equation}\label{eq: fnp_I N expansion}
	\hat{f}_{{\rm np}, I} = N^{ \hat{f}^{\rm (np)}_{{\rm log}, I} } \ e^{\hat{f}_{1/2, I}^{\rm (np)} \sqrt{N} } \ \sum_{i=0}^\infty \hat{f}^{\rm (np)}_{i,I} N^{-i/2},
	\end{equation}
	where the power series with coefficients $ \hat{f}^{\rm (np)}_{i,I} $ is understood to be an asymptotic expansion. Since only the sum $\hat{f}_0 + \hat{f}_{\rm np}$ is numerically available, we get rid of the constant map $\hat{f}_0$ by looking at the difference at some integer spacing $d$:
	\begin{equation}\label{eq: gI exp}
		g_I \equiv \log \left( \hat{f}_{{\rm np}, I}(N) - \hat{f}_{{\rm np}, I}(N+d) \right) =  g_{1/2, I} \sqrt{N} + g_{{\rm log}, I} \log N + \sum_{i=0}^\infty g_{i,I} N^{-i/2}.
	\end{equation}
	By expanding and matching coefficients, the first few coefficients in $\hat{f}_{{\rm np},I}$ are determined in terms of $g_{1/2,I}$, $g_{\log,I}$ and $g_{i,I}$ as
	\begin{equation}
		\begin{aligned}
			& \hat{f}_{1/2, I}^{\rm (np)} = g_{1/2, I},\quad \hat{f}^{\rm (np)}_{{\rm log}, I} = g_{{\rm log}, I} + \frac{1}{2},\\
			& \hat{f}^{\rm (np)}_{0,I} = \frac{2 e^{g_{0, I}}}{d g_{1/2, I}}
		,\quad \hat{f}^{\rm (np)}_{1,I} = \frac{ e^{g_{0, I}}}{2d g_{1/2, I}^2} \left( 4 + 4 g_{1/2, I}\, g_{1, I}  + d g_{1/2, I}^2 + 8 g_{\log, I} \right),\\
		& \hat{f}^{\rm (np)}_{2,I} = \frac{ e^{g_{0, I}}}{24 d g_{1/2, I}^3}\left[ g_{1/2, I}^2\left( 24 g_{1, I}^2 + 48 g_{2, I} + 12 d g_{1/2, I}\ g_{1, I} + d^2 g_{1/2, I}^2	 \right)\right. \\
        &\hspace{3cm} \left.+ 96 g_{\log , I} (1 + g_{1/2, I}\ g_{1 , I} ) + 192 g_{\log, I}^2 \right].
	\end{aligned}
	\end{equation}
	By fitting numerical data for many sets of $\hat{\Delta}_I$ to the ansatz \eqref{eq: gI exp}, we find that in the vast majority of cases, the leading exponent $f^{(\rm np)}_{1/2,I}$ controlling the size of the non-perturbative corrections is given by
	\begin{equation} \label{eq:BH-instanton}
		\begin{aligned}
			\hat{f}_{1/2, I}^{\rm (np)} &= - 2\sqrt{2}\pi \cdot \sqrt{ \frac{ \Delta_{\ll} \Delta_{<} }{ \Delta_{\gg} \Delta_{>} } }, \quad \Delta_{\gg} \ge \Delta_{>} \ge \Delta_{<} \ge \Delta_{\ll} \in \{ \hat{\Delta}_I\}.\\
		\end{aligned}
	\end{equation}
	The counterexamples to this general pattern are given in Table \ref{tbl:InstantonCounterExamples1}. As we will see in Section \ref{subsec:non-pert M2}, \eqref{eq:BH-instanton} matches $-S_{\text{M}2}$, where $S_{\text{M}2}$ is the action of a M$2$ brane wrapping the supersymmetric $S^3\subset S^7$ with the least volume or action\footnote{As we will see in Section \ref{subsec:non-pert M2}, the pullback of $C_3$ does not contribute for these wrappings.}. The counterexamples also match the action of M$2$ branes wrapping $S^3$'s, but they do not wrap the minimal cycle with the least volume.  
	
	\renewcommand{\arraystretch}{2}
	\begin{table}[!h] \small 
		\centering
        \hspace*{-1.5cm}
		\begin{tabular}{|c|c|c|}
			\hline
			$\hat{\Delta}_I$ &   $\hat{f}_{1/2, I}^{\rm (np)} $ & prediction  \\ \hline 
			$ \left(\frac{2\pi}{3}, \frac{2\pi}{3}, \Delta, \frac{2\pi}{3} - \Delta \right)$ & $ -2\sqrt{2}\pi \sqrt{\frac{{\rm min}(\Delta, \frac{2\pi}{3} - \Delta ) }{{\rm max}(\Delta, \frac{2\pi}{3} - \Delta )}} $ &  $ -2\sqrt{2}\pi \sqrt{\frac{ \Delta  (\frac{2\pi}{3} - \Delta )}{ \frac{2\pi}{3}  \cdot \frac{2\pi}{3}  }} $ \\ \hline 
			$ \left(\pi, \frac{\pi}{2},\Delta,\frac{\pi}{2} - \Delta \right)$ & $ -2\sqrt{2}\pi \cdot \left(  \sqrt{\frac{\Delta\left( \frac{\pi}{2} - \Delta \right)}{\pi \cdot \frac{\pi}{2} }}, \sqrt{\frac{\Delta\left( \frac{\pi}{2} - \Delta \right)}{\pi \cdot \frac{\pi}{2} }},  \sqrt{ \frac{\min\left( \Delta,  \frac{\pi}{2} - \Delta \right)}{2 \max\left( \Delta,  \frac{\pi}{2} - \Delta \right) } },  \sqrt{ \frac{\min\left( \Delta,  \frac{\pi}{2} - \Delta \right)}{2 \max\left( \Delta,  \frac{\pi}{2} - \Delta \right) } } \right) $ &  $ -2\sqrt{2}\pi  \cdot \sqrt{\frac{\Delta\left( \frac{\pi}{2} - \Delta \right)}{\pi \cdot \frac{\pi}{2} }} $ \\ \hline 
		\end{tabular}
        \caption{\rm Counterexamples to \eqref{eq:BH-instanton} for the coefficients $\hat{f}_{1/2, I}^{\rm (np)}$. Note that $\sqrt{\frac{{\rm min}(\Delta, \frac{2\pi}{3} - \Delta ) }{{\rm max}(\Delta, \frac{2\pi}{3} - \Delta )}}=\sqrt{\frac{\frac{2\pi}{3}{\rm min}(\Delta, \frac{2\pi}{3} - \Delta ) }{\frac{2\pi}{3}{\rm max}(\Delta, \frac{2\pi}{3} - \Delta )}}$, and $ \sqrt{ \frac{\min\left( \Delta,  \frac{\pi}{2} - \Delta \right)}{2 \max\left( \Delta,  \frac{\pi}{2} - \Delta \right) } } = \sqrt{\frac{\frac{\pi}{2} \min\left( \Delta,  \frac{\pi}{2} - \Delta \right) }{\pi \cdot \max \left( \Delta,  \frac{\pi}{2} - \Delta \right) }} $, so that both counterexamples still correspond to M$2$ branes wrapping $S^3\subset S^7$, whose action is given in \eqref{eq: M2 action gen wrapping}. The issue is that they do not wrap the $S^3$ with the least volume.}
		\label{tbl:InstantonCounterExamples1}
	\end{table}

    We find numerically that for chemical potentials $\hat{\Delta}_I^{(\epsilon)} = \left(\frac{2\pi}{3} + \epsilon, \frac{2\pi}{3} - \epsilon, \Delta, \frac{2\pi}{3} - \Delta  \right) $ with a small deformation $\epsilon$ away from the first counterexample in Table \ref{tbl:InstantonCounterExamples1}, the non-perturbative correction is schematically
    \begin{equation} \label{eq:one-instanton-BH}
    \hat{f}_{\rm np} = \epsilon^L e^{-s_0\sqrt{N}} + e^{-s_1\sqrt{N}} + \cdots\,,
    \end{equation}
    where $s_0$ is the generic leading-order coefficient \eqref{eq:BH-instanton}, $L$ is certain positive number, and $s_1 > s_0$ is the exceptional coefficient in the second column of Table \ref{tbl:InstantonCounterExamples1}. The suppression factor $\epsilon^L$ can be explained by a fermionic zero mode that is lifted when $\epsilon$ is turned on \cite{Cvetic:2012ts}. Pinning this down requires analyzing the quantization of the brane, which we leave to future work. 
	
	For the $\hat{f}_{{\rm log}, I}^{\rm (np)}$ coefficient, it takes different values depending on how many chemical potentials are equal:
	
	\begin{itemize}
		\item Case 1: all $\hat{\Delta}_I$ are distinct: 
		\begin{equation}
			 \hat{f}_{{\rm log}, I}^{\rm (np)} = \frac{1}{2},\quad {\rm if}\ \Delta_{\gg} > \Delta_{>} > \Delta_{<} > \Delta_{\ll}.
		\end{equation}
		
		\item Cases 2(a), 2(b), 2(c): only two chemical potentials are identical,
		\begin{equation}
		\begin{aligned}
			&(a):\;\; \hat{f}_{{\rm log}, I}^{\rm (np)} = \frac{1}{2}, \quad {\rm if}\ \Delta_{\gg} > \Delta_{>} > \Delta_{<} = \Delta_{\ll},\\
			&(b):\;\; \hat{f}_{{\rm log}, >}^{\rm (np)} = \hat{f}_{{\rm log}, <}^{\rm (np)} = \frac{1}{2}, \quad 	\hat{f}_{{\rm log}, \gg}^{\rm (np)} = \hat{f}_{{\rm log}, \ll}^{\rm (np)} = 1, \quad {\rm if}\ \Delta_{\gg} > \Delta_{>} = \Delta_{<} > \Delta_{\ll},\\
			&(c):\;\; \hat{f}_{{\rm log}, I}^{\rm (np)} = \frac{1}{2}, \quad 	\hat{f}_{{\rm log}, <}^{\rm (np)} = \hat{f}_{{\rm log}, \ll}^{\rm (np)} = 1, \quad {\rm if}\ \Delta_{\gg} = \Delta_{>} > \Delta_{<} > \Delta_{\ll}.\\
		\end{aligned}
		\end{equation}
		
		\item Case 3: $\{\hat{\Delta}_I\}$ are pairwise equal, then
		\begin{equation}
			\hat{f}_{{\rm log}, I}^{\rm (np)} = \frac{1}{2}, \quad {\rm if}\ \Delta_{\gg} = \Delta_{>} > \Delta_{<} = \Delta_{\ll}.
		\end{equation}
		
	\item Case 4: three chemical potentials are identical,
		\begin{equation}
			\begin{aligned}
				& \hat{f}_{{\rm log}, \gg}^{\rm (np)} = 1,\quad \hat{f}_{{\rm log}, {\rm other}}^{\rm (np)} = \frac{1}{2},\quad {\rm if}\ \Delta_{\gg} > \Delta_{>} = \Delta_{<} = \Delta_{\ll};\\
				& \hat{f}_{{\rm log}, I}^{\rm (np)} = \frac{3}{2},\quad {\rm if}\ \Delta_{\gg} = \Delta_{>} = \Delta_{<} > \Delta_{\ll}.\\
			\end{aligned}
		\end{equation}
		
	\item Case 5: all four chemical potentials are equal, which we call the ``superconformal'' point \cite{Azzurli:2017kxo}:
		\begin{equation}
			\hat{f}_{{\rm log}, I}^{\rm (np)} = 1,\quad \hat{\Delta}_I = \frac{\pi}{2}\equiv  \hat{\Delta}_{\rm s.c.}.
		\end{equation}
        The superconformal point is so-named because $\hat{\Delta}_{\rm s.c.}$ is proportional to the superconformal R-charges of ABJM/ADHM theory, and extremizes the free energy on $S^3$.
	\end{itemize}
	
	 Again, although most of our numerical data supports the formulas above, some counterexamples are already found in Table \ref{tbl:InstantonCounterExamples2}, and we do not claim to have exhausted all of them. It will be most interesting to understand this from the bulk perspective, where $\hat{f}_{{\rm log}, I}^{\rm (np)}$ should be reproduced by the one-loop quantum corrections in the membrane worldvolume theory. 
	
	\renewcommand{\arraystretch}{1.8}
	\begin{table}[!h] \small 
		\centering
		\begin{tabular}{|c|c|c|c|}
			\hline
	case &	$\hat{\Delta}_I$ &   $\hat{f}_{{\rm log}, I}^{\rm (np)} $ & prediction  \\ \hline 
    1 &	$ \left( \pi, \frac{\pi}{2}, \Delta , \frac{\pi}{2} - \Delta \right), \quad \Delta \ne \frac{\pi}{6}, \frac{\pi}{3} $ & $\Big(0,0,\frac{1}{2},\frac{1}{2} \Big)$ & $\Big( \frac{1}{2}, \frac{1}{2},\frac{1}{2},\frac{1}{2} \Big)$ \\ \hline 
    1 &	$ \left( \pi, \frac{\pi}{2}, \frac{\pi}{3} , \frac{\pi}{6} \right) $ & $\Big(0,1,0,0 \Big)$ & $\Big( \frac{1}{2}, \frac{1}{2},\frac{1}{2},\frac{1}{2} \Big)$ \\ \hline 
	2(a)&	$ \left(\pi , \frac{\pi}{2}, \frac{\pi}{4}, \frac{\pi}{4} \right) $ & $\Big(0,0,\frac{1}{2},\frac{1}{2} \Big)$ & $\Big(\frac{1}{2},\frac{1}{2} ,\frac{1}{2},\frac{1}{2} \Big)$ \\ \hline 
	2(b)&	$ \left(\frac{3\pi}{4}, \frac{\pi}{2}, \frac{\pi}{2}, \frac{\pi}{4} \right) $ & $\Big( \frac{1}{2},\frac{1}{2},\frac{1}{2},\frac{1}{2} \Big)$ & $\Big(1 ,\frac{1}{2},\frac{1}{2},1 \Big)$ \\ \hline 
	2(c)&	$ \left(\frac{2\pi}{3}, \frac{2\pi}{3}, \frac{4\pi}{9}, \frac{2\pi}{9} \right)  $ & $\Big( \frac{1}{2},\frac{1}{2},\frac{1}{2},\frac{1}{2} \Big) $ & $\Big( \frac{1}{2},\frac{1}{2},1,1 \Big)$ \\ \hline 
	3&	$ \left(\frac{2\pi}{3}, \frac{2\pi}{3}, \frac{\pi}{3}, \frac{\pi}{3} \right) $ & $ \Big( 1,1,1,1 \Big) $ & $\Big(\frac{1}{2},\frac{1}{2},\frac{1}{2},\frac{1}{2} \Big)$  \\ \hline 
	4&	$ \left( \frac{4\pi}{5}, \frac{2\pi}{5},\frac{2\pi}{5}, \frac{2\pi}{5} \right) $ & $\Big( \frac{1}{2},\frac{1}{2},\frac{1}{2},\frac{1}{2} \Big)$ & $\Big( 1, \frac{1}{2},\frac{1}{2},\frac{1}{2} \Big)$ \\ \hline
		\end{tabular}
        \caption{\rm Counterexamples for the coefficients $\hat{f}_{{\rm log}, I}^{\rm (np)}$ found so far. }
		\label{tbl:InstantonCounterExamples2}
	\end{table}
	
	Finally, we provide the higher expansion coefficients $\hat{f}_{0, I}^{\rm (np)}$, $\hat{f}_{1, I}^{\rm (np)}$, $\hat{f}_{2, I}^{\rm (np)}$, $\hat{f}_{3, I}^{\rm (np)}$ at the superconformal point $\hat{\Delta}_I = \frac{\pi}{2}$. These can be tested with high numerical precision:
	\begin{equation}
		\begin{aligned}
			& \hat{f}_{1/2, I}^{\rm (np)} = -2\pi \sqrt{2},\quad \hat{f}_{\log, I}^{\rm (np)} = 1,\quad \hat{f}_{0, I}^{\rm (np)} = - 26,\quad \hat{f}_{1, I}^{\rm (np)} =  \frac{ 5(13\pi^2 - 8) }{2\sqrt{2}\pi },\\
			& \hat{f}_{2, I}^{\rm (np)} = - \frac{ 325\pi^4 - 1096 \pi^2 + 160  }{32\pi^2},\quad \hat{f}_{3, I}^{\rm (np)} = \frac{ \pi \left(   1625 \pi^2 - 15390  \right) }{384 \sqrt{2} }.
		\end{aligned} 
	\end{equation}
	This is to be constrasted with the three-sphere partition function of $k=1$ ABJM theory, whose corresponding one-instanton coefficients (including membrane and worldsheet instantons) are: \footnote{ The seem-to-be minus overall sign difference comes from our definition of the free energy \eqref{eq:F=logI}, which differs from that of \cite{Hatsuda:2012dt} by a sign. }\cite{Hatsuda:2012dt}
		\begin{equation}
		\begin{aligned}
			& 	\hat{f}_{1/2, I}^{\rm (np)} = -2\pi \sqrt{2},\quad \hat{f}_{\log, I}^{\rm (np)} = 1,\quad \hat{f}_{0, I}^{\rm (np)} = 2,\quad \hat{f}_{1, I}^{\rm (np)} = - \frac{  13\pi^2 - 2 }{2\sqrt{2}\pi },\\
			& \hat{f}_{2, I}^{\rm (np)} = \frac{ 169 \pi^4 +116 \pi^2 + 8  }{32\pi^2},\quad \hat{f}_{3, I}^{\rm (np)} = \frac{ \pi \left( - 2197 \pi^2 - 4012  \right) }{ 384 \sqrt{2} }.
		\end{aligned} 
	\end{equation}
	In reference \cite{Hatsuda:2012dt}, with exact values of the $S^3$ partition function $Z(N)$ and an educated ansatz, the authors were able to obtain closed form expressions for the non-perturbative corrections in $Z(N)$ up to six-instanton contributions. Similarities in form between the large-$N$ expansion of $Z(N)$ and that of the TTI strongly suggests that a similar asymptotic expansion of the TTI instanton corrections in terms of ratios of Airy functions exists. Recently, the instanton corrections of the topologically twisted index is studied numerically with very high precision in \cite{Hosseini:2026lvj}, where a conjecture for instanton corrections has been made. Our results are in agreement with the conjectured formula. 
	
	%%%%%%%%%%%%%%%%%%%%%%%%
	\subsection{Dual dyonic black holes}
	\label{subsec:oneCut:bulk}
	%%%%%%%%%%%%%%%%%%%%%%%%

    In this subsection, we review the construction of the dyonic black holes that are dual to the one-cut solutions, and show that the solutions in BH$_\pm$ correspond to black holes in different branches. On the way, we complete the construction in \cite{Katmadas:2014faa} by providing a general solution to the flow equations in terms of symplectic invariants.

    \subsubsection*{FI gauging and the black hole ansatz}

    The black holes in question can be constructed in any $d=4$ $\cN=2$ FI gauged supergravity.  The bosonic field consists of $n_v$ complex scalars $z_{i=1,\ldots, n_v}$, $n_v+1$ gauge fields $A^{\Lambda =1,\ldots,n_v+1}$, and the metric $g_{\mu\nu}$. The scalars are parametrized by $n_v + 1$ redundant holomorphic sections $X^\Lambda$ as
    \be
    z_i = \frac{X^i}{X^{n_v+1}}\,.
    \ee
    The theory is specified by a choice of prepotential $F$ and FI parameters $\cG$. In particular, we are concerned with the STU model, which is a consistent $U(1)^4$ truncation of $d=4$, $\cN=8$ $SO(8)$ gauged supergravity, which in turn arises
    from the consistent reduction of eleven-dimensional supergravity
    on $S^7$ \cite{deWit:1986oxb,Cvetic:1999xp,Azizi:2016noi}. For the STU model with $n_v = 3$, the data specifying the theory is 
    \be\label{eq: def STU sqrt}
    F = -2i\sqrt{X^1X^2X^3X^4}\,,\quad \cG = g(0,0,0,0;\, 1,1,1,1)^T\,.
    \ee
    Since the black holes can be constructed in any FI gauged supergravity in a uniform way, we shall keep the discussion generic for now and only restrict to the STU model when comparing with the one-cut solution at the end.
    
    The static dyonically charged black holes that we are considering have the ansatz 
    \bea\label{eq: bh metric ansatz}
        &z_i(r)\,,\quad A^\Lambda = A^\Lambda_t(r)dt + p^\Lambda\omega_{\Sigma_\fg}\,, \quad d\omega_{\Sigma_\fg} = \text{vol}_{\Sigma_\fg}\,, \\
        &ds^2 = -e^{2U(r)}dt^2 + e^{-2U(r)}dr^2 + e^{2(\psi(r)-U(r))}d\Omega^2_{\Sigma_\fg}\,,\\
        &d\Omega^2_{\Sigma_\fg} = (dx^2+dy^2)\times{\begin{cases}\frac{4}{(1+x^2+y^2)^2}\,, & \fg = 0\,,\;(x,y)\in \bR^2 \\ 2\pi\,, & \fg = 1\,,\;(x,y)\in [0,1)^2 \\ \frac{1}{y^2}\,, & \fg > 1\,,\;(x,y)\in \bH/\Gamma\end{cases}}
    \eea
    where $p^\Lambda$ are its magnetic charges, $\text{vol}_{\Sigma_\fg}$ is the volume form on the Riemann surface $\Sigma_\fg$ of genus $\fg$ with unit radius, and $d\Omega^2_{\Sigma_\fg}$ is its squared line element. Note that $\fg > 1$ Riemann surfaces are described by quotienting the Poincar\'e half plane $\bH$ by a Fuchsian group $\Gamma \subset \PSL(2,\bR)$. By the Gauss-Bonnet theorem, the area of the unit Riemann surfaces are
    \be
    \int_{\Sigma_\fg} \text{vol}_{\Sigma_\fg} = 2\pi\eta \equiv 2\pi \begin{cases}2|\fg - 1|\,, & \fg \neq 1 \\ 1\,, & \fg = 1\end{cases}
    \ee
    The magnetic and electric charges are defined as
    \be
    p^\Lambda \equiv \frac{1}{2\pi\eta}\int_{\Sigma_\fg}F^\Lambda\,,\quad q_\Lambda \equiv \frac{1}{2\pi\eta}\int_{\Sigma_\fg}G_\Lambda\,, \quad G_\Lambda \equiv \frac{\delta \cL}{\delta F^\Lambda}\,,
    \ee
    where the definition of $G_\Lambda$ is understood as the derivative of a $4$-form with respect to a $2$-form. The charges are more conveniently organized into the symplectic vector 
    \be
    \cQ\equiv (p^\Lambda , q_\Lambda)^T\,.
    \ee
    Constructing the black hole then amounts to finding $U(r)$, $\psi(r)$ and $z_i(r)$ in terms of $\cQ$. This is done by solving the BPS flow equations, which we review in the following.

    \subsubsection*{BPS flow equations}
    
    Define the symplectic inner product
    \be
    \langle \cA,\cB\rangle = \cA_\Lambda \cB^\Lambda -\cA^\Lambda \cB_\Lambda\,. 
    \ee
    It is shown in \cite{DallAgata:2010ejj} that provided the following constraint holds
    \be
    \langle \cG,\cQ\rangle = -\kappa\,,\quad \kappa = \begin{cases} 1\,, & \fg = 0 \\ 0\,, & \fg = 1 \\ -1\,, & \fg > 1\end{cases}\,,
    \ee
    the bosonic action can be written as a sum of squares up to boundary terms. The BPS equations correspond to setting the bases of the squares to zero. They can be written entirely in terms of the unknown
    \be
    \im\wt\cV=e^{\psi - U}\im e^{-i\alpha}e^{\frac{K}{2}}\left(
        X^\Lambda\,, F_\Lambda\right)^T\,, 
    \ee
    where 
    \be
    F_\Lambda = \frac{\partial F}{\partial X^\Lambda}\,,\quad K = -\log\left[i(F_\Lambda\overline{X}^\Lambda - X^\Lambda\overline{F}_\Lambda)\right]\,,
    \ee
    and $\alpha$ is an auxiliary phase factor that also appears in the projectors satisfied by the Killing spinor. In addition, the equations transform covariantly under symplectic transformations and are written in terms of the quartic symplectic invariant $I_4$ and its derivative $I_4'$, which is a symplectic vector\footnote{See Appendix A.2 of \cite{Halmagyi:2014qza} for the definition of $I_4$ in the cubic symplectic frame where $F = -\frac{d_{ijk}X^iX^jX^k}{X^{n_v+1}}$. For the STU model, this the frame where $F=-\frac{X^1X^2X^3}{X^4}$ and $\cG = g(0,-1,-1,-1;\, 1,0,0,0)^T$. This is related to the square-root frame in \eqref{eq: def STU sqrt} via the symplectic transformation $S=\begin{psmallmatrix}A & B \\ -B &  A\end{psmallmatrix}\in \Sp(8,\bR)$, $A=\diag(1,0,0,0)$, $B=\diag(0,1,1,1)$. }. We consider the black holes with $2n_v - 1$ unconstrained dyonic charges constructed in \cite{Katmadas:2014faa}, which have the ansatz\footnote{There are in principle more general black holes with $2n_v$ unconstrained charges, constructed in \cite{Halmagyi:2014qza}, but their ansatz is considerably more complicated. We have stuck to the black holes of \cite{Katmadas:2014faa} for simplicity. }
    \be\label{eq: BH Vtilde ansatz}
    \im\wt\cV(r) = \cA r + \cB\,.
    \ee
    Via one of the BPS equations, this immediately implies that
    \be
    e^{\psi}=\langle\cG,\cA\rangle r^2 + 2\langle \cG,\cB\rangle r\,,
    \ee
    where the origin of the coordinate $r$ is chosen to be the horizon. With the ansatz \eqref{eq: BH Vtilde ansatz}, the remaining BPS equations reduce to the algebraic equations
    \bea\label{eq: BPS flow alg}
    I_4'(\cG,\cA,\cA)=2\langle\cG,\cA\rangle\cA\,,\quad I_4'(\cG,\cA,\cB)=2\langle\cG,\cB\rangle \cA\,,\quad I_4'(\cG,\cB,\cB)-\cQ = 0\,.
    \eea
    The first equation is solved by
    \be
    \cA = \frac{I_4'(\cG)}{4I_4(\cG)^\frac{3}{4}}\,,
    \ee
    where the normalization of $\cA$ is fixed by demanding that 
    \be
    e^{2\psi}|_{r=\infty} = \sqrt{I_4(\cG)}r^4 = \frac{r^4}{R^2_{\text{AdS}_4}}\,.
    \ee
    Consistency with the remaining equations gives a linear constraint on the charges:
    \be\label{eq: BH lin charge constr}
    I_4(\cG,\cG,\cG,\cQ)= 0\,,
    \ee
    which in the STU model is simply
    \be
    \sum_\Lambda q_\Lambda = 0\,.
    \ee
    One can then show that system \eqref{eq: BPS flow alg} is equivalent to
    \be\label{eq: BPS flow simp}
    I_4'(\cG,\cB,\cB)-\cQ = 0\,,\quad \langle I_4'(\cG),\cB\rangle = 0\,,\quad I_4(\cG,\cB,\cB,\cB) = 0\,.
    \ee

    \subsubsection*{Two branches of dyonic solutions}
    
    The exposition so far has just been a review of \cite{Katmadas:2014faa}. We now solve these equations in full generality, without restricting to a specific model and gauging $\cG$. Using the identities in Appendix A.3 of \cite{Halmagyi:2014qza}\footnote{The second equation under (86) in Appendix A.3 should be $I_4'(I_4'(\cG),I_4'(\cG,\cG,\cQ),\cG)=2I_4(\cG,\cG,\cG,\cQ)I_4'(\cG)+16I_4(\cG)\langle\cG,\cQ\rangle\cG$, where $\cG$ and $\cQ$ are arbitrary symplectic vectors.}, one can show that if $\cB$ is a linear combination of $\{I_4'(\cG)$, $I_4'(\cG,\cG,\cQ)$, $I_4'(\cG,\cQ,\cQ)$, $I_4'(\cQ)\}$, then $I_4'(\cG,\cB,\cB)$ is in the span of $\{\cG$, $\cQ$, $I_4'(\cG)$, $I_4'(\cG,\cG,\cQ)$, $I_4'(\cG,\cQ,\cQ)$, $I_4'(\cQ)\}$. It is therefore convenient to use the following ansatz for $\cB$:
    \be
    \cB = b_1 I_4'(\cG) + b_2 I_4'(\cG,\cG,\cQ) + b_3 I_4'(\cG,\cQ,\cQ) + b_4 I_4'(\cQ)\,.
    \ee
    Solving the first equation in \eqref{eq: BPS flow simp} is only possible if there is a second nonlinear constraint on the charges 
    \be\label{eq: BH nonlin charge constr}
    I_4(\cG,\cQ,\cQ,\cQ) = 0\,.
    \ee
    In this case, the solutions are\footnote{In principle, there are two more solutions where the sign in front of the square root in the numerator of $b_1$ is flipped. But these solutions are unphysical because the relative sign between the two terms in $e^\psi$ will also be flipped. This would imply that the horizon is at some positive value of $r$, which contradicts our assumption that the horizon is at $r=0$.}
    \bea\label{eq: BH B coeff sol}
    &b_1 = \frac{\pm I_4(\cG,\cG,\cQ,\cQ)+\sqrt{I_4(\cG,\cG,\cQ,\cQ)^2-64 I_4(\cG)I_4(\cQ)}}{16 I_4(\cG)\sqrt{-I_4(\cG,\cG,\cQ,\cQ)\kappa - 2\langle I_4'(\cG),I_4'(\cQ)\rangle}}\\
    &b_3 = \mp\frac{1}{4\sqrt{-I_4(\cG,\cG,\cQ,\cQ)\kappa - 2\langle I_4'(\cG),I_4'(\cQ)\rangle}}\,,\\
    & b_2 = b_4 = 0\,.
    \eea
    The remaining two equations in \eqref{eq: BPS flow simp} are then automatically satisfied due to the identity
    \be
    \langle I_4'(\cG,\cQ,\cQ),I_4'(\cG)\rangle = -\frac{2}{3}I_4(\cQ,\cG,\cG,\cG)\langle\cG,\cQ\rangle\,,
    \ee
    and the constraints \eqref{eq: BH lin charge constr}, \eqref{eq: BH nonlin charge constr}.  We therefore get two branches of black hole solutions. As will be shown shortly, in the STU model, the upper sign in \eqref{eq: BH B coeff sol} will be dual to the one-cut solution in BH$_+$, while the lower sign will be dual to the solution in BH$_-$. The origin of this sign is the fact that the first equation in \eqref{eq: BPS flow simp} is invariant under $\cB\mapsto -\cB$. The metric functions are then given by
    \bea
        &e^\psi = I_4(\cG)^\frac{1}{4}r^2+2\langle \cG,\cB\rangle r = r\left(I_4(\cG)^\frac{1}{4}r+\frac{\sqrt{I_4(\cG,\cG,\cQ,\cQ)^2-64 I_4(\cG)I_4(\cQ)}}{2\sqrt{-I_4(\cG,\cG,\cQ,\cQ)\kappa - 2\langle I_4'(\cG),I_4'(\cQ)\rangle}}\right)\,,\\
        & {e^{2(\psi - U)} = 4\sqrt{I_4(\cA r + \cB)}}
    \eea
    As shown in \cite{Katmadas:2014faa,Halmagyi:2014qza}, special geometry identities imply that
    \be
    \wt\cV = \re\wt\cV + i\im\wt\cV = -\frac{I_4'(\im\wt\cV)}{2\sqrt{I_4(\im\wt\cV)}} + i\im\wt\cV
    \ee
    after which the scalars are determined as
    \be
    z_i = \frac{X^i}{X^{n_v+1}} = \frac{\wt\cV^i}{\wt\cV^{n_v+1}}\,.
    \ee
    This completes the construction of static dyonic black holes with $2n_v-1$ charges in generic $d=4$ $\cN=2$ FI gauged supergravity.  

    \subsubsection*{Entropy in the STU model and physicality conditions}
    
    We now specify the solutions to the STU model defined by \eqref{eq: def STU sqrt}. We set $g = \frac{1}{\sqrt{2}}$ so that $R^2_{\text{AdS}_4} = 1$. Changing variables to the field theory fluxes and charges
    \be\label{eq: g q dict}
    \hat{\fn}_{\Lambda} =\frac{1}{2\pi}\int_{\Sigma_\fg} g F^\Lambda= \eta g p^\Lambda\,,\quad \hat{\fq}_{\Lambda} =\frac{1}{2\pi}\int_{\Sigma_\fg} g G_\Lambda= \eta g q_\Lambda\,,
    \ee
    the black hole entropy is
    \bea
    \frac{2\pi\eta R^2_{\Sigma_\fg}}{4G_N^{(4\text{d})}}&=\frac{2\pi\eta e^{2(\psi(0)-U(0))}}{4G_N^{(4\text{d})}}=\frac{\pi\eta\sqrt{I_4(\cG,\cG,\cQ,\cQ)\pm\sqrt{I_4(\cG,\cG,\cQ,\cQ)^2-64 I_4(\cG)I_4(\cQ)}}}{4\sqrt{2}G_N^{(4\text{d})}I_4(\cG)^\frac{1}{2}}\\
    &=\frac{\pi\sqrt{\Theta_2\pm\sqrt{\Theta_2^2-4\Theta_4}}}{2G_N^{(4\text{d})}}\,,
    \eea
    where the signs above are correlated with those in \eqref{eq: BH B coeff sol}, and we defined
    \bea
    &\Theta_2\equiv-\frac{1}{2}\sum_I\hat{\fn}_I^2+\sum_{I<J}\hat{\fn}_I\hat{\fn}_J+2\sum_{I<J}\hat{\fq}_I\hat{\fq}_J=\frac{\kappa^2\eta^2}{2}-\sum_I\hat{\fn}_I^2-\sum_I\hat{\fq}_I^2\,,\\
    &\Theta_4\equiv 2\sum_{I<J}\hat{\fn}_I\hat{\fn}_J\hat{\fq}_I\hat{\fq}_J-\sum_I\hat{\fn}_I^2\hat{\fq}_I^2+4\prod_I\hat{\fn}_I+4\prod_I\hat{\fq}_I\,.
    \eea
    In the above and what follows, we use the capital greek symplectic indices $\Lambda,\Sigma,\ldots$ in supergravity and the capital latin indices $I,J,\ldots$ labelling fluxes and charges in field theory interchangeably, since they are identified by \eqref{eq: g q dict}. Using the AdS/CFT dictionary $1/G_N^{(4\text{d})}=\frac{2\sqrt{2}}{3}N^\frac{3}{2}$, the entropy becomes
    \be\label{eq: BH entropy}
    S=N^\frac{3}{2}\frac{\pi\sqrt{2}}{3}\sqrt{\Theta_2\pm\sqrt{\Theta_2^2-4\Theta_4}}\,.
    \ee
    The black hole solution is only physical if the expressions under the square roots in $b_1$ and $b_3$ are positive, and $R^2_{\Sigma_\fg} > 0$, i.e.\footnote{We exclude the equality in the first constraint to ensure that the near-horizon geometry is AdS$_2\times S^2$. }
    \bea\label{eq: BH phys cond}
    &I_4(\cG,\cG,\cQ,\cQ)^2-64 I_4(\cG)I_4(\cQ) > 0\,,\quad -I_4(\cG,\cG,\cQ,\cQ)\kappa - 2\langle I_4'(\cG),I_4'(\cQ)\rangle > 0\,,\\
    &I_4(\cG,\cG,\cQ,\cQ)\pm\sqrt{I_4(\cG,\cG,\cQ,\cQ)^2-64 I_4(\cG)I_4(\cQ)} > 0\,.
    \eea
    This restricts the space of permissible charges. For the branch with the lower sign, solutions only exist at higher genus. For example, one can check that $\fg = 6, \hat{\fn} = (1,3,3,3), \hat{\fq} = 0$ satisfy these constraints with the lower sign. 

    \subsubsection*{Purely magnetic solutions and the match with 1-cut solutions}
    
    For our purposes, it will be sufficient to restrict to purely magnetic solutions with $\fq_\Lambda=0$. This is because we have restricted to real chemical potentials $\hat{\Delta}_I$. As we will see shortly, the horizon attractor values of the sections $2\pi X^I$ are identified with the extremal values of $[\hat{\Delta}_I]$ or $[\hat{\Delta}_I]'$ under the Legendre transform of the free energies at leading order in large $N$. Since the attractor values of $X^I$ are real for purely magnetic solutions, we are precisely restricting to this class of black holes. In this case, the charge constraints \eqref{eq: BH lin charge constr} and \eqref{eq: BH nonlin charge constr} are automatically satisfied. This restriction has the advantage of simplifying the black hole solution significantly. In particular, $I_4'(\im\wt\cV)$ does not have magnetic components, meaning that $\wt\cV^\Lambda$ are purely imaginary and the physical scalars $z^i$ are real. They are given by:
    \begin{gather}
    z_i = \frac{\im\wt\cV^i}{\im\wt\cV^4} = \frac{r+\frac{1}{\eta\sqrt{\Theta_3}}\left(\pm v_i + \sqrt{\Theta_2^2-4\Theta_4}\right)}{r+\frac{1}{\eta\sqrt{\Theta_3}}\left(\pm v_4 + \sqrt{\Theta_2^2-4\Theta_4}\right)}\,,\quad v_I \equiv 4\hat{\fn}_I^2 + 2\kappa\eta\hat{\fn}_I + \Theta_2\,,\notag\\
    \Theta_3 = \left(2\sum_I \hat{\fn}_I^2-\kappa^2\eta^2\right)\kappa-\frac{8}{\eta}\sum_{I<J<K}\hat{\fn}_I\hat{\fn}_J\hat{\fn}_K\,.\label{eq: scalars mag}
    \end{gather}
    If we fix the K\"ahler gauge by demanding that the sections $X^\Lambda$ are real, the BPS equations imply
    \be\label{eq: section constr}
    \sum_\Lambda X^\Lambda = \pm 1\,,
    \ee
    where this sign is correlated with the sign in \eqref{eq: BH B coeff sol} \cite{Hristov:2010ri}, and with a choice of spinor projection when solving the Killing spinor equations. This then determines
    \be
    X^{i} = \pm\frac{z_i}{1+\sum_iz_i}\,,\quad X^4 =\pm \frac{1}{1+\sum_iz_i}\,.
    \ee
    Due to the attractor mechanism, the values of the sections at the horizon must extremize \cite{Benini:2015eyy}
    \be\label{eq: attractor Z}
    \sqrt{X^1X^2X^3X^4}\sum_I \frac{\hat{\fn}_I}{X^I}\,,
    \ee
    subject to the constraint \eqref{eq: section constr}. By identifying
    \be\label{eq: iden X w Delta}
    2\pi X^I = \pm\frac{2\pi X^I}{\sum_{I}X^I} = \begin{cases} [\hat{\Delta}_I]\,, & +\;\text{branch} \\ [\hat{\Delta}_I]'\,, & -\;\text{branch}\end{cases}\,,
    \ee
    we observe that this extremization problem matches the extremization equations coming from the Legendre transform of \eqref{eq: logZ BH} for the $+$ branch, and \eqref{eq: logZ BH branch 2} for the $-$ branch. This is because an overall constant multiplying \eqref{eq: attractor Z} does not affect the extremization equations, and the constraint \eqref{eq: section constr} for the two branches matches the constraints $\sum_I[\hat{\Delta}_I]=2\pi$, $\sum_I[\hat{\Delta}_I]'=-2\pi$ for the two one-cut solutions. One can check explicitly that the sections evaluated at $r=0$ solve the constrained extremization equations, and evaluating \eqref{eq: logZ BH} or \eqref{eq: logZ BH branch 2} at their extremal values exactly reproduces the black hole entropy in \eqref{eq: BH entropy}, where the plus sign under the square root is reproduced by \eqref{eq: logZ BH}, and the minus sign by \eqref{eq: logZ BH branch 2}. In addition, for any set of charges such that the black hole physicality conditions \eqref{eq: BH phys cond} are satisfied with the $+$ sign, one finds that $2\pi X^\Lambda \in (0,2\pi)$, consistent with the range of $[\hat{\Delta}_I]$. Similarly, for a set of charges satisfying \eqref{eq: BH phys cond} with the $-$ sign, one can check that $2\pi X^\Lambda \in (-2\pi,0)$, consistent with the domain of $[\hat{\Delta}_I]'$. We can thus conclude that one-cut solutions in BH$_+$ are dual to black holes in the $+$ branch, while one-cut solutions in BH$_-$ are dual to black holes in the $-$ branch. 

    %%%%%%%%%%%%%%%%%%%%%%%%
	\subsection{Large $N$ expansion from equivariant localization}
	\label{subsec:equiv loc BH}
	%%%%%%%%%%%%%%%%%%%%%%%%

    In this subsection, we will show that the all-order perturbative corrections to $F_{\fg=0,\text{BH}_+}$ shown in \eqref{eq:TTI-BH} can be derived from equivariant localization in supergravity. Recently, the black hole free energy has also been computed in \cite{BenettiGenolini:2026kjy} for M-theory on AdS$_4\times S^7/\bZ_k$, and their result at $k=1$ coincides with ours here.

    In \cite{BenettiGenolini:2026cyc}, the authors provide a general formula for the partition function of M-theory on an 11d manifold $M$ of the form 
    \be
    Y_7 \hookrightarrow M \rightarrow M_4\,.
    \ee
    Consider the refined TTI, which is just the usual TTI on $\Sigma_\fg = S^2$ that we have been studying, but with an additional fugacity $q=e^{-\omega}$ that is conjugate to the angular momentum on $S^2$. We will eventually take $\omega\rightarrow 0$ to recover the unrefined index\footnote{The formula of \cite{BenettiGenolini:2026cyc} only applies when the Killing vector constructed out of the Killing spinor has isolated fixed points. This is true when $\omega\neq 0$, but not for $\omega = 0$ corresponding to the unrefined TTI, where the fixed locus is the entire horizon $S^2$. Thus we need to consider the refined TTI first and subsequently take $\omega\rightarrow 0$.}. In this case, $Y^7 = S^7$ and $M_4$ is the rotating generalization \cite{Hristov:2018spe} of the dyonic black hole spacetimes constructed in Section \ref{subsec:oneCut:bulk}. Equivariant localization gives the partition function in terms of the data $b_{p,i=1,\dots, 6}$ satisfying $\sum_i b_{p,i} = 0$, which are the tangent weights at the fixed points $p$ of the Killing vector constructed from the Killing spinor bilinear\footnote{There are $6$ tangent weights because the 11d manifold $M$ needs to be extended to a 12d manifold $W$ with $\partial W = M$ in order to define Chern-Simons terms.}. The last $4$ weights corresponding to $Y_7$ are denoted by $u_I \equiv b_{2+I}$, for $I=1,\ldots, 4$. The general formula for the M-theory partition function up to non-perturbative corrections is then
    \be\label{eq: M th Z}
    \cI_{0,\text{BH}_+}=e^A\prod_{p}C_p^{-\frac{1}{3}}\Ai\left[\frac{N-B_p}{C_p^\frac{1}{3}}\right]\,,
    \ee
    where $A$ is a constant in $N$, and we have defined
    \bea\label{eq: loc comb def}
    B_p\equiv\frac{1}{24 e_{p,4}}[e_{p,4}+(\hat{b}_{p,1}+\hat{b}_{p,2})e_{p,3}+\hat{b}_{p,1}\hat{b}_{p,2}e_{p,2}]\,,\quad \hat{b}_{p,i}\equiv\sigma_{p,i}b_{p,i}\,,\quad C_p\equiv\frac{(b_{p,1}b_{p,2})^2}{8\pi^2e_{p,4}}\,.
    \eea
    $e_{p,r}$ are the elementary symmetric polynomials of degree $r$ in the variables $u_I$, and $\sigma_{p,i}=\pm 1$ are signs associated with the spinor projections satisfied by the Killing spinor. 
    
    In the case of the rotating black hole contributing to the refined index, there are two fixed points at the north and south poles of the horizon $S^2$. The tangent weights and $\sigma_{p,i}$ at these points are identified as
    \bea\label{eq: ref TTI loc iden}
    (b_{1,1}\,,b_{1,2}) = (1\,,\omega)\,,\quad u_{1,I}&=-\frac{\hat{\Delta}_I}{2\pi}+\omega\frac{\hat{\fn}_I}{2}\,,\quad \sigma_{1,1} = \sigma_{1,2} = 1\,,\\
    (b_{2,1}\,,b_{2,2}) = (1\,,-\omega)\,,\quad u_{2,I}&=-\frac{\hat{\Delta}_I}{2\pi}-\omega\frac{\hat{\fn}_I}{2}\,,\quad \sigma_{2,1} = \sigma_{2,2} = 1\,.
    \eea
    We can now evaluate \eqref{eq: M th Z} in the strict unrefined limit $\omega\rightarrow 0$. Since $C_{1,2}\sim \omega^2$\,, we can use the asymptotic expansion of $\Ai(z)$ for $|z|\rightarrow \infty$, which is
    \be\label{eq: Ai asymp}
    \Ai(z) = \frac{1}{2\sqrt{\pi}z^\frac{1}{4}}e^{-\frac{2}{3}z^{3/2}}(1+\cO(z^{-3/2}))\,,
    \ee
    and keep only the leading term. By plugging the identifications \eqref{eq: ref TTI loc iden} into \eqref{eq: loc comb def} and simplifying using the constraints $\sum_I\hat{\Delta}_I = 2\pi$ and $\sum_I\hat{\fn}_I=-2$, one can show that
    \bea\label{eq: Bp omega exp}
    B_1 &= \frac{1}{24}-\frac{1}{12}\sum_{I=1}^4\frac{\pi}{\hat{\Delta}_I}-\frac{\omega}{24}\left(\sum_{I=1}^4\frac{2\pi^2\hat{\fn}_I}{\hat{\Delta}_I^2}+\sum_{I=1}^4\frac{2\pi}{\hat{\Delta}_I}-\sum_{I<J}\frac{4\pi^2}{\hat{\Delta}_I\hat{\Delta}_J}\right)+\cO(\omega^2)\\
    &=-\delta^{(\text{BH})} + \frac{\pi \omega}{3}\sum_{I=1}^4\frac{\fc_I^{(\rm BH)}\hat{\fn}_I}{\hat{\Delta}_I}+\cO(\omega^2)\,,\\
    B_2&=-\delta^{(\text{BH})} - \frac{\pi \omega}{3}\sum_{I=1}^4\frac{\fc_I^{(\rm BH)}\hat{\fn}_I}{\hat{\Delta}_I}+\cO(\omega^2)\,.
    \eea
    We also have\footnote{ The defining branch is
    \begin{equation}
    C_p^{-1/2} \equiv \frac{\sqrt{8\pi^2 e_{p,4}}}{b_{p,1}b_{p,2}},
    \end{equation}
    thus it is natural to have $C_1^{-1/2} >0, C_2^{-1/2}<0$. 
    }
    \bea\label{eq: Cp omega exp}
    C_1^{-\frac{1}{2}}=\frac{1}{\sqrt{2}\pi\omega}\sqrt{\prod_{I=1}^4\hat{\Delta}_I}\left(1-\frac{\pi\omega}{2}\sum_{I=1}^4\frac{\hat{\fn}_I}{\hat{\Delta}_I} +\cO(\omega^2) \right)\,,\\
    C_2^{-\frac{1}{2}}=-\frac{1}{\sqrt{2}\pi\omega}\sqrt{\prod_{I=1}^4\hat{\Delta}_I}\left(1+\frac{\pi\omega}{2}\sum_{I=1}^4\frac{\hat{\fn}_I}{\hat{\Delta}_I} +\cO(\omega^2) \right)\,.
    \eea
    After applying \eqref{eq: Ai asymp} to \eqref{eq: M th Z}, and using \eqref{eq: Bp omega exp} and \eqref{eq: Cp omega exp} to find the surviving terms as $\omega\rightarrow 0$, one exactly recovers \eqref{eq:TTI-BH} up to the non-perturbative terms.

    %%%%%%%%%%%%%%%%%%%%%%%%
	\subsection{Non-perturbative corrections from M2-branes}
	\label{subsec:non-pert M2}
	%%%%%%%%%%%%%%%%%%%%%%%%

    In this subsection, we shall review the uplift of the magnetically charged black holes considered in Section \ref{subsec:oneCut:bulk} to 11d, and compute the action of M$2$ branes wrapped on $S^3\subset S^7$ in the uplifted geometry. As anticipated, their action accounts for the non-perturbative corrections found numerically in Section \ref{subsec:oneCut:numerics}. We also provide an argument for why the wrappings that were considered are supersymmetric.

    Because the physical scalars and sections are real in the magnetic solutions, their uplift to 11d supergravity is also relatively simple \cite{Benini:2015eyy,Cvetic:1999xp}. The scalars are parametrized by the rescaled sections $L$ defined below:
    \be\label{eq: sec L def}
    L^\Lambda = 2\sqrt{2}e^{K/2}X^\Lambda\,, \quad e^{-K} = 8\sqrt{X^1X^2X^3X^4}\,,
    \ee
    where $K$ is the K\"ahler potential. $S^7$ is parametrized by the coordinates 
    \be
    \sum_{I=1}^4|Z_I|^2 = 1\,,\quad Z_I = \mu_I e^{i\varphi_I}\,,\quad \sum_{I=1}^4\mu_I^2 = 1\,.
    \ee
    The uplifted metric and 4-form flux are then given by
    \bea\label{eq: 11d uplifted g G}
    ds^2 &= \Delta^\frac{2}{3}ds_4^2 + \frac{2}{g^2\Delta^\frac{1}{3}}\sum_I\frac{1}{L_I}(d\mu_I^2 + \mu_I^2(d\varphi_I + gA_I)^2)\,,\quad \Delta \equiv \sum_I L_I \mu_I^2\,,\\
    G_4 &= \sqrt{2}g\sum_I(L_I^2\mu_I^2-\Delta L_I)\epsilon_4-\frac{1}{\sqrt{2}g}\sum_IL_I^{-1}(*_4dL_I)\wedge d\mu_I^2\\
    &\quad+\frac{\sqrt{2}}{g^2}\sum_IL_I^{-2}d\mu_I^2\wedge(d\varphi_I+gA_I)\wedge *_4F_I\,.
    \eea
    $ds^2_4$ is the black hole line element in $4$d, $\epsilon_4$ is the 4d volume form oriented as $dt\wedge dr\wedge dx\wedge dy$, and $*_4$ is the 4-dimensional hodge star defined with the convention $\alpha\wedge *_4\beta = \frac{1}{p!}\alpha_{\mu_1\ldots\mu_p}\beta^{\mu_1\ldots\mu_p}\epsilon_4$ for $p$-forms $\alpha$, $\beta$ in 4d. Using the scalar equations of motion, the Maxwell equation, and keeping in mind that $\sum_{I=1}^4d\mu_I^2 = 0$, one can show that $dG_4 = 0$ and $G_4=dC_3$, where
    \begin{gather}\label{eq: C3 uplift}
    C_3 = \frac{1}{\sqrt{2}g}\sum_{I=1}^4\mu_I^2L_I^{-1}*_4 dL_I + \frac{\sqrt{2}}{g^2}\sum_{I=1}^4\mu_I^2L_I^{-2}(d\varphi_I+gA_I)\wedge *_4 F_I + C_3^{\mathrm{ext}}\,,\notag\\
    dC_3^{\mathrm{ext}} = -\frac{1}{2\sqrt{2}g}\sum_{I=1}^4L_I^{-2}F_I\wedge *_4F_I-\frac{g}{\sqrt{2}}\sum_{\substack{I,J=1 \\ I<J}}^4 L_IL_J\epsilon_4\,.
    \end{gather}
    $C_3^{\mathrm{ext}}$ and $C_3$ are defined up to gauge transformations, and we can choose
    \be
    C_3^{\mathrm{ext}}=-\frac{1}{2\sqrt{2}g}\sum_{I=1}^4L_I^{-2}A_I\wedge *_4F_I-\frac{g}{\sqrt{2}}\sum_{\substack{I,J=1 \\ I<J}}^4L_IL_Jte^{\psi(r)-U(r)}dr\wedge \text{vol}_{\Sigma_\fg}\,.
    \ee
    For a Euclidean M2 brane whose worldvolume does not wrap the time direction, its bosonic action is \cite{Bergshoeff:1987cm,Bergshoeff:1987qx}
    \be\label{eq: M2 bos S}
    S_{\text{M}2}=\frac{1}{(2\pi)^2l_p^3}\int d^3\sigma\left( \sqrt{h} + \frac{i}{3!}\varepsilon^{ijk}\partial_ix^{M}\partial_jx^{N}\partial_kx^{P}(C_3)_{MNP}\right)\,.
    \ee
    $\sigma^{i=1,2,3}$ are worldvolume coordinates, $x^{M}$ are coordinates on 11d spacetime, $h_{ij}$ is the induced metric with determinant $h$, and $\varepsilon$ is the epsilon symbol. We consider a brane wrapping a particular $S^3\subset S^7$ at $r=0$, the radial position of the horizon. Since the metric component $g_{tt}=-e^{2U(r)}|_{r=0} = 0$, the coordinate $t$ degenerates at $r=0$ and is not a valid coordinate there. Instead of $t$, we use the ingoing Eddington-Finkelstein coordinate $v$ defined as
    \be\label{eq: EF coord def}
    v \equiv t + \int^r_{r_0}d\tilde{r}e^{-2U(\tilde{r})}\,,\quad r_0>0\,,\quad dv = dt + e^{-2U(r)}dr\,.
    \ee
    The brane embedding is given by
    \bea\label{eq: wrapped S3 def}
    r &= 0\,,\quad v = v_0, \quad x_{\Sigma_g} = x_0\,, \quad y_{\Sigma_\fg} = y_0\,,\\ 
    Z_3=Z_4=0,\quad &|Z_1|^2+|Z_2|^2=1\,,\quad Z_1 = \cos\frac{\theta}{2} e^{i\varphi_1}\,,\quad Z_2 = \sin\frac{\theta}{2} e^{i\varphi_2}\,,
    \eea
    where $(v_0,x_0,y_0)$ are arbitrary fixed values\footnote{Note that we could have used any other regular time coordinate $\tilde{v}$ at $r=0$ and set $\tilde{v} =\tilde{v}_0=\text{const}$. The family of embeddings parametrized by $(v_0,x_0,y_0)$ and $(\tilde{v}_0,x_0,y_0)$ are the same. This would just be a change of collective coordinates.}. The worldvolume is parametrized by $\theta\in (0,\pi)$ and $\varphi_{1,2}\in (0,2\pi)$, where $\mu_1 = \cos\frac{\theta}{2}$ and $\mu_2 = \sin\frac{\theta}{2}$. Note that $C_3$ does not have nonzero components along $(\theta,\varphi_1,\varphi_2)$, so the pullback term in \eqref{eq: M2 bos S} does not contribute. Choosing
    \be
    \sigma^1 = \theta\,,\quad \sigma^2 = \varphi_1\,,\quad \sigma^3 = \varphi_2\,,
    \ee
    and fixing $g=\frac{1}{\sqrt{2}}$ as in Section \ref{subsec:oneCut:bulk}, the induced metric is
    \be
    ds^2_{\text{M}2} = \frac{1}{\Delta^\frac{1}{3}}\left(\frac{1}{L_1}\sin^2\frac{\theta}{2}+\frac{1}{L_2}\cos^2\frac{\theta}{2}\right) d\theta^2 + \frac{4}{\Delta^\frac{1}{3}L_1}\cos^2\frac{\theta}{2}d\varphi_1^2 + \frac{4}{\Delta^\frac{1}{3}L_2}\sin^2\frac{\theta}{2}d\varphi_2^2\,. 
    \ee
    One can then compute 
    \bea\label{eq: M2 action 12 instanton}
    S_{\text{M}2}&=\frac{1}{(2\pi)^2l_p^3}\int d^3\sigma \sqrt{h} = \frac{1}{l_p^3}\int_0^\pi d\theta \frac{2\sin\theta}{L_1L_2} =\frac{4}{l_p^3 L_1L_2} = \frac{4}{l_p^3}\sqrt{\frac{X^3X^4}{X^1X^2}} \\
    &=2\sqrt{2}\pi\sqrt{N}\sqrt{\frac{[\hat{\Delta}_3][\hat{\Delta}_4]}{[\hat{\Delta}_1][\hat{\Delta}_2]}}\,.
    \eea
    We have used \eqref{eq: sec L def} to express $L_I$ in terms of the original sections $X^I$, and identified $2\pi X^I = [\hat{\Delta}_I]$ at the horizon $r=0$ where the branes are located, as in \eqref{eq: iden X w Delta}. In principle, the backgrounds contributing to the index are the Euclidean Wick rotated black saddles constructed in \cite{Bobev:2020pjk}, and not the Lorentzian black holes that we are considering. However, the ``UV-IR relation'' in \cite{Bobev:2020pjk} guarantees that the identification between the sections $X^I$ and $[\hat{\Delta}_I]$ is unchanged. The last step uses the AdS/CFT dictionary at $g=1/\sqrt{2}$, which is $1/l_p^3=\frac{\pi\sqrt{N}}{\sqrt{2}}$. 
    
    The computation goes through identically if we had chosen the M2 worldvolume to wrap the other $S^3$'s defined by $Z^I=Z^J=0$ for different pairs $(I,J)\neq (1,2)$. The result would just be
    \be\label{eq: M2 action gen wrapping}
    S_{\text{M}2}=2\sqrt{2}\pi\sqrt{N}\sqrt{\frac{[\hat{\Delta}_K][\hat{\Delta}_L]}{[\hat{\Delta}_I][\hat{\Delta}_J]}}\,,\quad K,L\neq I,J\,.
    \ee
    Our numerical procedure in Section \ref{subsec:oneCut:numerics} is only able to extract the largest non-perturbative correction. Thus it should only see the contribution of the M2-brane with the smallest action. If the chemical potentials $\{[\hat{\Delta}_I]\}$ are ordered as $\Delta_{\gg} > \Delta_{>} > \Delta_{<} > \Delta_{\ll}$, the smallest M2 brane action is
    \be
    S_{\text{M}2}=2\sqrt{2}\pi\sqrt{N}\sqrt{\frac{\Delta_{\ll}\Delta_{<}}{ \Delta_{>} \Delta_{\gg} }}\,,
    \ee
    which matches \eqref{eq:BH-instanton}. 
    
    In the Wick-rotated black saddles, the euclidean time coordinate is generated by the Killing vector constructed from a bilinear of the Killing spinor. The horizon is a fixed point of this Killing vector, and the time coordinate degenerates there \cite{Bobev:2020pjk}. Thus the position of the wrapped M2-branes in 4d are only specified by their coordinates along $\Sigma_\fg$. But we saw that the M2 action is independent of its position on $\Sigma_\fg$. This means that the $\Sigma_\fg$ position is a zero mode, or collective coordinate of the brane. If we assume that its measure in the path integral is the volume form on $\Sigma_\fg$, integrating over it gives
    \be
    \int_{\Sigma_\fg}d^2x e^{-S_{\text{M}2}} = 2\pi\eta e^{-S_{\text{M}2}}\,,
    \ee
    as noticed in \cite{Gautason:2025per}. This explains the genus dependence in 
    \be
    \sum_I\fn_I e^{-S_{\text{M}2}} = 2(\fg-1) e^{-S_{\text{M}2}}\,,
    \ee
    for the cases where \eqref{eq:BH-instanton} holds. For the exceptional cases where $\hat{f}^{(\rm np)}_{1/2,I}$ varies with $I$ in Table \ref{tbl:InstantonCounterExamples1}, we speculate that they may be explained by performing a full analysis of the quantum fluctuations of the brane around the classical configurations considered above, as in \cite{Beccaria:2023ujc,Gautason:2025per}. Our explanation for the non-perturbative corrections as wrapped M2 branes only makes sense if the wrappings considered in \eqref{eq: wrapped S3 def} are supersymmetric, so that these configurations contribute to the index. In Appendix \ref{App:M2 susy}, we shall present an argument in support of this.
    
	%%%%%%%%%%%%%%%%%%%%%%%%
	\section{Dressed one-cut solutions}
	\label{sec:dressed}
	%%%%%%%%%%%%%%%%%%%%%%%%

\begin{figure}[h!]
        \centering        
        \includegraphics[width=\textwidth]{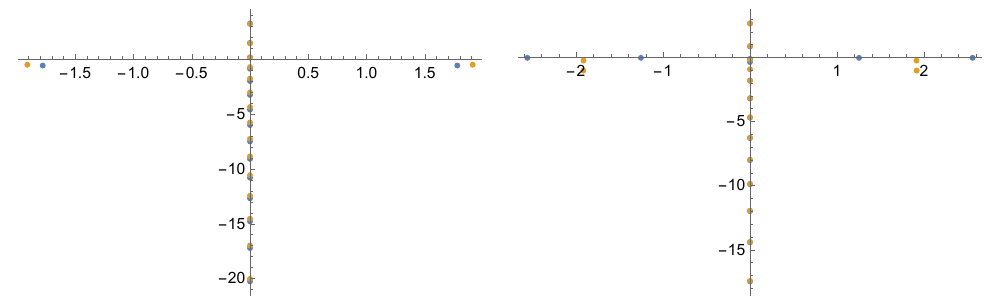}
        \caption{\small The left panel plots $u_a$ in the complex plane for two 1-pair solutions at $N=17$, $\Delta_1=\frac{\pi^2}{4}$, $\Delta_2=e$, $\Delta_C=\frac{\pi}{9}$, one in blue and the other in orange. The blue solution has the pair of outlying points at $\re u= \pm\left(\pi-\frac{\Delta_2}{2}\right)$, while the orange solution has $\re u =\pm\left(\pi-\frac{\Delta_1}{2}\right)$. The right panel shows two 2-pair solutions, also at $N=17$ and the same chemical potentials.}
        \label{fig: npair eg}
    \end{figure}

    In this section, we will study the $n$-pair ``dressed'' one-cut solutions, so called because their distribution of $u_a$ has $N-2n$ points along the imaginary axis, and $n$ pairs of points away from the imaginary axis. In each pair, the points are reflection symmetric about the imaginary axis. As examples, we have shown two $1$-pair and $2$-pair solutions in Figure \ref{fig: npair eg}. 

    A primary feature of these $n$-pair dressed solutions is that their free energy $F_{\fg,\lambda}$ at leading order in $N$ is equal to that of the black hole (i.e. one-cut solutions). This has been observed numerically. Analytically, one can isolate $n$ pairs of $u_a$ and apply the same one-cut ansatz \eqref{eq: BH ansatz} to the remaining variables. Since $n$ does not scale with $N$, it follows straightforwardly that the Bethe equations \eqref{eq: large N W'} at leading order in $N$ are unchanged, and so is the solution for $\rho$. Substituting into $F_{\fg,\lambda}$ in \eqref{eq: log Z} then gives the same result at order $N^\frac{3}{2}$. The free energy of the $n$-pair solutions only begins to differ from the black hole $F_{\fg,\text{BH}}$ at order $N^\frac{1}{2}$. Since the M2 brane tension $\sim 1/l_p^3\sim N^\frac{1}{2}$, this suggests that the bulk dual to the n-pair solutions are configurations of $n$ supersymmetric M2-branes wrapped around a black hole. We will provide evidence for this in the following subsections.
	
	%%%%%%%%%%%%%%%%%%%%%%%%
	\subsection{The one-pair dressed one-cut solutions}
	\label{subsec:dressed:1pair}
	%%%%%%%%%%%%%%%%%%%%%%%%
	
	%%%%%%%%%%%%%%%%%%%%%%%%
	\subsubsection{Regions of existence of the one-pair solutions}
	\label{subsubsec:dressed:1pair:Distribution}
\begin{figure}[!h]
		\centering 
		\includegraphics[width=\textwidth]{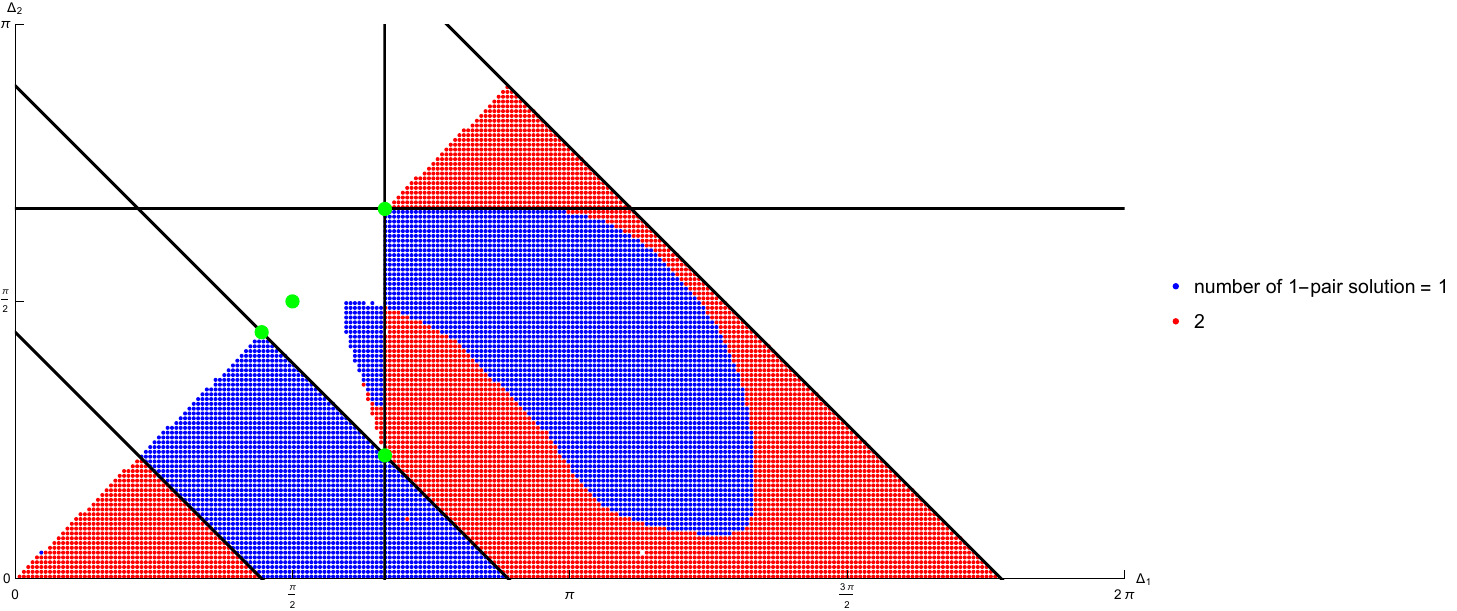}
        \caption{\rm The number of 1-pair solutions for different chemical potentials in the fundamental domain \eqref{eq: fund dom}, computed at $N = 5$ and fixed $\Delta_C = \frac{\pi}{9}$. There are 1 and 2 one-pair solutions at the blue and red dots respectively. The coordinates of the green dots are, from left to right, $(\frac{4\pi}{9}, \frac{4\pi}{9}), (\frac{\pi}{2}, \frac{\pi}{2}), (\frac{2\pi}{3}, \frac{2\pi}{9}) , (\frac{2\pi}{3}, \frac{2\pi}{3})$. }
		\label{fig:1PairRangeN=5}
	\end{figure}

	%%%%%%%%%%%%%%%%%%%%%%%%
    We begin by studying the one-pair dressed solutions. At a given chemical potential, there are at most two one-pair solutions. Let $\{u_1, u_N\}$ be the two outlying gauge holonomies taking values outside the imaginary axis. As $N$ increases, we observe numerically that the real parts of $\{u_1, u_N\}$ approach  
    \begin{equation} \label{eq:onepair-u1uN}
    	\pm \Big( \pi - \frac{\Delta_1}{2}  \Big)\quad {\rm and} \quad \pm \Big( \pi - \frac{\Delta_2}{2} \Big)
    \end{equation}
    for the two solutions respectively. An example is shown in the left panel of Figure \ref{fig: npair eg}. By sampling the fundamental domain of chemical potentials in \eqref{eq: fund dom} at a grid of points, and solving the Bethe equations using our algorithm at each point, we find that the regions where one-pair solutions exist for a fixed $\Delta_C$ are shown in Figure \ref{fig:1PairRangeN=5}. 

    Let us dissect this figure in detail. The blue region is where only one of the one-pair solutions exist, while the region in red is where both solutions exist. The rightmost sloped line is \[ \Delta_1 + \Delta_2 = 2\pi - 2 \Delta_C\, , \] which coincides with the boundary of BH$_+$ in \eqref{eq: BH reg exists}. There are no one-pair solutions to the right of this sloped line, which means that no one-pair solutions exist if the black hole (one-cut) solution does not exist. This is consistent with the proposition that the one-pair solution is a black hole perturbed by the addition of M2 branes. The other two sloped lines dividing the fundamental domain in Figure \ref{fig:1PairRangeN=5} are \[ \Delta_1 + \Delta_2 = \frac{2\pi}{3} \pm 2 \Delta_C \,. \] We observe that the one-pair solutions exist almost everywhere except for the white, almost triangular shaped region in the middle, where the triangle has vertices labeled by $3$ of the $4$ green dots. The remaining green dot on the edge of the triangle is at the ``superconformal'' point $\Delta_1=\Delta_2=\frac{\pi}{2}$\footnote{Presumably, this means that the dual M2 brane configurations cannot exist around the black hole constructed in minimal gauged supergravity.}. There are also curved boundaries between the regions which are difficult to quantify. The figure is plotted for $\Delta_C = \frac{\pi}{9}$.  As $\Delta_C \rightarrow 0$, the two sloped lines on the left coincide, and the white triangle is the largest at this point. As $\Delta_C \rightarrow \frac{\pi}{3}$, the white region vanishes, and the two sloped lines on the right coincide. If $\Delta_C\geq \frac{\pi}{2}$, it turns out that one-pair solutions cease to exist. 

    The regions of existence of two-pair dressed one-cut solutions are much more complicated. At $N = 5$ and a given chemical potential, there can be at most 7 two-pair solutions. Thus we leave an analysis for future work. 
    
	%%%%%%%%%%%%%%%%%%%%%%%%
	\subsubsection{TTI at finite $N$ }
	\label{subsubsec:dressed:1pair:TTIfiniteN}
	%%%%%%%%%%%%%%%%%%%%%%%%
	
	Given that the perturbative $1/N$ corrections to the free energy of the black saddle resums nicely into a compact expression in \eqref{eq:TTI-BH}, it is natural to ask if the same is true for the one-pair dressed solutions. We restrict our attention to the ``left'' and ``right'' regions, defined as subregions of the fundamental domain shown in Figure \ref{fig:1PairRangeN=5} by\footnote{In Figure \ref{fig:1PairRangeN=5}, the right region is the topmost small triangle bounded from below by the horizontal black line $\Delta_2 =\frac{2\pi}{3}$, and the left region is the leftmost small triangle bounded from the right by the sloped black line $\Delta_1+\Delta_2=\frac{2\pi}{3}-2\Delta_C$. This definition is only valid when $\Delta_C < \frac{\pi}{3}$. }
    \begin{equation}
        \begin{aligned}
            {\rm right\ region} &: \quad \hat\Delta_I>0,\qquad \hat\Delta_1,\hat\Delta_2>\frac{2\pi}{3}\,,\\
             {\rm left\ region} &: \quad \hat\Delta_I>0,\qquad \hat\Delta_3,\hat\Delta_4>\frac{2\pi}{3}\,.
        \end{aligned}
    \end{equation}
    The two regions transform into each other under joint symmetry 3 in \eqref{eq:Transformation3}. As seen from Figure \ref{fig:1PairRangeN=5}, there are two 1-pair solutions, which we label by $i=1,2$, at every point in the left and right regions. If we denote the contributions from the two 1-pair solutions to the index as $\cI_{\fg=0,1\text{-pair},i}$, with free energies $F_{\fg=0,1\text{-pair},i} = -\log\cI_{\fg=0,1\text{-pair},i}$, the free energies of both solutions fit the ansatz
    \begin{gather} 
    		F_{0,1\text{-pair},i} = - \frac{\sqrt{2\hat{\Delta}_1\hat{\Delta}_2\hat{\Delta}_3\hat{\Delta}_4}}{3} \sum_{I=1}^4 \frac{\hat{\fn}_I}{\hat{\Delta}_I} \left( (N + \delta^{({\rm 1-pair}, i)})^{\frac{3}{2}} + \fc_I^{({\rm 1-pair}, i)}(N + \delta^{({\rm 1-pair}, i)})^{\frac{1}{2}} \right)\notag\\
            + \frac{1}{2} \log (N + \delta^{({\rm 1-pair}, i)})  + \hat{f}_0 + \hat{f}_{{\rm np},i}\,.\label{eq:TTI-onePair}
    \end{gather}
    As in \eqref{eq:TTI-BH}, $\hat{f}_0\sim\cO(1)$ is $N$-independent while $\hat{f}_{{\rm np},i}$ are exponentially suppressed in $N$. In the right region with generic chemical potentials $\hat{\Delta}_1\neq\hat{\Delta}_2$, the fitted coefficients $\fc_I^{({\rm 1-pair}, i)}$ and $\delta^{({\rm 1-pair}, i)}$ are given by 
    \begin{equation}\label{eq: coeff 1p d12 uneq right}
        \begin{aligned}
            \fc_I^{\rm (1-pair, 1)} &= \fc_I^{\rm (BH)} +  \frac{6\pi}{\hat{\Delta}_1} \delta_{1I} \,, \qquad \delta^{\rm (1-pair,1)} = \delta^{\rm (BH)} - \frac{2\pi }{\hat{\Delta}_1} \,; \\
            \fc_I^{\rm (1-pair, 2)} &= \fc_I^{\rm (BH)} + \frac{6\pi}{\hat{\Delta}_2} \delta_{2I} \,,\qquad 
            \delta^{\rm (1-pair,2)} = \delta^{\rm (BH)} - \frac{2\pi }{\hat{\Delta}_2} \,.
        \end{aligned}
    \end{equation}
    Similarly, for the ``left region'' with $\hat{\Delta}_3 \ne \hat{\Delta}_4$, one has
    \begin{equation}\label{eq: coeff 1p d12 uneq left}
        \begin{aligned}
            \fc_I^{\rm (1-pair, 1)} &= \fc_I^{\rm (BH)} +  \frac{6\pi}{\hat{\Delta}_3} \delta_{3I} \,, \qquad \delta^{\rm (1-pair,1)} = \delta^{\rm (BH)} - \frac{2\pi }{\hat{\Delta}_3} \,; \\
            \fc_I^{\rm (1-pair, 2)} &= \fc_I^{\rm (BH)} + \frac{6\pi}{\hat{\Delta}_4} \delta_{4I} \,,\qquad 
            \delta^{\rm (1-pair,2)} = \delta^{\rm (BH)} - \frac{2\pi }{\hat{\Delta}_4} \,.
        \end{aligned}
    \end{equation}
    The two one-pair solutions are labeled by the real part of the dressing pairs $\{u_1,u_N\}$. The solution with coefficients $\{\fc_I^{\rm (1-pair, 1)}$, $\delta^{\rm (1-pair,1)}\}$ has real parts at $\pm \left( \pi - \frac{\Delta_{1}}{2} \right)$, while the other solution with coefficients $\{\fc_I^{\rm (1-pair, 2)}$, $\delta^{\rm (1-pair,2)}\}$ has real parts $\pm \left( \pi - \frac{\Delta_{2}}{2} \right)$. It is interesting to note that the shifts of the coefficients in \eqref{eq: coeff 1p d12 uneq right} relative to $\fc_I^{(\rm BH)}$ do not depend on $\Delta_C$.
    
    In the special cases of $\hat{\Delta}_1=\hat{\Delta}_2$ in the right region or $\hat{\Delta}_3=\hat{\Delta}_4$ in the left region, the general expressions do not apply, and the two solutions admit the same coefficients $\delta$ and $\fc_I$. In other words, the two 1-pair solutions have the same perturbative free energy in these cases. Explicitly, when $\hat{\Delta}_1 = \hat{\Delta}_2$ in the right region, one has 
	\begin{equation} \label{eq:coeExact1-pair right}
		\begin{aligned}
			\fc_I^{\rm (1-pair)}-\fc_I^{\rm (BH)} &=  \frac{3\pi}{2 \hat{\Delta}_1\hat{\Delta}_2 } \left( \hat{\Delta}_1, \hat{\Delta}_2, -\hat{\Delta}_3, -\hat{\Delta}_4 \right), \qquad \delta^{\rm (1-pair)}-\delta^{\rm (BH)} = - \frac{2\pi }{\hat{\Delta}_1}, \\ 
		\end{aligned}
	\end{equation}
    and when $\hat{\Delta}_3 = \hat{\Delta}_4$ in the left region, one has
    \begin{equation} \label{eq:coeExact1-pair left}
		\begin{aligned}
			\fc_I^{\rm (1-pair)}-\fc_I^{\rm (BH)} &= \frac{3\pi}{2 \hat{\Delta}_3\hat{\Delta}_4 } \left( -\hat{\Delta}_1, -\hat{\Delta}_2,  \hat{\Delta}_3,  \hat{\Delta}_4 \right), \qquad \delta^{\rm (1-pair)}-\delta^{\rm (BH)} = -\frac{2\pi }{\hat{\Delta}_3}.\\[.3cm]
		\end{aligned}
	\end{equation}
    
     We do not aim to obtain the coefficients $\{\fc_I,\delta\}$ for the one-pair dressed solutions across the whole fundamental domain beyond the left and right regions, as the regions of existence in Figure \ref{fig:1PairRangeN=5} are complicated and have non-linear boundaries that we do not understand. 

    Subtracting the perturbative expansion and constant map $\hat{f}_0$ from the free energy allows us to extract the leading non-perturbative correction numerically. For generic chemical potentials, the exponential rate $\hat{f}_{1/2,I}^{(\rm np)}$, defined via \eqref{eq: f0 fnp flux decomp} and \eqref{eq: fnp_I N expansion}, agrees with that of the black-hole solution in \eqref{eq:BH-instanton}:
    \be
    \hat{f}_{{\rm np},i}\sim e^{\hat{f}_{1/2,I}^{(\rm np)} N^\frac{1}{2}}\,.
    \ee
    The non-perturbative corrections for both 1-pair solutions are controlled by the same exponential rate. This can be attributed to the M2-branes wrapping $S^3\subset S^7$, as discussed in Section \ref{subsec:non-pert M2}. For $\hat{\Delta}_1=\hat{\Delta}_2$ in the right region, however, the leading non-perturbative corrections have half of the previous exponential rate $\hat{f}_{1/2,I}^{(\rm np)}$ in \eqref{eq:BH-instanton}, and the overall signs are opposite for the two solutions, i.e. 
    \begin{gather}
    \hat{f}_{{\rm np},1} = \hat{f}_{\rm 1/2-inst.} + \hat{f}_{\rm 1-inst.}\,,\quad \hat{f}_{{\rm np},2} = -\hat{f}_{\rm 1/2-inst.} + \hat{f}_{\rm 1-inst.}\,,\notag\\
    \hat{f}_{\rm 1/2-inst.} \sim e^{\frac{1}{2}\hat{f}_{1/2,I}^{(\rm np)} N^\frac{1}{2}}\,, \quad \hat{f}_{\rm 1-inst.} \sim e^{\hat{f}_{1/2,I}^{(\rm np)} N^\frac{1}{2}}\,.\label{eq: dressed half instanton expansion}
    \end{gather}
    Thus the two 1-pair solutions are distinguished by having different non-perturbative corrections to their free energies, even though the perturbative expansions are the same. Beyond the 1-pair solutions, such ``half-instanton'' corrections are also present for the $n$-pair dressed solutions with isolated points, that we will introduce below, when $\hat{\Delta}_1=\hat{\Delta}_2$. Examples of such solutions can be found in Figure \ref{fig:superposition_eg}. For the so-called non-isolated n-pair dressed solutions, also introduced below and plotted in Figure \ref{fig:noniso_eg_5pi/6}, the exponential rate controlling the leading non-perturbative correction is $1/(n+1)$ of the usual rate $\hat{f}_{1/2,I}^{(\rm np)}$ when $\hat{\Delta}_1=\hat{\Delta}_2$: 
    \be
    \hat{f}_{\rm np} \sim e^{\frac{1}{n+1}\hat{f}_{1/2,I}^{(\rm np)} N^\frac{1}{2}}\,. \label{eq: dressed 1/(n+1) instanton expansion}
    \ee
    Since these reduced exponential rates are only observed at $\hat{\Delta}_1=\hat{\Delta}_2$, we speculate that the enhancement of the isometries of the internal squashed $S^7$ at this special locus must play a role. A better understanding of the mechanism involved is left to future work. 
	
	%%%%%%%%%%%%%%%%%%%%%%%%
	\subsection{$n$-pair dressings and the superposition property}
	\label{subsec:dressed:superposition}
	%%%%%%%%%%%%%%%%%%%%%%%%

\begin{figure}[h!]
		\centering        
		\includegraphics[width=7cm]{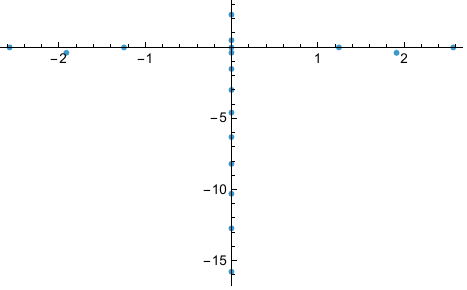} \hspace{0.5cm}
		\includegraphics[width=7cm]{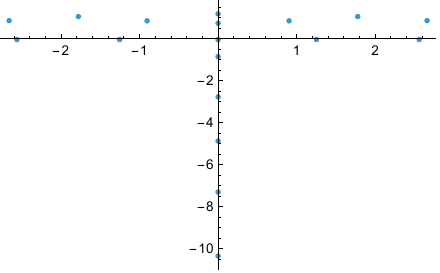}
		\caption{\small Plots of $u_a$ in the complex plane for solutions with isolated points at $N=18$, $\Delta_1=\frac{\pi^2}{4}$, $\Delta_2=e$, $\Delta_C=\frac{\pi}{9}$. }
		\label{fig:isolated_points_eg}
	\end{figure}

\begin{figure}[h!]
		\centering        
		\includegraphics[width=7cm]{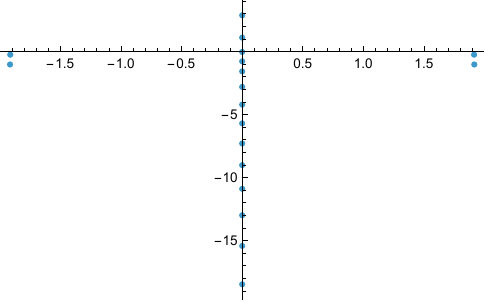} \hspace{0.5cm}
		\includegraphics[width=7cm]{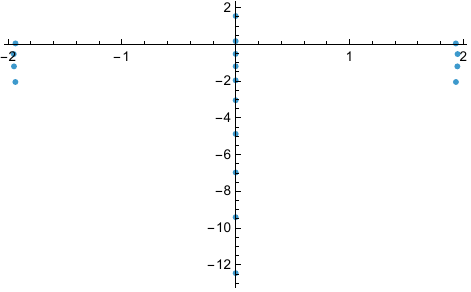}
		\caption{\small Plots of $u_a$ in the complex plane for solutions with non-isolated points at $N=18$, $\Delta_1=\frac{\pi^2}{4}$, $\Delta_2=e$, $\Delta_C=\frac{\pi}{9}$. }
		\label{fig:nonisolated_points_eg}
	\end{figure}

	By looking at the distribution of the $2n$ outlying points (that is, the points outside the imaginary axis) in an $n$-pair dressed solution on the complex plane, we notice that they are divided into two categories: isolated points and non-isolated points. An isolated point maintains a finite separation with other outlying points in the large $N$ limit, and tends to a fixed limit point as $N$ increases. On the other hand, a non-isolated point merges with other outlying points to form a cut in the large $N$ limit, where the separation between the points in the cut decreases as $\sim N^{-\frac{1}{2}}$. Figure \ref{fig:isolated_points_eg} shows examples of dressed solutions with isolated points, while Figure \ref{fig:nonisolated_points_eg} shows examples with non-isolated points. 

    We propose that the dressed solutions with isolated points are dual to configurations of M2-branes wrapped around the black hole, whereas the dressed solutions with non-isolated points will be further discussed in Section \ref{sec:multi-cut:mixing}. The latter interpolate between the black hole solution and its $\bZ_3$ orbifold, and their precise bulk dual is unclear. There could also be solutions with a mixture of isolated points and non-isolated points. A natural expectation is that their bulk dual are configurations of branes wrapped around the interpolating solutions. 
    
    One strong evidence for our proposal that solutions with isolated points correspond to branes wrapped on the black hole is the fact that their free energy satisfies an additive property. Let us examine $\hat{\Delta}_I=(\frac{5\pi}{6},\frac{5\pi}{6},\frac{\pi}{6},\frac{\pi}{6})$ as an example, where there are 6 3-pair solutions, 2 2-pair solutions and 2 1-pair solutions with isolated outlying points. Recall from the discussion around \eqref{eq:TTI-onePair} that when $\hat{\Delta}_1 = \hat{\Delta}_2$, the 2 1-pair solutions have outlying points which tend to the same limit points in the large $N$ limit. Correspondingly, the free energy of both solutions have the same perturbative expansion \eqref{eq:TTI-onePair} with identical coefficients $\{\fc^{(1\text{-pair})}, \delta^{(1\text{-pair})}\}$ given in \eqref{eq:coeExact1-pair right}. The situation is identical for the 2-pair and 3-pair solutions. For all of the 3-pair solutions, the 6 outlying points tend to the same 6 limit points in the large $N$ limit, and the 4 outlying points in the 2-pair solutions tend to the same 4 limit points. Accordingly, each class of solutions share the same perturbative expansion in \eqref{eq:TTI-onePair}, with coefficients $\{\fc^{(2\text{-pair})}, \delta^{(2\text{-pair})}\}$ and $\{\fc^{(3\text{-pair})}, \delta^{(3\text{-pair})}\}$ respectively. These coefficients are listed in Table \ref{tbl:CompositionTable}. Comparing the values with those of the black hole in \eqref{eq:coeExactBH}, we observe that
	\begin{equation}\label{eq: npair coeff diff lin n}
		\begin{aligned}
			\fc_I^{(n-{\rm pair})} = \fc_I^{\rm (BH)} + \Big( \frac{9}{5}n, \frac{9}{5}n, -\frac{9}{25}n, - \frac{9}{25}n \Big),\quad (\delta^{(n-{\rm pair})} + 2n) = \delta^{\rm (BH)} - \frac{2}{5}n\,.
		\end{aligned}
	\end{equation}
	The fact that the difference with the black hole coefficients is linear in $n$ implies that the free energy of these $n$-pair solutions has the form
    \be\label{eq: free E probe approx}
    F_{\fg,n\text{-pair}}(N) = F_{\fg,\text{BH}}(N) + n \delta F + \cO(N^{-\frac{1}{2}})\,,\quad \delta F \sim \cO(N^\frac{1}{2})\,.
    \ee
    This is consistent with the interpretation that the solutions are dual to wrapped branes on the black hole. When $n\ll N$, the probe brane approximation is valid, and one would expect that the $\cO(N^\frac{1}{2})$ correction to the black hole free energy is just $n$ times the DBI action $\delta F$ evaluated on the black hole solution. At order $\cO(N^{-\frac{1}{2}})$ and beyond, the backreaction due to the added branes start to contribute, and one expects that the linearity in $n$ is violated. By expanding \eqref{eq:TTI-onePair} with \eqref{eq: npair coeff diff lin n}, it is straightforward to check that this is indeed the case. 
	
	\renewcommand{\arraystretch}{1.8}
	\begin{table}[!h] \small 
		\centering
		\begin{tabular}{|c|c|c|}
			\hline
			solution & $\fc_I$ & $\delta $   \\ \hline 
			 black hole & $ \Big( -\frac{63}{5}, - \frac{63}{5}, - \frac{99}{25}, - \frac{99}{25} \Big) $ & $\frac{139}{120}$ \\ \hline 
			one-pair & $\Big(-\frac{54}{5} , -\frac{54}{5} , -\frac{108}{25} , -\frac{108}{25} \Big)$ & $ -\frac{149}{120} $\\ \hline 
			two-pair  & $\Big(- 9 , - 9 , -\frac{117}{25} , -\frac{117}{25} \Big)$ & $ -\frac{437}{120}  $\\ \hline 
			three-pair & $\Big(- \frac{36}{5} , - \frac{36}{5} , -\frac{126}{25} ,  -\frac{126}{25} \Big)$ & $ -\frac{145}{24} $\\ \hline 
		\end{tabular}
        \caption{\rm Expansion coefficients for the black hole, one-pair, isolated two-pair, and isolated three-pair dressed solutions, excluding those solutions whose outlying points form part of a cut at large $N$. The chemical potentials are $\hat{\Delta}_I =\Big( \frac{5\pi}{6}, \frac{5\pi}{6},  \frac{\pi}{6},  \frac{\pi}{6}  \Big)$. }
		\label{tbl:CompositionTable}
	\end{table}
	
\begin{figure}[h!]
		\centering        
		\includegraphics[width=7cm]{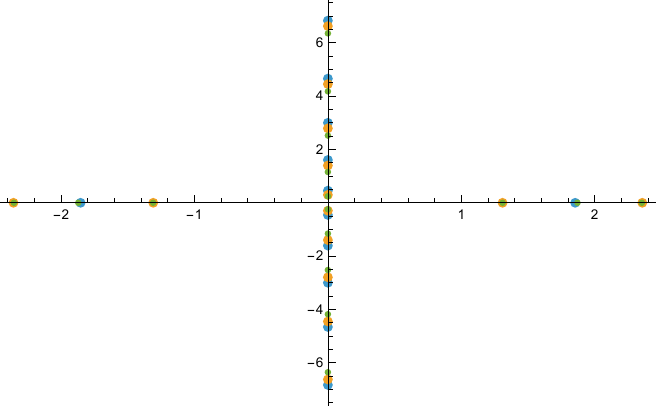} \hspace{0.5cm}
		\includegraphics[width=7cm]{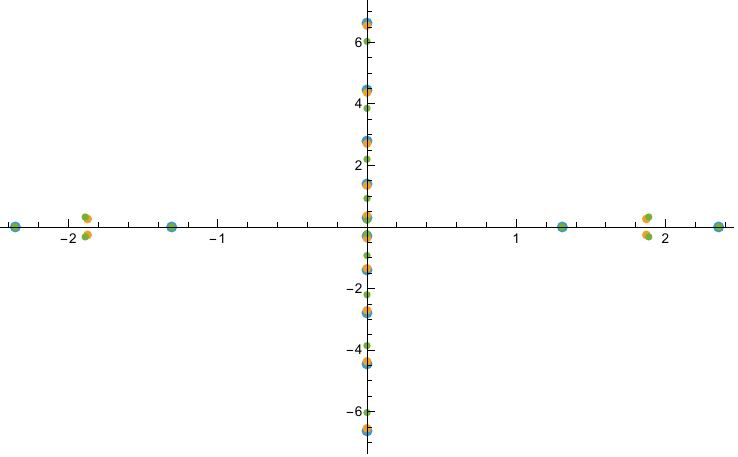}
		\caption{\small Plots of $u_a$ in the complex plane for $n$-pair dressed solutions with isolated points, found at $\Delta_1=\Delta_2=\frac{5\pi}{6}$, $\Delta_C= 0$, and $N = 10 + 2n$. {\it Left panel:} the 3-pair solution in green is a superposition of the 1-pair solution in blue and the 2-pair solution in orange. {\it Right panel:} the 4-pair solution in green is a superposition of two different 2-pair solutions in blue and orange. }
		\label{fig:superposition_eg}
	\end{figure}

	For more general cases where $\Delta_C \ne 0$ and $\Delta_1 \ne \Delta_2$, we expect (and checked in many cases) that the contributions of $n$-pair solutions also admit an expression of the form \eqref{eq:TTI-onePair}, which is specified by coefficients $\{\fc_I,\delta\}$. We thus make the conjecture
	\begin{conjecture}
		(Conjecture of index resummability) The all-order perturbative $1/N$ expansion of the free energy of the dressed black saddle is characterized by two sets of quantities up to constant maps independent of $N$ and exponentially suppressed instanton corrections: the shift of membrane charge $\delta$ and the subleading coefficients $\fc_I$. The leading $\cO(N^{3/2})$ coefficient is identical to that of the black hole.
	\end{conjecture}
    A notable characteristic of the solutions is the superposition of dressings. Taking $\hat{\Delta}_I=(\frac{5\pi}{6},\frac{5\pi}{6},\frac{\pi}{6},\frac{\pi}{6})$ as an example, the limit points of the outlying points in the 3-pair solution are precisely the union of the limit points of the 1-pair and 2-pair solutions, as shown in the left panel of Figure \ref{fig:superposition_eg}. Similarly, the outlying points of a specific 4-pair solution are the union of the outlying points of two different 2-pair solutions, as shown in the right panel of Figure \ref{fig:superposition_eg}. In such cases, we will say that one dressed one-cut solution is a superposition of the other two solutions:
    \begin{definition}
    A dressed solution A with $2n_A$ (not necessarily) isolated outlying points is said to be a superposition of the solution B with $2n_B$ outlying points and solution C with $2n_C$ outlying points if $n_A = n_B + n_C$, and the large $N$ limit of the $2n_A$ outlying points of A is identical to the union of the large $N$ limit of the $2n_B$ outlying points of B and the $2n_C$ outlying points of $C$.  
    \end{definition}
    As a generalization of \eqref{eq: npair coeff diff lin n} to $\Delta_C \ne 0$ and $\Delta_1 \ne \Delta_2$, we propose that whenever one solution is a superposition of two others, their coefficients are related in the following way.
	\begin{conjecture}
		(Conjecture of dressing superposition) If a dressed black saddle $A$ is a superposition of two dressed saddles $B$ and $C$, then
		\begin{equation}
			\begin{aligned}
				&\fc_I^{A} - \fc_I^{\rm (BH)} = (\fc_I^{B} -\fc_I^{\rm (BH)}) + (\fc_I^{C} - \fc_I^{\rm (BH)})\,,\\
				& \delta_A - \delta^{\rm (BH)}= (\delta_B- \delta^{\rm (BH)} ) + (\delta_C - \delta^{\rm (BH)}) \,,\\
			\end{aligned}
		\end{equation}
        where $\{\fc_I^A,\delta_A\}$, $\{\fc_I^B,\delta_B\}$, $\{\fc_I^C,\delta_C\}$ are the expansion coefficients of $A$, $B$, and $C$ respectively. 
	\end{conjecture}

    Note that the conjecture naturally reduces to the additivity of probe M2-brane actions at order $O(N^{1/2})$, but is much stronger as it predicts all perturbative orders in $1/N$ except for the constant term. 
	
\begin{figure}[h!]
		\centering        
		\includegraphics[width=7cm]{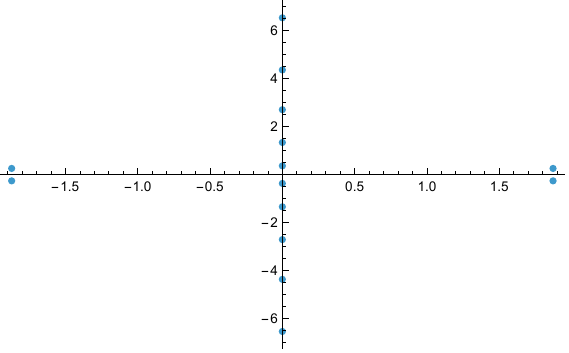} \hspace{0.5cm}
		\includegraphics[width=7cm]{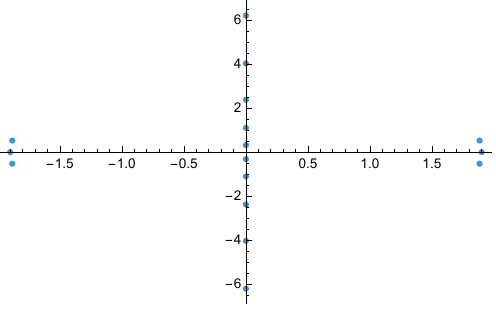}
		\caption{\small Plots of $u_a$ in the complex plane for $n$-pair dressed solutions with non-isolated points, computed at $\Delta_1=\Delta_2=\frac{5\pi}{6}$, $\Delta_C= 0$, and $N = 10 + 2n$. {\it Left panel:} a solution with $n=2$. {\it Right panel:} a solution with $n=3$.}
		\label{fig:noniso_eg_5pi/6}
	\end{figure}

	We contrast the discussion above about the superposition of isolated outlying points by briefly examining the solutions with only non-isolated points. For the same example with $\hat{\Delta}_I=(\frac{5\pi}{6},\frac{5\pi}{6},\frac{\pi}{6},\frac{\pi}{6})$, they form a family of $n$-pair dressed solutions where the outlying points are distributed vertically along $\re u \approx \pm\frac{2\pi}{3}$, as shown in Figure \ref{fig:noniso_eg_5pi/6}. As $N$ increases, the outlying points eventually form two cuts. The solutions with non-isolated outlying points also admit an expansion of the form \eqref{eq:TTI-onePair}, and their coefficients are listed in Table \ref{tbl:CompositionTable2}. We observe that the membrane charge shifts are exactly the same as those in Table \ref{tbl:CompositionTable} for the solutions with isolated points, but the other coefficients $\fc_I$ no longer add up linearly. Thus the free energy of these solutions do not have the form of \eqref{eq: free E probe approx}, and interpreting them as weakly-interacting branes on the black hole background would be inconsistent. As we will see in Section \ref{sec:multi-cut:mixing}, in a specific limit, they should be understood as distinct geometries which interpolate between the black hole and its orbifold.
	
	\renewcommand{\arraystretch}{1.8}
	\begin{table}[!h] \small 
		\centering
		\begin{tabular}{|c|c|c|}
			\hline
			solution & $\fc_I$ & $\delta $   \\ \hline 
			undressed & $ \Big( -\frac{63}{5}, - \frac{63}{5}, - \frac{99}{25}, - \frac{99}{25} \Big) $ & $\frac{139}{120}$ \\ \hline 
			one-pair & $\Big(-\frac{54}{5} , -\frac{54}{5} , -\frac{108}{25} , -\frac{108}{25} \Big)$ & $ -\frac{149}{120} $\\ \hline 
			two-pair (non-isolated) & $\Big(- \frac{39}{5} , -  \frac{39}{5}, -\frac{111}{25} , -\frac{111}{25} \Big)$ & $ -\frac{437}{120}  $\\ \hline 
			three-pair (non-isolated) & $\Big( -\frac{9}{2} , -\frac{9}{2}, -\frac{9}{2}, -\frac{9}{2} \Big)$ & $ -\frac{145}{24} $\\ \hline 
		\end{tabular}
        \caption{\rm Expansion coefficients for the black hole solution, the one-pair solution, and two-pair and three-pair dressed solutions with non-isolated outlying points. The chemical potentials are $\hat{\Delta}_I =\Big( \frac{5\pi}{6}, \frac{5\pi}{6},  \frac{\pi}{6},  \frac{\pi}{6}  \Big)$, as in Table \ref{tbl:CompositionTable}. }
		\label{tbl:CompositionTable2}
	\end{table}

	%%%%%%%%%%%%%%%%%%%%%%%%
	\subsection{Candidate brane configurations dual to isolated points}
	\label{subsec:dressed:DressVSinst}
	%%%%%%%%%%%%%%%%%%%%%%%%
	
	If we accept that the dressed solutions with isolated outlying points are dual to M2-branes wrapped around the black hole, based on the evidence presented in the previous subsection, the next step would be to figure out what specific cycle the M2-brane is wrapping. 

    For the first 1-pair solution in the ``right region'', with $\hat\Delta_1\neq\hat\Delta_2$, the coefficients in \eqref{eq: coeff 1p d12 uneq right} give the following free-energy difference at fixed $N$:
    \be\label{eq: F 1-pair corr}
    F_{\fg,1\text{-pair},1}(N)-F_{\fg,\text{BH}}(N)=\sqrt{N}\sqrt{2\prod_{I=1}^4\hat{\Delta}_I}\left[\frac{\pi}{\hat{\Delta}_1}\sum_{I=1}^4\frac{\hat{\fn}_I}{\hat{\Delta}_I}-\frac{2\pi\hat{\fn}_1}{\hat{\Delta}_1^2}\right] + \cO(N^{0})\,.
    \ee
    Unlike the instanton correction in \eqref{eq:BH-instanton}, the $\cO(N^\frac{1}{2})$ correction above is directly proportional to the magnetic fluxes. This is expected for the free energy, due to \eqref{eq:general-g}. If the correction and its flux dependence were to be reproduced by the action of a brane, a natural choice would be for the brane to wrap $\Sigma_\fg$ at the horizon. Strictly speaking, we should be working in Euclidean signature, where the Euclidean time circle degenerates at the horizon and the geometry caps off. Thus the only possible choice is for the remaining worldvolume direction to wrap an internal $S^1\subset S^7$\footnote{Since the Euclidean time circle $S^1_\tau$ shrinks, any brane wrapping $\Sigma_\fg\times S^1_\tau$ would have zero volume. In the pullback of $C_3$, the first term in \eqref{eq: C3 uplift} has the correct index structure to contribute, but due to $*_4$, it is proportional to the 4d measure $\propto r$ and vanishes at the horizon. Therefore a brane wrapping $\Sigma_\fg\times S^1_\tau$ has zero action and cannot give a finite $\cO(N^\frac{1}{2})$ correction. }. The worldvolume is parametrized by $\sigma^{1,2,3}$, with the embedding
    \bea
    x = \sigma^1\,,\quad y=\sigma^2\,,\quad \varphi_I = c_I\sigma^3\,,\quad \sigma^3 \in (0,2\pi)\,,\quad c_I\in\bZ\,,\quad \mu_I = \text{const.}\,,\quad \sum_{I=1}^4\mu_I^2 = 1\,.
    \eea     
    The $c_I$ must be integers in order for the brane to wrap a genuine circle. We have kept the positions $\mu_I$ generic for now. One can check using \eqref{eq: C3 uplift} that the pullback of $C_3$ does not have the correct index structure to contribute, so we only need to compute the volume term in the brane action. From \eqref{eq: 11d uplifted g G}, the induced metric is
    \bea
    h=\begin{pmatrix}\Delta^\frac{2}{3}e^{2(\psi-U)}g_{xx}+\frac{2}{\Delta^\frac{1}{3}}\sum_I\frac{\mu_I^2(A_I)_x^2}{L_I} & \frac{2}{\Delta^\frac{1}{3}}\sum_I\frac{\mu_I^2(A_I)_x(A_I)_y}{L_I} & \frac{2}{g\Delta^\frac{1}{3}}\sum_I\frac{c_I\mu_I^2(A_I)_x}{L_I} \\ \frac{2}{\Delta^\frac{1}{3}}\sum_I\frac{\mu_I^2(A_I)_x(A_I)_y}{L_I} & \Delta^\frac{2}{3}e^{2(\psi-U)}g_{yy}+\frac{2}{\Delta^\frac{1}{3}}\sum_I\frac{\mu_I^2(A_I)_y^2}{L_I} & \frac{2}{g\Delta^\frac{1}{3}}\sum_I\frac{c_I\mu_I^2(A_I)_y}{L_I} \\ \frac{2}{g\Delta^\frac{1}{3}}\sum_I\frac{c_I\mu_I^2(A_I)_x}{L_I} & \frac{2}{g\Delta^\frac{1}{3}}\sum_I\frac{c_I\mu_I^2(A_I)_y}{L_I} & \frac{2}{g^2\Delta^\frac{1}{3}}\sum_I\frac{c_I^2\mu_I^2}{L_I}\end{pmatrix}\,,
    \eea
	where $g_{xx}$, $g_{yy}$ are the metric components of the unit Riemann surface $\Sigma_\fg$ given in the third line of \eqref{eq: bh metric ansatz}. We shall focus on $\fg = 0$ for now. Simplifying using \eqref{eq: sec L def} and 
    \be
    e^{2(\psi(0)-U(0))}=\sqrt{\prod_{I=1}^4X^I}\sum_{I=1}^4\frac{\hat{\fn}_I}{X^I}\,,\quad A_I = \frac{\hat{\fn}_I(xdy-ydx)}{g(1+x^2+y^2)}\,,
    \ee
    we obtain
    \bea
    \sqrt{\det h} &= \frac{8\sqrt{\prod_IX^I}}{(1+x^2+y^2)^2}\left[\left(\sum_I\frac{\hat{\fn}_I}{X^I}\right)^2\sum_J\mu_J^2X^J\sum_K\frac{c_K^2\mu_K^2}{X^K}\right.\\
    &\qquad \left.+(x^2+y^2)\sum_I\frac{\hat{\fn}_I}{X^I}\left(\sum_J\frac{\mu_J^2\hat{\fn}_J^2}{X^J}\sum_K\frac{\mu_K^2c_K^2}{X^K}-\left(\sum_J\frac{c_J\mu_J^2\hat{\fn}_J}{X^J}\right)^2\right)\right]^\frac{1}{2}\,.
    \eea
    The M2-brane action is then given by
    \be
    S_{\text{M}2}=\frac{1}{(2\pi)^2 l_p^3}\int d^3\sigma\sqrt{\det h} = \left\{ \begin{aligned}
    		& 2\pi\sqrt{N}\sqrt{2\prod_IX^I}\left(\sqrt{a}+\frac{b\sec^{-1}\sqrt{\frac{b}{a}}}{\sqrt{b-a}}\right),\quad b>a,\\
    		& 2\pi\sqrt{N}\sqrt{2\prod_IX^I}\left(  \sqrt a+ \frac{b\, {\rm tanh}^{-1} \sqrt{\frac{a-b}{a}}}{\sqrt{a-b}} \right),\quad b<a,\\
    	\end{aligned} \right.
    \ee
    where we have used $1/l_p^3 = \pi\sqrt{N}/\sqrt{2}$ and
    \bea
        &a\equiv \left(\sum_I\frac{\hat{\fn}_I}{X^I}\right)^2\sum_J\mu_J^2X^J\sum_K\frac{c_K^2\mu_K^2}{X^K}\,,\quad b\equiv \sum_I\frac{\hat{\fn}_I}{X^I}\left(\sum_J\frac{\mu_J^2\hat{\fn}_J^2}{X^J}\sum_K\frac{\mu_K^2c_K^2}{X^K}-\left(\sum_J\frac{c_J\mu_J^2\hat{\fn}_J}{X^J}\right)^2\right)\,.
    \eea
    
    If we choose $c_I = \mu_I= (1,0,0,0)$, then $b=0$. Taking the continuous $b\to0^+$ limit of the $b<a$ expression, the action simplifies to
    \be
    S_{\text{M}2}=\sqrt{N}\sqrt{2\prod_{I=1}^4\hat{\Delta}_I} \sum_I \frac{\hat{\fn}_I}{\hat{\Delta}_I}\,,
    \ee
    after identifying $2\pi X^I=\hat{\Delta}_I$ as in \eqref{eq: iden X w Delta}. The other choices $c_I=\mu_I=\delta_{I2},\delta_{I3},\delta_{I4}$ give the same result. We need to keep in mind that since the brane is at the horizon, the sections $X^I$ must take their horizon attractor values. The extremization of \eqref{eq: attractor Z} under the constraint $\sum_I X^I=1$, and its degree-one homogeneity imply that
    \[
    \frac{2\pi\hat{\fn}_I}{\hat{\Delta}_I^2}=\left(\frac{\pi}{\hat{\Delta}_I}-1\right)\sum_{J=1}^4\frac{\hat{\fn}_J}{\hat{\Delta}_J}\,,\qquad I=1,\ldots,4\,.
    \]
    Substituting this relation into \eqref{eq: F 1-pair corr}, we obtain
    \be\label{eq: eucl DGG match}
    F_{0,1\text{-pair},1}(N)-F_{0,\text{BH}}(N) \overset{!}{=} S_{\text{M}2}+\cO(N^0)\,.
    \ee
    Thus the wrapped membrane action reproduces the full $\cO(N^{1/2})$ free-energy difference at fixed $N$. We explicitly check in Appendix \ref{App:M2 susy} that these candidate wrappings are supersymmetric.

    These wrappings are the AdS$_4$ analogue of the supersymmetric D3-branes wrapping $(S^3\subset\mathrm{AdS}_5)\times (S^1\subset S^5)$ at the horizon of uplifted AdS$_5$ Kerr-Newman black holes, constructed in Appendix C.3.3 of \cite{Aharony:2021zkr}. In a loose sense, they are the Euclidean analogues of the Lorentzian dual giant gravitons considered in \cite{Grisaru:2000zn,Basu:2006id,Aharony:2021zkr}. The spatial worldvolume directions of the Lorentzian dual giant graviton wraps $\Sigma_\fg$ at an arbitrary fixed radius $r_0$, but as time advances, the brane moves at a BPS velocity along an angular direction in $S^7$. Naively, if we analytically continue this configuration to Euclidean signature and take $r_0\rightarrow 0$, we recover the wrappings considered in this subsection, because the Euclidean time direction degenerates at $r=0$. However, unlike the Lorentzian dual giant gravitons, which are domain walls in the Freund-Rubin flux $N$ such that the region $r < r_0$ has flux $N-1$ while the exterior region $r >r_0$ has flux $N$\footnote{We thank Francesco Benini for clarifying this to us.}, the Euclidean analogues considered here do not alter this flux. An obvious reason why they cannot be domain walls is that they have codimension 2 in the AdS$_4$ directions, unlike the Lorentzian dual giants, which have codimension 1. More precisely, the Freund-Rubin flux is computed by an integral of the Page current over the internal $S^7$ fibres, but the source current due to the M2-brane worldvolume wrapping $\Sigma_\fg\times (S^1\subset S^7)$ integrates to zero over the internal $S^7$. This is why it makes sense that the match in \eqref{eq: eucl DGG match} is made with the difference with respect to a black hole with flux $N$, and not $N-1$. 

    \subsection{Bulk interpretation of the Bethe ansatz formula}
    \label{subsec:inst_vs_dress}

    By comparing how the instanton corrections \eqref{eq:BH-instanton} and the $\cO(N^\frac{1}{2})$ corrections from the dressed solutions enter the index, one can make a simple observation about how the Bethe ansatz formula organizes saddles in the gravitational path integral. Namely, it automatically sums instanton corrections due to M2 branes wrapping $S^3\subset S^7$ into the contribution of a single Bethe root, but corrections due to M2 branes wrapping $\Sigma_\fg \times S^1$ or small changes in geometry are treated as distinct Bethe roots. 

    As we saw, the free energy of the $n$-pair dressed solutions is the same as that of the black hole solution to leading order in $N$, but starts to differ at $\cO(N^\frac{1}{2})$. If we accept that the dressed solutions with isolated outlying points are dual to M2-branes wrapping $\Sigma_\fg \times S^1$ in the black hole, and dressed solutions with non-isolated points are dual to a family of geometries interpolating between a black hole and its orbifold, we see that these gravity configurations are treated as distinct Bethe roots. Their contribution to the index has the form
    \be
    \cI_{\fg, n\text{-pair}}=e^{-F_{\fg,\text{BH}}+N^\frac{1}{2} \delta F + \cO(N^0)}\,,
    \ee
    where $\delta F$ is a function of chemical potentials and fluxes. In the case of dressed solutions with isolated outlying points, $N^\frac{1}{2} \delta F$ is provided by the action of a M2 brane on $\Sigma_\fg\times S^1$, where $S^1\subset S^7$. We contrast this with how the instanton action $\sqrt{N}\hat{f}^{(\text{np})}_{1/2,I}$ given in \eqref{eq:BH-instanton}, which is the action of a M2-brane on $S^3\subset S^7$, contributes to the index:
    \be\label{eq: bh contrib inst form}
    \cI_{\fg,\text{BH}}=e^{-F_{\fg,\text{BH},\text{pert.}} + \sum_I\hat{\fn}_Ie^{\sqrt{N}\hat{f}^{(\text{np})}_{1/2,I}+\ldots}} = e^{-F_{\fg,\text{BH},\text{pert.}}}\left(1+\sum_I\hat{\fn}_Ie^{\sqrt{N}\hat{f}^{(\text{np})}_{1/2,I}+\ldots}+\cO\left(e^{2\sqrt{N}\hat{f}^{(\text{np})}_{1/2,I}}\right)\right)\,.
    \ee 
    Here, $F_{\fg,\text{BH},\text{pert.}}$ denotes the full perturbative $1/N$ expansion of the black hole free energy.  Note that the instanton action is exponentiated as $e^{\sqrt{N}\hat{f}^{(\text{np})}_{1/2,I}}$ in the free energy. Since $\sqrt{N}\hat{f}^{(\text{np})}_{1/2,I}<0$, this is exponentially small, and the black hole contribution to the index can be expanded as the sum of the perturbative series, the 1-instanton sector, and higher instanton sectors of order $\cO(e^{2\sqrt{N}\hat{f}^{(\text{np})}_{1/2,I}})$ and so on in \eqref{eq: bh contrib inst form}. We clearly see that the contribution to the index of a single Bethe root---the one-cut solution dual to the black hole---already contains the sum over all corrections due to M2 branes wrapping $S^3\subset S^7$. They are not treated as separate Bethe roots. 

    This is in some sense a physically obvious and natural result. The M2-brane wrapping $S^3\subset S^7$ has no dynamical 4d degrees of freedom from the point of view of M-theory dimensionally reduced on $S^7$. It only has the constant collective coordinates parametrizing its position on $\Sigma_\fg$. By contrast, the M2-brane wrapping $\Sigma_\fg\times S^1$, with $S^1\subset S^7$, has dynamical embedding modes which describe how its shape fluctuates in the directions normal to $\Sigma_\fg$ in the 4d geometry. Since the Bethe ansatz formula follows from localizing the 3d CFT dual to M theory reduced on $S^7$, it is natural that configurations which cannot be differentiated by looking at the dynamical degrees of freedom in 4d, such as those with additional M2 branes on $S^3\subset S^7$, are automatically summed into a single Bethe root, while configurations which do differ in 4d are treated as separate Bethe roots. 
	
	%%%%%%%%%%%%%%%%%%%%%%%%
	\section{Multi-cut solutions}
	\label{sec:multi-cut}
	%%%%%%%%%%%%%%%%%%%%%%%%

   Besides the one-cut solution and dressed one-cut solutions which have already been considered, a broad class of solutions with the required large $N$ scaling $F_{\fg,\lambda}\sim N^\frac{3}{2}\sim 1/G_N$ are multi-cut solutions in which the gauge holonomies $u_a$ are distributed equally across $m$ cuts in the complex plane. In particular, the 2-cut and 3-cut solutions will be studied in Sections \ref{subsec:multi-cut:2cut} and \ref{subsec:multi-cut:3cut} respectively, where we will identify the regions of chemical potentials where they exist, find their corresponding large $N$ densities, and compute their free energy. Later, in Section \ref{subsec:multi-cut:orb grav}, it is shown that the free energies of the 2-cut and 3-cut solutions at leading $\cO(N^\frac{3}{2})$ order exactly match the on-shell actions of the $\bZ_2$ and $\bZ_3$ orbifolds of the black saddle solutions considered by \cite{Bobev:2020pjk}. In addition, the BPS condition on the orbifolds exactly restricts their existence to the regions of chemical potentials found in Sections \ref{subsec:multi-cut:2cut} and \ref{subsec:multi-cut:3cut}. Moreover, the coefficients of $\log N$ in the free energies can also be reproduced by a bulk computation, once additional degrees of freedom on certain fixed loci are taken into account. These calculations show that the 2-cut and 3-cut Bethe solutions are exactly dual to the $\bZ_2$ and $\bZ_3$ orbifolds of Euclidean black saddles. In general, it is reasonable to expect that the $m$-cut Bethe roots are dual to $\bZ_m$ orbifolds.

	%%%%%%%%%%%%%%%%%%%%%%%%
	\subsection{Two-cut solutions}
	\label{subsec:multi-cut:2cut}
	%%%%%%%%%%%%%%%%%%%%%%%%

\begin{figure}[!h]
    	\centering 
        \includegraphics[width=\textwidth]{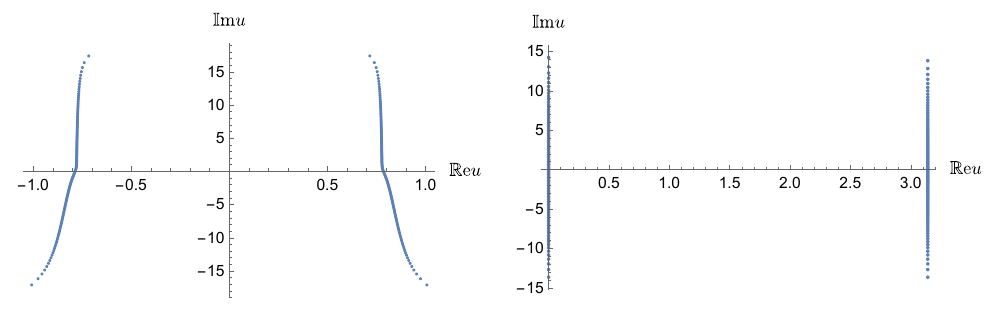}
    	\caption{\rm Examples of the two types of 2-cut solutions. The left panel plots the positions of $u_a$ for a 	solution of the first type with cuts that are reflection symmetric about the imaginary axis. Its parameters are $N=	300$, $\Delta_C=\frac{\pi}{2}$, $\Delta_1=\Delta_2=\frac{\pi^2}{4}$. The right panel plots a solution of the second 	type with one cut along the imaginary axis and another at $\re u_a = \pi$. The corresponding parameters are $N=200$, $\Delta_C=\frac{\pi}{2}$, $\Delta_1=\Delta_2=\frac{3\pi}{4}$.}
    	\label{fig: 2cut eg}
    \end{figure}

    Numerically, we observe that there are two types of 2-cut solutions: the first has cuts which are symmetric under reflections about the imaginary axis, as shown in the left panel of Figure \ref{fig: 2cut eg}, while the second type has one cut along the imaginary axis and the other cut at $\re u = \pi$, as shown in the right panel of the same figure.

    By sampling the fundamental domain \eqref{eq: fund dom} at various points, we find that the 2-cut solutions only exist in the subregions $\tilde{1}$, $\tilde{2}$, $\tilde{3}$, and $\tilde{3}'$ as defined in the second column of Table \ref{tab: 2col reg}. There is a single 2-cut solution at every point in regions $\tilde{1}$, $\tilde{3}$, while there are two distinct 2-cut solutions at every point in $\tilde{2}$ and $\tilde{3}'$. The type of 2-cut solution that is present in each region is written in the third column, where ``symm.'' stands for the first type with reflection symmetric cuts, and ``$0$-$\pi$'' stands for the second type with cuts at $\re u = 0,\pi$. It turns out that the free energy of the 2-cut solution is most conveniently expressed in terms of $[2\hat{\Delta}_I]$ in regions $\tilde{1}$, $\tilde{3}$, and $\tilde{3}'$, while it is naturally written in terms of $[2\hat{\Delta}_I]'$ in region $\tilde{2}$. Here, $[\cdot]$ and $[\cdot]'$ are the functions defined in \eqref{eq: bracket gp G def} and \eqref{eq: bracket prime def} respectively. In the fourth column of Table \ref{tab: 2col reg}, we also keep track of the integer shifts $s_I$ in $[2\hat{\Delta}_I]$ and $[2\hat{\Delta}_I]'$, where
    \be
    s_I\equiv\begin{cases}\left\lfloor\frac{\hat{\Delta}_I}{\pi}\right\rfloor\,, & \textrm{regions }\tilde{1}\,,\tilde{3}\,,\tilde{3}' \\[.2cm] \left\lceil\frac{\hat{\Delta}_I}{\pi}\right\rceil\,, & \text{region }\tilde{2}\end{cases}\,.
    \ee  
    As we will see in Section \ref{subsec:multi-cut:orb grav}, these integers specify the $\bZ_2$ action that is used to construct the dual orbifolds in supergravity.

    \begin{table}[h]
        \centering
        \centerline{
        \begin{tabularx}{0.9\textwidth}{|c|
        >{\centering\arraybackslash\hsize=2.2\hsize\linewidth=\hsize}X|>{\centering\arraybackslash\hsize=0.3\hsize\linewidth=\hsize}X|>{\centering\arraybackslash\hsize=0.5\hsize\linewidth=\hsize}X|}
        \hline
        Reg. & Definition & Type & $s_I$ \\
        \hline
        $\tilde{1}$ & $0<\Delta_C<\pi\,,\;\;$ $0<\Delta_2<\min(\Delta_C,\pi-\Delta_C)\,,$ \par $\Delta_2<\Delta_1<\min(2\Delta_C,2\pi-2\Delta_C)-\Delta_2$ & symm. & $\{0,0,1,0\}$ \\
        \hline
        $\tilde{2}$ & $0<\Delta_C<\pi\,,\;\;$ $\max(\Delta_C,\pi-\Delta_C)<\Delta_1<\pi\,,$ \par $ \max(2\Delta_C,2\pi-2\Delta_C)-\Delta_1<\Delta_2<\Delta_1$ & $0$-$\pi$ & $\{1,1,1,0\}$ \\
        \hline
        $\tilde{3}$ & $0<\Delta_C<\frac{\pi}{2}\,,\;\;$ $\pi<\Delta_1<2\pi-2\Delta_C\,,\;\;$ \par $0<\Delta_2<2\pi-2\Delta_C-\Delta_1$  & symm. & $\{1,0,0,0\}$ \\
        \hline
        $\tilde{3}'$ & $\frac{\pi}{2}<\Delta_C<\pi\,,\;\;$ $\pi<\Delta_1<2\Delta_C\,,\;\;$ \par $0<\Delta_2<2\Delta_C-\Delta_1$  & symm. & $\{1,0,1,-1\}$ \\
        \hline
        \end{tabularx}
        }
        \caption{Regions where 2-cut solutions exist, together with the type of solution in each region (column 3) and the integers $s_I$ specifying the orbifold action (column 4).}
        \label{tab: 2col reg}
    \end{table}

    In the following, Sections \ref{subsubsec:multi-cut:2cut:largeN} and \ref{subsubsec:multi-cut:2cut:0pi cont} will analytically study the 2-cut solutions with reflection symmetric cuts and cuts at $\re = 0,\pi$ respectively. The analysis will show that imposing necessary positivity conditions on the solutions exactly restricts their regions of existence to those listed in Table \ref{tab: 2col reg}. Section \ref{subsubsec:multi-cut:2cut:contrib index} then computes the free energies of the 2-cut solutions in each region at leading $\cO(N^\frac{3}{2})$ order. It is already apparent there that the most natural variables are $[2\hat{\Delta}_I]$ in regions $\tilde{1}$, $\tilde{3}$, $\tilde{3}'$, and $[2\hat{\Delta}_I]'$ in region $\tilde{2}$. Finally, in Section \ref{subsubsec:multi-cut:2cut:numerics}, we will attempt to numerically compute the free energy at the ``superconformal'' points $[2\hat{\Delta}_I]=\frac{\pi}{2}$ and $[2\hat{\Delta}_I]'=-\frac{\pi}{2}$. 
    
	%%%%%%%%%%%%%%%%%%%%%%%%
	\subsubsection{Large $N$ ansatz with reflection symmetric cuts}
	\label{subsubsec:multi-cut:2cut:largeN}
	%%%%%%%%%%%%%%%%%%%%%%%%

    Motivated by numerical solutions where the two cuts are reflection symmetric about the $\im u$ axis, as in the left panel of Figure \ref{fig: 2cut eg}, we divide the original $N$ gauge holonomies into two groups of $M\equiv N/2$ each: $\{u_{a=1,\ldots,M}\}$ and $\{v_{a=1,\ldots, M}\}$, and consider the ansatz
    \be\label{eq: large N 2 col ansatz}
    u_a=iM^\frac{1}{2}t_a + f(t_a)\,, \quad v_a = iM^\frac{1}{2}t_a - f(t_a)\,,\quad a=1,\ldots,M\,.
    \ee
    Defining 
    \be
    \rho(t) = \frac{1}{M}\frac{da}{dt}\,,
    \ee
    we suppose that $\rho$ tends to a piecewise smooth function as $N\rightarrow\infty$. In the large $N$ limit, the Bethe equations \eqref{eq: W'} become
    \bea\label{eq: large N W' 2 col}
    &(M^\frac{1}{2}-if'(t))\rho(t)\sum_{I=1}^3\left[g_+'([\Delta_I+2f(t)])-g_+'([\Delta_I-2f(t)])\right]\\
    &-i\rho'(t)\sum_{I=1}^3\left[2g_+([\Delta_I])+g_+([\Delta_I+2f(t)])+g_+([\Delta_I-2f(t)])\right]\\
    &-i\pi\left[1+\sum_{I=1}^3\left(\left\lfloor\frac{\Delta_I+2f(t)}{2\pi}\right\rfloor+\left\lfloor\frac{\Delta_I-2f(t)}{2\pi}\right\rfloor\right)\right]-i\Delta_C-i\sgn t\frac{\Delta_3}{2}\\
    &-2\pi i\sgn t\left\lfloor\frac{\frac{\Delta_1+\Delta_2}{2}-\sgn(t)f(t)}{2\pi}\right\rfloor=0\,.
    \eea
    The steps taken to derive this can be found in Appendix \ref{App: large N Bethe 2cut}. It is worth noting that unlike the analysis of the one-cut ansatz, it is important here to account for $\cO(1)$ corrections when approximating sums by integrals, using the Euler-Maclaurin formula. One can find an exposition in Appendix \ref{App: E-M formula}. The real part of \eqref{eq: large N W' 2 col} is
    \be\label{eq: large N W' 2 col re}
    M^\frac{1}{2}\rho(t)\sum_{I=1}^3\left[g_+'([\Delta_I+2f(t)])-g_+'([\Delta_I-2f(t)])\right]=0\,,
    \ee
    while the imaginary part is
    \bea\label{eq: large N W' 2 col im}
    &2\rho'(t)G(\Delta,f)+\Delta_C+\sgn t\frac{\Delta_3}{2}+\pi\sgn(t)\wt n(t)=0\,,\\
    &G(\Delta,f)\equiv \frac{1}{2}\sum_{I=1}^3\left[2g_+([\Delta_I])+g_+([\Delta_I+2f(t)])+g_+([\Delta_I-2f(t)])\right]\,,\\
    &\wt n(t)\equiv 2\left\lfloor\frac{\frac{\Delta_1+\Delta_2}{2}-\sgn(t)f(t)}{2\pi}\right\rfloor+\sgn t\left[1+\sum_{I=1}^3\left(\left\lfloor\frac{\Delta_I+2f(t)}{2\pi}\right\rfloor+\left\lfloor\frac{\Delta_I-2f(t)}{2\pi}\right\rfloor\right)\right]\,,
    \eea
    where \eqref{eq: large N W' 2 col re} has been used to remove the term proportional to $f'(t)$.
    
    We would like to impose \eqref{eq: large N W' 2 col re} first. Note that the bracket functions $[\cdot]$ in the arguments of $g_+'$ are the same ones defined in \eqref{eq: bracket gp G def}, and contain the integers
    \be\label{eq: 2 col int def}
    m_{+,I}\equiv \left\lfloor\frac{\Delta_I+2f(t)}{2\pi}\right\rfloor\,,\quad m_{-,I}\equiv \left\lfloor\frac{\Delta_I-2f(t)}{2\pi}\right\rfloor\,. 
    \ee
    As observed numerically, we assume that $m_{\pm,I}$ are piecewise constant functions of $t$, with one possible discontinuity at $t=0$. In other words, $f(t)$ only varies slightly with $t$ such that $m_{\pm,I}$ takes the same constant value for $t>0$, and a possibly different value for $t<0$. Consequently, $\tilde{n}(t)$ in \eqref{eq: large N W' 2 col im} is also piecewise constant, with an allowed discontinuity at $t=0$. We shall denote its values as $\tilde{n}_+\equiv \tilde{n}(t>0)$ and $\tilde{n}_-\equiv \tilde{n}(t<0)$. In the fundamental domain \eqref{eq: fund dom}, and given that $f\in(0,\pi)$, the possible values of $m_{\pm,I}$ are
    \be\label{eq: 2 col int ranges}
    m_{+,I}=0,1\,,  \qquad m_{-,I}=-1,0\,.
    \ee
    Since $g_+'$ is a quadratic polynomial and \eqref{eq: large N W' 2 col re} is odd in $f$, the dependence on $f$ is at most linear. Thus if $f$ does not cancel out completely, imposing \eqref{eq: large N W' 2 col re} will fix $f$ in terms of $\Delta_I$ uniquely. In order to find the allowed values of $m_{\pm,I}$, we performed a sweep across all possible values in \eqref{eq: 2 col int ranges}. For each value of $m_{\pm,I}$, it is directly kept if \eqref{eq: large N W' 2 col re} holds for any $f$ and there exists $\Delta_{1,2},\;f$ such that \eqref{eq: 2 col int def} is satisfied. Otherwise, if \eqref{eq: large N W' 2 col re} has non-trivial $f$ dependence, we solve for $f$. Then if there exists $\Delta_{1,2}$ such that \eqref{eq: 2 col int def} is consistent with the determined value of $f$, this value of $m_{\pm,I}$ is kept; otherwise it is discarded\footnote{We allow edge cases where for example $m_{+,3}=0$ is deemed to be consistent with $f=\frac{2\pi-\Delta_3}{2}$, even though $\Delta_3+2f=2\pi$. This is to allow for the fact that in the actual numerical solution, $f<\frac{2\pi-\Delta_3}{2}$ but approaches $\frac{2\pi-\Delta_3}{2}$ at large $N$. }. The allowed integers $\{m_{+,I},m_{-,I}\}$, their corresponding values of $f$, and the regions in which they are consistent are listed in Table \ref{tab: 2 col regions}. For later reference, we have also included the corresponding values of $\{\tilde{n}_+,\tilde{n}_-\}$.
    
    \begin{table}[h]
    \centering
    \footnotesize
    \begin{tabularx}{\textwidth}{|c|
    >{\centering\arraybackslash\hsize=1.8\hsize\linewidth=\hsize}X|>{\centering\arraybackslash\hsize=1.1\hsize\linewidth=\hsize}X|>{\centering\arraybackslash\hsize=0.5\hsize\linewidth=\hsize}X|>{\centering\arraybackslash\hsize=0.6\hsize\linewidth=\hsize}X|}
        \hline
         & Region & $\{m_{+,I},m_{-,I}\}$ & $f$ & $\{\wt n_+,\wt n_-\}$ \\[.2em]
        \hline
        1 & $\quad 0<\Delta_2<\frac{\pi}{2},\quad  \Delta_2<\Delta_1<\pi-\Delta_2$ & ${\{0,0,1,-1,-1,0\}}$ & ${\text{free}}$ & ${\{-2,0\}}$ \\
         & & ${\{0,0,0,-1,-1,0\}}$ & ${\quad\frac{2\pi-\Delta_3}{2}\quad}$ & ${\{-1,1\}}$ \\
         & & ${\{0,0,1,-1,-1,-1\}}$ & ${\frac{\Delta_3}{2}}$ & ${\{-3,1\}}$\\[.2em]\hline
        2 & $\quad \frac{\pi}{2}<\Delta_1<\pi,\quad \pi-\Delta_1<\Delta_2<\Delta_1$ & ${\{0,0,0,-1,-1,-1\}}$ & ${\frac{\pi}{2}}$ & ${\{-2,2\}}$ \\[.2em] \hline
        3 & $\quad \pi<\Delta_1<2\pi,\quad 0<\Delta_2<2\pi-\Delta_1$ & ${\{1,0,0,0,-1,-1\}}$ & ${\text{free}}$ & ${\{0,0\}}$ \\
         & & ${\{0,0,0,0,-1,-1\}}$ & ${\frac{2\pi-\Delta_1}{2}}$ & ${\{-1,1\}}$ \\
         & & $\{1,0,0,-1,-1,-1\}$ & ${\frac{\Delta_1}{2}}$ & ${\{-1,1\}}$\\[.2em] \hline
        % 4 & $\quad 0<\Delta_1<\pi,\quad -\Delta_1<\Delta_2<0$ & ${\{0,0,1,0,-1,0\},\;f=\frac{\Delta_1}{2}}$ ${\{1,0,1,-1,-1,0\},\;f=\frac{2\pi-\Delta_1}{2}}$ ${\{0,0,1,-1,-1,0\},\;f\;\text{undetermined}}$ \\[.2em] \hline
        % 5 & ${\pi<\Delta_1<\frac{3\pi}{2},\quad \Delta_1-2\pi<\Delta_2<\pi-\Delta_1}$ & ${\{1,0,1,0,-1,0\},\;f=\frac{\pi}{2}}$ \\[.2em] \hline
        % 6 & ${-\frac{\pi}{2}<\Delta_2<0,\quad \pi-\Delta_2<\Delta_1<\Delta_2+2\pi}$ & ${\{1,0,0,0,-1,0\},\;f=\frac{\Delta_3}{2}}$ ${\{1,0,1,0,-1,-1\},\;f=\frac{2\pi-\Delta_3}{2}}$ ${\{1,0,0,0,-1,-1\},\;f\;\text{undetermined}}$ \\[.2em] \hline
    \end{tabularx}
    \normalsize
    \caption{\small Allowed integers $\{m_{+,I},m_{-,I}\}$, the corresponding values of $f$, and the regions in which they are consistent. $\wt n_\pm$ is the integer $\tilde{n}$ defined in \eqref{eq: large N W' 2 col im}, evaluated at $t>0$ and $t<0$.}
    \label{tab: 2 col regions}
    \end{table}

\begin{figure}[h]
    \centering        
    \begin{tikzpicture}[x=1.6cm, y=1.6cm]
        \def\PI{3.14159}

        % region 1
        \fill[blue!30] (0, 0) -- (\PI, 0) -- (\PI/2, \PI/2) -- cycle;
        \draw[blue!75!black, thick] (0, 0) -- (\PI, 0) -- (\PI/2, \PI/2) -- cycle;
        \node[blue!75!black] at (\PI/2, \PI/4-0.18) {$1$};

        % region 1 tilde
        \fill[red!30] (0, 0) -- (2*\PI/9, 0) -- (\PI/9, \PI/9) -- cycle;
        \draw[red!60, thick] (0, 0) -- (2*\PI/9, 0) -- (\PI/9, \PI/9) -- cycle;
        \node[red!75!black] at (\PI/9, \PI/18-0.02) {$\tilde{1}$};

        % region 2
        \fill[orange!30] (\PI, 0) -- (\PI, \PI) -- (\PI/2, \PI/2) -- cycle;
        \draw[orange!75!black, thick] (\PI, 0) -- (\PI, \PI) -- (\PI/2, \PI/2) -- cycle;
        \node[orange!75!black] at (3*\PI/4+0.2, \PI/2) {$2$};

        % region 2 tilde
        \fill[teal!30] (8*\PI/9, 8*\PI/9) -- (\PI, 7*\PI/9) -- (\PI, \PI) -- cycle;
        \draw[teal!75!black, thick] (8*\PI/9, 8*\PI/9) -- (\PI, 7*\PI/9) -- (\PI, \PI) -- cycle;
        \node[teal!75!black] at (17*\PI/18+0.05, 8*\PI/9) {$\tilde{2}$};

        % region 3
        \fill[green!50] (\PI, 0) -- (2*\PI, 0) -- (\PI, \PI) -- cycle;
        \draw[green!80!black, thick] (\PI, 0) -- (2*\PI, 0) -- (\PI, \PI) -- cycle;
        \node[green!60!black] at (3*\PI/2-0.12, \PI/2-0.2) {$3$};

        % region 3 tilde
        \fill[purple!30] (\PI, 0) -- (16*\PI/9, 0) -- (\PI, 7*\PI/9) -- cycle;
        \draw[purple!60,thick] (\PI, 0) -- (16*\PI/9, 0) -- (\PI, 7*\PI/9) -- cycle;
        \node[purple!75!black] at (4*\PI/3-0.25, \PI/3-0.25) {$\tilde{3}$};

        % axes
        \draw[->] (0,0) -- (2.2*\PI,0) node[right] {$\Delta_1$};
        \draw[->] (0,0) -- (0,1.2*\PI) node[above] {$\Delta_2$};
        
        % ticks
        \draw (\PI,-0.06)--(\PI,0.06) node[below=5pt] {$\pi$};
        \draw (2*\PI,-0.06)--(2*\PI,0.06) node[below=4pt] {$2\pi$};
        \draw (-0.06,\PI)--(0.06,\PI) node[left=3pt] {$\pi$};
        
    \end{tikzpicture}
    \caption{\small Plots of the regions listed in Table \ref{tab: 2 col regions}, as well as the regions $\tilde{1}$, $\tilde{2}$, $\tilde{3}$ in which 2-cut solutions exist. We have used an arbitrary value $\Delta_C=\frac{\pi}{9}$ for illustration. }
    \label{fig: 2 col regions}
    \end{figure}
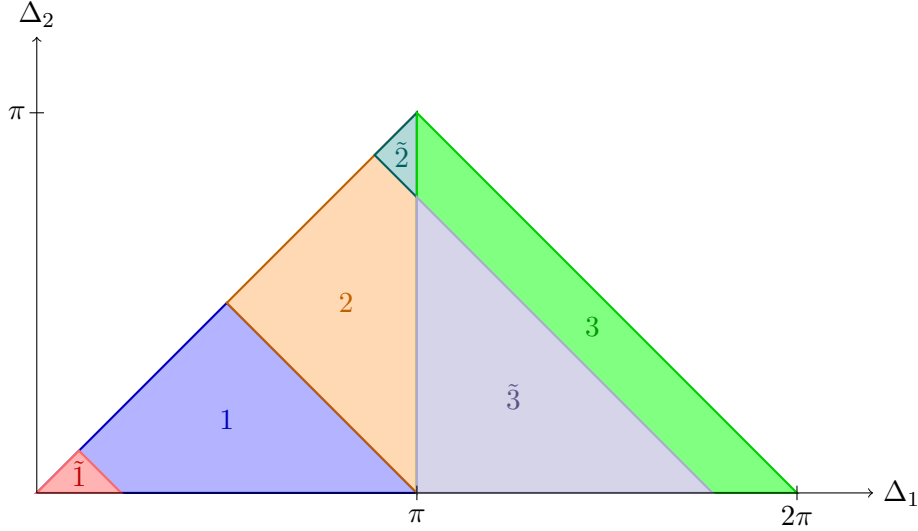

    As a visual aid, we plot the regions $1,2,3$ of Table \ref{tab: 2 col regions} in Figure \ref{fig: 2 col regions}, as well as the subregions $\tilde{1}$, $\tilde{2}$, $\tilde{3}$ where 2-cut solutions exist when $\Delta_C = \frac{\pi}{9}$.

    Now, solving \eqref{eq: large N W' 2 col im} fixes
    \be\label{eq: rho gen 2-col}
    \rho(t)=\frac{\mu-\left[\frac{\Delta_3}{2}+\sgn t\Delta_C+\pi\wt n(t)\right]|t|}{2G}\,,
    \ee
    up to a constant $\mu$ to be determined. Like in Section \ref{subsec:oneCut:largeN}, we assume that $\rho$ is a piecewise linear function that is positive and has finite support between two zeros. Therefore its gradient must be positive for $t<0$ and negative for $t>0$. We shall examine this condition in each region of Table \ref{tab: 2 col regions} to see if a solution exists. 
    
    In region 1, $G=\Delta_1\Delta_2(\Delta_3-\pi)>0$ when evaluated on the determined values of $f$, regardless of the choice of integers $\{m_{+,I},m_{-,I}\}$. One can check that this is true generally: the expression for $G$ only depends on the region of chemical potentials in Table \ref{tab: 2 col regions}, and not on the choices of integers that are consistent within each region. For $t>0$, we need $\frac{\Delta_3}{2}+\Delta_C+\pi\tilde{n}_+>0$. Considering the $3$ choices of integers, one finds that only $\{0,0,0,-1,-1,0\}$ and $\tilde{n}_+=-1$ is consistent with being in region 1. For $t<0$ on the other hand, $\frac{\Delta_3}{2}-\Delta_C+\pi\tilde{n}_->0$ with all $3$ choices of $\tilde{n}_-$ are consistent with being in region 1. Assuming that $f(t)$ is continuous, it cannot jump from $f=\frac{2\pi-\Delta_3}{2}$ to $f=\frac{\Delta_3}{2}$ at $t=0$, so we can choose either $\{0,0,0,-1,-1,0\}$ or $\{0,0,1,-1,-1,0\}$ for $t<0$. Numerically, only solutions of the latter type are found. It is perhaps possible to explain this fact by solving the Bethe equations at subleading order in $N$, but here we shall be content with describing the solutions that we do find. The constraints $\frac{\Delta_3}{2}\pm\Delta_C+\pi\tilde{n}_\pm>0$ for $\tilde{n}_+ = -1$ and $\tilde{n}_-=0$ restrict region $1$ such that solutions only exist in the subregion $\tilde{1}$ defined in Table \ref{tab: 2col reg}. Normalizing $\rho$ fixes $\mu$, and the density is given by 
    \bea\label{eq: 2col reg1 rho sol}
    \rho &= \begin{cases}
        \frac{\mu-(\frac{\Delta_3}{2}+\Delta_C-\pi)t}{2\Delta_1\Delta_2(\Delta_3-\pi)}\,, & t > 0\,, \\
         \frac{\mu+(\frac{\Delta_3}{2}-\Delta_C)t}{2\Delta_1\Delta_2(\Delta_3-\pi)}\,, & t < 0\,,
    \end{cases}\quad t\in \left(-\frac{\mu}{\frac{\Delta_3}{2}-\Delta_C}\,,\frac{\mu}{\frac{\Delta_3}{2}+\Delta_C-\pi}\right)\,,\\
    \mu &= \sqrt{\Delta_1\Delta_2(\Delta_3+2\Delta_C-2\pi)(\Delta_3-2\Delta_C)}\,.
    \eea
    
    In region 2, $G(\Delta,\frac{\pi}{2})=(\pi-\Delta_1)(\pi-\Delta_2)(\Delta_3-\pi)<0$ and $\wt n_\pm=\mp 2$, so the positivity of $\rho$ requires $\frac{\Delta_3}{2}\pm(\Delta_C-2\pi)<0$. For the lower sign, this requires $\Delta_1+\Delta_2>6\pi-2\Delta_C$, which is impossible in region 2, so there are no solutions.
    
    In region 3, $G=(\Delta_1-\pi)\Delta_2\Delta_3>0$, and we need $\frac{\Delta_3}{2}\pm\Delta_C+\pi\wt n_\pm>0$. Considering $0<\Delta_C<\frac{\pi}{2}$ first, only $\{1,0,0,0,-1,-1\}$ with $\tilde{n}_+ = 0$ is consistent with region 3 for $t>0$, while all $3$ choices of integers are consistent for $t<0$. In principle, we could pick a different set of integers for $t<0$, but the only solutions that were found numerically in region 3 have the same set of integers for both signs of $t$, and we shall restrict to this case in the following. The conditions $\frac{\Delta_3}{2}\pm\Delta_C>0$ imply that solutions only exist in the subregion $\tilde{3}$ defined in Table \ref{tab: 2col reg}. As before, normalizing $\rho$ fixes $\mu$, and the solution is
    \bea\label{eq: 2col reg3 rho sol}
    \rho &= \frac{\mu-(\frac{\Delta_3}{2}+\sgn t\Delta_C)|t|}{2(\Delta_1-\pi)\Delta_2\Delta_3}\,,\quad t \in \left(-\frac{\mu}{\frac{\Delta_3}{2}-\Delta_C},\frac{\mu}{\frac{\Delta_3}{2}+\Delta_C}\right)\,,\\
    \mu &= \sqrt{(\Delta_1-\pi)\Delta_2(\Delta_3+2\Delta_C)(\Delta_3-2\Delta_C)}\,.
    \eea
    
    If we are in region 3 with $\frac{\pi}{2}<\Delta_C<\pi$, $\{1,0,0,0,-1,-1\}$ is no longer consistent for $t<0$, while the other $2$ choices of integers are consistent for both signs of $t$. Thus we expect $2$ solutions in this range of $\Delta_C$; one has $f(t)\approx \frac{\Delta_1}{2}$, while the other has $f\approx\frac{2\pi-\Delta_1}{2}$. The conditions $\frac{\Delta_3}{2}\pm(\Delta_C-\pi)>0$ imply that the subregion of region 3 where the solutions exist is $\tilde{3}'$ in Table \ref{tab: 2col reg}. The normalized density in this case is
    \bea\label{eq: 2col reg3p rho sol}
    \rho &= \frac{\mu-(\frac{\Delta_3}{2}+\sgn t(\Delta_C-\pi))|t|}{2(\Delta_1-\pi)\Delta_2\Delta_3}\,,\quad t\in\left(-\frac{\mu}{\frac{\Delta_3}{2}-\Delta_C+\pi}\,,\frac{\mu}{\frac{\Delta_3}{2}+\Delta_C-\pi}\right)\,, \\
    \mu&=\sqrt{(\Delta_1-\pi)\Delta_2(\Delta_3+2\Delta_C-2\pi)(\Delta_3-2\Delta_C+2\pi)}\,.
    \eea
    One can check that the solutions \eqref{eq: 2col reg1 rho sol}, \eqref{eq: 2col reg3 rho sol}, \eqref{eq: 2col reg3p rho sol} in regions $\tilde{1}$, $\tilde{3}$, $\tilde{3}'$ can all be uniformly written as
    \bea\label{eq: 2cut reg 1 3 sol uniform}
    \rho &= \begin{cases}\frac{\sqrt{[2\hat{\Delta}_1][2\hat{\Delta}_2][2\hat{\Delta}_3][2\hat{\Delta}_4]}-[2\hat{\Delta}_3]t}{\frac{1}{2}[2\hat{\Delta}_1][2\hat{\Delta}_2]([2\hat{\Delta}_3]+[2\hat{\Delta}_4])}\,, & t>0 \\[.2cm] \frac{\sqrt{[2\hat{\Delta}_1][2\hat{\Delta}_2][2\hat{\Delta}_3][2\hat{\Delta}_4]}+[2\hat{\Delta}_4]t}{\frac{1}{2}[2\hat{\Delta}_1][2\hat{\Delta}_2]([2\hat{\Delta}_3]+[2\hat{\Delta}_4])}\,, & t<0\end{cases}\\[.2cm]
    t&\in\left(-\frac{\sqrt{[2\hat{\Delta}_1][2\hat{\Delta}_2][2\hat{\Delta}_3][2\hat{\Delta}_4]}}{[2\hat{\Delta}_4]}\,,\frac{\sqrt{[2\hat{\Delta}_1][2\hat{\Delta}_2][2\hat{\Delta}_3][2\hat{\Delta}_4]}}{[2\hat{\Delta}_3]}\right)\,.
    \eea
    Note that $s_I=\left\lfloor\frac{\hat{\Delta}_I}{\pi}\right\rfloor= \{0,0,1,0\}$, $\{1,0,0,0\}$, $\{1,0,1,-1\}$ in regions $\tilde{1}$, $\tilde{3}$, $\tilde{3}'$ respectively. These integers coincide with those listed for regions $\tilde{1}$, $\tilde{3}$, $\tilde{3}'$ in Table \ref{tab: 2col reg}.

    %%%%%%%%%%%%%%%%%%%%%%%%
    \subsubsection{Large $N$ ansatz with cuts at $\re u =0,\pi$}
    \label{subsubsec:multi-cut:2cut:0pi cont} 
    %%%%%%%%%%%%%%%%%%%%%%%%
    
    Motivated by the second type of 2-cut solutions in the right panel of Figure \ref{fig: 2cut eg}, we also consider the ansatz
    \be\label{eq: 2 col ansatz 0 pi}
    u_a = iM^\frac{1}{2}t_a\,,\quad v_b = iM^\frac{1}{2}\tilde{t}_b + \pi\,, \quad a,b=1,\ldots,M\,,
    \ee
    where again we have split the gauge holonomies into two groups $\{u_a\}$, $\{v_b\}$ of $M=N/2$ each. We order the variables such that $\{t_a\}$ and $\{\tilde{t}_b\}$ are increasing sequences in $a$ and $b$. As observed numerically, we assume that $t_a\neq \tilde{t}_b$ for all $a,b=1,\ldots, M$ and both $t$ and $\tilde{t}$ are identically distributed in the large $N$ limit. That is,
    \be
    \rho(t) = \frac{1}{M}\frac{da}{dt}\,,\quad \rho(\tilde{t}) = \frac{1}{M}\frac{db}{d\tilde{t}}\,.
    \ee
    This implies that $a(t)=b(t)+C$ where $C$ is piecewise constant, and therefore $\left\lceil b(t_a)\right\rceil=b(t_a)+r$, where $r\in (0,1)$ is piecewise constant in $t$. We allow $r$ to have a discontinuity at $t = 0$, and denote its values for $t>0$ and $t<0$ as $r_+$ and $r_-$ respectively. As shown in Appendix \ref{App: large N Bethe 2cut 0pi}, the Bethe equations at leading order in $N$ can be simplified to
    \bea\label{eq: 2cut large N W' 0 pi}
    &2\rho'(t)\sum_{I=1}^3\left[g_+([\Delta_I])+g_+([\Delta_I+\pi])\right]+\Delta_C+\sgn(t)\frac{\Delta_3}{2}+2\pi\sgn(t)\left\lfloor\frac{\Delta_1+\Delta_2}{4\pi}\right\rfloor\\
    &\hspace{3cm}+2\pi(2r-1)\left(1-\sum_{I=1}^3\left\lfloor\frac{\Delta_I+\pi}{2\pi}\right\rfloor\right)-2\pi=0\,, \\
    &\sgn(t)\left(\left\lfloor\frac{\Delta_1+\Delta_2}{4\pi}\right\rfloor - \left\lfloor\frac{\frac{\Delta_1+\Delta_2}{2}-\sgn(t)\pi}{2\pi}\right\rfloor\right)+2(2r-1)\left(1-\sum_{I=1}^3\left\lfloor\frac{\Delta_I+\pi}{2\pi}\right\rfloor\right)-2=0\,,
    \eea
    where $r$ enters the equations via the $\cO(1)$ corrections in the Euler-Maclaurin formula when approximating sums by integrals. The first equation in \eqref{eq: 2cut large N W' 0 pi} determines $\rho$ to be
    \bea
    &\rho = \frac{\mu-(\frac{\Delta_3}{2}+\sgn(t)\Delta_C+\pi\tilde{n})|t|}{2G}\,,\quad G \equiv \sum_{I=1}^3\left[g_+([\Delta_I])+g_+([\Delta_I+\pi])\right]\,,\\
    &{\tilde{n}\equiv 2\left\lfloor\frac{\Delta_1+\Delta_2}{4\pi}\right\rfloor+\sgn(t)\left[2(2r-1)\left(1-\sum_{I=1}^3\left\lfloor\frac{\Delta_I+\pi}{2\pi}\right\rfloor\right)-2\right]\,,}
    \eea
    where $\mu$ is a constant to be determined. Like $r$, $\tilde{n}$ is also piecewise constant, with a possible discontinuity at $t=0$. We denote its values for $t>0$ and $t<0$ by $\tilde{n}_+$ and $\tilde{n}_-$ respectively.
    
    Recall that the fundamental domain \eqref{eq: fund dom} can be divided into regions $1$, $2$, and $3$ in Table \ref{tab: 2 col regions}. In regions $1$ and $3$, one can check that $\lfloor\frac{\Delta_1+\Delta_2}{4\pi}\rfloor = 0$, $\left\lfloor\frac{\frac{\Delta_1+\Delta_2}{2}-\sgn(t)\pi}{2\pi}\right\rfloor=-\Theta(t)$, and $1-\sum_{I=1}^3\lfloor\frac{\Delta_I+\pi}{2\pi}\rfloor=0$. This implies that the second equation of \eqref{eq: 2cut large N W' 0 pi} is independent of $r$ and cannot be satisfied. Thus no solutions of this type exist in regions 1 and 3. 
    
\begin{figure}[h!]
        \centering        
        \includegraphics[width=.7\textwidth]{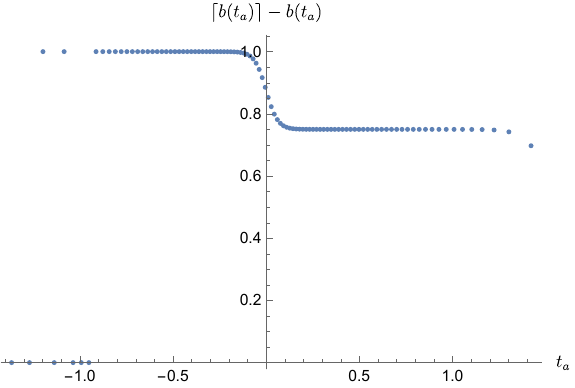}
        \caption{\small Plot of $\lceil b(t_a)\rceil - b(t_a)$ against $t_a$ for the 2-cut solution at $\Delta_1=\Delta_2=\frac{3\pi}{4}$, $\Delta_C=\frac{\pi}{2}$, $N=200$. }
        \label{fig: 2cut 0-pi r}
    \end{figure}

    In region 2 however, the second equation of \eqref{eq: 2cut large N W' 0 pi} becomes $2(2r_+-1)-1=0$ for $t>0$ and $2(2r_--1)-2=0$ for $t<0$, which are solved by $r_+ = \frac{3}{4}$ and $r_-=1$. In region $2$, $G=(\Delta_1-\pi)(\Delta_2-\pi)(\Delta_3-\pi) < 0$ and $\tilde{n}_+=-1$, $\tilde{n}_-=0$. For $\rho$ to have support between two zeros and be positive, we therefore need $\frac{\Delta_3}{2}+\Delta_C-\pi < 0$ and $\frac{\Delta_3}{2}-\Delta_C<0$. This implies that solutions only exist in subregion $\tilde{2}$, defined in Table \ref{tab: 2col reg}. After determining $\mu$ by normalizing $\rho$, we get
    \bea\label{eq: rho 2 col 0 pi}
    &\rho(t) = \begin{cases} \frac{\mu-\left(\frac{\Delta_3}{2}+\Delta_C-\pi\right)t}{2(\Delta_1-\pi)(\Delta_2-\pi)(\Delta_3-\pi)}\,, & t>0 \\[0.2cm] \frac{\mu+\left(\frac{\Delta_3}{2}-\Delta_C\right)t}{2(\Delta_1-\pi)(\Delta_2-\pi)(\Delta_3-\pi)}\,, & t<0\end{cases}\quad t\in\left(-\frac{\mu}{\frac{\Delta_3}{2}-\Delta_C},\frac{\mu}{\frac{\Delta_3}{2}+\Delta_C-\pi}\right)\,,\\[0.3cm]
    &\mu = -2\sqrt{(\Delta_1-\pi)(\Delta_2-\pi)\left(\frac{\Delta_3}{2}+\Delta_C-\pi\right)\left(\frac{\Delta_3}{2}-\Delta_C\right)}\,.
    \eea
    Our determination that $r_+ = \frac{3}{4}$ and $r_-=1$ is indeed borne out by the numerical solutions, as shown in Figure \ref{fig: 2cut 0-pi r}. Given a numerical solution with $\tilde{t}_b = \im v_b/M^\frac{1}{2}$, an interpolating function $b(\tilde{t})$ is created from the data $\{\tilde{t}_b,b\}$, which can then be used to evaluate $\lceil b(t_a)\rceil - b(t_a)$. 
    
    Note that due to the inequalities satisfied by region $\tilde{2}$ in Table \ref{tab: 2col reg}, the density can also be written as
    \bea\label{eq: 2cut reg 2t sol uniform}
    &{\rho(t) = \begin{cases}\frac{-\sqrt{[2\hat{\Delta}_1]'[2\hat{\Delta}_2]'[2\hat{\Delta}_3]'[2\hat{\Delta}_4]'}-[2\hat{\Delta}_3]'t}{\frac{1}{2}[2\hat{\Delta}_1]'[2\hat{\Delta}_2]'([2\hat{\Delta}_3]'+[2\hat{\Delta}_4]')}\,, & t > 0 \\[0.2cm] \frac{-\sqrt{[2\hat{\Delta}_1]'[2\hat{\Delta}_2]'[2\hat{\Delta}_3]'[2\hat{\Delta}_4]'}+[2\hat{\Delta}_4]'t}{\frac{1}{2}[2\hat{\Delta}_1]'[2\hat{\Delta}_2]'([2\hat{\Delta}_3]'+[2\hat{\Delta}_4]')}\,, & t < 0\end{cases}}\\[0.2cm] &t\in\left(\frac{\sqrt{[2\hat{\Delta}_1]'[2\hat{\Delta}_2]'[2\hat{\Delta}_3]'[2\hat{\Delta}_4]'}}{[2\hat{\Delta}_4]'}\,,-\frac{\sqrt{[2\hat{\Delta}_1]'[2\hat{\Delta}_2]'[2\hat{\Delta}_3]'[2\hat{\Delta}_4]'}}{[2\hat{\Delta}_3]'}\right)\,,
    \eea
    where the functions $[\cdot]'$ were defined in \eqref{eq: logZ BH branch 2}, and $s_I=\left\lceil\frac{\hat{\Delta}_I}{\pi}\right\rceil=\{1,1,1,0\}$ are exactly the integers associated with region $\tilde{2}$ in Table \ref{tab: 2col reg}.

    %%%%%%%%%%%%%%%%%%%%%%%%
    \subsubsection{Leading contribution of 2-cut solutions to the index}
    \label{subsubsec:multi-cut:2cut:contrib index}
    %%%%%%%%%%%%%%%%%%%%%%%%

    Substituting the reflection symmetric ansatz \eqref{eq: large N 2 col ansatz} into \eqref{eq: log Z} and simplifying to leading order in $N$ gives
    \begin{gather}
    -\frac{F_{\fg,2-\text{cut},+}}{M^\frac{3}{2}}=\int dt\rho(t)^2\bigg\{\sum_{I=1}^3(\fn_I+1-\fg)\left[g_+'([\Delta_I+2f(t)])+g_+'([\Delta_I-2f(t)])+2g_+'([\Delta_I])\right]\notag\\
    -2(1-\fg)[g_+'([2f(t)])+g_+'(0)]\bigg\}+\fn_3\int dt\rho(t)|t|-2\fn_C\int dt\rho(t)t+\cO\left(M^{-\frac{1}{2}}\right)\,.
    \end{gather}
    Further substituting the integers $\{m_{+,I},m_{-,I}\}$ and $f$ in regions 1 and 3, together with the solution \eqref{eq: 2cut reg 1 3 sol uniform} gives
    \be\label{eq: 2cut LO logZ plus}
    -\frac{F_{\fg,2-\text{cut},+}}{N^\frac{3}{2}}=\frac{1}{2\cdot 3}\sqrt{2[2\hat{\Delta}_1][2\hat{\Delta}_2][2\hat{\Delta}_3][2\hat{\Delta}_4]}\sum_{I=1}^4\frac{\hat{\fn}_I}{[2\hat{\Delta}_I]}+\cO\left(N^{-\frac{1}{2}}\right)\,.
    \ee
    On the other hand, if we substitute the ansatz \eqref{eq: 2 col ansatz 0 pi} with real parts at $0$ and $\pi$ into \eqref{eq: log Z}, the result to leading order in $N$ is
    \begin{gather}
    -\frac{F_{\fg,2-\text{cut},-}}{M^\frac{3}{2}} = 2\left\{\sum_{I=1}^3(\fn_I+1-\fg)[g_+'([\Delta_I])+g_+'([\Delta_I+\pi])]- (1-\fg)[g_+'(0)+g_+'(\pi)]\right\}\int dt\rho(t)^2\notag\\
    +\fn_3\int dt\rho(t)|t|-2\fn_C\int dt\rho(t)t+\cO\left(M^{-\frac{1}{2}}\right)\,.
    \end{gather}
    Further substituting the solution \eqref{eq: 2cut reg 2t sol uniform} gives
    \be\label{eq: 2cut LO logZ minus}
    -\frac{F_{\fg,2-\text{cut},-}}{N^\frac{3}{2}}=-\frac{1}{2\cdot 3}\sqrt{2[2\hat{\Delta}_1]'[2\hat{\Delta}_2]'[2\hat{\Delta}_3]'[2\hat{\Delta}_4]'}\sum_{I=1}^4\frac{\hat{\fn}_I}{[2\hat{\Delta}_I]'}+\cO\left(N^{-\frac{1}{2}}\right)\,.
    \ee
    As we shall see in Section \ref{subsec:multi-cut:orb grav}, \eqref{eq: 2cut LO logZ plus} and \eqref{eq: 2cut LO logZ minus} exactly match the on-shell action of the $\bZ_2$ orbifolds of black saddles in the $+$ and $-$ branches respectively.
	
	%%%%%%%%%%%%%%%%%%%%%%%%
	\subsubsection{Numerics at the superconformal points}
	\label{subsubsec:multi-cut:2cut:numerics}
	%%%%%%%%%%%%%%%%%%%%%%%%

    Much like Section \ref{subsec:oneCut:numerics}, we want to use the densities $\rho(t)$ found in Sections \ref{subsubsec:multi-cut:2cut:largeN} and \ref{subsubsec:multi-cut:2cut:0pi cont} to generate 2-cut solutions for a range of large $N$, and fit the corresponding free energies to an ansatz. To be as general as possible, this should be done for each of the regions $\tilde{1}$, $\tilde{2}$, $\tilde{3}$, and $\tilde{3}'$. However, it turns out that joint symmetry 3 in \eqref{eq:Transformation3} exchanges regions $\tilde{1}$ and $\tilde{3}$, while the combination of joint symmetry 1 in \eqref{eq: joint symm 1} and joint symmetry 3 exchanges regions $\tilde{2}$ and $\tilde{3}'$\footnote{Note that joint symmetries 1 and 3 commute.}. Therefore it is sufficient to consider regions $\tilde{1}$ and $\tilde{2}$. For simplicity, we will focus on the ``superconformal'' points in the two regions, where $[2\hat{\Delta}_I]=\frac{\pi}{2}$ in region $\tilde{1}$ and $[2\hat{\Delta}_I]'=-\frac{\pi}{2}$ in region $\tilde{2}$. These are the points $(\Delta_1,\Delta_2,\Delta_C)=(\frac{\pi}{4},\frac{\pi}{4},\frac{\pi}{2})$ and $(\frac{3\pi}{4},\frac{3\pi}{4},\frac{\pi}{2})$ respectively.
	
	At the superconformal point in region $\tilde{2}$, there are two distinct 2-cut solutions with the same distribution $\rho(t)$ of $t_a= \im u_a/M^\frac{1}{2}$ and therefore the same leading order free energies. At subleading orders in $N$, the free energies can be fitted to the ansatz 
    \begin{gather}
    	F_{0,\textrm{2-cut},\tilde{2}}  = \frac{\sqrt{2[2\hat{\Delta}_1]'[2\hat{\Delta}_2]'[2\hat{\Delta}_3]'[2\hat{\Delta}_4]'}}{6} \sum_{I=1}^4 \frac{\hat{\fn}_I}{[2\hat{\Delta}_I]'} \Big[ (N + \delta^{\rm (2-cut)})^{\frac{3}{2}} \notag\\
    	+ \fc_I^{\rm (2-cut)} (N + \delta^{\rm (2-cut)})^{\frac{1}{2}} \Big] + \frac{1}{2} \log (N + \delta^{\rm (2-cut)})  + \hat{f}_0 + \hat{f}_{\rm np}\,,\label{eq:TTI-twocut}
    \end{gather}
    where $\hat{f}_0$ is constant in $N$, and $\hat{f}_{\rm np}$ is exponentially small in $N$. The coefficients $\fc_I^{\rm (2-cut)}$ and $\delta^{\rm (2-cut)}$ for the two solutions are listed in Table \ref{tbl:2colRegion4SC}. In the fourth and fifth columns of Table \ref{tbl:2colRegion4SC}, $\hat{f}^{\rm (np)}_{1/2,I}$ and $\hat{f}^{\rm (np)}_{{\rm log},I}$ are still defined via \eqref{eq: f0 fnp flux decomp} and \eqref{eq: fnp_I N expansion}. They control the non-perturbative corrections $\hat{f}_{\rm np}$. Apart from replacing $[\hat{\Delta}_I]'\mapsto [2\hat{\Delta}_I]'$ and dividing the $(N+\delta)^\frac{3}{2}$ and $(N+\delta)^\frac{1}{2}$ terms by an overall factor of 2, the ansatz above is identical to the black hole free energy in \eqref{eq:TTI-BH}. Like for the black hole, we only examine the case $\fg=0$ since the free energy at generic $\fg\neq 1$ can be obtained via \eqref{eq:general-g}. 
	
	\begin{table}[!h]
		\centering
        \renewcommand{\arraystretch}{1.5}
		\begin{tabular}{|c|c|c|c|c|}
			\hline
			solution \# & $\fc_I$ & $\delta $ & $\hat{f}^{\rm (np)}_{1/2,I} $ & $\hat{f}^{\rm (np)}_{{\rm log},I} $ \\ \hline  
			\rule{0pt}{15pt} 1 & $\Big( -3, -3,  -3 ,\frac{3}{2} \Big)$ & $ - \frac{7}{8} $ & $-\sqrt{2}\pi$ & $\frac{3}{2}$  \\ \hline 
			2 & $\Big( -3, -3,-3, - \frac{9}{2} \Big)$ & $  \frac{9}{8} $ & $-\sqrt{2}\pi$ &  $\frac{3}{2}$ \\ \hline 
			black hole & $\Big( -3, -3, -3, -3 \Big)$ & $\frac{5}{8}$ & $-2\sqrt{2}\pi$ & $1$ \\\hline 
		\end{tabular}
        \caption{\rm The coefficients $\fc_I^{(\textrm{2-cut})}$ and $\delta^{(\textrm{2-cut})}$ controlling the $1/N$ expansion of the free energies of the two 2-cut solutions in region $\tilde{2}$ at the superconformal point $(\Delta_1, \Delta_2, \Delta_C) = \Big(\frac{3\pi}{4},\frac{3\pi}{4},\frac{\pi}{2} \Big)$. We also include the coefficients $\hat{f}^{\rm (np)}_{1/2,I}$ and $\hat{f}^{\rm (np)}_{{\rm log},I}$, which determine the non-perturbative corrections $\hat{f}_{\rm np}$. For comparison, we have listed the corresponding coefficients for the black hole at its superconformal point $\hat{\Delta}_I=\frac{\pi}{2}$. }
		\label{tbl:2colRegion4SC}
	\end{table}
	
	Since the free energies of the two 2-cut solutions in region $\tilde{2}$ are equal at leading order in $N$, we expect them to be dual to the same classical geometry in the bulk, that is a $\bZ_2$ orbifold of the Euclidean black saddle, as we will discuss in Section \ref{subsec:multi-cut:orb grav}. At subleading orders in $N$, their free energies differ since the coefficients $\fc_I$ and $\delta$ determining the $1/N$ expansion in Table \ref{tbl:2colRegion4SC} are different. This difference must come from the higher-derivative and quantum corrections in the bulk, which should be reproduced by equivariant localization as in Section \ref{subsec:equiv loc BH}. As we will explain in Section \ref{subsubsec:holography:log}, the coefficient of $\frac{1}{2}$ multiplying $\log N$ in \eqref{eq:TTI-twocut}, which is the same as in the black hole free energy in \eqref{eq:TTI-BH}, can be understood as the contribution of a universal zero mode. Lastly, we observe in Table \ref{tbl:2colRegion4SC} that the exponent $\hat{f}^{\rm (np)}_{1/2,I}$ in the non-perturbative corrections for the 2-cut solutions are halved with respect to that of the black hole. This is simply due to the fact that the range of one of the angular coordinates in $S^7$ is halved in the $\bZ_2$ orbifold, and therefore the M2-brane worldvolume wrapping $S^3\subset S^7$ is halved.   

    For the superconformal point in region $\tilde{1}$, the situation is markedly different. If we suppose that the free energy has the form
    \begin{gather}
        F_{0,\textrm{2-cut},\tilde{1}} \stackrel{?}{=} -\frac{\sqrt{2[2\hat{\Delta}_1][2\hat{\Delta}_2][2\hat{\Delta}_3][2\hat{\Delta}_4]}}{6} \sum_{I=1}^4 \frac{\hat{\fn}_I}{[2\hat{\Delta}_I]} \Big[ (N + \delta)^{\frac{3}{2}} + \fc_I (N + \delta)^{\frac{1}{2}} \Big]\notag\\
         + c_{\log}\log (N + \delta)  + \hat{f}_0 + \hat{f}_{\rm np}\notag\\
        =- \frac{\sqrt{2[2\hat{\Delta}_1][2\hat{\Delta}_2][2\hat{\Delta}_3][2\hat{\Delta}_4]}}{6} \sum_I  \frac{\hat{\fn}_I}{[2\hat{\Delta}_I]} N^{3/2} - \frac{\sqrt{2[2\hat{\Delta}_1][2\hat{\Delta}_2][2\hat{\Delta}_3][2\hat{\Delta}_4]}}{6} \sum_I  \frac{\hat{\fn}_I}{[2\hat{\Delta}_I]} \left( \frac{3}{2} \delta  + \fc_I \right) N^{1/2} \notag\\
    	+ c_{\log}\log N  - \frac{\sqrt{2[2\hat{\Delta}_1][2\hat{\Delta}_2][2\hat{\Delta}_3][2\hat{\Delta}_4]}}{6} \sum_I  \frac{\hat{\fn}_I}{[2\hat{\Delta}_I]} \left( \frac{3}{8} \delta^2 + \frac{1}{2} \fc_I\delta  \right) N^{-1/2} + c_{\log} \delta  N^{-1} + \cdots\,,\label{eq: 1 tilde naive/wrong ansatz}
    \end{gather}
    and attempt to fit the numerically computed free energy to this ansatz, the fitted coefficients $c_{\log}$ and $\frac{3}{2}\delta + \fc_I$ at $\cO(N^\frac{1}{2})$ are
    \begin{equation} \label{eq:CI}
    		c_{\log}=2\,,\quad C_I\equiv \frac{3}{2} \delta  + \fc_I = \left( -\frac{93}{16}, -\frac{93}{16}, -\frac{69}{16}, -\frac{93}{16}  \right)\,,
    \end{equation}
    while the coefficients $\frac{3}{8}\delta^2+\frac{1}{2}\fc_I\delta$ at $\cO(N^{-\frac{1}{2}})$ are irrational. Given this set of fitted coefficients, we have an overdetermined set of 8 equations for 5 unknowns $\delta$ and $\fc_I$. It turns out that there are no consistent solutions in this case, and the ansatz \eqref{eq: 1 tilde naive/wrong ansatz} cannot be true. A plausible way out is to suppose that in addition  to the terms in \eqref{eq: 1 tilde naive/wrong ansatz} with rational coefficients $\delta$, $\fc_I$, there are extra contributions starting from $\cO(N^{-\frac{1}{2}})$ which contain irrational functions of $[2\hat{\Delta}_I]$. This is natural from the bulk point of view, since 3 of the 4 integers in $s_I=\{0,0,1,0\}$ corresponding to region $\tilde{1}$ are zero. As we will explain in Section \ref{subsubsec:holography:orbifold:ADE}, there is a 7d SU(2) super-Yang-Mills theory living on the fixed locus $\Sigma_\fg\times S^5$ of the $\bZ_2$ orbifold action in this case, whose action also contributes to the free energy. In particular, the contribution of this 7d gauge theory is also needed to reproduce the logarithmic coefficient $c_{\log}=2$, as we will demonstrate in Section \ref{subsubsec:holography:log}.
	
	%%%%%%%%%%%%%%%%%%%%%%%%
	\subsection{Three-cut solutions}
	\label{subsec:multi-cut:3cut}
	%%%%%%%%%%%%%%%%%%%%%%%%

\begin{figure}[!h]
    	\centering 
        \includegraphics[width=\textwidth]{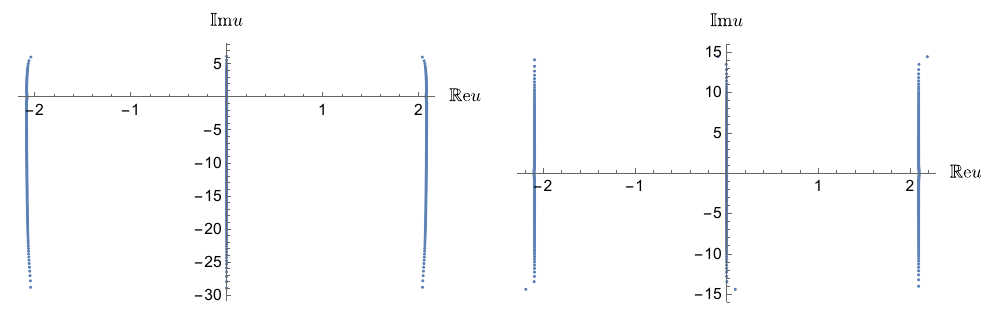}
    	\caption{\rm Examples of the first type of 3-cut solution with one cut along the imaginary axis. The left panel shows a solution of the ``symmetric'' type, at $N=501$, $\Delta_C=\frac{\pi}{9}$, $\Delta_1=\frac{\pi^2}{4}$, $\Delta_2=e$, while the right panel shows a ``deformed'' solution at $N=447$, $\Delta_C=0$, $\Delta_1=\Delta_2=\frac{\pi}{2}$. }
    	\label{fig: 3cut 0 eg}
    \end{figure}

\begin{figure}[!h]
    	\centering 
        \includegraphics[width=\textwidth]{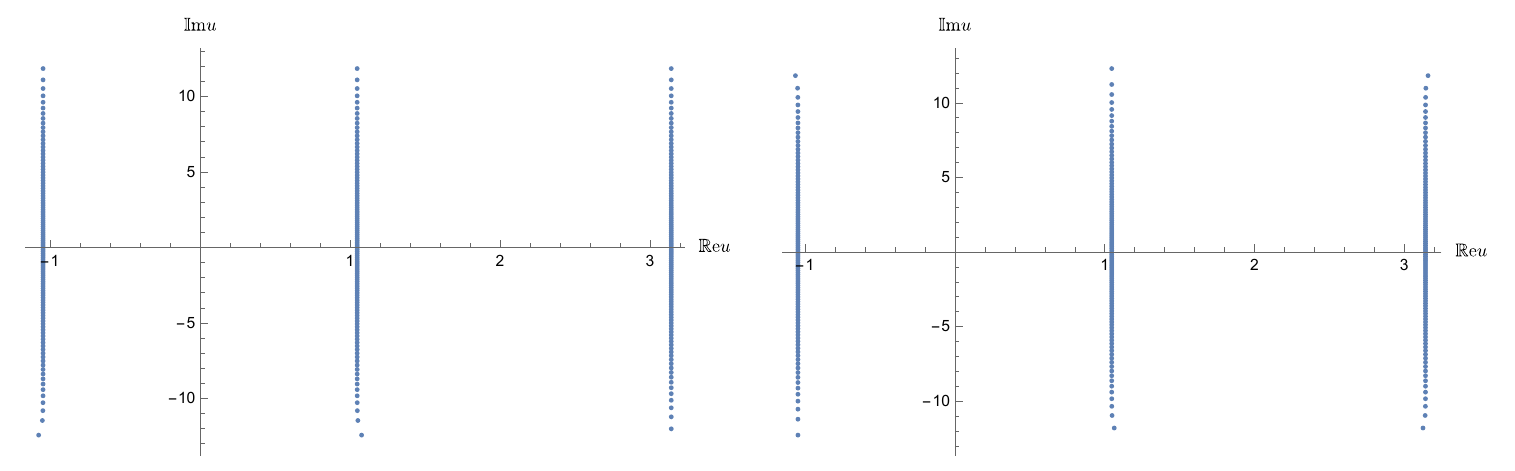}
    	\caption{\rm Examples of the second type of 3-cut solution with one cut along $\re u=\pi$. The left panel shows a solution of the ``symmetric'' type, at $N=339$, $\Delta_C=\frac{\pi}{3}$, $\Delta_1=\frac{7\pi}{6}$, $\Delta_2=\frac{\pi}{2}$, while the right panel shows a ``deformed'' solution at $N=324$, $\Delta_C=\pi$, $\Delta_1=\frac{7\pi}{6}$, $\Delta_2=\frac{\pi}{2}$. }
    	\label{fig: 3cut pi eg}
    \end{figure}

\begin{figure}[!h]
    	\centering 
        \includegraphics[width=0.6\textwidth]{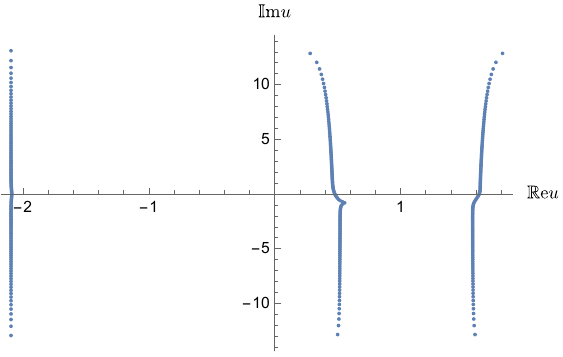}
    	\caption{\rm Example of a solution of the third type, at $N=381$, $\Delta_C=\frac{2\pi}{3}$, $\Delta_1=\Delta_2=\frac{5\pi}{6}$. }
    	\label{fig: 3cut odd eg}
    \end{figure}

    We now turn our attention to the 3-cut solutions. Numerically, we observe that there are 3 broad types of solutions, distinguished by the positions of the cuts. The first has one cut along the imaginary axis, and two other cuts placed symmetrically on the left and right of the axis, as shown in Figure \ref{fig: 3cut 0 eg}. The second type in Figure \ref{fig: 3cut pi eg} has one cut along $\re u = \pi$ and, similar to the first, two additional cuts placed symmetrically about the imaginary axis. The last type has one cut along $\re u=-\frac{2\pi}{3}$ and two cuts to the left and right of $\re u =\frac{\pi}{3}$, as seen in Figure \ref{fig: 3cut odd eg}. 

    Within the first and second types of solutions, we also differentiate between ``symmetric'' and ``deformed'' solutions. In the former, the two cuts to the left and right of the imaginary axis are exact reflections of each other about the axis. In other words, if we label the gauge holonomies in the cut on the right as $v_a$, and those in the left cut as $w_a$, with $a=1,\ldots, M\equiv N/3$, they exactly satisfy $v_a = -w_a^*$, or $\re v_a = -\re w_a$ and $\im v_a = \im w_a$. By contrast, in the ``deformed'' solutions, the imaginary parts of $v_a$ and $w_a$ are always slightly displaced, and $\re v_a$ is only approximately equal to $-\re w_a$. The positions of the holonomies in the remaining cut also deviate slightly away from $\re u =0$ and $\re u =\pi$, which does not happen in the ``symmetric'' solutions. In Figures \ref{fig: 3cut 0 eg} and \ref{fig: 3cut pi eg}, the left and right panels show the ``symmetric'' and ``deformed'' versions respectively of each type of solution. In the right panels, the violations of $v_a = -w_a^*$ are most obvious at the upper and lower limits of $\im v_a$, that is, at the top-most and bottom-most points of the cuts. 

    By sampling the fundamental domain \eqref{eq: fund dom} at a grid of points, we find that there are 12 different regions of chemical potentials where 3-cut solutions exist, numbered as $\tilde{1}$ to $\tilde{6}$ and $\tilde{1}'$ to $\tilde{6}'$. The definitions of these regions are in the second column of Table \ref{tab: 3cut reg}, while the third column records the type of solution that is found in each region. The first and second types of solutions are denoted as ``0-symm.'', ``0-def.'' and ``$\pi$-symm.'', ``$\pi$-def.'' respectively, where ``symm.'' stands for symmetric and ``def.'' stands for deformed, while the third type of solution is labelled as $-\frac{2\pi}{3}$, according to the positions of the cuts. As we will observe both analytically and numerically in the following sections, the free energies of the 3-cut solutions in regions $\tilde{5}$, $\tilde{5}'$, $\tilde{6}$, $\tilde{6}'$ are most naturally expressed in terms of $[3\hat{\Delta}_I]'$, while the natural variable in the other regions is $[3\hat{\Delta}_I]$. We denote the integer shifts in these variables as
    \begin{equation}
    s_I=\begin{cases} \left\lfloor \frac{3\hat{\Delta}_I}{2\pi} \right\rfloor\,, & \text{regions }\tilde{1}\text{--}\tilde{4}\,,\tilde{1}'\text{--}\tilde{4}' \\[.2cm] \left\lceil \frac{3\hat{\Delta}_I}{2\pi} \right\rceil\,, & \text{regions }\tilde{5},\tilde{5}',\tilde{6},\tilde{6}' \end{cases}
    \end{equation}
    As we will show in Section \ref{subsec:multi-cut:orb grav}, the integers $s_I$ specify the $\bZ_3$ action that is used to construct the orbifolds which are holographically dual to the 3-cut solutions. For future reference, the $s_I$ are listed in the last column of Table \ref{tab: 3cut reg}. 

    \begin{table}[h]
    \centering
    \footnotesize
    \begin{tabularx}{0.9\textwidth}{|
    >{\centering\arraybackslash\hsize=0.15\hsize\linewidth=\hsize}X|>{\centering\arraybackslash\hsize=2.4\hsize\linewidth=\hsize}X|>{\centering\arraybackslash\hsize=0.8\hsize\linewidth=\hsize}X|>{\centering\arraybackslash\hsize=0.65\hsize\linewidth=\hsize}X|}
        \hline
         Reg. & Definition & Type & $\{s_I\}$ \\[.2em]
        \hline
        $\tilde{1}$ & $0<\Delta_C<\frac{\pi}{3}\,,\;$ $0<\Delta_2<\frac{\pi}{3}-\Delta_C\,,\;$ $\Delta_2<\Delta_1<\frac{2\pi}{3}-2\Delta_C-\Delta_2$ & 0-symm. & $\{0,0,1,1\}$ \\[.2em]
        \hline
        $\tilde{1}'$ & $\frac{\pi}{3}<\Delta_C<\pi\,,\;$ $0<\Delta_2<\frac{\pi}{3}-|\Delta_C-\frac{2\pi}{3}|\,,\;$ $\Delta_2<\Delta_1<\frac{2\pi}{3}-2|\Delta_C-\frac{2\pi}{3}|-\Delta_2$ & 0-symm. & $\{0,0,2,0\}$ \\[.2em]
        \hline
        $\tilde{2}$ & $0<\Delta_C<\frac{2\pi}{3}\,,\;$ $\frac{2\pi}{3}<\Delta_1<\frac{4\pi}{3}-2|\Delta_C-\frac{\pi}{3}|\,,\;$ $0<\Delta_2<\frac{4\pi}{3}-2|\Delta_C-\frac{\pi}{3}|-\Delta_1$ & $\pi$-symm. & $\{1,0,1,0\}$ \\[.2em]
        \hline
        $\tilde{2}'$ & $\frac{2\pi}{3}<\Delta_C<\pi\,,\;$ $\frac{2\pi}{3}<\Delta_1<-\frac{2\pi}{3}+2\Delta_C\,,\;$ $0<\Delta_2<-\frac{2\pi}{3}+2\Delta_C-\Delta_1$ & $\pi$-symm. & $\{1,0,2,-1\}$ \\[.2em]
        \hline
        $\tilde{3}$ & $0<\Delta_C<\frac{\pi}{3}\,,\;$ $\frac{2\pi}{3}<\Delta_2<\pi-\Delta_C\,,\;$ $\Delta_2<\Delta_1<2\pi-2\Delta_C-\Delta_2$ & 0-symm. & $\{1,1,0,0\}$ \\[.2em]
        \hline
        $\tilde{3}'$ & $\frac{\pi}{3}<\Delta_C<\pi\,,\;$ $\frac{2\pi}{3}<\Delta_2<\pi-|\Delta_C-\frac{2\pi}{3}|\,,\;$ $\Delta_2<\Delta_1<2\pi-2|\Delta_C-\frac{2\pi}{3}|-\Delta_2$ & $-\frac{2\pi}{3}$ & $\{1,1,1,-1\}$ \\[.2em]
        \hline
        $\tilde{4}$ & $0<\Delta_C<\frac{\pi}{3}\,,\;$ $\frac{4\pi}{3}<\Delta_1<2\pi-2\Delta_C\,,\;$ $0<\Delta_2<2\pi-\Delta_1-2\Delta_C$ & 0-symm. & $\{2,0,0,0\}$\\[.2em]
        \hline
        $\tilde{4}'$ & $\frac{\pi}{3}<\Delta_C<\pi\,,\;$ $\frac{4\pi}{3}<\Delta_1<2\pi-2|\Delta_C-\frac{2\pi}{3}|\,,\;$ $0<\Delta_2<2\pi - 2|\Delta_C-\frac{2\pi}{3}|-\Delta_1$ & 0-symm. & $\{2,0,1,-1\}$ \\[.2em]
        \hline
        $\tilde{5}$ & $0<\Delta_C<\frac{\pi}{3}\,,\;$ $\frac{\pi}{3}+\Delta_C<\Delta_1<\frac{2\pi}{3}\,,\;$ $\frac{2\pi}{3}+2\Delta_C-\Delta_1<\Delta_2<\Delta_1$ & 0-def. & $\{1,1,1,1\}$\\[.2em]
        \hline
        $\tilde{5}'$ & $\frac{\pi}{3}<\Delta_C<\pi\,,\;$ $\frac{\pi}{3}+|\Delta_C-\frac{2\pi}{3}|<\Delta_1<\frac{2\pi}{3}\,,\;$ $\frac{2\pi}{3}+2|\Delta_C-\frac{2\pi}{3}|-\Delta_1<\Delta_2<\Delta_1$ & 0-symm. & $\{1,1,2,0\}$\\[.2em]
        \hline
        $\tilde{6}$ & $0<\Delta_C<\frac{2\pi}{3}\,,\;$ $\frac{2\pi}{3}+2|\Delta_C-\frac{\pi}{3}|<\Delta_1<\frac{4\pi}{3}\,,\;$ $\frac{4\pi}{3}+2|\Delta_C-\frac{\pi}{3}|-\Delta_1<\Delta_2<\frac{2\pi}{3}$ & $\pi$-symm. & $\{2,1,1,0\}$ \\[.2em]
        \hline
        $\tilde{6}'$ & $\frac{2\pi}{3}<\Delta_C<\pi\,,\;$ $\frac{8\pi}{3}-2\Delta_C<\Delta_1<\frac{4\pi}{3}\,,\;$ $\frac{10\pi}{3}-2\Delta_C-\Delta_1<\Delta_2<\frac{2\pi}{3}$ & $\pi$-def. & $\{2,1,2,-1\}$ \\[.2em]
        \hline
    \end{tabularx}
    \normalsize
    \caption{\small Regions where the 3-cut solutions exist, together with the type of solution in each region and the integers $s_I$ associated with the solution.}
    \label{tab: 3cut reg}
    \end{table}

    The rest of Section \ref{subsec:multi-cut:3cut} is organized as follows. In Sections \ref{subsubsec:multi-cut:3cut:largeN} and \ref{subsubsec:multi-cut:3cut:largeNpi}, we will analytically study the first (0-symm., 0-def.) and second ($\pi$-symm., $\pi$-def.) types of 3-cut solutions respectively using a large $N$ ansatz. Apart from a few regions that we will specify where an approximation that we used during the large $N$ limit fails, this study determines the density $\rho(t)$ describing the distribution of gauge holonomies along the cuts, and the positivity of $\rho$ exactly limits the region of existence of the solution to one of the regions listed in Table \ref{tab: 3cut reg}. For the sake of brevity, we will not analytically study the third type of 3-cut solution in Figure \ref{fig: 3cut odd eg}. In Section \ref{subsubsec:multi-cut:3cut:LO index}, we then use the densities $\rho$ obtained in each region to determine the free energy of the 3-cut solution at leading order in $N$. This is supplemented by a numerical study in Section \ref{subsubsec:multi-cut:3cut:numerics}, where we compute the free energy to all perturbative orders in $N$.

    %%%%%%%%%%%%%%%%%%%%%%%%
	\subsubsection{Large $N$ ansatz with one cut at $\re u=0$}
	\label{subsubsec:multi-cut:3cut:largeN}
	%%%%%%%%%%%%%%%%%%%%%%%%

    Motivated by the first type of numerical solutions in Figure \ref{fig: 3cut 0 eg}, we split the gauge holonomies into $3$ groups $\{u_{a=1,\ldots,M}\}$, $\{v_{b=1,\ldots,M}\}$, $\{w_{c=1,\ldots,M}\}$, each of size $M\equiv N/3$, and consider the ansatz
    \be\label{eq: large N 3 col ansatz}
    u_a = i M^\frac{1}{2} t_{u,a}\,, \quad v_b = i M^\frac{1}{2} t_{v,b} + f(t_{v,b})\,, \quad w_c = i M^\frac{1}{2} t_{w,c} - f(t_{w,c})\,,
    \ee
    where $f\in (0,\pi]$. This ansatz is meant to encapsulate both the ``symmetric'' and ``deformed'' solutions in the left and right panels of Figure \ref{fig: 3cut 0 eg}. In the symmetric solutions, since $\{v_b\}$ and $\{w_c\}$ are reflections of each other about the imaginary axis, we should restrict \eqref{eq: large N 3 col ansatz} to $\{t_{v,b}\}=\{t_{w,c}\}$. In the deformed solutions however, this symmetry is not present and $\{t_{u,a}\}$, $\{t_{v,b}\}$, $\{t_{w,c}\}$ are distinct. In addition, for the deformed solution displayed in the right panel of Figure \ref{fig: 3cut 0 eg}, $\re u_a$ deviate slightly away from $0$, and $\re w_c$ also differ slightly from $-\re v_b$. These variations in the real parts are negligible in the large $N$ limit, and they have been discarded in the ansatz \eqref{eq: large N 3 col ansatz}.
    
    We assume that $t_{u,a}$, $t_{v,b}$, and $t_{w,c}$ are all distributed according to the same density $\rho$:
    \be\label{eq: 3cut rho def}
    \rho(t_u) = \frac{1}{M}\frac{da}{dt_u}\,, \quad \rho(t_v) = \frac{1}{M}\frac{db}{dt_v}\,, \quad \rho(t_w) = \frac{1}{M}\frac{dc}{dt_w}\,. 
    \ee
    This implies that
    \be\label{eq: 3cut rv rw def}
    \lceil b(t_{u,a})\rceil = b(t_{u,a}) + r_v\,,\quad \lceil c(t_{u,a})\rceil = c(t_{u,a}) + r_w\,,\quad \forall a=1\,,\ldots,M\,,  
    \ee
    where $r_{v,w}\in (0,1)$ are piecewise constants in $t$, with a possible discontinuity at $t=0$. For the ``symmetric'' solutions, since $\{t_{v,b}\}$ and $\{t_{w,c}\}$ are identical, we have $r_v=r_w$. On the other hand, the ``deformed'' solutions have $r_v\neq r_w$. As explained in Appendix \ref{App: E-M formula}, it is important to keep track of $r_{v,w}$ because they contribute to the large $N$ Bethe equations via corrections in the Euler-Maclaurin formula when approximating sums as integrals. In the large $N$ limit, the Bethe equations \eqref{eq: W'} simplify to
    \begin{align}\label{eq: 3col large N BAE LO and rho}
    &\sum_{I=1}^3[g_+'([\Delta_I+f(t)]) - g_+'([\Delta_I-f(t)]) + g_+'([\Delta_I+2f(t)]) - g_+'([\Delta_I-2f(t)])] = 0\,,\notag\\
    &\sum_{I=1}^3[g_+([\Delta_I+f(t)]) + g_+([\Delta_I-f(t)]) -g_+([\Delta_I+2f(t)]) - g_+([\Delta_I-2f(t)])]=0\,,\notag\\
    &2\partial_t\left\{\rho(t)\sum_{I=1}^3\left[g_+([\Delta_I])+g_+([\Delta_I+f(t)])+g_+([\Delta_I-f(t)])\right]\right\}+\Delta_C+\sgn(t)\frac{\Delta_3}{2}\notag\\
    &+2\pi(1-r_v-r_w)\left(1+\sum_{I=1}^3\left\lfloor\frac{\Delta_I+f(t)}{2\pi}\right\rfloor+\sum_{I=1}^3\left\lfloor\frac{\Delta_I-f(t)}{2\pi}\right\rfloor\right)-2\pi\Theta\left(\Delta_C-\frac{\pi}{3}\right)=0\,,
    \end{align}
    and
    \begin{align}\label{eq: 3col large N BAE phases}
    &(r_v+r_w-1)\left(1+\sum_{I=1}^3\left\lfloor\frac{\Delta_I+f(t)}{2\pi}\right\rfloor+\sum_{I=1}^3\left\lfloor\frac{\Delta_I-f(t)}{2\pi}\right\rfloor\right)+\Theta\left(\Delta_C-\frac{\pi}{3}\right)\notag\\
    &=-\sgn(t)\left\lfloor\frac{\frac{\Delta_1+\Delta_2}{2}-\sgn(t)f(t)}{2\pi}\right\rfloor-\left(r_v-\frac{1}{2}\right)\left(1+\sum_{I=1}^3\left\lfloor\frac{\Delta_I+f(t)}{2\pi}\right\rfloor+\sum_{I=1}^3\left\lfloor\frac{\Delta_I-f(t)}{2\pi}\right\rfloor\right)\notag\\
    &+\begin{cases}\frac{1}{2}\sum_{I=1}^3\left(1+\left\lfloor\frac{\Delta_I+2f(t)}{2\pi}\right\rfloor-\left\lfloor\frac{\Delta_I-2f(t)}{2\pi}\right\rfloor\right)-4\,, & r_v = r_w \\ \sgn(r_w-r_v)\left(|r_w-r_v|-\frac{1}{2}\right)\left(1+\sum_{I=1}^3\left\lfloor\frac{\Delta_I+2f(t)}{2\pi}\right\rfloor+\sum_{I=1}^3\left\lfloor\frac{\Delta_I-2f(t)}{2\pi}\right\rfloor\right)\,, & r_v \neq r_w \end{cases}\notag\\
    &=-\sgn(t)\left\lfloor\frac{\frac{\Delta_1+\Delta_2}{2}+\sgn(t)f(t)}{2\pi}\right\rfloor-\left(r_w-\frac{1}{2}\right)\left(1+\sum_{I=1}^3\left\lfloor\frac{\Delta_I+f(t)}{2\pi}\right\rfloor+\sum_{I=1}^3\left\lfloor\frac{\Delta_I-f(t)}{2\pi}\right\rfloor\right)\notag\\
    &+\begin{cases}\frac{1}{2}\sum_{I=1}^3\left(1+\left\lfloor\frac{\Delta_I-2f(t)}{2\pi}\right\rfloor-\left\lfloor\frac{\Delta_I+2f(t)}{2\pi}\right\rfloor\right)+1\,, & r_v = r_w \\ \sgn(r_v-r_w)\left(|r_v-r_w|-\frac{1}{2}\right)\left(1+\sum_{I=1}^3\left\lfloor\frac{\Delta_I+2f(t)}{2\pi}\right\rfloor+\sum_{I=1}^3\left\lfloor\frac{\Delta_I-2f(t)}{2\pi}\right\rfloor\right)\,, & r_v \neq r_w \end{cases}
    \end{align}
    The derivations of \eqref{eq: 3col large N BAE LO and rho} and \eqref{eq: 3col large N BAE phases} can be found in Appendix \ref{App: 3cut w 1cut 0}. We have also used the fact that $\lfloor\frac{\Delta_1+\Delta_2}{4\pi}\rfloor = 0$ in the fundamental domain \eqref{eq: fund dom}. 
    
\begin{figure}[h]
    \centering        
    \begin{tikzpicture}[x=1.6cm, y=1.6cm]
        \def\PI{3.14159}

        % region 1
        \fill[customOrange!30] (0, 0) -- (2*\PI/3, 0) -- (\PI/3, \PI/3) -- cycle;
        \draw[customOrange!75!black, thick] (0, 0) -- (2*\PI/3, 0) -- (\PI/3, \PI/3) -- cycle;
        \node[customOrange!75!black] at (\PI/3, \PI/6-0.1) {$1$};

        % region 1 tilde
        \fill[skyBlue!30] (0, 0) -- (\PI/3, 0) -- (\PI/6, \PI/6) -- cycle;
        \draw[skyBlue!60, thick] (0, 0) -- (\PI/3, 0) -- (\PI/6, \PI/6) -- cycle;
        \node[skyBlue!75!black] at (\PI/7+0.05, \PI/16) {$\tilde{1}$};

        % region 2
        \fill[bluishGreen!30] (2*\PI/3, 0) -- (4*\PI/3, 0) -- (2*\PI/3, 2*\PI/3) -- cycle;
        \draw[bluishGreen!75!black, thick] (2*\PI/3, 0) -- (4*\PI/3, 0) -- (2*\PI/3, 2*\PI/3) -- cycle;
        \node[bluishGreen!75!black] at (\PI-0.2, \PI/3-0.3) {$2$};

        % region 2 tilde
        \fill[reddishPurple!30] (2*\PI/3, 0) -- (\PI, 0) -- (2*\PI/3, \PI/3) -- cycle;
        \draw[reddishPurple!60, thick] (2*\PI/3, 0) -- (\PI, 0) -- (2*\PI/3, \PI/3) -- cycle;
        \node[reddishPurple!75!black] at (5*\PI/6-0.17, \PI/6-0.2) {$\tilde{2}$};

        % region 3
        \fill[darkBlue!30] (2*\PI/3, 2*\PI/3) -- (4*\PI/3, 2*\PI/3) -- (\PI, \PI) -- cycle;
        \draw[darkBlue!75!black, thick] (2*\PI/3, 2*\PI/3) -- (4*\PI/3, 2*\PI/3) -- (\PI, \PI) -- cycle;
        \node[darkBlue!75!black] at (\PI, 5*\PI/6-0.1) {$3$};

        % region 3 tilde
        \fill[vermillion!30] (2*\PI/3, 2*\PI/3) -- (\PI, 2*\PI/3) -- (5*\PI/6, 5*\PI/6) -- cycle;
        \draw[vermillion!60, thick] (2*\PI/3, 2*\PI/3) -- (\PI, 2*\PI/3) -- (5*\PI/6, 5*\PI/6) -- cycle;
        \node[vermillion!75!black] at (5*\PI/6-0.02, 9*\PI/12-0.02) {$\tilde{3}$};

        % region 4
        \fill[charcoal!30] (4*\PI/3, 0) -- (2*\PI, 0) -- (4*\PI/3, 2*\PI/3) -- cycle;
        \draw[charcoal!75!black, thick] (4*\PI/3, 0) -- (2*\PI, 0) -- (4*\PI/3, 2*\PI/3) -- cycle;
        \node[charcoal!75!black] at (5*\PI/3-0.3, \PI/3-0.2) {$4$};

        % region 4 tilde
        \fill[yellow!30] (4*\PI/3, 0) -- (5*\PI/3, 0) -- (4*\PI/3, \PI/3) -- cycle;
        \draw[yellow!60, thick] (4*\PI/3, 0) -- (5*\PI/3, 0) -- (4*\PI/3, \PI/3) -- cycle;
        \node[yellow!75!black] at (9*\PI/6-0.24, \PI/6-0.18) {$\tilde{4}$};

        % region 5
        \fill[lightGray!30] (\PI/3, \PI/3) -- (2*\PI/3, 0) -- (2*\PI/3, 2*\PI/3) -- cycle;
        \draw[lightGray!75!black, thick] (\PI/3, \PI/3) -- (2*\PI/3, 0) -- (2*\PI/3, 2*\PI/3) -- cycle;
        \node[lightGray!75!black] at (\PI/2+0.1, \PI/3-0.1) {$5$};

        % region 5 tilde
        \fill[purple!30] (\PI/2, \PI/2) -- (2*\PI/3, \PI/3) -- (2*\PI/3, 2*\PI/3) -- cycle;
        \draw[purple!60, thick] (\PI/2, \PI/2) -- (2*\PI/3, \PI/3) -- (2*\PI/3, 2*\PI/3) -- cycle;
        \node[purple!75!black] at (7*\PI/12+0.04, \PI/2) {$\tilde{5}$};

        % region 6
        \fill[lightGreen!30] (2*\PI/3, 2*\PI/3) -- (4*\PI/3, 0) -- (4*\PI/3, 2*\PI/3) -- cycle;
        \draw[lightGreen!75!black, thick] (2*\PI/3, 2*\PI/3) -- (4*\PI/3, 0) -- (4*\PI/3, 2*\PI/3) -- cycle;
        \node[lightGreen!75!black] at (\PI+0.25, \PI/3+0.25) {$6$};

        % region 6 tilde
        \fill[terracotta!30] (\PI, 2*\PI/3) -- (4*\PI/3, \PI/3) -- (4*\PI/3, 2*\PI/3) -- cycle;
        \draw[terracotta!60, thick] (\PI, 2*\PI/3) -- (4*\PI/3, \PI/3) -- (4*\PI/3, 2*\PI/3) -- cycle;
        \node[terracotta!75!black] at (7*\PI/6+0.2, \PI/2+0.2) {$\tilde{6}$};

        % axes
        \draw[->] (0,0) -- (2.2*\PI,0) node[right] {$\Delta_1$};
        \draw[->] (0,0) -- (0,1.2*\PI) node[above] {$\Delta_2$};
        
        % ticks
        \draw (\PI,-0.06)--(\PI,0.06) node[below=5pt] {$\pi$};
        \draw (2*\PI,-0.06)--(2*\PI,0.06) node[below=4pt] {$2\pi$};
        \draw (-0.06,\PI)--(0.06,\PI) node[left=3pt] {$\pi$};
        
    \end{tikzpicture}
    \caption{\small Plots of the regions 1 to 6 listed in Table \ref{tab: 3 col regions}, as well as the subregions $\tilde{1}$ to $\tilde{6}$ where 3-cut solutions exist, for the illustrative value of $\Delta_C=\frac{\pi}{6}$.}
    \label{fig: 3cut wot t regions}
    \end{figure}
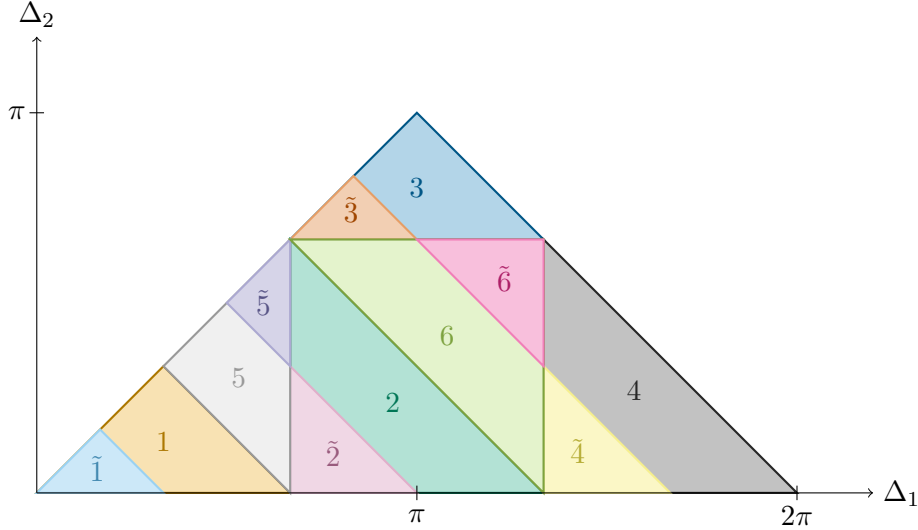

    We begin by solving the first two equations in \eqref{eq: 3col large N BAE LO and rho}. The arguments of $g_+'$ and $g_+$ depend on the integers
    \be\label{eq: 3 col ints def}
    m_{\pm,I}\equiv\left\lfloor\frac{\Delta_I\pm f(t)}{2\pi}\right\rfloor\,,\quad \tilde{m}_{\pm,I}\equiv\left\lfloor\frac{\Delta_I\pm 2f(t)}{2\pi}\right\rfloor\,.
    \ee
    As before, we assume that $m_{\pm,I}$ and $\tilde{m}_{\pm,I}$ are constant except for a possible discontinuity at $t = 0$. In the fundamental domain \eqref{eq: fund dom}, keeping in mind that $f\in(0,\pi)$, the integers can take the following range of values
    \begin{align}\label{eq: 3-col ints rang}
    m_{+,1}=0,1\,,\quad m_{+,2}&=0\,,\quad m_{+,3}=0,1\,,\quad m_{-,I}=-1,0\notag\\
    \tilde{m}_{+,I}&=0,1\,,\quad \tilde{m}_{-,I}=-1,0\,.
    \end{align}
    Examining all possible combinations of the integers in \eqref{eq: 3-col ints rang}, we try to solve the first two equations in \eqref{eq: 3col large N BAE LO and rho} for $f$. For the allowed values of $f$, if there exist $\Delta_I$, such that \eqref{eq: 3 col ints def} is consistent, this choice of integers is kept as a possibility. If not, it is discarded. In this way, we obtain the following values of $\{m_{+,I},m_{-,I}\}$, $\{\tilde{m}_{+,I},\tilde{m}_{-,I}\}$ in Table \ref{tab: 3 col regions}, together with their required values of $f$, and the regions of $\Delta_{1,2}$ in which they are consistent. As a visual aid, we plot the regions 1 to 6 of Table \ref{tab: 3 col regions} in Figure \ref{fig: 3cut wot t regions}. In the same figure, we also show the subregions $\tilde{1}$ to $\tilde{6}$ of regions 1 to 6 where the 3-cut solutions exist, at an example value of $\Delta_C = \frac{\pi}{6}$. 
    
    \begin{table}[h]
    \centering
    \footnotesize
    \centerline{
    \begin{tabularx}{\textwidth}{|c|
    >{\centering\arraybackslash\hsize=1.8\hsize\linewidth=\hsize}X|>{\centering\arraybackslash\hsize=0.83\hsize\linewidth=\hsize}X|>{\centering\arraybackslash\hsize=0.91\hsize\linewidth=\hsize}X|>{\centering\arraybackslash\hsize=0.46\hsize\linewidth=\hsize}X|}
        \hline
         & Region & $\{m_{+,I},m_{-,I}\}$ & $\{\tilde{m}_{+,I},\tilde{m}_{-,I}\}$ & $f$ \\[.2em]
        \hline
        1 & $0<\Delta_2<\frac{\pi}{3}\,,\quad\Delta_2<\Delta_1<\frac{2\pi}{3}-\Delta_2$ & ${\{0,0,0,-1,-1,0\}}$ & $\{0,0,1,-1,-1,0\}$ & ${2\pi-\Delta_3}$ \\
         & & ${\{0,0,1,-1,-1,0\}}$ & $\{0,0,1,-1,-1,0\}$ & free \\
         & & ${\{0,0,1,-1,-1,0\}}$ & $\{0,0,1,-1,-1,-1\}$ & $\frac{\Delta_3}{2}$ \\[.2em]\hline
        2 & $\frac{2\pi}{3}<\Delta_1<\frac{4\pi}{3}\,,\quad 0<\Delta_2<\frac{4\pi}{3}-\Delta_1$ & $\{0,0,0,0,-1,0\}$ & $\{1,0,1,-1,-1,-1\}$ & $\frac{2\pi}{3}$ \\[.2em]\hline
        3 & $\frac{2\pi}{3}<\Delta_2<\pi\,,\quad \Delta_2<\Delta_1<2\pi-\Delta_2$ & $\{0,0,0,0,0,-1\}$ & $\{1,1,0,-1,-1,-1\}$ & $\frac{2\pi}{3}$ \\[.2em]\hline
        4 & $\frac{4\pi}{3}<\Delta_1<2\pi\,,\quad 0<\Delta_2<2\pi-\Delta_1$ & $\{0,0,0,0,-1,-1\}$ & $\{1,0,0,0,-1,-1\}$ & $2\pi-\Delta_1$ \\
         & & $\{1,0,0,0,-1,-1\}$ & $\{1,0,0,0,-1,-1\}$ & free \\
         & & $\{1,0,0,0,-1,-1\}$ & $\{1,0,0,-1,-1,-1\}$ & $\frac{\Delta_1}{2}$\\[.2em]\hline
        5 & $\frac{\pi}{3}<\Delta_1<\frac{2\pi}{3}\,,\quad \frac{2\pi}{3}-\Delta_1<\Delta_2<\Delta_1$ & $\{0,0,0,-1,-1,0\}$ & $\{0,0,1,-1,-1,-1\}$ & $\frac{2\pi}{3}$ \\[.2em]\hline
        6 & $\frac{2\pi}{3}<\Delta_1<\frac{4\pi}{3}\,,\quad\frac{4\pi}{3}-\Delta_1<\Delta_2<\frac{2\pi}{3}$ & $\{0,0,0,0,-1,-1\}$ & $\{1,0,0,-1,-1,-1\}$ & $\frac{2\pi}{3}$ \\[.2em]\hline
    \end{tabularx}
    }
    \normalsize
    \caption{\small Allowed integers $\{m_{+,I},m_{-,I}\}$ and $\{\tilde{m}_{+,I},\tilde{m}_{-,I}\}$, the corresponding values of $f$, and the regions that they are consistent with.}
    \label{tab: 3 col regions}
    \end{table}

    Now, the third equation in \eqref{eq: 3col large N BAE LO and rho} determines $\rho$ to be
    \begin{align}\label{eq: 3 col rho gen}
    &\hspace{3cm}\rho(t) = \frac{\mu-\left(\frac{\Delta_3}{2}+\sgn t\Delta_C+2\pi p\right)|t|}{2G}\,,\notag\\
    &\hspace{1.45cm}G\equiv \sum_{I=1}^3\left[g_+([\Delta_I])+g_+([\Delta_I+f(t)])+g_+([\Delta_I-f(t)])\right]\,,\notag\\
    &p\equiv \sgn(t)(1-r_v-r_w)\left(1+\sum_{I=1}^3m_{+,I}+ \sum_{I=1}^3m_{-,I}\right)-\sgn(t)\Theta\left(\Delta_C-\frac{\pi}{3}\right)\,.
    \end{align}
    We must then solve \eqref{eq: 3col large N BAE phases} for $r_v$ and $r_w$ in each region of Table \ref{tab: 3 col regions}. Note that in regions $1$ and $4$, the values of $f=2\pi -\Delta_{1,3}$, $\frac{\Delta_{1,3}}{2}$ are such that the summands in \eqref{eq: 3cut uv sum} and \eqref{eq: 3cut vw int rem} have logarithmic singularities at the endpoints. As explained in Appendix \ref{App: E-M formula}, this invalidates the Euler-Maclaurin formula that we have used to take the large $N$ limit, and \eqref{eq: 3col large N BAE phases} is not reliable in these regions. One would have to consider generalized Euler-Maclaurin formulae such as those in \cite{Sidi2012} to get the correct large $N$ limit, and we leave this analysis to future work. In the following, we will consider the remaining regions $2,3,5,6$ in turn, determine $r_{v,w}$ in each region, and examine if solutions exist.  
    
    In region 2, since $1+\sum_{I=1}^3\lfloor\frac{\Delta_I+f}{2\pi}\rfloor + \sum_{I=1}^3\lfloor\frac{\Delta_I-f}{2\pi}\rfloor=0$ and $1+\sum_{I=1}^3\lfloor\frac{\Delta_I+2f}{2\pi}\rfloor + \sum_{I=1}^3\lfloor\frac{\Delta_I-2f}{2\pi}\rfloor=0$, \eqref{eq: 3col large N BAE phases} is independent of $r_{v,w}$, and the left hand side (first line) of \eqref{eq: 3col large N BAE phases} is independent of $t$. However, because $\lfloor\frac{\frac{\Delta_1+\Delta_2}{2}-\sgn(t)f}{2\pi}\rfloor = -\Theta(t)$, the right hand sides of \eqref{eq: 3col large N BAE phases} are not independent of $t$, and \eqref{eq: 3col large N BAE phases} cannot be satisfied. This is true for any $\Delta_C\in (0,\pi)$. 
    
    In region 3, for $0<\Delta_C<\frac{\pi}{3}$, one can check that \eqref{eq: 3col large N BAE phases} is satisfied for both the symmetric ($r_v=r_w$) and deformed ($r_v\neq r_w$) ansatz at any value of $r_{v,w}$. In the numerics, however, only symmetric solutions are found. Since $2G=\frac{1}{3}(3\Delta_1-2\pi)(3\Delta_2-2\pi)\Delta_3>0$ and $p = 0$, for $\rho$ to be positive and have support between two zeros, we need $\frac{\Delta_3}{2}\pm\Delta_C>0$. Therefore the solution only exists in the subregion $\tilde{3}$ defined in Table \ref{tab: 3cut reg}. Normalizing $\rho$ fixes $\mu$, and we get
    \begin{align}\label{eq: 3 col rho}
    &\rho(t) = \frac{\mu - \Delta_C t - \frac{\Delta_3}{2}|t|}{\frac{1}{3}(3\Delta_1-2\pi)(3\Delta_2-2\pi)\Delta_3}\,\quad t\in\left(-\frac{\mu}{\frac{\Delta_3}{2}-\Delta_C},\frac{\mu}{\frac{\Delta_3}{2}+\Delta_C}\right)\,,\notag\\
    &\mu = \sqrt{\frac{1}{6}(3\Delta_1-2\pi)(3\Delta_2-2\pi)\left(\Delta_3+2\Delta_C\right)\left(\Delta_3-2\Delta_C\right)}\,.
    \end{align}
    For $\frac{\pi}{3}<\Delta_C<\pi$, there are no solutions since \eqref{eq: 3col large N BAE phases} is violated for both the symmetric ($r_v=r_w$) and deformed ($r_v\neq r_w$) ansatz.
    
\begin{figure}[h!]
        \centering        
        \includegraphics[width=\textwidth]{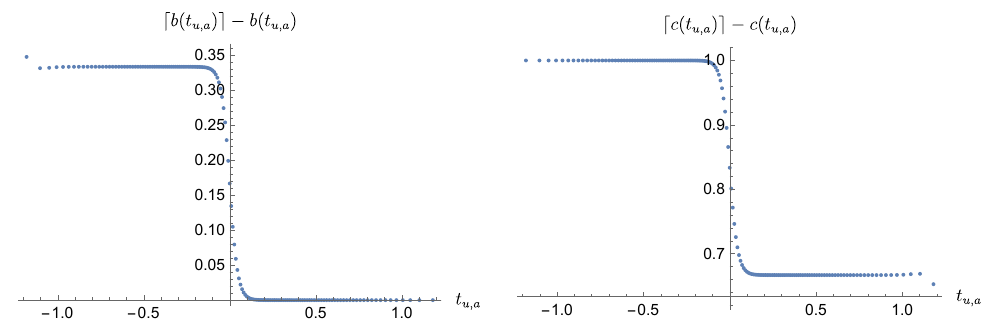}
        \caption{\small Plots of $\lceil b(t_{u,a})\rceil - b(t_{u,a})$ and $\lceil c(t_{u,a})\rceil - c(t_{u,a})$ against $t_{u,a}$ for the 3-cut solution at $\Delta_1=\Delta_2=\frac{\pi}{2}$, $\Delta_C=0$, $N=447$. $b(t)$ and $c(t)$ were computed by interpolating the data $\{t_{v,b}, b\}$ and $\{t_{w,c},c\}$ for $b,c=1,\ldots,M$.}
        \label{fig: 3cut 5t r}
    \end{figure}

    In region $5$, for $0<\Delta_C<\frac{\pi}{3}$, solving \eqref{eq: 3col large N BAE phases} in the symmetric case gives $r_v=r_w=\Theta(t)\frac{1}{3}+\Theta(-t)\frac{2}{3}$, while solving it in the deformed case gives $r_v = \Theta(-t)\frac{1}{3}$, $r_w = \Theta(t)\frac{2}{3}+\Theta(-t)$. In the numerics, only deformed solutions are found, and the values of $r_{v,w}$ computed from the interpolated functions $b(t)$, $c(t)$ indeed match the values obtained by solving \eqref{eq: 3col large N BAE phases}, as shown in figure \ref{fig: 3cut 5t r}. Substituting the values of $r_{v,w}$, one finds that $p=-\frac{1}{3}$. Since $2G=\frac{1}{9}(3\Delta_1-2\pi)(3\Delta_2-2\pi)(3\Delta_3-4\pi)<0$, the positivity of $\rho$ between two zeros requires $\frac{\Delta_3}{2}\pm\Delta_C-\frac{2\pi}{3}<0$. Therefore solutions only exist in the region $\tilde{5}$ in Table \ref{tab: 3cut reg}. Normalizing $\rho$ fixes $\mu$, and the solution is
    \begin{align}\label{eq: rho 3cut 5t}
    &\rho(t) = \frac{-\mu - \Delta_C t - \left(\frac{\Delta_3}{2}-\frac{2\pi}{3}\right)|t|}{\frac{1}{9}(3\Delta_1-2\pi)(3\Delta_2-2\pi)(3\Delta_3-4\pi)}\,\quad t\in\left(\frac{\mu}{\frac{\Delta_3}{2}-\Delta_C-\frac{2\pi}{3}},-\frac{\mu}{\frac{\Delta_3}{2}+\Delta_C-\frac{2\pi}{3}}\right)\,,\notag\\
    &\mu = \sqrt{\frac{2}{3}(3\Delta_1-2\pi)(3\Delta_2-2\pi)\left(\frac{\Delta_3}{2}+\Delta_C-\frac{2\pi}{3}\right)\left(\frac{\Delta_3}{2}-\Delta_C-\frac{2\pi}{3}\right)}\,.
    \end{align}
    
    For $\frac{\pi}{3}<\Delta_C<\pi$, if we consider the deformed ansatz, the only solution to \eqref{eq: 3col large N BAE phases} when $t<0$ is $r_v=r_w=1$, which contradicts the premise that $r_v\neq r_w$. For the symmetric ansatz, solving \eqref{eq: 3col large N BAE phases} gives $r_v=r_w=\Theta(t)\frac{2}{3}+\Theta(-t)$ and $p=-\Theta(t)\frac{2}{3}$. The positivity of $\rho$ between two zeros requires $\frac{\Delta_3}{2}+\Delta_C-\frac{4\pi}{3}<0$ and $\frac{\Delta_3}{2}-\Delta_C<0$, implying that the solution only exists in the region $\tilde{5}'$ in Table \ref{tab: 3cut reg}. The normalized density is
    \begin{align}\label{eq: rho 3cut 5tp}
    &\rho(t) = \frac{-\mu - \left(\frac{\Delta_3}{2}+\sgn(t)\Delta_C-\Theta(t)\frac{4\pi}{3}\right)|t|}{\frac{1}{9}(3\Delta_1-2\pi)(3\Delta_2-2\pi)(3\Delta_3-4\pi)}\,\quad t\in\left(\frac{\mu}{\frac{\Delta_3}{2}-\Delta_C},-\frac{\mu}{\frac{\Delta_3}{2}+\Delta_C-\frac{4\pi}{3}}\right)\,,\notag\\
    &\mu = \sqrt{\frac{2}{3}(3\Delta_1-2\pi)(3\Delta_2-2\pi)\left(\frac{\Delta_3}{2}+\Delta_C-\frac{4\pi}{3}\right)\left(\frac{\Delta_3}{2}-\Delta_C\right)}\,.
    \end{align}
    
    In region 6, for $0<\Delta_C<\frac{\pi}{3}$, solving \eqref{eq: 3col large N BAE phases} for the symmetric ansatz requires $r_w=r_v-1$, which contradicts the premise that $r_v=r_w$. If we consider the deformed ansatz, \eqref{eq: 3col large N BAE phases} is solved by $r_v=\frac{1}{3}$, $r_w=\frac{2}{3}$, or $r_w=\frac{1}{3}$, $r_v=\frac{2}{3}$. For either set of values, $p=0$. Since $2G=\frac{1}{9}(3\Delta_1-4\pi)(3\Delta_2-2\pi)(3\Delta_3-2\pi)<0$, the positivity of $\rho$ requires $\frac{\Delta_3}{2}\pm\Delta_C<0$, which is inconsistent with region 6. For $\frac{\pi}{3}<\Delta_C<\pi$, \eqref{eq: 3col large N BAE phases} for the symmetric ansatz still requires $r_w=r_v-1$, which is inconsistent. Considering the deformed ansatz, solving \eqref{eq: 3col large N BAE phases} gives either $r_v=\frac{2}{3}$, $r_w=1$, or $r_w=\frac{2}{3}$, $r_v=1$. For both sets of values, $p=-\frac{1}{3}\sgn(t)$, and the positivity of $\rho$ requires $\frac{\Delta_3}{2}+\Delta_C-\frac{2\pi}{3}<0$ and $\frac{\Delta_3}{2}-\Delta_C+\frac{2\pi}{3}<0$. The first inequality requires $\frac{\pi}{3}<\Delta_C<\frac{2\pi}{3}$ to be compatible with region 6, while the second inequality instead requires $\frac{2\pi}{3}<\Delta_C<\pi$. Thus they are incompatible and there are no solutions.

    %%%%%%%%%%%%%%%%%%%%%%%%
    \subsubsection{Large $N$ ansatz with one cut at $\re u=\pi$}
    \label{subsubsec:multi-cut:3cut:largeNpi}
	%%%%%%%%%%%%%%%%%%%%%%%%

    Mimicking the solutions in Figure \ref{fig: 3cut pi eg}, we also consider a 3-cut ansatz where one cut is along the vertical line $\re u=\pi$, and the other two cuts are reflections of each other about the $\im u$ axis:
    \be\label{eq: large N 3 col ansatz pi}
    u_a = i M^\frac{1}{2} t_{u,a} + \pi\,, \quad v_b = i M^\frac{1}{2} t_{v,b} + f(t_{v,b})\,, \quad w_c = i M^\frac{1}{2} t_{w,c} - f(t_{w,c})\,,\quad f\in (0,\pi)\,.
    \ee
    Exactly as in \eqref{eq: large N 3 col ansatz}, \eqref{eq: large N 3 col ansatz pi} describes both ``symmetric'' solutions such as the left panel of Figure \ref{fig: 3cut pi eg}, which conforms to \eqref{eq: large N 3 col ansatz pi} exactly and where $\{t_{v,b}\}=\{t_{w,c}\}$, and ``deformed'' solutions such as the right panel of Figure \ref{fig: 3cut pi eg}, where $\re u_a$ differ slightly from $\pi$, $\re v_b$ have small corrections away from $-\re w_c$, and $\{t_{v,b}\}\neq\{t_{w,c}\}$. In the large $N$ limit, the deviations in the real parts of $u,v,w$ in the deformed solutions are negligible. Like in \eqref{eq: 3cut rho def}, we assume that $t_{u,v,w}$ are identically distributed according to a single density $\rho$ at large $N$. This implies that \eqref{eq: 3cut rv rw def} holds for piecewise constants $r_{v,w}\in (0,1)$, which can have a discontinuity at $t=0$. We have $r_v=r_w$ for the symmetric solutions with $\{t_{v,b}\}=\{t_{w,c}\}$, while the deformed solutions have $r_v\neq r_w$. Following the analysis in Appendix \ref{App: 3cut w 1cut pi}, one can show that the Bethe equations are solved if
    \begin{align}\label{eq: bae 3cut pi LO rho' consis}
    &\sum_{I=1}^3\left[g_+'([\Delta_I+f(t)-\pi])-g_+'([\Delta_I-f(t)+\pi])+g_+'([\Delta_I+2f(t)])-g_+'([\Delta_I-2f(t)])\right]=0\notag\\
    &\sum_{I=1}^3\left[g_+([\Delta_I+f(t)-\pi])+g_+([\Delta_I-f(t)+\pi])-g_+([\Delta_I+2f(t)])-g_+([\Delta_I-2f(t)])\right] = 0\notag\\
    &2\partial_t\left\{\rho(t)\sum_{I=1}^3\left[g_+([\Delta_I])+g_+([\Delta_I+f(t)-\pi])+g_+([\Delta_I-f(t)+\pi])\right]\right\}+\Delta_C+\sgn(t)\frac{\Delta_3}{2}\notag\\
    &+2\pi\sgn(t)\left\lfloor\frac{\frac{\Delta_1+\Delta_2}{2}-\sgn(t)\pi}{2\pi}\right\rfloor+2\pi(1-r_v-r_w)\left(1+\sum_{I=1}^3\left\lfloor\frac{\Delta_I+f(t)-\pi}{2\pi}\right\rfloor\right.\notag\\
    &\left.+\sum_{I=1}^3\left\lfloor\frac{\Delta_I-f(t)+\pi}{2\pi}\right\rfloor\right)+2\pi\Theta\left(\frac{2\pi}{3}-\Delta_C\right)=0\,,
    \end{align}
    and
    \begin{align}\label{eq: bae 3cut pi rho' numer}
    &-\sgn(t)\left\lfloor\frac{\frac{\Delta_1+\Delta_2}{2}-\sgn(t)\pi}{2\pi}\right\rfloor+(r_v+r_w-1)\left(1+\sum_{I=1}^3\left\lfloor\frac{\Delta_I+f(t)-\pi}{2\pi}\right\rfloor\right.\notag\\
    &\left.+\sum_{I=1}^3\left\lfloor\frac{\Delta_I-f(t)+\pi}{2\pi}\right\rfloor\right)-\Theta\left(\frac{2\pi}{3}-\Delta_C\right)\notag\\
    &=-\sgn(t)\left\lfloor\frac{\frac{\Delta_1+\Delta_2}{2}-\sgn(t)f(t)}{2\pi}\right\rfloor+\left(\frac{1}{2}-r_v\right)\left(1+\sum_{I=1}^3\left\lfloor\frac{\Delta_I+f(t)-\pi}{2\pi}\right\rfloor+\sum_{I=1}^3\left\lfloor\frac{\Delta_I-f(t)+\pi}{2\pi}\right\rfloor\right)\notag\\
    &+\begin{cases}\frac{1}{2}\sum_{I=1}^3\left(1+\left\lfloor\frac{\Delta_I+2f(t)}{2\pi}\right\rfloor-\left\lfloor\frac{\Delta_I-2f(t)}{2\pi}\right\rfloor\right)-2\,, & r_v=r_w \\[.2cm] \left(r_w-r_v-\frac{\sgn(r_w-r_v)}{2}\right)\left(1+\sum_{I=1}^3\left\lfloor\frac{\Delta_I+2f(t)}{2\pi}\right\rfloor+\sum_{I=1}^3\left\lfloor\frac{\Delta_I-2f(t)}{2\pi}\right\rfloor\right)+\Theta\left(\Delta_C-\frac{2\pi}{3}\right)\,, & r_v\neq r_w\end{cases}\notag\\
    &=-\sgn(t)\left\lfloor\frac{\frac{\Delta_1+\Delta_2}{2}+\sgn(t)f(t)}{2\pi}\right\rfloor+\left(\frac{1}{2}-r_w\right)\left(1+\sum_{I=1}^3\left\lfloor\frac{\Delta_I+f(t)-\pi}{2\pi}\right\rfloor+\sum_{I=1}^3\left\lfloor\frac{\Delta_I-f(t)+\pi}{2\pi}\right\rfloor\right)\notag\\
    &+\begin{cases}\frac{1}{2}\sum_{I=1}^3\left(1+\left\lfloor\frac{\Delta_I-2f(t)}{2\pi}\right\rfloor-\left\lfloor\frac{\Delta_I+2f(t)}{2\pi}\right\rfloor\right)\,, & r_v=r_w \\ \left(r_v-r_w-\frac{\sgn(r_v-r_w)}{2}\right)\left(1+\sum_{I=1}^3\left\lfloor\frac{\Delta_I+2f(t)}{2\pi}\right\rfloor+\sum_{I=1}^3\left\lfloor\frac{\Delta_I-2f(t)}{2\pi}\right\rfloor\right)\,, & r_v\neq r_w\end{cases}
    \end{align}
    hold. These equations depend on the integers
    \be\label{eq: 3cut pi int def}
    m_{\pm,I}\equiv\left\lfloor\frac{\Delta_I\pm f(t)\mp\pi}{2\pi}\right\rfloor\,, \quad \tilde{m}_{\pm,I}\equiv\left\lfloor\frac{\Delta_I\pm 2f(t)}{2\pi}\right\rfloor\,,
    \ee
    which are assumed to be piecewise constant, with an allowed discontinuity at $t=0$. Note that the argument of the floor function in $m_{\pm,I}$ is shifted with respect to what was previously defined in \eqref{eq: 3 col ints def}. Since $f\in(0,\pi)$, the possible integer values in the fundamental domain \eqref{eq: fund dom} are
    \begin{align}
    &m_{+,I}=-1,0\,,\quad m_{-,1}=0,1\,,\quad m_{-,2}=0\,,\quad m_{-,3}=0,1\notag \\
    &\hspace{2.5cm}\tilde{m}_{+,I}=0,1\,,\quad \tilde{m}_{-,I}=-1,0\,.
    \end{align}
    For every possible combination of integer values, we solve the first two equations in \eqref{eq: bae 3cut pi LO rho' consis} for $f$. The solution is only consistent if there exists a region in the fundamental domain \eqref{eq: fund dom} where \eqref{eq: 3cut pi int def} holds for the given set of integers and $f$. The possibilities are listed in Table \ref{tab: 3 col regions pi}.
    
    \begin{table}[h]
    \centering
    \footnotesize
    \centerline{
    \begin{tabularx}{\textwidth}{|c|
    >{\centering\arraybackslash\hsize=1.8\hsize\linewidth=\hsize}X|>{\centering\arraybackslash\hsize=0.83\hsize\linewidth=\hsize}X|>{\centering\arraybackslash\hsize=0.91\hsize\linewidth=\hsize}X|>{\centering\arraybackslash\hsize=0.46\hsize\linewidth=\hsize}X|}
        \hline
         & Region & $\{m_{+,I},m_{-,I}\}$ & $\{\tilde{m}_{+,I},\tilde{m}_{-,I}\}$ & $f$ \\[.2em]
        \hline
        1 & $0<\Delta_2<\frac{\pi}{3}\,,\quad\Delta_2<\Delta_1<\frac{2\pi}{3}-\Delta_2$ & ${\{-1,-1,0,0,0,1\}}$ & $\{0,0,1,-1,-1,0\}$ & free \\
         & & ${\{-1,-1,0,0,0,0\}}$ & $\{0,0,1,-1,-1,0\}$ & $\Delta_3-\pi$ \\
         & & ${\{-1,-1,0,0,0,1\}}$ & $\{0,0,0,-1,-1,0\}$ & $\frac{2\pi-\Delta_3}{2}$ \\[.2em]\hline
        2 & $\frac{2\pi}{3}<\Delta_1<\frac{4\pi}{3}\,,\quad 0<\Delta_2<\frac{4\pi}{3}-\Delta_1$ & $\{0,-1,0,0,0,0\}$ & $\{0,0,0,0,-1,0\}$ & $\frac{\pi}{3}$ \\[.2em]\hline
        3 & $\frac{2\pi}{3}<\Delta_2<\pi\,,\quad \Delta_2<\Delta_1<2\pi-\Delta_2$ & $\{0,0,-1,0,0,0\}$ & $\{0,0,0,0,0,-1\}$ & $\frac{\pi}{3}$ \\[.2em]\hline
        4 & $\frac{4\pi}{3}<\Delta_1<2\pi\,,\quad 0<\Delta_2<2\pi-\Delta_1$ & $\{0,-1,-1,1,0,0\}$ & $\{1,0,0,0,-1,-1\}$ & free \\
         & & $\{0,-1,-1,0,0,0\}$ & $\{1,0,0,0,-1,-1\}$ & $\Delta_1-\pi$ \\
         & & $\{0,-1,-1,1,0,0\}$ & $\{0,0,0,0,-1,-1\}$ & $\frac{2\pi-\Delta_1}{2}$\\[.2em]\hline
        5 & $\frac{\pi}{3}<\Delta_1<\frac{2\pi}{3}\,,\quad \frac{2\pi}{3}-\Delta_1<\Delta_2<\Delta_1$ & $\{-1,-1,0,0,0,0\}$ & $\{0,0,0,-1,-1,0\}$ & $\frac{\pi}{3}$ \\[.2em]\hline
        6 & $\frac{2\pi}{3}<\Delta_1<\frac{4\pi}{3}\,,\quad\frac{4\pi}{3}-\Delta_1<\Delta_2<\frac{2\pi}{3}$ & $\{0,-1,-1,0,0,0\}$ & $\{0,0,0,0,-1,-1\}$ & $\frac{\pi}{3}$ \\[.2em]\hline
    \end{tabularx}
    }
    \normalsize
    \caption{\small Allowed integers $\{m_{+,I},m_{-,I}\}$ and $\{\tilde{m}_{+,I},\tilde{m}_{-,I}\}$, the corresponding values of $f$ solving the first two equations of \eqref{eq: bae 3cut pi LO rho' consis}, and the regions that they are consistent with.}
    \label{tab: 3 col regions pi}
    \end{table}
    Note that the regions 1 to 6 in Table \ref{tab: 3 col regions pi} have the exact same definitions as in Table \ref{tab: 3 col regions}. The remaining equation in \eqref{eq: bae 3cut pi LO rho' consis} fixes
    \begin{align}
    &\hspace{3.7cm}\rho = \frac{\mu-\left(\frac{\Delta_3}{2}+\sgn(t)\Delta_C+2\pi p\right)|t|}{2G}\,,\notag\\
    &\hspace{1.4cm}G\equiv\sum_{I=1}^3\left[g_+([\Delta_I])+g_+([\Delta_I+f(t)-\pi])+g_+([\Delta_I-f(t)+\pi])\right]\,,\notag\\
    &p\equiv-\Theta(t)+\sgn(t)(1-r_v-r_v)\left(1+\sum_{I=1}^3m_{+,I}+\sum_{I=1}^3m_{-,I}\right)+\sgn(t)\Theta\left(\frac{2\pi}{3}-\Delta_C\right)\,,
    \end{align}
    where we have used the fact that $\left\lfloor\frac{\frac{\Delta_1+\Delta_2}{2}-\sgn(t)\pi}{2\pi}\right\rfloor=-\Theta(t)$ in the fundamental domain \eqref{eq: fund dom}. The next step is to solve \eqref{eq: bae 3cut pi rho' numer} for $r_{v,w}$ in each region. Like in Section \ref{subsubsec:multi-cut:3cut:largeN}, we cannot treat regions 1 and 4 because the summands in \eqref{eq: 3cut pi uv sum} and \eqref{eq: 3cut vw sum} have logarithmic singularities at the endpoints, rendering the Euler-Maclaurin formula \eqref{eq: Euler-Mac simp} and, by extension, \eqref{eq: bae 3cut pi rho' numer} incorrect. We will therefore only consider the regions $2,3,5,6$ in the following. 
    
    In region 2, \eqref{eq: bae 3cut pi rho' numer} requires 
    \be\label{eq: 3cut pi reg 2 cond}
    \Theta(t) -\Theta\left(\frac{2\pi}{3}-\Delta_C\right)=\begin{cases}0\,, & r_v=r_w \\ \Theta\left(\Delta_C-\frac{2\pi}{3}\right)\,, & r_v\neq r_w\end{cases}
    \ee
    This is impossible to satisfy since the left hand side has a discontinuity at $t = 0$ while the right hand side does not. However, we do find numerical 3-cut solutions with the form \eqref{eq: large N 3 col ansatz pi} in region 2. As explained in Appendix \ref{App: 3cut w 1cut pi}, the caveat is that the second equation of \eqref{eq: bae 3cut pi LO rho' consis} and the equations in \eqref{eq: bae 3cut pi rho' numer} come from the assumption \eqref{eq: rho u v consis num denom} that the numerators and denominators of \eqref{eq: rho u v consis} are independently equal. It turns out that this assumption is false for the numerical solution in region 2. It violates the second equation in \eqref{eq: bae 3cut pi LO rho' consis} and instead satisfies \eqref{eq: rho u v consis}, which is less restrictive. Since the value of $f=\frac{\pi}{3}$ for region 2 in Table \ref{tab: 3 col regions pi} comes from solving the first two equations in \eqref{eq: bae 3cut pi LO rho' consis}, this is also violated, and the numerical solution instead has $f=\pi-\Delta_1$. With this value of $f$, the summand in \eqref{eq: 3cut pi uv sum} has a logarithmic singularity near $\im v_b=\im u_a$, and our large $N$ analysis fails because the Euler-Maclaurin formula \eqref{eq: Euler-Mac simp} does not apply. Like the excluded regions 1 and 4, treating this would require examining whether generalized Euler-Maclaurin formulae such as \cite{Sidi2012} can be applied, which we leave for future work.  
    
    In region 3, solving \eqref{eq: bae 3cut pi rho' numer} would also require \eqref{eq: 3cut pi reg 2 cond}, which is impossible. Consistently, there are no 3-cut solutions in the form of \eqref{eq: large N 3 col ansatz pi}. The 3-cut solutions which are present in region 3 instead look like Figure \ref{fig: 3cut odd eg}, where one cut is along $\re u=-\frac{2\pi}{3}$ while the other two cuts are reflection symmetric about $\re u = \frac{\pi}{3}$.
    
    In region 5, we first consider $0<\Delta_C<\frac{2\pi}{3}$. For the symmetric ansatz ($r_v=r_w$), solving \eqref{eq: bae 3cut pi rho' numer} gives $r_v=r_w=\frac{1}{3}\Theta(t)$ and $p=-\frac{1}{3}\Theta(t)$. For $\rho$ to be positive between two zeros, since $2G=\frac{1}{9}(3\Delta_1-2\pi)(3\Delta_2-2\pi)(3\Delta_3-4\pi)<0$, we need $\frac{\Delta_3}{2}+\Delta_C-\frac{2\pi}{3}<0$ and $\frac{\Delta_3}{2}-\Delta_C<0$. The two inequalities are compatible with region 5 if $0<\Delta_C<\frac{\pi}{3}$ and $\frac{\pi}{3}<\Delta_C<\frac{2\pi}{3}$, which cannot both be true. For the deformed ansatz ($r_v\neq r_w$), solving \eqref{eq: bae 3cut pi rho' numer} gives $r_v=\frac{1}{3}$, $r_w=\frac{2}{3}$ or $r_v=\frac{2}{3}$, $r_w=\frac{1}{3}$ at $t>0$, and $r_v=0$, $r_w=\frac{1}{3}$ or $r_v=\frac{1}{3}$, $r_w=0$ at $t<-0$. Regardless of which solution for $r_{v,w}$ we choose, one can check that $p=-\frac{1}{3}\Theta(-t)$. The positivity of $\rho$ thus requires $\frac{\Delta_3}{2}+\Delta_C<0$ and $\frac{\Delta_3}{2}-\Delta_C-\frac{2\pi}{3}<0$. The latter inequality is equivalent to $\Delta_1+\Delta_2>2\pi+2\Delta_C$, which is impossible in the fundamental domain \eqref{eq: fund dom}. Moving on to $\frac{2\pi}{3}<\Delta_C<\pi$, solving \eqref{eq: bae 3cut pi rho' numer} gives $r_v=r_w=\frac{2}{3}\Theta(t)+\frac{1}{3}\Theta(-t)$ for the symmetric ansatz, and $r_v=\frac{1}{3}\Theta(t)$, $r_w=\Theta(t)+\frac{2}{3}\Theta(-t)$ for the deformed ansatz. In both cases, $p=-\frac{2}{3}\Theta(t)+\frac{1}{3}\Theta(-t)$. The positivity of $\rho$ requires $\frac{\Delta_3}{2}+\Delta_C-\frac{4\pi}{3}<0$ and $\frac{\Delta_3}{2}-\Delta_C+\frac{2\pi}{3}<0$. The latter inequality is incompatible with region 5. Therefore, no solutions---whether symmetric or deformed---exist in region 5 for any value of $0<\Delta_C<\pi$.  
    
    In region 6, it turns out that the equations \eqref{eq: bae 3cut pi rho' numer} are identical to those in region 5. When $0<\Delta_C<\frac{2\pi}{3}$, the symmetric ansatz gives $r_v=r_w=\frac{1}{3}\Theta(t)$ and $p=-\frac{1}{3}\Theta(t)$. Since $2G=\frac{1}{9}(3\Delta_1-4\pi)(3\Delta_2-2\pi)(3\Delta_3-2\pi)<0$, the positivity of $\rho$ requires $\frac{\Delta_3}{2}+\Delta_C-\frac{2\pi}{3}<0$ and $\frac{\Delta_3}{2}-\Delta_C<0$. This restricts the solution to the subregion $\tilde{6}$ in Table \ref{tab: 3cut reg}. The normalized density is
    \bea\label{eq: rho 3cut 6t}
    &\rho(t) = \frac{-\mu - \left(\frac{\Delta_3}{2}+\sgn(t)\Delta_C-\Theta(t)\frac{2\pi}{3}\right)|t|}{\frac{1}{9}(3\Delta_1-4\pi)(3\Delta_2-2\pi)(3\Delta_3-2\pi)}\,\quad t\in\left(\frac{\mu}{\frac{\Delta_3}{2}-\Delta_C},-\frac{\mu}{\frac{\Delta_3}{2}+\Delta_C-\frac{2\pi}{3}}\right)\,,\\
    &\mu = \sqrt{\frac{2}{3}(3\Delta_1-4\pi)(3\Delta_2-2\pi)\left(\frac{\Delta_3}{2}+\Delta_C-\frac{2\pi}{3}\right)\left(\frac{\Delta_3}{2}-\Delta_C\right)}\,.
    \eea
    On the other hand, the deformed ansatz would give $p=-\frac{1}{3}\Theta(-t)$ as in region 5. Like before, the positivity of $\rho$ implies $\Delta_1+\Delta_2>2\pi+2\Delta_C$, which is incompatible with the fundamental domain. When $\frac{2\pi}{3}<\Delta_C<\pi$, both the symmetric and deformed ansatz give $p=-\frac{2}{3}\Theta(t)+\frac{1}{3}\Theta(-t)$, which implies $\frac{\Delta_3}{2}+\Delta_C-\frac{4\pi}{3}<0$ and $\frac{\Delta_3}{2}-\Delta_C+\frac{2\pi}{3}<0$ due to the positivity of $\rho$. This restricts the solution to region $\tilde{6}'$ in Table \ref{tab: 3cut reg}, where the density is
    \bea\label{eq: rho 3cut 6tp}
    &\rho(t) = \frac{-\mu - \left(\frac{\Delta_3}{2}+\sgn(t)\Delta_C-\Theta(t)\frac{4\pi}{3}+\Theta(-t)\frac{2\pi}{3}\right)|t|}{\frac{1}{9}(3\Delta_1-4\pi)(3\Delta_2-2\pi)(3\Delta_3-2\pi)}\,\quad t\in\left(\frac{\mu}{\frac{\Delta_3}{2}-\Delta_C+\frac{2\pi}{3}},-\frac{\mu}{\frac{\Delta_3}{2}+\Delta_C-\frac{4\pi}{3}}\right)\,,\\
    &\mu = \sqrt{\frac{2}{3}(3\Delta_1-4\pi)(3\Delta_2-2\pi)\left(\frac{\Delta_3}{2}+\Delta_C-\frac{4\pi}{3}\right)\left(\frac{\Delta_3}{2}-\Delta_C+\frac{2\pi}{3}\right)}\,.
    \eea
    
    One can check that the density \eqref{eq: 3 col rho} in region $\wt 3$ can be written as
    \begin{align}
    &\hspace{1.3cm}\rho = \begin{cases}\frac{\sqrt{\frac{2}{3}[3\hat{\Delta}_1][3\hat{\Delta}_2][3\hat{\Delta}_3][3\hat{\Delta}_4]}-[3\hat{\Delta}_3]t}{\frac{1}{3}[3\hat{\Delta}_1][3\hat{\Delta}_2]([3\hat{\Delta}_3]+[3\hat{\Delta}_4])}\,, & t>0 \\[.2cm] \frac{\sqrt{\frac{2}{3}[3\hat{\Delta}_1][3\hat{\Delta}_2][3\hat{\Delta}_3][3\hat{\Delta}_4]}+[3\hat{\Delta}_4]t}{\frac{1}{3}[3\hat{\Delta}_1][3\hat{\Delta}_2]([3\hat{\Delta}_3]+[3\hat{\Delta}_4])}\,, & t<0\end{cases}\notag\\[.2cm]
    t\in &\left(-\frac{\sqrt{\frac{2}{3}[3\hat{\Delta}_1][3\hat{\Delta}_2][3\hat{\Delta}_3][3\hat{\Delta}_4]}}{[3\hat{\Delta}_4]}\,,\frac{\sqrt{\frac{2}{3}[3\hat{\Delta}_1][3\hat{\Delta}_2][3\hat{\Delta}_3][3\hat{\Delta}_4]}}{[3\hat{\Delta}_3]}\right)\label{eq: 3cut rho plus uniform}\,,
    \end{align}
    while the densities \eqref{eq: rho 3cut 5t}, \eqref{eq: rho 3cut 5tp}, \eqref{eq: rho 3cut 6t}, \eqref{eq: rho 3cut 6tp} in regions $\wt 5$, $\wt 5'$, $\wt 6$, $\wt 6'$ can be uniformly written as
    \begin{align}
    &\hspace{1.3cm}\rho = \begin{cases}\frac{-\sqrt{\frac{2}{3}[3\hat{\Delta}_1]'[3\hat{\Delta}_2]'[3\hat{\Delta}_3]'[3\hat{\Delta}_4]'}-[3\hat{\Delta}_3]'t}{\frac{1}{3}[3\hat{\Delta}_1]'[3\hat{\Delta}_2]'([3\hat{\Delta}_3]'+[3\hat{\Delta}_4]')}\,, & t>0 \\[.2cm] \frac{-\sqrt{\frac{2}{3}[3\hat{\Delta}_1]'[3\hat{\Delta}_2]'[3\hat{\Delta}_3]'[3\hat{\Delta}_4]'}+[3\hat{\Delta}_4]'t}{\frac{1}{3}[3\hat{\Delta}_1]'[3\hat{\Delta}_2]'([3\hat{\Delta}_3]'+[3\hat{\Delta}_4]')}\,, & t<0\end{cases}\notag\\[.2cm]
    t\in &\left(\frac{\sqrt{\frac{2}{3}[3\hat{\Delta}_1]'[3\hat{\Delta}_2]'[3\hat{\Delta}_3]'[3\hat{\Delta}_4]'}}{[3\hat{\Delta}_4]'}\,,-\frac{\sqrt{\frac{2}{3}[3\hat{\Delta}_1]'[3\hat{\Delta}_2]'[3\hat{\Delta}_3]'[3\hat{\Delta}_4]'}}{[3\hat{\Delta}_3]'}\right)\label{eq: 3cut rho minus uniform}\,.
    \end{align}

    %%%%%%%%%%%%%%%%%%%%%%%%
    \subsubsection{Leading contribution of 3-cut solutions to the index}
    \label{subsubsec:multi-cut:3cut:LO index}
    %%%%%%%%%%%%%%%%%%%%%%%%

    Substituting the ansatz \eqref{eq: large N 3 col ansatz} into \eqref{eq: log Z}, the contribution of 3-cut solutions of the form \eqref{eq: large N 3 col ansatz} to the index is
    \begin{gather}
    -\frac{F_{\fg,\textrm{3-cut},\re u=0}}{N^\frac{3}{2}}= \frac{1}{3\sqrt{3}}\bigg\{\int dt\rho(t)^2\bigg[\sum_{I=1}^3(\fn_I+1-\fg)(3g_+'([\Delta_I]) + 2g_+'([\Delta_I+f(t)])\notag\\
    +2g_+'([\Delta_I-f(t)])+g_+'([\Delta_I+2f(t)])+g_+'([\Delta_I-2f(t)]))-(1-\fg)(4g_+'([f(t)])\notag\\
    +2g_+'([2f(t)])+\pi^2)\bigg]-3\fn_C\int dt\rho(t)t + \frac{3\fn_3}{2}\int dt\rho(t)|t|\bigg\}+\cO\left(N^{-\frac{1}{2}}\right)\notag\\
    =\frac{1}{3\cdot 3}\sqrt{2[3\hat{\Delta}_1][3\hat{\Delta}_2][3\hat{\Delta}_3][3\hat{\Delta}_4]}\sum_{I=1}^4\frac{\hat{\fn}_I}{[3\hat{\Delta}_I]}+\cO\left(N^{-\frac{1}{2}}\right)\label{eq: 3cut 0 logZ}\,.
    \end{gather}
    If we further substitute the solution \eqref{eq: 3 col rho} in region $\tilde{3}$, \eqref{eq: 3cut 0 logZ} simplifies to
    \be\label{eq: 3cut logZ plus}
    -\frac{F_{\fg,\textrm{3-cut},+}}{N^\frac{3}{2}}=\frac{1}{3\cdot 3}\sqrt{2[3\hat{\Delta}_1][3\hat{\Delta}_2][3\hat{\Delta}_3][3\hat{\Delta}_4]}\sum_{I=1}^4\frac{\hat{\fn}_I}{[3\hat{\Delta}_I]}+\cO\left(N^{-\frac{1}{2}}\right)\,.
    \ee
    On the other hand, substituting the solutions \eqref{eq: rho 3cut 5t}, \eqref{eq: rho 3cut 5tp} in regions $\tilde{5}$, $\tilde{5}'$ into \eqref{eq: 3cut 0 logZ} gives
    \be\label{eq: 3cut logZ minus}
    -\frac{F_{\fg,\textrm{3-cut},-}}{N^\frac{3}{2}}=-\frac{1}{3\cdot 3}\sqrt{2[3\hat{\Delta}_1]'[3\hat{\Delta}_2]'[3\hat{\Delta}_3]'[3\hat{\Delta}_4]'}\sum_{I=1}^4\frac{\hat{\fn}_I}{[3\hat{\Delta}_I]'}+\cO\left(N^{-\frac{1}{2}}\right)\,.
    \ee
    Similarly, if we substitute the ansatz \eqref{eq: large N 3 col ansatz pi} with one cut at $\re u=\pi$ into \eqref{eq: log Z}, and evaluate it on the solutions \eqref{eq: rho 3cut 6t}, \eqref{eq: rho 3cut 6tp} in regions $\tilde{6}$ and $\tilde{6}'$, the result simplifies to \eqref{eq: 3cut logZ minus}. As we shall see in Section \ref{subsec:multi-cut:orb grav}, \eqref{eq: 3cut logZ plus} and \eqref{eq: 3cut logZ minus} exactly match the on-shell action of the $\bZ_3$ orbifolds of black saddles in the $+$ and $-$ branches respectively.
    
	%%%%%%%%%%%%%%%%%%%%%%%%
	\subsubsection{Numerics at finite $N$}
	\label{subsubsec:multi-cut:3cut:numerics}
	%%%%%%%%%%%%%%%%%%%%%%%%

	Firstly, we compute the free energy of the 3-cut solution in region $\tilde{3}$ using the same method as \cite{Bobev:2022jte, Bobev:2022eus} and the previous Sections \ref{subsec:oneCut:numerics}, \ref{subsubsec:dressed:1pair:TTIfiniteN}, \ref{subsubsec:multi-cut:2cut:numerics}. Namely, the numerical solution is found to very high precision for a wide range of large $N$ using the density $\rho(t)$ found in \eqref{eq: 3 col rho}, and the free energy computed using the solution is fitted to the ansatz
    \begin{gather}
    	F_{\fg=0,\text{3-cut},+}\left(N,\fc_I^{\rm (3-cut)},\delta^{\rm (3-cut)}\right)  = - \frac{\sqrt{2[3\hat{\Delta}_1][3\hat{\Delta}_2][3\hat{\Delta}_3][3\hat{\Delta}_4]}}{9} \sum_{I=1}^4 \frac{\hat{\fn}_I}{[3\hat{\Delta}_I]} \Big[ (N + \delta^{\rm (3-cut)})^{\frac{3}{2}}\notag\\
    	+ \fc_I^{\rm (3-cut)} (N + \delta^{\rm (3-cut)})^{\frac{1}{2}} \Big] + c_{\log}\log (N + \delta^{\rm (3-cut)})  + \hat{f}_0 + \hat{f}_{\rm np}\,.\label{eq:TTI-threecut}
    \end{gather}
    By repeating the fit for many different values of chemical potentials in region $\tilde{3}$, we are able to extrapolate an analytical formula for the coefficients $\fc_I^{\rm (3-cut)}$, $\delta^{\rm (3-cut)}$, and determine that $c_{\log}=\frac{1}{2}$. The coefficients are very similar to those of the black hole in \eqref{eq:coeExactBH}, once the chemical potentials $\hat{\Delta}_I=[\hat{\Delta}_I]$ in \eqref{eq:coeExactBH} are replaced with $[3\hat{\Delta}_I]$. The differences between $\fc_I^{\rm (3-cut)}$, $\delta^{\rm (3-cut)}$ and those of the black hole are given by
    \begin{gather} 
    	\fc_{1,2}^{\rm (3-cut)} = \fc_{1,2}^{\rm (BH)}([3\hat{\Delta}_I]) - \frac{2\pi[3\hat{\Delta}_{1,2}]}{[3\hat{\Delta}_3][3\hat{\Delta}_4]},\quad \fc_{3,4}^{\rm (3-cut)} = \fc_{3,4}^{\rm (BH)}([3\hat{\Delta}_I]) - \frac{2\pi([3\hat{\Delta}_3] + [3\hat{\Delta}_4])}{[3\hat{\Delta}_3][3\hat{\Delta}_4]}\,,\notag\\
    	\delta^{\rm (3-cut)} =  \delta^{\rm (BH)}([3\hat{\Delta}_I]) + \frac{2\pi}{3} \left( \frac{1}{[3\hat{\Delta}_3]} + \frac{1}{[3\hat{\Delta}_4]} \right)\,.\label{eq:coeExact3Cut}
    \end{gather}
	
	Having obtained the full perturbative expansion in $1/N$, the non-perturbative corrections can also be studied. Expanding $\hat{f}_{\rm np}$ as 
	\begin{equation}
		\hat{f}_{\rm np}  = \sum_I  \hat{f}^{\rm (np)}_{ I} (\Delta_I) \wtd{\fn}_I,\quad \log \hat{f}^{\rm (np)}_{I} = \hat{f}_{1/2, I}^{\rm (np)} \sqrt{N} + \hat{f}^{\rm (np)}_{{\rm log}, I} \log N + \hat{f}^{\rm (np)}_{0, I} + \hat{f}^{\rm (np)}_{-1/2, I} N^{-1/2} + \cdots,
	\end{equation}
	the coefficient $\hat{f}_{1/2, I}^{\rm (np)}$ is given by
    \begin{equation} \label{eq:Z3instanton}
		\begin{aligned}
			\hat{f}_{1/2, I}^{\rm (np)} &= - \frac{2\sqrt{2}\pi}{3} \cdot \min_{I\ne J\ne K\ne L} \left( \sqrt{ \frac{[3\hat{\Delta}_I] [3\hat{\Delta}_J]}{[3\hat{\Delta}_K],[3\hat{\Delta}_L]}}\Big|_{(I,J)\ne (1,2)}, \quad 3 \sqrt{ \frac{[3\hat{\Delta}_1] [3\hat{\Delta}_2]}{[3\hat{\Delta}_3],[3\hat{\Delta}_4]}} \right).\\
		\end{aligned}
	\end{equation}
	Recall that $-\hat{f}_{1/2, I}^{\rm (np)}$ in \eqref{eq:BH-instanton}, which controls the leading non-perturbative correction in the black hole free energy, exactly matches the action of M2-branes wrapping the $S^3\subset S^7$ with the least volume in the black hole geometry, neglecting counterexamples at special values of $\hat{\Delta}_I$. Similarly, we expect that the coefficients $-\hat{f}_{1/2, I}^{\rm (np)}$ above should be reproduced by the action of M2-branes wrapping the minimal $S^3\subset S^7$ in the $\bZ_3$ orbifold. Based on the discussion in Section \ref{subsec:multi-cut:orb grav}, the M2-brane action should simply be given by $\frac{1}{3}$ of \eqref{eq:BH-instanton}, where every $\hat{\Delta}_I$ in \eqref{eq:BH-instanton} is replaced with $[3\hat{\Delta}_I]$. The factor of $\frac{1}{3}$ arises because the range of one of the angular coordinates in the orbifold is a third of what it was in the black hole, and hence the volume of $S^3\subset S^7$ should be divided by $3$. The special case given in \eqref{eq:Z3instanton} corresponds to M2-branes wrapping the $S^3\subset S^7$ defined by $\mu_1=\mu_2=0$. From the analysis in Section \ref{subsec:multi-cut:orb grav} and the values of $s_I$ in Table \ref{tab: 3cut reg}, we know that the orbifolds which are dual to the 3-cut solutions in region $\tilde{3}$ are constructed using the $\bZ_3$ action that acts as $(\phi_1,\phi_2,\phi_3,\phi_4)\mapsto (\phi_1+2\pi/3,\phi_2+2\pi/3,\phi_3,\phi_4)$ on the angular coordinates of $S^7$. The $S^3$ defined by $\mu_1=\mu_2=0$ is precisely the fixed locus of this action, and hence its volume and corresponding M2-brane action is enhanced by a factor of $3$ when compared with the other choices of $S^3\subset S^7$. We leave a systematic study of counterexamples as well as the subleading coefficient $\hat{f}_{{\rm log}, I}^{\rm (np)}$ to the future.
	
    \paragraph{The free energy at ``superconformal'' points}
    
    In addition, we also study the free energy at the ``superconformal'' points in each of the 12 regions in Table \ref{tab: 3cut reg}, which are defined as
    \be
    \begin{cases}[3\hat{\Delta}_I] = 3\hat{\Delta}_I - 2\pi s_I=\frac{\pi}{2}\,, & \text{Regions }\tilde{1}\text{--}\tilde{4},\tilde{1}'\text{--}\tilde{4}'\,, \\ [3\hat{\Delta}_I]' = 3\hat{\Delta}_I - 2\pi s_I=-\frac{\pi}{2}\,, & \text{Regions }\tilde{5},\tilde{5}',\tilde{6},\tilde{6}'\end{cases}
    \ee
    It turns out that at 10 of the 12 superconformal points, excluding the points in regions $\tilde{1}'$ and $\tilde{4}$, the free energy of the 3-cut solution can either be fitted to the ansatz in \eqref{eq:TTI-threecut} for regions $\tilde{1}$--$\tilde{3}$ and $\tilde{2}'$--$\tilde{4}'$, or to the ansatz
    \begin{gather}
        	F_{\fg=0,\text{3-cut},-}  = \frac{\sqrt{2[3\hat{\Delta}_1]'[3\hat{\Delta}_2]'[3\hat{\Delta}_3]'[3\hat{\Delta}_4]'}}{9} \sum_{I=1}^4 \frac{\hat{\fn}_I}{[3\hat{\Delta}_I]'} \Big[ (N + \delta^{\rm (3-cut)})^{\frac{3}{2}}\notag\\
        	+ \fc_I^{\rm (3-cut)} (N + \delta^{\rm (3-cut)})^{\frac{1}{2}} \Big] + c_{\log}\log (N + \delta^{\rm (3-cut)})  + \hat{f}_0 + \hat{f}_{\rm np}\,.\label{eq:TTI-threecut minus}
    \end{gather}
    for regions $\tilde{5}$,$\tilde{5}'$,$\tilde{6}$,$\tilde{6}'$. The fitted values for the coefficients $\fc_I$, $\delta$, and $c_{\log}$ at each point are listed in Table \ref{tbl:SC3colCoe}. When the fit for $\fc_I$ and $\delta$ fails in regions $\tilde{1}'$ and $\tilde{4}$, the situation is identical to region $\tilde{1}$\footnote{It is important to note that the regions $\tilde{1}$--$\tilde{3}$ are defined differently for the 2-cut and 3-cut solutions.} for the 2-cut solution in Section \ref{subsubsec:multi-cut:2cut:numerics}, where there is no consistent solution if the coefficients $C_I=\frac{3}{2}\delta +\fc_I$ at $\cO(N^\frac{1}{2})$ and $\frac{3}{8}\delta^2+\frac{1}{2}\fc_I\delta$ at $\cO(N^{-\frac{1}{2}})$ are used to solve for $\fc_I$ and $\delta$. In particular,  $C_I$ is rational while $\frac{3}{8}\delta^2+\frac{1}{2}\fc_I\delta$ is irrational. From Table \ref{tab: 3cut reg}, notice that $\tilde{1}'$ and $\tilde{4}$ are the only two regions where 3 of the 4 $s_I$'s vanish. As explained in \ref{subsubsec:holography:orbifold:ADE}, this is precisely the case where extra contributions from a 7d $\SU(3)$ gauge theory living on the fixed locus $\Sigma_\fg\times S^5$ are expected at $\cO(N^{-\frac{1}{2}})$, so we expect that \eqref{eq:TTI-threecut} should be supplemented by these contributions. One evidence for this is that $c_{\log}=\frac{9}{2}$ in regions $\tilde{1}'$ and $\tilde{4}$ can be accounted for by the contributions from the 7d gauge theory, as we will show in Section \ref{subsubsec:holography:log}. For now, we list the coefficients $C_I=\frac{3}{2}\delta +\fc_I$ in regions $\tilde{1}'$ and $\tilde{4}$ in Table \ref{tbl:SC3colCoe}. They are unaffected by the presence of the 7d gauge theory since the latter does not contribute at $\cO(N^\frac{1}{2})$.  
	
	\renewcommand{\arraystretch}{1.5}
	\begin{table}[!h] \small 
		\centering
		\caption{\rm Fitted values of the coefficients $\fc_I$, $\delta$, $c_{\log}$ at the superconformal points of each region $\tilde{1}$--$\tilde{6}$, $\tilde{1}'$--$\tilde{6}'$. The third column indicates whether the fit for $\fc_I$ and $\delta$ is successful (Y) or unsuccessful (N). When the fit fails, as in regions $\tilde{1}'$ and $\tilde{4}$, we list the values of $C_I=\frac{3}{2}\delta +\fc_I$ in place of $\fc_I$. }
        \vspace{10pt}
		\begin{tabular}{|c|c|c|c|c|}
			\hline
			region & $c_{\rm log}$ & resum. & $\fc_I$ or $C_I$ & $\delta$  \\ \hline
			$\tilde{3}'$, $\tilde{5}$, $\tilde{6}'$ &   $  \frac{1}{2} $ & Y & $\Big(  -3,-3,-3,-3 \Big)$ & $ \frac{5}{8}  $ \\ \hline  \hline 
			$\tilde{2}', \tilde{4}'$ &   $ \frac{1}{2}  $ & Y  & $\Big( -3, -7,  -3, -3 \Big) $ & $ \frac{47}{24} $ \\ \hline 
			$\tilde{5}', \tilde{6}$ &  $ \frac{1}{2}  $ & Y & $\Big( -3, -3, -3, -7  \Big)$ & $ \frac{47}{24}  $ \\ \hline  \hline 
			$\tilde{1}$ & $ \frac{1}{2}  $ & Y & $\Big(-11, -11, -7, -7\Big)$ & $\frac{79}{24}$ \\ \hline  
			$\tilde{2}$ & $\frac{1}{2}$ & Y & $\Big(-7, -11, -7, -11\Big)$ & $\frac{79}{24}$ \\ \hline
            $\tilde{3}$ & $\frac{1}{2}$ & Y & $\Big(-7, -7, -11, -11\Big)$ & $\frac{79}{24}$ \\ \hline \hline 
			$\tilde{1}'$  & $ \frac{9}{2}  $ & N & $\Big( -\frac{193}{16}, -\frac{193}{16}, -\frac{129}{16}, -\frac{193}{16}\Big)$ & $ $ \\ \hline 
            $\tilde{4}$ & $\frac{9}{2}$ & N & $\Big(-\frac{129}{16}, -\frac{193}{16}, -\frac{193}{16}, -\frac{193}{16}\Big)$ & $$ \\ \hline
		\end{tabular}
		\label{tbl:SC3colCoe}
	\end{table}
    
	From the data in Table \ref{tbl:SC3colCoe}, we notice that the values of $\fc_I$ and $\delta$ for a given region only depend on how many of the corresponding integers $s_I$ in Table \ref{tab: 3cut reg} are zero, and on which component in particular is zero. Specifically, we extract the following rules and patterns: 
	\begin{itemize}
		\item For superconformal points where $s_I \ne 0\; \forall I$, $\fc_I$ and $\delta$ are exactly the same as the black hole values \eqref{eq:coeExactBH}. 
		
		\item For superconformal points where one component of $s_I$ vanishes, when compared to the black hole coefficients, the corresponding $\fc_I$ differs by $4$, other $\fc_I$ are the same, and $\delta$ differs by $\frac{4}{3}$. 
		
		\item For superconformal points where two components of $s_I$ vanish (such as region $\tilde{3}$ studied above), all four components of $\fc_I$ are different from the black hole values, and $\delta$ differs by $\frac{8}{3}$.  More precisely, the $\fc_I$ are shifted by $4$ with respect to the black hole values if the corresponding $s_I=1$, while they are shifted by $8$ if the corresponding $s_I=0$.
		
		\item If three components of $s_I$ vanish, the free energy cannot be fitted to \eqref{eq:coeExact3Cut}, and extra terms must be included in the ansatz. 
	\end{itemize}
	
		\renewcommand{\arraystretch}{1.8}
	\begin{table}[!h] 
		\centering
		\caption{\rm Conjectured analytic expressions for $\fc_I$ and $\delta$ in the 12 regions where 3-cut solutions exist. }
        \vspace{10pt}
		\begin{tabular}{|c|c|c|}
			\hline
			region &  $\fc_I - \fc_I^{\rm (BH)}$ & $\delta - \delta^{\rm (BH)} $  \\ \hline 
			$\tilde{3}'$, $\tilde{5}$, $\tilde{6}'$ & $0$ & $ 0  $ \\[.5em] \hline \hline
			$\tilde{2}', \tilde{4}'$ &  $-\delta_{I2}\frac{2\pi}{[3\hat{\Delta}_2]}$ & $\frac{2\pi}{3[3\hat{\Delta}_2]}$ \\[.5em] \hline  
			$\tilde{5}', \tilde{6}$ &  $-\delta_{I4}\frac{2\pi}{[3\hat{\Delta}_4]'}$ & $\frac{2\pi}{3[3\hat{\Delta}_4]'}$ \\[.5em] \hline   \hline 
			$\tilde{1}$ &  $-\frac{2\pi[3\hat{\Delta}_I]}{\prod_{J\neq 3,4}[3\hat{\Delta}_J]}-\delta_{I1}\frac{2\pi}{[3\hat{\Delta}_1]}-\delta_{I2}\frac{2\pi}{[3\hat{\Delta}_2]}$ & $\frac{2\pi}{3}\sum_{J\neq 3,4}\frac{1}{[3\hat{\Delta}_J]}$ \\[.5em] \hline  
			$\tilde{2}$ & $-\frac{2\pi[3\hat{\Delta}_I]}{\prod_{J\neq 1,3}[3\hat{\Delta}_J]}-\delta_{I2}\frac{2\pi}{[3\hat{\Delta}_2]}-\delta_{I4}\frac{2\pi}{[3\hat{\Delta}_4]}$ & $\frac{2\pi}{3}\sum_{J\neq 1,3}\frac{1}{[3\hat{\Delta}_J]}$ \\[.5em] \hline
            $\tilde{3}$ & $-\frac{2\pi[3\hat{\Delta}_I]}{\prod_{J\neq 1,2}[3\hat{\Delta}_J]}-\delta_{I3}\frac{2\pi}{[3\hat{\Delta}_3]}-\delta_{I4}\frac{2\pi}{[3\hat{\Delta}_4]}$ & $\frac{2\pi}{3}\sum_{J\neq 1,2}\frac{1}{[3\hat{\Delta}_J]}$
            \\[.5em] \hline \hline 	  
			$ \tilde{1}'$ & $-\frac{2\pi[3\hat{\Delta}_I]\sum_{J\neq 3}[3\hat{\Delta}_J]}{\prod_{J\neq 3}[3\hat{\Delta}_J]}-(1-\delta_{I3})\frac{2\pi}{[3\hat{\Delta}_I]}$ & $\frac{2\pi}{3}\sum_{J\neq 3}\frac{1}{[3\hat{\Delta}_J]}$ \\[.5em] \hline  
            $\tilde{4}$ & $-\frac{2\pi[3\hat{\Delta}_I]\sum_{J\neq 1}[3\hat{\Delta}_J]}{\prod_{J\neq 1}[3\hat{\Delta}_J]}-(1-\delta_{I1})\frac{2\pi}{[3\hat{\Delta}_I]}$ & $\frac{2\pi}{3}\sum_{J\neq 1}\frac{1}{[3\hat{\Delta}_J]}$ \\[.5em] \hline
		\end{tabular}
		\label{tbl:SC3colConj}
	\end{table}
	
	Based on these observations, we conjecture that the coefficients $\fc_I$ and $\delta$ in each region have the analytic expressions given in Table \ref{tbl:SC3colConj}. These conjectured expressions pass all of the numerical tests that we have performed, in which different points are sampled in every region except $\tilde{1}'$ and $\tilde{4}$, and the free energy is fitted to either \eqref{eq:TTI-threecut} or \eqref{eq:TTI-threecut minus}. On the other hand, since the ansatz \eqref{eq:TTI-threecut} fails in regions $\tilde{1}'$ and $\tilde{4}$ as mentioned above, the conjectures in the last two rows of Table \ref{tbl:SC3colConj} are not supported by numerical checks, and should be understood as follows. We assume that the correct ansatz for the free energy is the sum of the gravitational free energy, which is still given by \eqref{eq:TTI-threecut}, and an unknown contribution coming from the 7d $\SU(3)$ gauge theory living on the fixed locus of the $\bZ_3$ group action. The $\fc_I$ and $\delta$ listed for regions $\tilde{1}'$ and $\tilde{4}$ are then the coefficients appearing in the gravitational part of the total free energy. They are obtained by extrapolating the coefficients in the other regions where the numerical fit is successful. Based on the values of $\delta$ in the 10 regions where the ansatz \eqref{eq:TTI-threecut} or \eqref{eq:TTI-threecut minus} is valid, it is clear that the difference $\delta-\delta^{(\textrm{BH})}$ should be given by $\frac{2\pi}{3}$ times the sum $\sum_I 1/[3\hat{\Delta}_I]$, where $I$ only ranges over the indices for which $s_I=0$. If we extend this pattern to regions $\tilde{1}'$ and $\tilde{4}$, the values of $C_I=\frac{3}{2}\delta +\fc_I$ in Table \ref{tbl:SC3colCoe} can then be used to fix $\fc_I$, which gives us the conjectures for $\tilde{1}'$ and $\tilde{4}$ in Table \ref{tbl:SC3colConj}. By subtracting the gravitational free energy from the numerical free energy, we can potentially study the unknown gauge theory contribution systematically. We will leave this to future work.

    %%%%%%%%%%%%%%%%%%%%%%%%
    \subsection{Orbifolds of black saddles}
    \label{subsec:multi-cut:orb grav}
    %%%%%%%%%%%%%%%%%%%%%%%%

    Recall that in Section \ref{subsec:oneCut:bulk}, we have shown that there is an exact match between the Legendre transform of the free energies $F_{\fg, \textrm{BH}_\pm}$ in \eqref{eq: logZ BH}, \eqref{eq: logZ BH branch 2} at leading order in $N$ and the entropy of Lorentzian BPS black holes, which come in two corresponding branches $+$ and $-$. This establishes a correspondence between the one-cut solutions in regions BH$_\pm$ and Lorentzian black holes in branches $\pm$. More properly, however, the AdS/CFT dictionary prescribes that the topologically twisted index is computed by the gravitational path integral in Euclidean signature, which is dominated by the contributions of Euclidean supergravity solutions in the large $N$ or semiclassical limit. The free energies $F_{\fg,\lambda}$ of Bethe solutions $\lambda$ with $F_{\fg,\lambda}=\cO(N^\frac{3}{2})=\cO(G_N^{-1})$ should match the on-shell actions of the corresponding Euclidean saddles. Indeed, in \cite{Bobev:2020pjk}, it was found that there is a class of ``Euclidean black saddles'' whose on-shell action exactly matches \eqref{eq: logZ BH} and \eqref{eq: logZ BH branch 2}. In this context, no Legendre transform is necessary, and the chemical potentials $\hat{\Delta}_I$ in field theory correspond to holonomies of the bulk gauge fields along the Euclidean time direction. The Lorentzian BPS black holes correspond to special points in parameter space where the Euclidean black saddles have a physically sensible Lorentzian continuation. In this section, we will review the black saddles constructed in \cite{Bobev:2020pjk} and show that the on-shell action of their $\bZ_2$ and $\bZ_3$ orbifolds exactly match the free energies of the 2-cut and 3-cut solutions respectively, at leading order in $N$. Furthermore, imposing supersymmetry on the $\bZ_2$ and $\bZ_3$ orbifolds precisely restricts their regions of existence in $\hat{\Delta}_I$ to the regions introduced in Sections \ref{subsec:multi-cut:2cut} and \ref{subsec:multi-cut:3cut} respectively. This provides strong evidence that the 2-cut and 3-cut solutions are holographically dual to $\bZ_2$ and $\bZ_3$ orbifolds of the Euclidean black saddles. More generally, we expect that Bethe solutions with $m$ cuts in the complex $u$ plane are dual to $\bZ_m$ orbifolds.

    The black saddle solutions found in \cite{Bobev:2020pjk} are solutions of the Euclidean STU model, which is the Wick rotation of the Lorentzian STU model considered in Section \ref{subsec:oneCut:bulk}. The bosonic field content contains the metric, $4$ gauge fields $A^{I=1,\ldots,4}$, and complex scalars $z_{i=1,2,3}$, $\tilde{z}_{i=1,2,3}$. Note that in Euclidean signature, the gauge fields $A^I$ are generically complex, while $z_i$ and $\tilde{z}_i$ are independent complex scalars rather than complex conjugates of each other. Following the notation in \cite{Bobev:2020pjk}, the metric and gauge fields of the black saddle solutions take the form     
    \be\label{eq: black saddle ansatz}
    ds^2 = e^{2f_1(r)}d\tau^2 + e^{2f_2(r)}dr^2 + e^{2f_3(r)}d\Omega^2_{\Sigma_\fg}\,,\quad A^I = e^I(r)d\tau + p^I\omega_{\Sigma_\fg}\,.
    \ee
    As in \eqref{eq: bh metric ansatz}, $d\Omega^2_{\Sigma_\fg}$ is the constant curvature metric on a Riemann surface of genus $\fg$, and $\omega_{\Sigma_\fg}$ is the local $1$-form such that $d\omega_{\Sigma_\fg}$ is the volume form on $\Sigma_\fg$. $p^I$ are precisely the magnetic charges of the solutions which satisfy an analogue of the topological twist condition. The black saddle interpolates between an asymptotically AdS$_4$ region when $r\rightarrow\infty$, and caps off smoothly at $r=r_0$ into $\bR^2\times\Sigma_\fg$. We shall denote the limiting values of $e^I$ and $f_{1,2,3}$ as for example $e^I(r\rightarrow\infty)=e^I|_\text{UV}$ and $e^I(r_0)=e^I|_\text{IR}$. For the geometry to be smooth at $r=r_0$, the periodicity of $\tau$ must be $\beta_\tau\equiv \frac{2\pi e^{f_2-f_1}}{f_1'}|_\text{IR}$. The BPS equations in supergravity impose that
    \be\label{eq: sugra BPS orig}
    \sum_{I=1}^4e^I|_\text{UV} = 0\,,\quad \xi g\sum_{I=1}^4 e^I|_\text{IR} =\frac{2\pi}{\beta_\tau}\,,
    \ee
    where $\xi=\pm1$ is a parameter in the spinor projections used to solve the BPS equations, and $g$ is the gauge coupling\footnote{In the terminology of \cite{Bobev:2020pjk}, we have chosen to consider solutions in ``branch 1'', but the results which follow apply equally well to solutions in ``branch 2'' with minor modifications in various sign conventions.}. However, since $\tau$ parametrizes a vanishing cycle at $r=r_0$ and $e^I|_\text{IR}\neq 0$, the gauge field is singular at $r=r_0$ in this gauge. We can perform a gauge transformation by $e^{-ige^I|_\text{IR}\tau}$, after which $e^I|_\text{IR} = 0$ and the BPS constraint becomes
    \be\label{eq: wilson line BPS constr}
    \xi g\sum_{I=1}^4 e^I|_\text{UV} = -\frac{2\pi}{\beta_\tau}\,.
    \ee
    According to \cite{Bobev:2020pjk}, the preserved Killing spinors are antiperiodic under $\tau\mapsto\tau +\beta_\tau$ in the frame satisfying \eqref{eq: sugra BPS orig}. After transforming by $e^{-\frac{i}{2}g\sum_Ie^I|_\text{IR}\tau}=e^{\mp i\xi\frac{\pi\tau}{\beta_\tau}}$ under the gauge transformation, they are now periodic in $\tau$. The holographic dictionary then dictates that the black saddles have the correct boundary conditions to contribute to the topologically twisted index on $\Sigma_\fg\times S^1$, where the chemical potentials are identified as 
    \be\label{eq: Delta grav iden}
    \hat{\Delta}_I = -\beta_\tau g e^I|_\text{UV} +2\pi n_I\,,\quad n_I\in\bZ\,, \quad \sum_{I=1}^4n_I=\textrm{even}.
    \ee
    The shifts by $2\pi n_I$ come from $e^I|_\text{UV}\mapsto e^I|_\text{UV}-\frac{2\pi n_I}{\beta_\tau g}$, which are large gauge transformations at the boundary that do not affect the asymptotic boundary conditions \cite{Aharony:2021zkr}. Note that $\sum_{I=1}^4n_I=\text{even}$ is required, otherwise the gauge transformation would change the periodicity of spinors. The shifted solutions are geometrically distinct in the bulk because they have different values of $e^I|_\text{UV}$. We therefore have an infinite family of black saddle solutions contributing to the same partition function. Apart from $e^I|_\textrm{UV}$, the boundary conditions at $r\rightarrow\infty$ also require specifying the real masses $\sigma_i\sim z_i|_\textrm{UV}-\tilde{z}_i|_\textrm{UV}$, which are set to zero in our field theory computation. In this case, the UV-IR relation in equation (3.83) of \cite{Bobev:2020pjk} states that the normalized sections $\hat{u}^I$ appearing in the on-shell action only contain $e^I|_\textrm{UV}$:
    \be
    \hat{u}^I = \left.\pm\frac{2\pi X^I}{\sum_{I=1}^4X^I}\right|_\textrm{IR} = -\beta_\tau ge^I|_\textrm{UV}\,,
    \ee
    where the $\pm$ sign above corresponds to the $\pm$ branches of Lorentzian black holes discussed in Section \ref{subsec:oneCut:bulk}. Recall from \eqref{eq: iden X w Delta} there that the same normalized section $\hat{u}^I$ must be identified with $[\hat{\Delta}_I]$ in the $+$ branch and $[\hat{\Delta}_I]'$ in the $-$ branch. Therefore the allowed integer shifts in \eqref{eq: Delta grav iden} are 
    \be\label{eq: delta shift iden}
    -\beta_\tau g e^I|_\text{UV} = \hat{u}^I = \begin{cases}  [\hat{\Delta}_I]\,, & +\textrm{ branch} \\  [\hat{\Delta}_I]'\,, & -\textrm{ branch}  \end{cases}\,.
    \ee
    As we saw in \eqref{eq: M2 action 12 instanton}, this identification ensures that the action of wrapped M2 branes is real and positive, meaning that the allowed gravitational saddles are stable with respect to brane condensation, as was found in \cite{Aharony:2021zkr}. \eqref{eq: delta shift iden} also implies that the BPS constraint \eqref{eq: wilson line BPS constr} matches the field theory relations
    \be\label{eq: branch bracket constr}
    \sum_{I=1}^4 [\hat{\Delta}_I] = 2\pi\,,\quad \sum_{I=1}^4 [\hat{\Delta}_I]'= -2\pi\,
    \ee
    in \eqref{eq: logZ BH}, \eqref{eq: logZ BH branch 2}, provided that
    \be
    \xi = \begin{cases}+1\,, & +\textrm{ branch} \\ -1\,, & -\textrm{ branch}\end{cases}\,.
    \ee
    This can also be derived independently because as \cite{Hristov:2010ri} has shown, the $+$ and $-$ branches of black holes differ by a sign in the projections satisfied by the Killing spinor, which are given in equation (4.7) of \cite{Hristov:2010ri}. This exactly corresponds to the choice of sign controlled by $\xi$ in the Killing spinor projections in equation (3.24) of \cite{Bobev:2020pjk}.
    
    We now proceed to construct orbifolds of the black saddles, following the same logic as \cite{Aharony:2021zkr}. Using the result of \cite{Azizi:2016noi}, any solution of the STU model can be uplifted to $11$d supergravity. Denoting the coordinates of $\bR^8$ as $x^{1,\ldots,8}$, we parametrize $S^7$ by the embedding
    \bea
    x^1+ix^2=\mu_1e^{i\phi_1}\,,\quad x^3+ix^4=\mu_2e^{i\phi_2}\,,&\quad x^5+ix^6=\mu_3e^{i\phi_3}\,,\quad x^7+ix^8=\mu_4 e^{i\phi_4}\,,\\
    &\sum_{I=1}^4\mu_I^2=1\,,
    \eea
    where $\phi_I$ have periods $2\pi$. From the uplift ansatz in \cite{Azizi:2016noi}, one can observe that there is no dependence on the angular coordinates $\phi_I$ in any component of the metric or $3$-form potential. We now consider the uplift of a black saddle whose parameters include the period $\tilde{\beta}_\tau$ of $\tau$, and $\tilde{e}^I|_\text{UV}$. In the following, parameters of the ``parent'' black saddle will be denoted with a tilde, while those of its orbifold will not contain tildes. Since the components of the metric and gauge fields are independent of $\tau$, the uplifted solution will also be independent of $\tau$. The coordinates are identified as
    \be\label{eq: BH coord iden}
    (\tau,\phi_I)\cong (\tau+\tilde{\beta}_\tau,\phi_I)\,,\quad (\tau,\phi_I)\cong(\tau,\phi_I+2\pi n_I)\,,\quad n_I\in\bZ\,.
    \ee
    As such,
    \be
    (\tau,\phi_I)\mapsto \left(\tau +\frac{\tilde{\beta}_\tau}{m},\phi_I+\frac{2\pi s_I}{m}\right)\,,\quad s_I=0,\ldots,m-1
    \ee
    is a $\bZ_m$ isometry of the solution that we can orbifold by to get another solution. In the orbifold, the identification along $\tau$ is modified to
    \be\label{eq: orbifold coord iden}
    (\tau,\phi_I)\cong \left(\tau +\frac{\tilde{\beta}_\tau}{m},\phi_I+\frac{2\pi s_I}{m}\right)\,,\quad s_I=0,\ldots,m-1\,.
    \ee
    In the uplifted black saddle, the connection $\tilde{e}^I$ along $\tau$ is encoded in the off-diagonal components of the metric. This is because the gauge fields only appear in the metric through the covariant derivative $D\phi_I\equiv d\phi_I-gA^I$. To eliminate the $\phi$-$\tau$ components of the metric at $r\rightarrow\infty$ and read off $\tilde{e}^I|_\text{UV}$ from the coordinate identifications, we can perform the change of coordinates
    \be\label{eq: coord change to shift frame}
    \hat{\phi}_I\equiv\phi_I-g\tilde{e}^I|_\text{UV}\tau\,,
    \ee
    after which \eqref{eq: BH coord iden} becomes
    \be\label{eq: BH coord iden shift}
    (\tau,\hat{\phi}_I)\cong (\tau+\tilde{\beta}_\tau,\hat{\phi}_I - g\tilde{e}^I|_\text{UV}\tilde{\beta}_\tau)
    \ee
    If we perform the same change of coordinates \eqref{eq: coord change to shift frame} in the orbifold, the identification \eqref{eq: orbifold coord iden} becomes
    \be\label{eq: orbifold coord iden shift}
    (\tau,\hat{\phi}_I)\cong \left(\tau+\frac{\tilde{\beta}_\tau}{m},\hat{\phi}_I-\frac{g\tilde{e}^I|_\text{UV}\tilde{\beta}_\tau}{m}+\frac{2\pi s_I}{m}\right)\,.
    \ee
    Comparing \eqref{eq: BH coord iden shift} with \eqref{eq: orbifold coord iden shift}, we see that the $\tau$ periodicity $\beta_\tau$ and connection $e^I|_\text{UV}$ of the orbifold are
    \be
    \beta_\tau = \frac{\tilde{\beta}_\tau}{m}\,,\quad e^I|_\text{UV} = \tilde{e}^I|_\text{UV} - \frac{2\pi s_I}{g\tilde{\beta}_\tau}\,.
    \ee
    Substituting into \eqref{eq: Delta grav iden}, one concludes that the orbifold contributes to the index with chemical potentials
    \be
    \hat{\Delta}_I = -\frac{\tilde{\beta}_\tau g}{m}\left(\tilde{e}^I|_\text{UV} - \frac{2\pi s_I}{g\tilde{\beta}_\tau}\right) = \begin{cases} \frac{[\tilde{\Delta}_I] + 2\pi s_I}{m} \,, & +\textrm{ branch} \\ \frac{[\tilde{\Delta}_I]' + 2\pi s_I}{m} \,, & -\textrm{ branch} \end{cases}\,, 
    \ee
    where we have used the fact that the black saddle before the orbifold contributes to the index with chemical potentials $\tilde{\Delta}_I$ such that
    \be
    -\tilde{\beta}_\tau g\tilde{e}^I|_\textrm{UV}=\begin{cases} [\tilde{\Delta}_I]\,, & +\textrm{ branch}\,,\quad \sum_{I=1}^4[\tilde{\Delta}_I]=2\pi \\ [\tilde{\Delta}_I]'\,, & -\textrm{ branch}\,,\quad \sum_{I=1}^4[\tilde{\Delta}_I]'=-2\pi\end{cases}
    \ee
    Depending on whether we started with a black saddle in the $+$ or $-$ branch, the on-shell action of its orbifold should be the same function of $[\tilde{\Delta}_I]$ or $[\tilde{\Delta}_I]'$ respectively, divided by $m$. We therefore get that the on-shell action of the orbifolds are
    \be\label{eq: orbifold logZ}
    \begin{cases} \frac{1}{m\cdot 3}\sqrt{2[m\hat{\Delta}_1][m\hat{\Delta}_2][m\hat{\Delta}_3][m\hat{\Delta}_4]}\sum_{I=1}^4\frac{\hat{\fn}_I}{[m\hat{\Delta}_I]}\,,\qquad\hspace{.5cm} [m\hat{\Delta}_I] \equiv m\hat{\Delta}_I - 2\pi s_I\,, & +\text{ branch}\\[.2cm]
    -\frac{1}{m\cdot 3}\sqrt{2[m\hat{\Delta}_1]'[m\hat{\Delta}_2]'[m\hat{\Delta}_3]'[m\hat{\Delta}_4]'}\sum_{I=1}^4\frac{\hat{\fn}_I}{[m\hat{\Delta}_I]'}\,,\quad [m\hat{\Delta}_I]' \equiv m\hat{\Delta}_I - 2\pi s_I\,, & -\text{ branch}\end{cases}
    \ee
    In addition, due to the ranges of $m\hat{\Delta}_I-2\pi s_I=[\tilde{\Delta}_I]\in(0,2\pi)$ for the $+$ branch and $m\hat{\Delta}_I-2\pi s_I=[\tilde{\Delta}_I]'\in(-2\pi,0)$ for the $-$ branch, we must have $s_I=\left\lfloor\frac{m\hat{\Delta}_I}{2\pi}\right\rfloor$ and $s_I=\left\lceil\frac{m\hat{\Delta}_I}{2\pi}\right\rceil$ for the $+$ and $-$ branches respectively. Under this identification, there is an exact match between \eqref{eq: orbifold logZ} and the large $N$ free energies of the $\bZ_2$ and $\bZ_3$ orbifolds in \eqref{eq: 2cut LO logZ plus}, \eqref{eq: 2cut LO logZ minus} and \eqref{eq: 3cut logZ plus}, \eqref{eq: 3cut logZ minus} respectively, when $m$ is set to $2$ and $3$. Since our identification between the field theory fluxes and supergravity fluxes in \eqref{eq: g q dict} differs from the identification in \cite{Bobev:2020pjk} by a factor of $\xi$, this gives rise to the relative sign difference in \eqref{eq: orbifold logZ} between the two branches.
    
    For the orbifolds to be supersymmetric in the fundamental domain, we need $\sum_{I=1}^4\hat{\Delta}_I=2\pi$, i.e. 
    \be
    \sum_{I=1}^4\hat{\Delta}_I = \frac{2\pi}{m}\left(\pm 1 + \sum_{I=1}^4 s_I\right) = 2\pi\,,
    \ee
    where the upper and lower signs hold for the $+$ and $-$ branches in that order, and we have used $\sum_I[\tilde{\Delta}_I]=2\pi$, $\sum_I[\tilde{\Delta}_I]'=-2\pi$. For orbifolds of black saddles in the $+$ branch, this requires $\sum_{I=1}^4s_I=m-1$, while orbifolds of black saddles in the $-$ branch require $\sum_{I=1}^4s_I=m+1$.  Therefore, the supersymmetry of the orbifolds require
    \be\label{eq: orbifold susy cond}
    \begin{cases}+\text{ branch:} & \sum_{I=1}^4s_I = m-1\,,\quad s_I=\left\lfloor\frac{m\hat{\Delta}_I}{2\pi}\right\rfloor \\[3ex] -\text{ branch:} & \sum_{I=1}^4s_I = m+1\,,\quad s_I=\left\lceil\frac{m\hat{\Delta}_I}{2\pi}\right\rceil\end{cases}
    \ee
    We shall show that the regions in $\Delta_{1,2}$, $\Delta_C$ which are compatible with \eqref{eq: orbifold susy cond} for $m=2,3$ exactly match the regions in which the 2-cut and 3-cut solutions were found to exist numerically. 
    
    For $m=2$, let us consider orbifolds of black saddles in the $+$ branch first. In the fundamental domain, the possible values of $s_I=\left\lfloor\frac{m\hat{\Delta}_I}{2\pi}\right\rfloor$ are $s_1 = 0,1$, $s_2 = 0$, $s_3=0,1$ and $s_4 = -1,0$. The $3$ combinations of $s_I$ which sum to $1$ are $s_I=\{1,0,0,0\}$,  $\{0,0,1,0\}$, or $\{1,0,1,-1\}$. One can check that the regions which are compatible with these $3$ choices are $\wt 3$, $\wt 1$ and $\wt 3'$ respectively, as defined in Table \ref{tab: 2col reg}. For orbifolds of black saddles in the $-$ branch, the possible values of $\left\lceil\frac{m\hat{\Delta}_I}{2\pi}\right\rceil$ are $s_1 = 1,2$, $s_2 = 1$, $s_3 = 1,2$, and $s_4 = 0,1$. One can check that the only combination that is consistent with the fundamental domain, and whose sum is $3$ is $\{1,1,1,0\}$. This corresponds to region $\tilde{2}$ in Table \ref{tab: 2col reg}. Thus, the solutions to \eqref{eq: orbifold susy cond} when $m=2$ precisely reproduce all the regions $\tilde{1}$, $\tilde{2}$, $\tilde{3}$, and $\tilde{3}'$ where 2-cut solutions exist.   
    
    For $m = 3$, considering orbifolds of black saddles in the $+$ branch, the possible values of $s_I=\left\lfloor\frac{m\hat{\Delta}_I}{2\pi}\right\rfloor$ in the fundamental domain are $s_1 = 0,1,2$, $s_2 = 0,1$, $s_3 = 0,1,2$, and $s_4 = -2,-1,0,1$. On the other hand, for orbifolds of black saddles in the $-$ branch, the possible values of $s_I=\left\lceil\frac{m\hat{\Delta}_I}{2\pi}\right\rceil$ are $s_1 = 1,2,3$, $s_2 = 1,2$, $s_3 = 1,2,3$, and $s_4 = -1,0,1,2$. If we test each combination of $s_I$'s, keeping only those which are consistent with \eqref{eq: orbifold susy cond}, one can check that we precisely get the options in Table \ref{tab: 3cut reg}. 

    %%%%%%%%%%%%%%%%%%%%%%%%
	\subsubsection{ADE enhancement and 7d YM theory}
	\label{subsubsec:holography:orbifold:ADE}
	%%%%%%%%%%%%%%%%%%%%%%%%
	
	For orbifolds whose corresponding $s_I$ have only one nonzero component, the supersymmetry condition \eqref{eq: orbifold susy cond} implies that either $s_I=(m-1,0,0,0)$ or $s_I=(m+1,0,0,0)$ up to permutations of $I$, corresponding to orbifolds in the $+$ or $-$ branch respectively. Since the $\bZ_2$ and $\bZ_3$ orbifolds which exist in the fundamental domain only contain instances of $s_I=(m-1,0,0,0)$ and permutations thereof, as seen in Tables \ref{tab: 2col reg} and \ref{tab: 3cut reg}, we will mainly focus on this case in the following discussion. 

    Denoting the complex coordinate of the near-horizon cigar geometry in the black saddle as $w\equiv re^{2\pi i\tau/\tilde{\beta}_\tau}$, and defining $z_I\equiv \mu_I e^{i\phi_I}$, one can see from \eqref{eq: orbifold coord iden} that the orbifold action corresponding to $s_I=(m-1,0,0,0)$ is 
	\begin{equation}\label{eq: ADE Zm action}
		(w, z_1) \rightarrow (e^{2\pi i/m}w, e^{-2\pi i/m} z_1)\,.
	\end{equation}
	Its fixed locus is $w=z_1=0$, where $w=0$ restricts to the horizon at $r=0$, while $z_1=0$ defines a $S^5\subset S^7$. We therefore get a seven-dimensional fixed locus $\Sigma_\fg\times S^5$ at the horizon. The transverse space to the fixed locus is the singularity $\bC^2/\bZ_m$ of type $A_{m-1}$, where the $\bZ_m$ action is defined by \eqref{eq: ADE Zm action}. The resolution of $\bC^2/\bZ_m$ contains $m-1$ exceptional 2-cycles $C_a\cong\bP^1$, $a=1,2,\cdots, m-1$. The three-form of M-theory wraps these 2-cycles to give $m-1$ gauge fields along the seven-dimensional fixed locus on which the surface singularity is fibered. Away from the singular limit, this is a gauge theory with abelian gauge group $\U(1)^{m-1}$. At the orbifold point the exceptional cycles collapse, the states created by M2-branes wrapping the exceptional cycles become massless, and the gauge group enhances to the simply-connected ADE group $\SU(m)$. This is the standard ADE enhancement mechanism in M-theory on K3/ALE singularities \cite{Witten:1995ex, Aspinwall:1995zi, Douglas:1996xg, Sen:1997kz}. 
	
	The gauge theory living on the seven-dimensional fixed locus contains a gauge field $A_\alpha$, three scalars $\phi^{u= 1,2,3} $, and a symplectic-Majorana spinor $\lambda$, all in the adjoint representation of $\SU(m)$. The theory couples to 11d supergravity non-trivially \cite{Horava:1996ma, Anderson:2006pb, Barrett:2006sn}, with the standard Yang-Mills coupling scaling as
	\begin{equation}\label{eq: 7d YM coupling}
		g_7^2 \sim \kappa_{11}^{2/3} \sim \ell_p^3 \propto N^{-1/2}.
	\end{equation}
	The classical action vanishes as there are no background fields turned on, but the one-loop determinant contributes non-trivially at orders $O({\rm log}\, N)$ and $O(1)$, the two-loop corrections contribute at $O(N^{-1/2})$, and so on. This explains why the free energies of 2-cut and 3-cut solutions cannot be resummed into forms such as \eqref{eq:TTI-twocut} or \eqref{eq:TTI-threecut} when the corresponding $s_I$ have 3 zero components and 1 nonzero component, such as in region $\tilde{1}$ in Table \ref{tab: 2col reg} and regions $\tilde{1}'$ and $\tilde{4}$ in Table \ref{tab: 3cut reg}. There are extra contributions to the free energy coming from the 7d $\SU(m)$ gauge theory living on the fixed locus, while \eqref{eq:TTI-twocut} or \eqref{eq:TTI-threecut} only accounts for the free energy of M-theory in the bulk.

    %%%%%%%%%%%%%%%%%%%%%%%%
	\subsubsection{Logarithmic corrections}
	\label{subsubsec:holography:log}
	%%%%%%%%%%%%%%%%%%%%%%%%

    To begin with, we compute the $\cO(\log N)$ correction to the free energy of the orbifolds due to corrections in M-theory. For the black hole background, this has already been considered in \cite{Liu:2017vbl}, which we now review. By dimensional analysis, $\log N$ terms in the free energy can only come from the 1-loop determinant, which can be evaluated in terms of a heat kernel expansion \cite{Vassilevich:2003xt}. As we are working in odd dimensions ($d=11$), the relevant heat kernel coefficients are zero, and we only need to consider zero mode contributions. It turns out that the only contribution comes from the zero mode of the two-form ghost arising in the gauge fixing of the M-theory three-form $C_3$, and it evaluates to
	\begin{equation} \label{eq:logNbulk}
		c_{\rm log} = \frac{1}{2}(1-\fg) .
	\end{equation} 
	This zero mode is invariant under shifts in $\tau$ and $\phi_I$, and thus it survives the orbifold projection. For the examples of 2-cut and 3-cut solutions in Tables \ref{tab: 2col reg} and \ref{tab: 3cut reg} where at least two components of $s_I$ are nonzero, the gravitational contribution \eqref{eq:logNbulk} is consistent with the coefficients $c_{\log}$ computed numerically in field theory, as can be seen from \eqref{eq:TTI-twocut} and Table \eqref{tbl:SC3colCoe}. Otherwise, if $s_I=(m-1,0,0,0)$ or permutations thereof, the coefficients $c_{\log}$ obtained numerically differ from \eqref{eq:logNbulk}, but this is precisely the case where the contributions from the 7d $\SU(m)$ gauge theory living on the fixed locus of the orbifold action must be accounted for, as explained in Section \ref{subsubsec:holography:orbifold:ADE}. We will show in the following that the total $\cO(\log N)$ correction coming from M-theory and the 7d gauge theory exactly reproduces the numerical values of $c_{\log}$ in these cases. More generally, for $m>3$, the fundamental domain also includes cases where $s_I=(m+1,0,0,0)$, in which the contribution of the fixed locus must also be considered. This case, and the contributions of a general fixed locus can be found in Appendix \ref{App:Log}.

    Due to the scaling of the 7d gauge coupling in \eqref{eq: 7d YM coupling}, the $\log N$ contribution in the free energy of the gauge theory can only come from the 1-loop determinant, as it was in the gravitational free energy. Since we are still working in an odd dimension ($d=7$), the heat kernel coefficients are zero, and the 1-loop determinant only receives contributions from the zero modes of the 7d fields on the fixed locus $\Sigma_\fg\times S^5$. Recall from Section \ref{subsubsec:holography:orbifold:ADE} that the field content consists of 3 scalars $\phi^u$, a symplectic-Majorana gaugino $\lambda$, and a gauge field $A$, all in the adjoint representation of $\SU(m)$. We will consider each of these in turn. 

    \paragraph{Scalar and fermion zero modes}
	
	Naively, the zeroth Betti number $b_0(\Sigma_\fg \times S^5) = 1$ gives the number of zero modes $n_\phi^0$ of a scalar, but this local argument is not enough in the uplift of the dyonic black hole. The fixed locus has the form of $S^5$ fibered over $\Sigma_\fg$, and the 7d action contains kinetic terms with covariant derivatives and flux-induced mass terms schematically of the form
	\begin{equation}
		\widehat{D}_\mu \phi \widehat{D}^\mu \phi,\quad M_{\rm flux}^2\phi^2 \sim F^I_{\mu\nu} F^{J\mu\nu} \phi^2\,.
	\end{equation}
	The 3 scalars transform as a triplet under the $\SU(2)_R$ isometry group of the transverse space $\bC^2/\bZ_m$, and thus the covariant derivative $\widehat{D}$ contains an induced background $\SU(2)_R$ connection. Due to the topological twist in the dyonic black hole, the transverse space $\bC^2/\bZ_m$ in its uplift is fibered nontrivially over $\Sigma_\fg$, and the $\SU(2)_R$ connection in $\widehat{D}$ has nontrivial flux $F_I$, as written in the second term above. Since the scalars are sections of this $\SU(2)_R$ bundle, their zero modes must be covariantly constant and also annihilated by the flux-induced algebraic mass matrix:
	\begin{equation}
		\widehat{D}\phi = 0,\quad M^2_{\rm flux} \phi = 0.
	\end{equation}
	For generic magnetic fluxes, the $\SU(2)_R$ bundle has nonzero curvature and has no globally covariantly constant triplet section. We thus have no scalar zero modes:
	\begin{equation}
		n_\phi^0 = 0.
	\end{equation}
	
	As for the fermion, a bare product Dirac operator on $\Sigma_\fg \times S^5$ has no zero modes because the Dirac operator on a round $S^5$ has a gap. In the uplifted dyonic black hole, the actual operator is twisted by the normal-bundle $SU(2)_R$ connection. Such a twist can change the kernel, but if the scalar blow-up modes are lifted, their modulino partners are lifted as well due to supersymmetry. We therefore do not include an additional gaugino zero-mode contribution.

    \paragraph{Zero modes in the gauge sector}
    
	For the gauge field $A$, the finite configurations on which the Yang-Mills action is zero are flat connections satisfying $F=dA+A\wedge A=0$. Expanding around the trivial connection $A=0$ as
    \be
    A = a_1\delta + a_2\delta^2 + \ldots\,,\quad \delta \ll 1\,,
    \ee
    the zero modes that we want to count are then the infinitesimal fluctuations $a_1$ which can be completed by $a_2$, $a_3$ etc. to form a flat connection. To first order in $\delta$, flatness requires $da_1 = 0$, which means that $a_1$ is a closed one-form. One must also take into account that $a_1$ includes degrees of freedom which are pure gauge. If we similarly expand the gauge transformation parameter as $\epsilon = \epsilon_1 \delta + \epsilon_2 \delta^2 + \ldots$, the gauge transformation of $A$ at $\cO(\delta)$ sends $a_1\mapsto a_1 + d\epsilon_1$. Thus the space of closed one-forms modulo gauge redundancies is precisely $H^1(\Sigma_\fg\times S^5, \su(m))$. By the K\"unneth formula, and since $H^1(S^5)=0$, its dimension is 
	\begin{equation} \label{eq:1formZMnumberNaive}
		\dim H^1(\Sigma_\fg \times S^5, \su(m)) = \dim H^1(\Sigma_\fg , \su(m)) \otimes H^0(S^5) = 2\fg\, \dim\su(m) = 2\fg (m^2 - 1).
	\end{equation}
    At $\cO(\delta^2)$, the flatness condition requires $da_2 + a_1 \wedge a_1=0$, which implies that $a_1\wedge a_1$ is cohomologically trivial. From the first equality in \eqref{eq:1formZMnumberNaive}, it is clear that $a_1\in H^1(\Sigma_\fg\times S^5, \su(m)) \cong H^1(\Sigma_\fg,\su(m))$ is simply an $\su(m)$-valued cohomology class on $\Sigma_\fg$. We choose a standard basis of harmonic one-forms $\{\alpha_{i=1,\ldots,\fg}, \beta_{i=1,\ldots,\fg}\}$ for $H^1(\Sigma_\fg)$, where $\alpha_i$ and $\beta_i$ are dual to the canonical $A_i$ and $B_i$ cycles of $\Sigma_\fg$ under the natural pairing defined by integration. In addition, $\alpha_i\wedge\beta_j = \delta_{ij}d\text{vol}_{\Sigma_\fg}$, where $d\text{vol}_{\Sigma_\fg}$ is the volume form on $\Sigma_\fg$. Expanding $a_1$ as
    \be
    a_1 = \sum_{i=1}^{\fg}(X_i\alpha_i+Y_i\beta_i)\,,\quad X_i\,,Y_i\in\su(m)\,,
    \ee  
    the flatness condition becomes
	\begin{equation}\label{eq: quad flat restr}
		\sum_{i=1}^{\fg} [X_i, Y_i] = 0\,,
	\end{equation}
	because $d\text{vol}_{\Sigma_\fg}$ is cohomologically non-trivial. This imposes $m^2 - 1$ constraints on $a_1$. Once \eqref{eq: quad flat restr} is satisfied, one can always find a one-form $a_2$ such that $da_2=-a_1\wedge a_1$. At this point, the number of gauge-inequivalent zero modes in $a_1$ is
	  \begin{equation}\label{eq: num 1-form zero modes}
	  	n_A^0={\rm dim}\, H^1 (\Sigma_\fg \times S^5, {\rm ad}\, SU(m))-(m^2-1) = (2\fg - 1)(m^2-1).
	  \end{equation}
    In principle, one should also examine the flatness condition at successive orders $\cO(\delta^3)$, $\cO(\delta^4)$, etc., which could place further constraints on $a_1$. However, Theorem 1 in \cite{Goldman:1988dem} guarantees that the higher order equations can all be satisfied by appropriately choosing $a_2$, $a_3$, etc., and do not constrain $a_1$. Thus the counting of one-form zero modes in \eqref{eq: num 1-form zero modes} is complete. For a bosonic one-form zero mode on a seven-dimensional manifold, its contribution to the 1-loop determinant is \footnote{The constant $\beta_p=\frac{D}{2}-p$ comes from correctly normalizing the integration measure of a $p$-form field on a $D$-dimensional base. } \cite{Bhattacharyya:2012ye,Banerjee:2010qc}
	\begin{equation}
		\Delta F_A^{(1)} = -(\beta_1 - 1) \log L = - \frac{3}{2} \log L,\quad \beta_1 = \frac{7}{2} - 1 = \frac{5}{2}\,,
	\end{equation}
	where $L$ is the AdS radius. The total contribution from the zero modes in $a_1$ is therefore
	\begin{equation}
		\Delta F_{A}  =  n_A^0 \cdot \left( - \frac{3}{2} \log L  \right) = -\frac{3}{2} (2\fg - 1)(m^2-1) \log L.
	\end{equation}
    
	In the gauge sector, we also need to consider the contribution of ghost zero modes. Recall that the gauge transformations generated by non-constant $\epsilon_1$ have already been fixed when restricting $a_1$ to an equivalence class under $a_1 \sim a_1 + d\epsilon_1$ in \eqref{eq:1formZMnumberNaive}. The ghost zero modes exactly correspond to residual constant gauge transformations, which are generated by constant adjoint valued scalars. Their number is 
	\begin{equation}
		n_{\rm ghost}^0={\rm dim}\, H^0(\Sigma_\fg \times S^5, \su(m)) = {\rm dim} \, \su(m) = m^2 - 1.
	\end{equation}
	So the one-loop correction from the scalar ghosts is
	\begin{equation}
		\Delta F_{\rm ghost} = + (\beta_0 - 2) (m^2 - 1) \log L = \frac{3}{2} (m^2-1) \log L. 
	\end{equation}
    Combining the contributions of the gauge field and ghosts, we obtain that the total logarithmic correction from the $\SU(m)$ gauge sector is
	  \begin{equation} \label{eq:ADElog}
	  	\Delta F_{\rm gauge} = 3( 1 - \fg )(m^2-1) \log L.
	  \end{equation}
	  
    Using the holographic dictionary $L \propto N^{1/6}$, and summing the gravitational contribution in \eqref{eq:logNbulk} and the gauge theory contribution in \eqref{eq:ADElog}, the total logarithmic correction is given by 
	  \begin{equation}
	  	c_{\rm log, total} = \frac{1}{2}(1-\fg) m^2  ,
	  \end{equation}
	  exactly reproducing the results in \eqref{eq:CI} and for regions $\tilde{1}'$ and $\tilde{4}$ in Table \ref{tbl:SC3colCoe}. 

    %%%%%%%%%%%%%%%%%%%%%%%%
	\subsubsection{Comments on the logarithmic corrections from the 4d viewpoint}
	\label{subsubsec:holography:log:4dpers}
	%%%%%%%%%%%%%%%%%%%%%%%%
	
	We now adopt the four-dimensional KK supergravity description of \cite{Bobev:2023dwx}. From the four-dimensional perspective, the field content does not contain massless two-form fields. Thus the non-local contribution to the logarithmic correction vanishes. The bootstrap constraints on the total heat-kernel coefficients then give
	\begin{equation}
		C_{\mathrm{non\text{-}local}}=0,\qquad
		c^{\mathrm{tot}}=b_1^{\mathrm{tot}}=b_2^{\mathrm{tot}}=0.
	\end{equation}
	The remaining local contribution is proportional to the Euler density,
	\begin{equation}
		\label{eq:4d-Euler-density}
		(4\pi)^2 a_4^{\mathrm{tot}}=-a_E^{\mathrm{tot}}E_4,\qquad
		E_4=R_{\mu\nu\rho\sigma}R^{\mu\nu\rho\sigma}
		-4R_{\mu\nu}R^{\mu\nu}+R^2.
	\end{equation}
	Here the total coefficients include the full KK spectrum and its regularization. With the conjectured normalization $a_E^{\mathrm{tot}}=1/6$ of \cite{Bobev:2023dwx}, the logarithmic term in $F=-\log Z$ is
	\begin{equation}
		\label{eq:4d-log-Euler}
		F_{\log}(X)=a_E^{\mathrm{tot}}\chi(X)\log(L^2/G_4)
		=\frac{\chi(X)}{4}\log N,
	\end{equation}
	where we used $L^2/G_4\propto N^{3/2}$ at fixed discrete parameters. The dependence on the Euler characteristic is also consistent with the four-dimensional localization analysis of \cite{Hristov:2021zai}. For the orbifold backgrounds considered below, we assume that this prescription extends by evaluating the Euler functional with its conical Gauss--Bonnet completion. This extension is the physical input; our task here is to compute its geometric factor.
	
	Let $X$ be the smooth black saddle of \eqref{eq: black saddle ansatz}, whose Euclidean time circle contracts at the bolt $\Sigma_\fg$. The $\bZ_m$ quotient rotates the normal plane to the bolt by $2\pi/m$. The quotient $X_m=X/\bZ_m$ has the local metric near the cap
	\begin{equation}
		\label{eq:4d-orbifold-cone}
		ds^2=d\rho^2+m^{-2}\rho^2d\theta^2
		+h_{ab}(y)dy^a dy^b+\cdots,\qquad
		\theta\sim\theta+2\pi,
	\end{equation}
	where $h$ is the induced metric on $\Sigma_\fg$. The normal cone has deficit angle $2\pi(1-m^{-1})$. Since the bolt is a fixed surface of the circle isometry, its extrinsic curvatures vanish.
	
	To evaluate the Euler contribution at the cone tip, replace the normal cone by a smooth cap with metric
	\begin{equation}
		ds_{\perp,\epsilon}^2=d\rho^2+f_\epsilon(\rho)^2d\theta^2,\qquad
		f_\epsilon(0)=0,\quad f_\epsilon'(0)=1,\quad
		f_\epsilon'(\epsilon)\longrightarrow m^{-1}.
	\end{equation}
	The last condition matches the cap to the cone as $\epsilon\to0$. Its scalar curvature is $R_\perp=-2f_\epsilon''/f_\epsilon$, so
	\begin{equation}
		\label{eq:4d-cone-curvature}
		\lim_{\epsilon\to0}
		\int_{\mathrm{cap}_\epsilon}d^2x\sqrt{g_\perp}\,R_\perp
		=-4\pi\lim_{\epsilon\to0}
		\bigl[f_\epsilon'(\epsilon)-f_\epsilon'(0)\bigr]
		=4\pi(1-m^{-1}).
	\end{equation}
	For a product of two two-dimensional metrics, the Euler density is $E_4=2R_\perp R_\Sigma$. This also determines the singular part at the bolt: the antisymmetric structure of the Euler density removes terms quadratic in the singular normal curvature, while the vanishing extrinsic curvatures identify the remaining tangential curvature with $R_\Sigma$. All bounded contributions vanish in the shrinking-cap integral. Hence
	\begin{equation}
		\label{eq:4d-bolt-Euler}
		\begin{aligned}
			\Delta_\Sigma\chi
			&=\frac{1}{32\pi^2}\lim_{\epsilon\to0}
			\int_{\mathrm{cap}_\epsilon\times\Sigma_\fg}d^4x\sqrt g\,E_4\\
			&=\frac{8\pi(1-m^{-1})}{32\pi^2}
			\int_{\Sigma_\fg}d^2y\sqrt h\,R_\Sigma
			=(1-m^{-1})\chi(\Sigma_\fg).
		\end{aligned}
	\end{equation}
	In the last step we used the two-dimensional Gauss--Bonnet theorem. This is the conical Euler contribution derived in \cite{Fursaev:1995ef}; it is exact at finite deficit angle.
	
	To include the asymptotic boundary, first introduce a radial cutoff and denote the resulting spaces by $X_R$ and $X_{m,R}$. Let $\mathcal B_{\mathrm{GB}}(\partial X_{m,R})$ be the normalized outer Gauss--Bonnet boundary term. Away from the bolt, the local geometry is unchanged and the time-circle integration range is reduced by $m^{-1}$. Both the regular bulk integral and the outer boundary term therefore scale by $m^{-1}$. Adding \eqref{eq:4d-bolt-Euler} gives
	\begin{equation}
		\label{eq:4d-conical-GB}
		\begin{aligned}
			\chi_{\mathrm{top}}(X_{m,R})
			&=\frac{1}{32\pi^2}
			\int_{X_{m,R}\setminus\Sigma_\fg}d^4x\sqrt g\,E_4
			+\mathcal B_{\mathrm{GB}}(\partial X_{m,R})
			+(1-m^{-1})\chi(\Sigma_\fg)\\
			&=m^{-1}\chi(X_R)+(1-m^{-1})\chi(\Sigma_\fg).
		\end{aligned}
	\end{equation}
	Here $\chi_{\mathrm{top}}$ is the ordinary topological Euler characteristic, including the cone-tip contribution. The identity holds at finite $R$ and remains valid when the cutoff is removed with the corresponding Euler boundary completion. Since $X_R$ has topology $D^2\times\Sigma_\fg$, we obtain
	\begin{equation}
		\label{eq:4d-orbifold-Euler}
		\chi_{\mathrm{top}}(X/\bZ_m)
		=\frac{1}{m}\,2(1-\fg)
		+\left(1-\frac{1}{m}\right)2(1-\fg)
		=2(1-\fg).
	\end{equation}
	As a topological check, the underlying quotient space is $(D^2/\bZ_m)\times\Sigma_\fg\simeq D^2\times\Sigma_\fg$.
	
	Substituting \eqref{eq:4d-orbifold-Euler} into \eqref{eq:4d-log-Euler} yields
	\begin{equation}
		\label{eq:4d-orbifold-log}
		F_{\log}(X/\bZ_m)=\frac{1-\fg}{2}\log N.
	\end{equation}
	The bolt contribution to the four-dimensional Euler term thus supplies the part omitted by simply dividing the smooth result by $m$. The agreement with the numerical results supports the assumed conical extension of the four-dimensional prescription. This conclusion concerns the sector described by that prescription; additional ADE enhancement sectors require their own contributions.

    %%%%%%%%%%%%%%%%%%%%%%%%
	\subsection{General expectations for $m$-cut solutions}
	\label{subsec:multi-cut:mcut}
	%%%%%%%%%%%%%%%%%%%%%%%%

    %%%%%%%%%%%%%%%%%%%%%%%%
	\subsubsection{Regions of existence of $m$-cut solutions}
	\label{subsubsec:multi-cut:mcut:distribution}
	%%%%%%%%%%%%%%%%%%%%%%%%
	
	Recall that the conditions for supersymmetric $\bZ_m$ orbifolds of black saddles to exist are given in \eqref{eq: orbifold susy cond}. They are analysed in great detail in Appendix \ref{App:multicut}, but we will highlight some important conclusions here. 

    We first describe the regions of existence of the $m$-cut solutions in the $(\Delta_1,\Delta_2)$ plane at a given $m$ and $\Delta_C$. They can be periodically described in terms of square cells of length $\frac{2\pi}{m}$ numbered by $(p,q)=(\lfloor\frac{m\Delta_1}{2\pi}\rfloor,\lfloor\frac{m\Delta_2}{2\pi}\rfloor)$:
    \be
    (\Delta_1,\Delta_2)\in \left[\frac{2\pi p}{m},\frac{2\pi(p+1)}{m}\right) \times \left[\frac{2\pi q}{m},\frac{2\pi(q+1)}{m}\right)\,.
    \ee  
    Each cell is parametrized by the coordinates $(a,b)\equiv (\{\frac{m\Delta_1}{2\pi}\}, \{\frac{m\Delta_2}{2\pi}\})\in [0,1)\times[0,1)$, where $\{\cdot\}$ is the fractional part. We also define $h\equiv (m-p-q)\mod 2$, which is fixed for a given cell, and
    \be
    \lambda_h \equiv\begin{cases}2\min\left(\left\{\frac{m\Delta_C}{2\pi}\right\},1-\left\{\frac{m\Delta_C}{2\pi}\right\}\right)\,, & h=0 \\ 1-2\min\left(\left\{\frac{m\Delta_C}{2\pi}\right\},1-\left\{\frac{m\Delta_C}{2\pi}\right\}\right)\,, & h = 1\end{cases}\,, \qquad \lambda\in[0,1]\,.
    \ee
    Then the cell is always divided into 3 regions, where orbifolds in the $+$ branch exist in a triangular region in the lower left corner, orbifolds in the $-$ branch exist in a triangular region in the upper right corner, and no $\bZ_m$ orbifolds exist in the rest of the cell. The triangular regions where solutions exist are defined by
    \be
    \begin{cases}+\text{ branch:} & 0\leq a+b\leq \lambda_h \\ -\text{ branch:} & 2-\lambda_h<a+b<2 \end{cases}
    \ee
    To visualize the division of regions within a square cell, one can refer to Figure \ref{fig:microcell}, where the left and right panels plot the cells with $h=0$ and $h=1$ respectively, at the fixed value of $2\min\left(\left\{\frac{m\Delta_C}{2\pi}\right\},1-\left\{\frac{m\Delta_C}{2\pi}\right\}\right)=0.32$. Orbifolds in the $+$ and $-$ branches exist in the black and gray regions respectively. By its definition, the value of $h$ alternates between $0$ and $1$ in neighbouring square cells. Thus the $(\Delta_1,\Delta_2)$ plane is divided into a checkerboard pattern of squares alternating between the left and right panels of Figure \ref{fig:microcell}, as seen in Figure \ref{fig:irrational-regions}.
    
    For a given point in the space of chemical potentials, it is also natural to ask what is the set of $m$ for which the corresponding $\bZ_m$ orbifold exists at that point. It turns out that for generic chemical potentials away from some co-dimension one hyperplanes whose positions are studied in Section \ref{Appsubsec:multicut:exceptional-points}, there exists infinitely many multi-cut solutions. As a result, the large $N$ phase diagram in the ensemble where the magnetic fluxes are fixed is very complicated and involves infinitely many competing saddles, including all the multi-cut solutions, dressed multi-cut solutions, and others. 

    \begin{figure}[h!]
            \centering        
            \includegraphics[width=\textwidth]{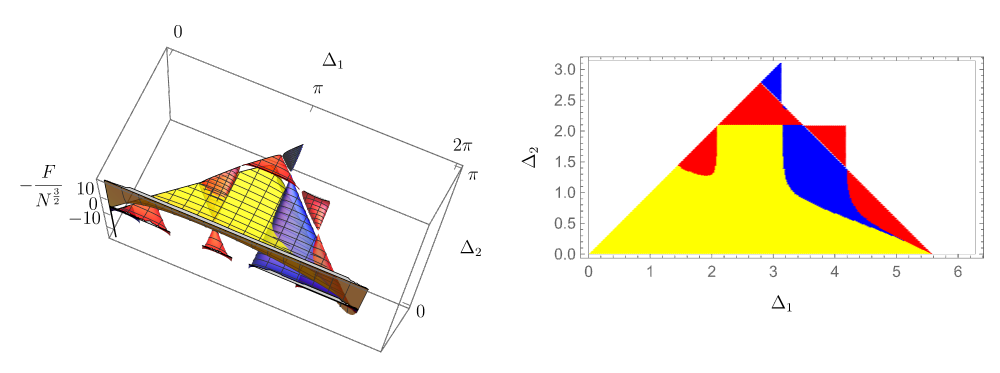}
            \caption{\small Phase structure between the black hole, $\bZ_2$ orbifold, and $\bZ_3$ orbifold. The left panel plots the leading free energy $\lim_{N\rightarrow\infty}-\frac{F}{N^{3/2}}$ of the 3 solutions in the fundamental domain, with the black hole in yellow, $\bZ_2$ orbifold in blue, and $\bZ_3$ orbifold in red. The right panel plots the phase diagram in the fundamental domain. The phases where the black hole, $\bZ_2$, or $\bZ_3$ orbifold dominate are colored yellow, blue, or red respectively. In both panels, we have chosen $\Delta_C=\frac{\pi}{9}$, $\fg=0$, $\fn_1=-2$, $\fn_2=4$, $\fn_C=0$.} 
            \label{fig: mleq3 phase diag}
    \end{figure}
    A competition between saddles can already be seen If we restrict our attention to the black hole and its $\bZ_2$ and $\bZ_3$ orbifolds. In the left panel of Figure \ref{fig: mleq3 phase diag}, the leading $\cO(N^\frac{3}{2})$ free energies of these 3 saddles are plotted as functions of $\Delta_{1,2}$ at a fixed value of the magnetic fluxes and $\Delta_C$, where the black hole, $\bZ_2$, and $\bZ_3$ orbifold are in yellow, blue and red respectively. In the right panel of the same figure, we plot a phase diagram for these 3 phases, where a point $(\Delta_1,\Delta_2)$ is colored yellow, blue, or red if the dominant phase with the smallest free energy at that point is the black hole, $\bZ_2$, or $\bZ_3$ orbifolds respectively. It is clear from the phase diagram that there are curves of first-order phase transitions where the values of competing free energies cross each other, and also straight phase boundaries where the free energy of a phase diverges and it cannot be continued past the boundary.   
    
\begin{figure}[!h]
		\centering 
		\caption{\rm The free energy $-\re F$ for all the BAE solutions at $N = 8$ as a function of $\Delta_1 \in (0, 2\pi)$ for fixed $\Delta_2 = \frac{e}{3}, \Delta_C = \frac{\pi}{9}$ and $\hat{\fn}_I = (-2,-2,1,1)$.}
		\includegraphics[width=12cm]{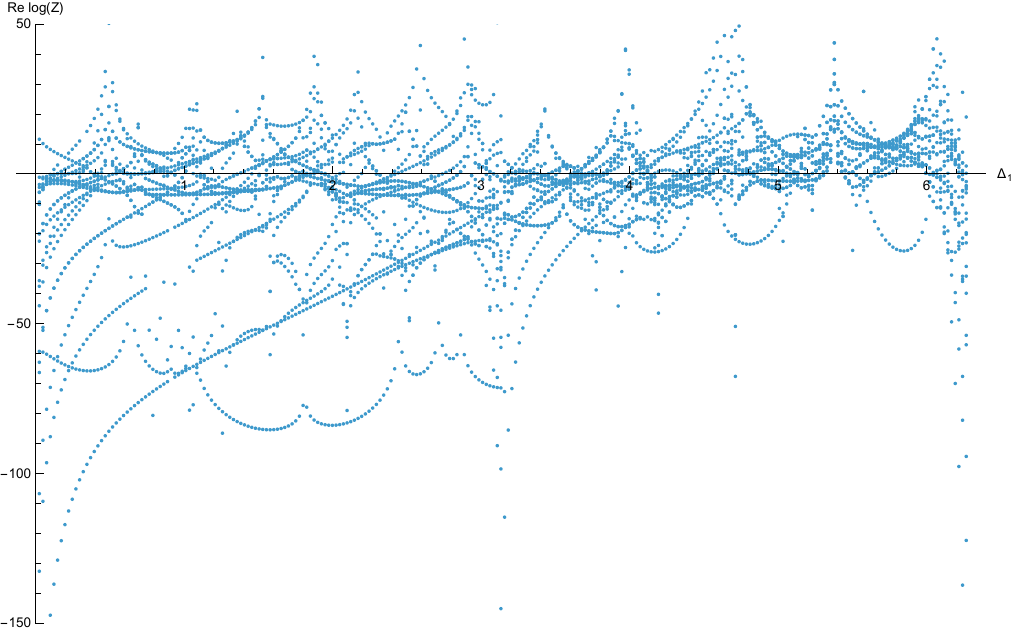}
		\label{fig:DomainWall}
	\end{figure}

    More generally, instead of restricting to 3 phases, we can examine the free energy of all the Bethe roots at finite $N$ as a function of the chemical potentials. For instance, in Figure \ref{fig:DomainWall}, we plot the free energies of all the $p(8)=22$ Bethe roots at $N=8$ as a function of $\Delta_1$, where the other parameters are kept fixed. The individual roots or phases can be roughly identified as smooth curves in $\Delta_1$, which have vertical asymptotes at phase boundaries. One can observe that the curves cross and intersect in a very complex manner, making the phase diagram extremely complicated.  
	
	%%%%%%%%%%%%%%%%%%%%%%%%
    \subsubsection{Conjectured densities for generic $m$-cut solutions}
    \label{subsubsec:multi-cut:mcut:density}
    %%%%%%%%%%%%%%%%%%%%%%%%

    Examining the sequence of one-cut densities in \eqref{eq: BH+ rho uniform}, \eqref{eq: BH- rho uniform}, 2-cut densities in \eqref{eq: 2cut reg 1 3 sol uniform}, \eqref{eq: 2cut reg 2t sol uniform}, and 3-cut densities in \eqref{eq: 3cut rho plus uniform}, \eqref{eq: 3cut rho minus uniform}, it is natural to extrapolate that the density of $m$-cut solutions with $m>3$ can either be written as
    \begin{gather}
    \rho(t) = \begin{cases}\frac{\mu-[m\hat{\Delta}_3]t}{\frac{1}{m}[m\hat{\Delta}_1][m\hat{\Delta}_2]([m\hat{\Delta}_3]+[m\hat{\Delta}_4])}\,, & 0<t<\frac{\mu}{[m\hat{\Delta}_3]} \\[.2cm] \frac{\mu+[m\hat{\Delta}_4]t}{\frac{1}{m}[m\hat{\Delta}_1][m\hat{\Delta}_2]([m\hat{\Delta}_3]+[m\hat{\Delta}_4])}\,, & -\frac{\mu}{[m\hat{\Delta}_4]}<t<0\end{cases}\quad \mu\equiv\sqrt{\frac{2}{m}[m\hat{\Delta}_1][m\hat{\Delta}_2][m\hat{\Delta}_3][m\hat{\Delta}_4]}
    \end{gather}
    or
    \begin{gather}
    \rho(t) = \begin{cases}\frac{-\mu-[m\hat{\Delta}_3]'t}{\frac{1}{m}[m\hat{\Delta}_1]'[m\hat{\Delta}_2]'([m\hat{\Delta}_3]'+[m\hat{\Delta}_4]')}\,, & 0<t<-\frac{\mu}{[m\hat{\Delta}_3]'} \\[.2cm] \frac{-\mu+[m\hat{\Delta}_4]'t}{\frac{1}{m}[m\hat{\Delta}_1]'[m\hat{\Delta}_2]'([m\hat{\Delta}_3]'+[m\hat{\Delta}_4]')}\,, & \frac{\mu}{[m\hat{\Delta}_4]'}<t<0\end{cases}\quad \mu\equiv\sqrt{\frac{2}{m}[m\hat{\Delta}_1]'[m\hat{\Delta}_2]'[m\hat{\Delta}_3]'[m\hat{\Delta}_4]'}
    \end{gather}
    depending on whether the solution is in the $+$ or $-$ branch. One can check that the densities above are correctly normalized as $\int dt\rho(t)=1$.

    %%%%%%%%%%%%%%%%%%%%%%%%
    \subsubsection{Free energy of generic $m$-cut solutions}
    \label{subsubsec:multi-cut:mcut:free energy}
    %%%%%%%%%%%%%%%%%%%%%%%%
    
	Based on the free energy of generic $\bZ_m$ orbifolds to leading order in $N$ given in \eqref{eq: orbifold logZ}, and the finite-$N$ free energies of $\bZ_2$ and $\bZ_3$ orbifolds in \eqref{eq:TTI-twocut}, \eqref{eq:TTI-threecut}, and \eqref{eq:TTI-threecut minus}, it is natural to extrapolate that the finite-$N$ free energy of a generic $m$-cut solution in the $+$ branch is
	\begin{gather} 
		F_{\fg=0,m-{\rm cut},+}  = - \frac{\sqrt{2[m\hat{\Delta}_1][m\hat{\Delta}_2][m\hat{\Delta}_3][m\hat{\Delta}_4]}}{3m} \sum_{I=1}^4 \frac{\hat{\fn}_I}{[m\hat{\Delta}_I]} \Big[ (N + \delta^{(m-{\rm cut})})^{\frac{3}{2}}\notag\\
		+ \fc_I^{(m-{\rm cut})} (N + \delta^{(m-{\rm cut})})^{\frac{1}{2}} \Big] + \frac{1}{2} \log (N + \delta^{(m-{\rm cut})}) + \hat{f}_0 + \hat{f}_{\rm np}+ \hat{f}_{\rm ADE}\,,\label{eq:TTI-mcut}
	\end{gather}
    while that of a $m$-cut solution in the $-$ branch is
    \begin{gather} 
		F_{\fg=0,m-{\rm cut},-}  = \frac{\sqrt{2[m\hat{\Delta}_1]'[m\hat{\Delta}_2]'[m\hat{\Delta}_3]'[m\hat{\Delta}_4]'}}{3m} \sum_{I=1}^4 \frac{\hat{\fn}_I}{[m\hat{\Delta}_I]'} \Big[ (N + \delta^{(m-{\rm cut})})^{\frac{3}{2}}\notag\\
		+ \fc_I^{(m-{\rm cut})} (N + \delta^{(m-{\rm cut})})^{\frac{1}{2}} \Big] + \frac{1}{2} \log (N + \delta^{(m-{\rm cut})}) + \hat{f}_0 + \hat{f}_{\rm np}+ \hat{f}_{\rm ADE}\,.\label{eq:TTI-mcut minus}
	\end{gather}
	In the above, $\hat{f}_{\rm ADE} \sim\#\log N+ \#(1+\cO(N^{-\frac{1}{2}})) $ is the contribution from the ADE gauge theory supported on the fixed locus of the orbifold action, if such a fixed locus exists, as discussed in Section \ref{subsubsec:holography:orbifold:ADE}. We have tested these formulae against the numerical free energy of many $m$-cut solutions, up to $m=8$. This high-precision numerical study gives evidence for the following properties:
	\begin{property}
		For a $m$-cut solution, its finite-$N$ free energy can be expressed as the compact formulae in \eqref{eq:TTI-mcut} or \eqref{eq:TTI-mcut minus}, excluding $\hat{f}_{\rm ADE}$, if and only if the coefficient of $\log N$ in the free energy is $\frac{1}{2}$. 
	\end{property}
	This is because the orbifold fixed locus, if present, contributes simultaneously to the $\log N$ coefficient and the expansion $\#(1+\cO(N^{-\frac{1}{2}}))$, as discussed in Section \ref{subsubsec:holography:orbifold:ADE}. A compact expression for the latter is unknown as of now, and deserves further study. 
	\begin{property}
		At the superconformal points defined as $[m\hat{\Delta}_I] = \frac{\pi}{2}$ or $[m\hat{\Delta}_I]' = -\frac{\pi}{2}$, the exponent of the leading non-perturbative correction of an $m$-cut solution is given by
		\begin{equation}
			\hat{f}^{\rm np}_{1/2, I} = - \frac{2\sqrt{2}\pi }{m}\,,
		\end{equation}
        where $\hat{f}^{\rm np}_{1/2, I}$ is defined as in \eqref{eq: f0 fnp flux decomp}, \eqref{eq: fnp_I N expansion}. 
	\end{property}
    This is consistent with the action of M2-branes wrapping $S^3\subset S^7$ in the bulk, whose action is $1/m$ of what it was in the black hole since an angular coordinate has been orbifolded by a $\bZ_m$ identification.

	%%%%%%%%%%%%%%%%%%%%%%%%
	\section{Solutions with unequal filling fractions}
	\label{sec:multi-cut:mixing}
	%%%%%%%%%%%%%%%%%%%%%%%%

\begin{figure}[h!]
        \centering        
        \includegraphics[width=\textwidth]{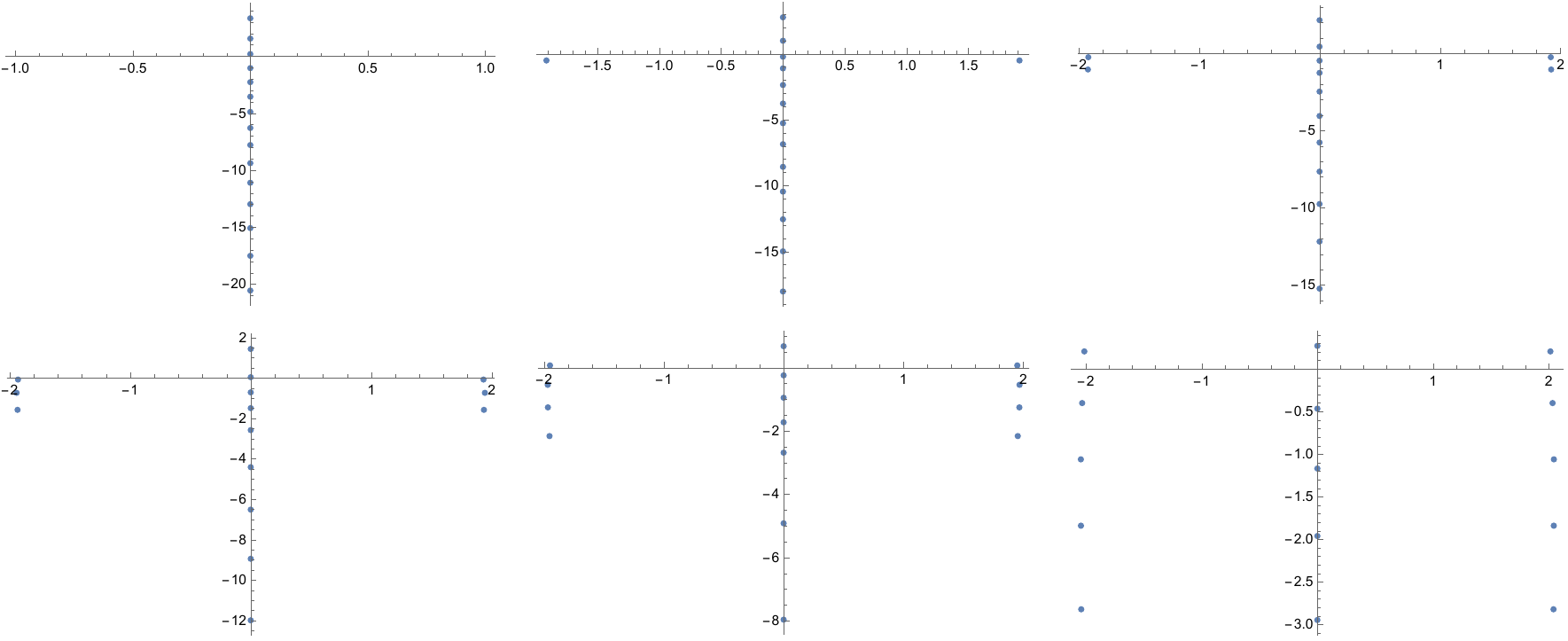}
        \caption{\small The family of $6$ interpolating solutions at $N=15$ in order of increasing $i=0,\ldots,M$ from top left to bottom right. They are computed at $\Delta_1=\frac{\pi^2}{4}$, $\Delta_2=e$, $\Delta_C=\frac{\pi}{9}$. }
        \label{fig: interpolating egs}
    \end{figure}

    At certain values of the chemical potentials where both 1-cut and 3-cut solutions exist, we observe that there is a discrete family of numerical solutions which have the following form. Let $N=3M$ where $M\in\bN$. In each solution, the $N$ gauge holonomies are still spread among 3 cuts, but the number of holonomies in each cut are $(i, N-2i, i)$, where $i=0,\ldots,M$. We have plotted the family of 6 solutions at $N=15$ in Figure \ref{fig: interpolating egs}. In the terminology used to describe matrix models, the filling fraction of a cut is the number of eigenvalues (i.e. holonomies) in the cut divided by the total number of eigenvalues. Here, the cuts generically have unequal filling fractions $(\frac{i}{N},\frac{N-2i}{N},\frac{i}{N})$, hence the name of this family of solutions. At $i=0$, we recover the familiar one-cut solution with all $N$ points along the central cut, while at $i=M$, we get the 3-cut solution with equal filling fractions $\frac{1}{3}$. At intermediate values of $i$, the solution interpolates between the one-cut and 3-cut solutions, and hence we will also refer to the family of solutions as interpolating solutions. It is worth noting that for $i\ll M$, we obtain i-pair dressed solutions with non-isolated points, which were discussed in Section \ref{subsec:dressed:superposition}.

    In Section \ref{subsubsec:multi-cut:mixing:largeN}, we analytically solve for the large $N$ distributions of the holonomies in the cuts, and compute the free energy to leading order in $N$. Numerically, this is extended to all perturbative orders in $1/N$ in Section \ref{subsubsec:multi-cut:mixing:numerics}. The bulk dual of these solutions are unknown, and it would be extremely interesting to construct them in supergravity. 

    It is worth mentioning that there are many other solutions with free energies of order $N^\frac{3}{2}$ that is beyond the scope of this paper. In regions $\tilde{1}$ and $\tilde{3}$ of Table \ref{tab: 2col reg} where the one-cut and 2-cut solutions coexist, we also find a discrete family of solutions with 3 cuts where one cut is along the imaginary axis, and two other cuts are placed symmetrically on the left and right of the axis. Within this family, the number of points on the imaginary axis varies between $0$ and $N$, and the solutions correspondingly interpolate between the 2-cut solution and 1-cut solution. This is very similar to the family of solutions interpolating between 1-cut and 3-cut solutions that we will study below. In addition, in regions $\tilde{2}$ and $\tilde{3}'$ of Table \ref{tab: 2col reg}, we also find ``dressed'' 2-cut solutions where there are pairs of gauge holonomies whose positions are removed from the cuts. These are analogous to the dressed one-cut solutions in Section \ref{sec:dressed}.
	
	%%%%%%%%%%%%%%%%%%%%%%%%
	\subsection{The analytic large $N$ solution}
	\label{subsubsec:multi-cut:mixing:largeN}
	%%%%%%%%%%%%%%%%%%%%%%%%

    Let $\nu=\frac{i}{N-2i}\in\bQ$ be the ratio of the number of points in one of the cuts at $\re u=\pm f$ to the number of points in the central cut. In an abuse of terminology, we will sometimes refer to $\nu$ as the filling fraction. We then consider the large $N$ limit of the solutions with fixed $\nu\sim\cO(1)$, where there are $M$ points in the central cut and $\nu M\in\bN$ points in each of the other cuts. $\nu = 0$ and $\nu = 1$ then correspond to the 1-cut and 3-cut solutions respectively. In the large $M$ limit, $\nu$ is approximately a continuous parameter in $(0,1)$. Denoting the holonomies in the central cut as $u_a$ and those in the cuts on either side as $v_b$ and $w_b$, the ansatz that we shall consider is 
    \begin{gather}
    u_a = iM^\frac{1}{2} t_a\,,\quad v_b = i(\nu M)^\frac{1}{2}\tilde{t}_b + f(\tilde{t}_b)\,,\quad w_b = i(\nu M)^\frac{1}{2}\tilde{t}_b - f(\tilde{t}_b)\notag\\
    a=1,\ldots,M\,,\qquad b=1,\ldots,\nu M\,.\label{eq: cont sol ansatz}
    \end{gather}
    In the large $N$ limit, we assume that $\{t_a\}$ and $\{\tilde{t}_b\}$ are distributed according to the densities $\rho$ and $\tilde{\rho}$ respectively, where
    \be\label{eq: cont rho rhot def}
    \rho(t)=\frac{1}{M}\frac{da}{dt}\,,\quad \tilde{\rho}(\tilde{t})=\frac{1}{\nu M}\frac{db}{d\tilde{t}}\,.
    \ee
    The family of solutions with unequal filling fractions are consistently found in region $3$ of Table \ref{tab: 3 col regions}, and we shall restrict our attention to this region in particular. Since \eqref{eq: cont sol ansatz} is very similar to the 3-cut ansatz in \eqref{eq: large N 3 col ansatz}, the large $N$ limit of the Bethe equations can be taken following the exact same steps as in Appendix \ref{App: 3cut w 1cut 0}. We shall thus omit the intermediate steps, and only report the result. The Bethe equations corresponding to $u$ simplify to
    \begin{gather}
    2\partial_t\left\{\rho(t)\sum_{I=1}^3g_+([\Delta_I])+\nu^\frac{1}{2}\tilde{\rho}(\nu^{-\frac{1}{2}}t)\sum_{I=1}^3\big[g_+([\Delta_I+f(\nu^{-\frac{1}{2}}t)])+g_+([\Delta_I-f(\nu^{-\frac{1}{2}}t)])\big]\right\}\notag\\
    +\Delta_C+\sgn t\frac{\Delta_3}{2} = 0\label{eq: cont u bae}\,,
    \end{gather}
    while the equations corresponding to $v$ become\footnote{Since $\rho$ and $\tilde{\rho}$ are independent in general, we cannot assume that $a(t)$ and $b(\tilde{t})$ are related via an equation like \eqref{eq: 3cut rv rw def}. Consequently, the $\cO(1)$ corrections when approximating sums of the form \eqref{eq: EM form sum} by integrals are not under analytic control. In deriving \eqref{eq: cont u bae} and \eqref{eq: cont v bae}, we have assumed that these $\cO(1)$ corrections are zero. This can be checked a posteriori in region $3$ of Table \ref{tab: 3 col regions}, and it is also confirmed numerically.}
    \begin{gather}
    iM^\frac{1}{2}\rho(\nu^\frac{1}{2}\tilde{t})\sum_{I=1}^3\left[g_+'([\Delta_I+f(\tilde{t})])-g_+'([\Delta_I-f(\tilde{t})])\right]+i(\nu M)^\frac{1}{2}\tilde{\rho}(\tilde{t})\sum_{I=1}^3[g_+'([\Delta_I+2f(\tilde{t})])\notag\\
    -g_+'([\Delta_I-2f(\tilde{t})])]+2\tilde{\rho}'(\tilde{t})\sum_{I=1}^3g_+([\Delta_I])+\rho'(\nu^\frac{1}{2}\tilde{t})\sum_{I=1}^3[g_+([\Delta_I+f(\tilde{t})])+g_+([\Delta_I-f(\tilde{t})])]\notag\\
    + \tilde{\rho}'(\tilde{t})\sum_{I=1}^3[g_+([\Delta_I+2f(\tilde{t})])+g_+([\Delta_I-2f(\tilde{t})])]+f'(\tilde{t})\tilde{\rho}(\tilde{t})\sum_{I=1}^3[g_+'([\Delta_I+2f(\tilde{t})])\notag\\
    -g_+'([\Delta_I-2f(\tilde{t})])]+\Delta_C+\sgn(\tilde{t})\frac{\Delta_3}{2}+2\pi\sgn(\tilde{t})\left\lfloor\frac{\frac{\Delta_1+\Delta_2}{2}-\sgn(\tilde{t})f(\tilde{t})}{2\pi}\right\rfloor\notag\\
    -\pi\sum_{I=1}^3\left(1+\left\lfloor\frac{\Delta_I+2f(\tilde{t})}{2\pi}\right\rfloor - \left\lfloor\frac{\Delta_I-2f(\tilde{t})}{2\pi}\right\rfloor\right)+8\pi=0\label{eq: cont v bae}\,.
    \end{gather}
    We have fixed the integer ambiguities in the Bethe equations by comparing \eqref{eq: cont u bae} and \eqref{eq: cont v bae} with \eqref{eq: W'} evaluated on numerical solutions in region $3$. The Bethe equations corresponding to $v$ can be obtained by taking $f\mapsto -f$ in \eqref{eq: cont v bae}, and do not give any new constraints. 
    
\begin{figure}[h]
        \centering        
        \includegraphics[width=.7\textwidth]{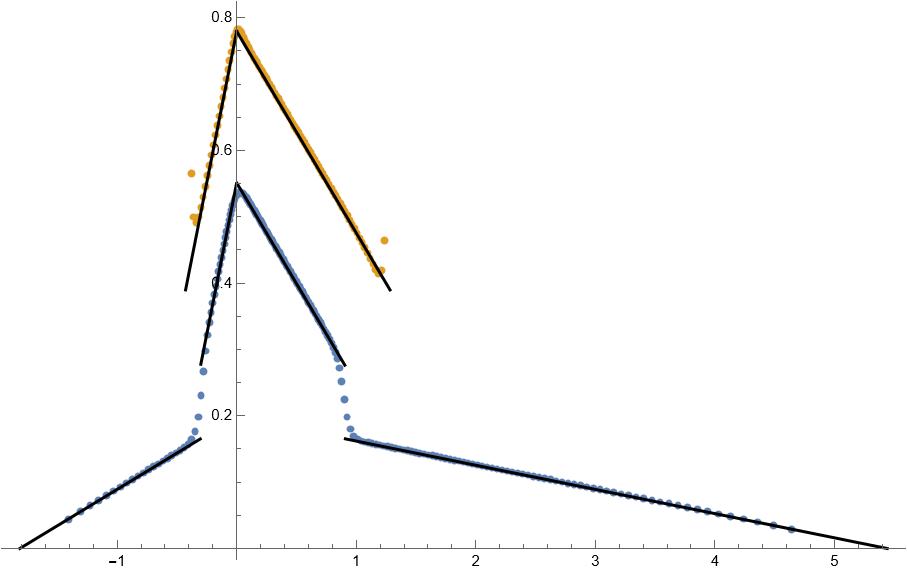}
        \caption{\small Plots of the distributions of $\{t_a\}$ (blue) and $\{\tilde{t}_b\}$ (yellow) of the $\nu=\frac{1}{2}$ solution at $N=400$, where $\Delta_1=\Delta_2=\frac{5\pi}{6}$, $\Delta_C=-\frac{\pi}{12}$. The black lines plot the analytic solutions for $\rho$ and $\tilde{\rho}$ in \eqref{eq: cont inn rho sol} and \eqref{eq: cont outer rho sol}. }
        \label{fig: cont sol eg}
    \end{figure}

    We start by solving the leading $\cO(M^\frac{1}{2})$ part of \eqref{eq: cont v bae}, which is
    \begin{gather}
    \rho(\nu^\frac{1}{2}\tilde{t})\sum_{I=1}^3[g_+'([\Delta_I+f(\tilde{t})])-g_+'([\Delta_I-f(\tilde{t})])]+\nu^\frac{1}{2}\tilde{\rho}(\tilde{t})\sum_{I=1}^3[g_+'([\Delta_I+2f(\tilde{t})])\notag\\
    -g_+'([\Delta_I-2f(\tilde{t})])] = 0\label{eq: cont v bae re LO}\,.
    \end{gather}
    In region $3$, \eqref{eq: cont v bae re LO} simplifies to
    \be\label{eq: rho = rescaled rho tilde}
    \rho(\nu^\frac{1}{2}\tilde{t}) = \nu^\frac{1}{2}\tilde{\rho}(\tilde{t})\,.
    \ee
    Using this, \eqref{eq: cont v bae} can be rewritten as
    \begin{gather}
    \partial_{\tilde{t}}\left\{2\tilde{\rho}(\tilde{t})\sum_{I=1}^3g_+([\Delta_I]) + \tilde{\rho}(\tilde{t})\sum_{I=1}^3[g_+([\Delta_I+f(\tilde{t})])+g_+([\Delta_I-f(\tilde{t})])]\right.\notag\\
    \left.+\tilde{\rho}(\tilde{t})\sum_{I=1}^3[g_+([\Delta_I+2f(\tilde{t})])+g_+([\Delta_I-2f(\tilde{t})])]\right\}+\Delta_C+\sgn(\tilde{t})\frac{\Delta_3}{2}= 0\,.\label{eq: cont v bae im inner}
    \end{gather}
    If $\tilde{\rho}$ has support $(\tilde{t}_-,\tilde{t}_+)$, we assume that $(\nu^\frac{1}{2}\tilde{t}_-,\nu^\frac{1}{2}\tilde{t}_+)$ is a subset of the support $(t_-,t_+)$ of $\rho$. In this subset $(\nu^\frac{1}{2}\tilde{t}_-,\nu^\frac{1}{2}\tilde{t}_+)$, $\tilde{\rho}(\nu^{-\frac{1}{2}}t)\neq 0$, and \eqref{eq: cont u bae} can be written as 
    \begin{gather}
    2\partial_t\left\{\rho(t)\sum_{I=1}^3g_+([\Delta_I])+\rho(t)\sum_{I=1}^3\big[g_+([\Delta_I+f(\nu^{-\frac{1}{2}}t)])+g_+([\Delta_I-f(\nu^{-\frac{1}{2}}t)])\big]\right\}\notag\\
    +\Delta_C+\sgn (t)\frac{\Delta_3}{2}= 0\,.\label{eq: cont u bae inner}
    \end{gather}
    For \eqref{eq: cont v bae im inner} and \eqref{eq: cont u bae inner} to be consistent with \eqref{eq: rho = rescaled rho tilde}, we need
    \be
    \sum_{I=1}^3\left[g_+([\Delta_I+f(\tilde{t})])+g_+([\Delta_I-f(\tilde{t})])\right] = \sum_{I=1}^3\left[g_+([\Delta_I+2f(\tilde{t})])+g_+([\Delta_I-2f(\tilde{t})])\right]\,, 
    \ee
    which fixes $f = \frac{2\pi}{3}$. Integrating either \eqref{eq: cont v bae im inner} or \eqref{eq: cont u bae inner} and imposing \eqref{eq: rho = rescaled rho tilde} then gives
    \begin{align}
    &\tilde{\rho}(\tilde{t})=\frac{\tilde{\mu}-\left(\Delta_C+\sgn \tilde{t}\frac{\Delta_3}{2}\right)\tilde{t}}{\frac{1}{3}(3\Delta_1-2\pi)(3\Delta_2-2\pi)\Delta_3}\,,\quad \tilde{t}\in(\tilde{t}_-,\tilde{t}_+)\,,\notag\\
    &\rho(t)=\frac{\nu^\frac{1}{2}\tilde{\mu}-\left(\Delta_C+\sgn t\frac{\Delta_3}{2}\right)t}{\frac{1}{3}(3\Delta_1-2\pi)(3\Delta_2-2\pi)\Delta_3}\,,\quad t\in(\nu^\frac{1}{2}\tilde{t}_-,\nu^\frac{1}{2}\tilde{t}_+)\subset (t_-,t_+)\,.\label{eq: cont inn rho sol}
    \end{align}
    The solution for $\tilde{\rho}$ is complete, but we have not yet solved for $\rho$ in the region $(t_-,t_+)\setminus (\nu^\frac{1}{2}\tilde{t}_-,\nu^\frac{1}{2}\tilde{t}_+)$. In this remaining region, the term proportional to $\tilde{\rho}(\nu^{-\frac{1}{2}}t)$ in \eqref{eq: cont u bae} evaluates to zero because $\nu^{-\frac{1}{2}}t$ is outside the support of $\tilde{\rho}$, and integrating \eqref{eq: cont u bae} gives
    \be\label{eq: cont outer rho sol}
    \rho(t) = \frac{\mu-\left(\Delta_C+\sgn t\frac{\Delta_3}{2}\right)t}{\Delta_1\Delta_2\Delta_3}\,,\quad t\in (t_-,t_+)\setminus (\nu^\frac{1}{2}\tilde{t}_-,\nu^\frac{1}{2}\tilde{t}_+)\,,\quad t_\pm =\pm\frac{\mu}{\frac{\Delta_3}{2}\pm\Delta_C}\,.
    \ee
    It remains to fix the constants $\tilde{\mu}$, $\tilde{t}_\pm$, and $\mu$. Imposing that $\rho$ and $\tilde{\rho}$ are normalized is not enough to fix them completely. By fitting with numerical results to very high precision, we can determine that
    \be
    \tilde{\mu} = \frac{\sqrt{1+2\nu}}{\sqrt{3\nu}}\sqrt{\frac{2}{3}(3\Delta_1-2\pi)(3\Delta_2-2\pi)\left(\frac{\Delta_3}{2}+\Delta_C\right)\left(\frac{\Delta_3}{2}-\Delta_C\right)}\,.
    \ee
    Motivated by the shape of the numerical densities computed via \eqref{eq: cont rho rhot def}, we further assume that $\tilde{\rho}(\tilde{t}_-)=\tilde{\rho}(\tilde{t}_+)\neq 0$, then $-(\frac{\Delta_3}{2}-\Delta_C)\tilde{t}_- = (\frac{\Delta_3}{2}+\Delta_C)\tilde{t}_+$, and normalizing $\tilde{\rho}$ fixes
    \be
    \tilde{t}_\pm = \pm\left(1-\sqrt{\frac{1-\nu}{1+2\nu}}\right)\frac{\tilde{\mu}}{\frac{\Delta_3}{2}\pm\Delta_C}\,. 
    \ee
    Finally, normalizing $\rho$ fixes
    \be
    \mu = \sqrt{2(1-\nu)\Delta_1\Delta_2\left(\frac{\Delta_3}{2}+\Delta_C\right)\left(\frac{\Delta_3}{2}-\Delta_C\right)} +\nu^\frac{1}{2}\left(\frac{\Delta_3}{2}+\Delta_C\right)\tilde{t}_+\,.
    \ee
    As seen in figure \ref{fig: cont sol eg}, there is a very close match between the numerical densities computed using \eqref{eq: cont rho rhot def} and the analytic solutions found above. It is worth noting that $\rho(t)$ has discontinuities at $t=\nu^\frac{1}{2}\tilde{t}_\pm$. 
    
    \paragraph{Leading contribution of interpolating solutions to the index}

    Substituting \eqref{eq: cont sol ansatz} into \eqref{eq: log Z} and retaining the leading $\cO(N^\frac{3}{2})$ term gives
    \begin{gather}
    -\frac{F_{\fg,\nu}}{M^\frac{3}{2}}= \sum_{I=1}^3[(\fn_I+1-\fg)g_+'([\Delta_I])-(1-\fg)g_+'(0)]\left(\int_{t_-}^{\nu^\frac{1}{2}\tilde{t}_-}dt\rho(t)^2+\int^{t_+}_{\nu^\frac{1}{2}\tilde{t}_+}dt\rho(t)^2\right)\notag\\
    +\nu^\frac{3}{2}\int_{\tilde{t}_-}^{\tilde{t}_+}d\tilde{t}\tilde{\rho}(\tilde{t})^2\left\{\sum_{I=1}^3(\fn_I+1-\fg)\left[3g_+'([\Delta_I])+2g_+'([\Delta_I+f(\tilde{t})])+2g_+'([\Delta_I-f(\tilde{t})])\right.\right.\notag\\
    \left.\left.+g_+'([\Delta_I+2f(\tilde{t})])+g_+'([\Delta_I-2f(\tilde{t})])\right]-(1-\fg)\left[3g_+'(0)+4g_+'([f(\tilde{t})])+2g_+'([2f(\tilde{t})])\right]\vphantom{\sum_I^3}\right\}\notag\\
    +\frac{\fn_3}{2}\left(\int_{t_-}^{\nu^\frac{1}{2}\tilde{t}_-}dt\rho(t)|t|+\int^{t_+}_{\nu^\frac{1}{2}\tilde{t}_+}dt\rho(t)|t|+3\nu^\frac{3}{2}\int_{\tilde{t}_-}^{\tilde{t}_+}d\tilde{t}\tilde{\rho}(\tilde{t})|\tilde{t}|\right)-\fn_C\left(\int_{t_-}^{\nu^\frac{1}{2}\tilde{t}_-}dt\rho(t)t\right.\notag\\
    \left.+\int^{t_+}_{\nu^\frac{1}{2}\tilde{t}_+}dt\rho(t)t+3\nu^\frac{3}{2}\int_{\tilde{t}_-}^{\tilde{t}_+}d\tilde{t}\tilde{\rho}(\tilde{t})\tilde{t}\right)+\cO\left(M^{-\frac{1}{2}}\right).
    \end{gather}
    After plugging in the densities \eqref{eq: cont inn rho sol} and \eqref{eq: cont outer rho sol}, as well as $f=\frac{2\pi}{3}$, we get
    \begin{gather}
    -\frac{F_{\fg,\nu}}{N^\frac{3}{2}} = \alpha^\frac{3}{2}\frac{1}{3}\sqrt{2[\hat{\Delta}_1][\hat{\Delta}_2][\hat{\Delta}_3][\hat{\Delta}_4]}\sum_{I=1}^4\frac{\hat{\fn}_I}{[\hat{\Delta}_I]}\notag\\
    +\left(1-\alpha^\frac{3}{2}\right)\frac{1}{3\cdot 3}\sqrt{2[3\hat{\Delta}_1][3\hat{\Delta}_2][3\hat{\Delta}_3][3\hat{\Delta}_4]}\sum_{I=1}^4\frac{\hat{\fn}_I}{[3\hat{\Delta}_I]}+\cO\left(N^{-\frac{1}{2}}\right)\label{eq: LO interp F}\,,\quad \alpha\equiv\frac{1-\nu}{1+2\nu}\,.
    \end{gather}
    Suggestively, this is a linear combination of the free energies of a black hole and a $\bZ_3$ orbifold, where the relative size of the contributions are controlled by $\nu$.

    \paragraph{Similarities with saddles of the 4d superconformal index}

    The solutions with unequal filling fractions are very similar to a class of large $N$ saddle points found in the matrix model of the superconformal index of $\cN=4$ super-Yang-Mills (SYM) by the authors of \cite{Choi:2021rxi}. As in our set-up, the variables in their matrix model are gauge holonomies $u_{a=1,\ldots,N}$ parametrizing the Cartan subalgebra of $\U(N)$. For the class of solutions in question, they are distributed along $2m$ cuts in the complex plane, divided into two groups of $m$ cuts each. Much like our solutions, the filling fractions of the cuts is controlled by a parameter $\nu\in[0,1]$. When $\nu = 0$, all of the holonomies are in the first group of $m$ cuts, which gives a $m$-cut saddle. As $\nu$ increases, the share of holonomies in the first group correspondingly increases. When $\nu=\frac{1}{2}$, the number of holonomies in the two groups are equal, and we get a $2m$-cut saddle with equal filling fractions. When $\nu$ reaches $\nu=1$, all of the holonomies are in the second group of $m$ cuts, which gives another $m$-cut saddle. The way that the saddle interpolates between $m$ cuts and $2m$ cuts as $\nu$ varies is exactly analogous to how our solution interpolates between 1 cut and 3 cuts. Under the AdS$_5$/CFT$_4$ duality between $\cN=4$ SYM and IIB string theory on AdS$_5\times S^5$, the $m$-cut saddles are dual to $\bZ_m$ orbifolds of rotating electrically charged black holes. Thus the leading order free energy $F$ in $-\log\cI_{\text{SCI}}/N^2=F+\cO(N^{-1})$ for the 4d superconformal index $\cI_{\text{SCI}}$ evaluated in the large $N$ saddle point approximation correspondingly interpolates between the on-shell actions $S_{\bZ_m}$ and $S_{\bZ_{2m}}$ of the $\bZ_m$ and $\bZ_{2m}$ orbifolds respectively:
    \be
        F=\alpha^2 S_{\bZ_m}([m\Delta]) + (1-\alpha^2)S_{\bZ_{2m}}([2m\Delta])\,,\quad \alpha\equiv 1-2\nu\in [-1,1]\,.
    \ee         
    Here $\Delta$ collectively denote the 3 chemical potentials in the 4d superconformal index, which were written as $\delta_{I=1,2,3}$ in \cite{Choi:2021rxi}. This is exactly analogous to \eqref{eq: LO interp F}, where the free energy interpolates between those of the black hole and $\bZ_3$ orbifold. In our set-up, we also find numerical solutions which interpolate the 1-cut and 2-cut solutions, which are perhaps more directly related to the saddles of $\cN=4$ SYM discussed above, but it proved difficult to study them analytically. In general, we expect there to exist many more classes of Bethe roots interpolating between orbifolds of different orders. It would be interesting to study them more systematically, and we leave this to future work.
    
	%%%%%%%%%%%%%%%%%%%%%%%%
	\subsection{Finite $N$ numerics}
	\label{subsubsec:multi-cut:mixing:numerics}
	%%%%%%%%%%%%%%%%%%%%%%%%
	
	For simplicity, we focus on the chemical potentials $(\Delta_1,\Delta_2,\Delta_C)=\Big( \frac{5\pi}{6}, \frac{5\pi}{6}, 0 \Big)$, i.e. $\hat{\Delta}_I = \left(\frac{5\pi}{6}, \frac{5\pi}{6}, \frac{\pi}{6}, \frac{\pi}{6} \right)$, which is the superconformal point of region $\tilde{3}$ shown in Table \ref{tab: 3cut reg}. At this point, the one-cut and three-cut solutions coexist together with the family of interpolating solutions. At a fixed $\nu\sim\cO(1)$, we can generate the interpolating solution for a range of large $N$, and compute the free energy numerically. Based on the leading $\cO(N^{3/2})$ coefficient of the free energy found in \eqref{eq: LO interp F}, we continue to compute the subleading coefficients numerically for different values of $\nu$. We observe that at each order, the coefficient depends on $\nu$ only through the combination $\alpha(\nu)\equiv\frac{1-\nu}{1+2\nu}\in[0,1]$, and the power of $\alpha$ is identical to the power of $N$. This motivates us to define the effective membrane charge $N_\alpha\equiv \alpha N+\delta'$, where $\delta'$ is a constant in $N$. Interestingly, the free energy resums into a compact expression in terms of $N$ and $N_\alpha$:
	\begin{equation} \label{eq:TTI-mixing}
		\begin{aligned}
			F_{\fg=0,\nu} & = F_{0,\rm 3-cut,+}\left(N,\fc_I^{\rm (3-cut)},\delta^{\rm (3-cut)}\right) - \frac{\sqrt{2 \hat{\Delta}_1 \hat{\Delta}_2 \hat{\Delta}_3 \hat{\Delta}_4 }}{3} \sum_I \frac{\hat{\fn}_I}{ \hat{\Delta}_I } \left( N_\alpha^{3/2} + \tilde{\fc}_I^{(\rm BH)} N_\alpha^{1/2} \right)  \\
			& \textcolor{red}{+} \frac{\sqrt{2 [3 \hat{\Delta}_1] [3 \hat{\Delta}_2] [3 \hat{\Delta}_3] [3 \hat{\Delta}_4] } }{9}  \sum_I \frac{\hat{\fn}_I}{ [3 \hat{\Delta}_I] } \left( N_\alpha^{3/2} + \tilde{\fc}_I^{(\rm 3-cut)} N_\alpha^{1/2} \right) + \frac{1}{2} \log N_\alpha + \hat{f}_0 + \hat{f}_{\rm np},\\
			& \qquad N_\alpha \equiv \alpha N + \delta'\,,\\
		\end{aligned}
	\end{equation}
	where $F_{0,\rm 3-cut,+}(N,\fc_I^{\rm (3-cut)},\delta^{\rm (3-cut)})$ is given in \eqref{eq:TTI-threecut}. This has been tested at the superconformal point $\hat{\Delta}_I = \left(\frac{5\pi}{6}, \frac{5\pi}{6}, \frac{\pi}{6}, \frac{\pi}{6} \right)$, where the numerical free energy of interpolating solutions of various $\nu$ can be successfully fitted to the above ansatz, with coefficients
	\begin{equation}\label{eq: interp coeff}
		\delta' = \frac{3}{4},\quad \tilde{\fc}_{1,2}^{(\rm BH)} - \tilde{\fc}_{1,2}^{(\rm 3-cut)} = -12,\quad 5 \tilde{\fc}_{3,4}^{(\rm BH)} - \tilde{\fc}_{3,4}^{(\rm 3-cut)} = -12.
	\end{equation}
	
	Some comments on the result are in order.
	\begin{itemize}
		\item The ansatz only depends on $\tilde{c}_I^\text{(BH)}$ and $\tilde{c}_I^{(3\text{-cut})}$ through the linear combinations
        \be
        -\frac{\sqrt{2\hat{\Delta}_1\hat{\Delta}_2\hat{\Delta}_3\hat{\Delta}_4}}{3}\frac{\tilde{c}_I^\text{(BH)}}{\hat{\Delta}_I}+\frac{\sqrt{2[3\hat{\Delta}_1][3\hat{\Delta}_2][3\hat{\Delta}_3][3\hat{\Delta}_4]}}{9}\frac{\tilde{c}_I^{(3\text{-cut})}}{[3\hat{\Delta}_I]}\,,
        \ee 
        which evaluate to
        \be
        -\frac{\pi}{9\sqrt{2}}\left(\tilde{c}_1^\text{(BH)}-\tilde{c}_1^{(3\text{-cut})},\tilde{c}_2^\text{(BH)}-\tilde{c}_2^{(3\text{-cut})}, 5\tilde{c}_3^\text{(BH)}-\tilde{c}_3^{(3\text{-cut})}, 5\tilde{c}_4^\text{(BH)}-\tilde{c}_4^{(3\text{-cut})}\right)
        \ee
        at the superconformal point. Hence matching the numerical coefficients of $N_\alpha^{1/2}$, $N_\alpha^{-1/2}$, etc. can only determine these linear combinations, not $\tilde{c}_I^\text{(BH)}$ and $\tilde{c}_I^{(3\text{-cut})}$ individually. Perhaps understanding the bulk dual of these solutions better, or finding the analytical solution to subleading order will help to fix these coefficients.
		
		\item All dependence on the filling fraction $\nu$ is absorbed into the effective membrane charge $N_\alpha$.
		
		\item The ansatz in \eqref{eq:TTI-mixing} has the form
        \begin{gather}
        F_{\fg=0,\nu}=F_{0,\rm{3-cut},+}\left(N, \fc_I^{(\rm{3-cut})},\delta^{(\rm{3-cut})}\right) - F_{0,\rm{3-cut},+}\left(\alpha N, \tilde{\fc}_I^{(\rm{3-cut})}, \delta'\right) \notag\\
        + F_{0,BH_+}\left(\alpha N, \tilde{\fc}_I^{(\rm{BH})}, \delta'\right) + \ldots\,,
        \end{gather}
        up to logarithmic, constant, and non-perturbative corrections in $N$. The fact that the perturbative corrections in $1/N$ have this structure suggests, schematically, that the interpolating solution is constructed by ``removing'' $\alpha N$ units of M2-brane charge from the $\bZ_3$ orbifold and ``giving'' it instead to the black hole. Perhaps this might shed some light on the bulk dual of these solutions.   

	\end{itemize}
	Since the family of solutions that we are considering reduces to the one-cut solution when $\alpha \rightarrow 1$ or $\nu \rightarrow 0$, and to the 3-cut solution when $\alpha \rightarrow 0$ or $\nu \rightarrow 1$, it is natural to study the free energy under these simplifying limits. When $\alpha \rightarrow 0$, one can immediately observe that \eqref{eq:TTI-mixing} reduces to $F_{0,\rm 3-cut,+}(N,\fc_I^{\rm (3-cut)},\delta^{\rm (3-cut)})$ up to constant and non-perturbative corrections, as expected. In the opposite limit $\alpha \rightarrow 1$, however, one can check that the coefficients of $N^\frac{1}{2}$ and $\log N$ in \eqref{eq:TTI-mixing} are all different from those in the black hole free energy in \eqref{eq:TTI-BH}. This discrepancy can be explained by a subtlety in the order of limits. The formula in \eqref{eq:TTI-mixing} applies when the large $N$ limit is taken at fixed $\nu\sim\cO(1)$. Subsequently taking $\nu\rightarrow 0$ then naively gives the one-cut solution. On the other hand, the black hole free energy in \eqref{eq:TTI-BH} was obtained by first restricting to one-cut solutions with $\nu = 0$ before taking the large $N$ limit. There is no guarantee that the two limits $N\rightarrow\infty$ and $\nu\rightarrow 0$ commute. One might also be tempted to suggest that since $\nu = \frac{n}{N-2n}$ with $n\sim\cO(1)$ corresponds to $n$-pair dressed solutions, substituting this value of $\nu$ into \eqref{eq:TTI-mixing} will give the free energy of the dressed solutions. However, testing this directly with the examples discussed in Section \ref{subsec:dressed:superposition} shows that the comparison does not hold. This failure is expected because one of the assumptions when obtaining \eqref{eq:TTI-mixing} is $\nu\sim\cO(1)$. This is violated here since $\nu = \frac{n}{N-2n}\sim \cO(N^{-1})\rightarrow 0$ in the large $N$ limit. Also, the fact that both $n$-pair solutions and one-cut solutions in the large $N$ limit correspond to interpolating solutions with $\nu=0$ is further evidence for the proposal that the limits $N\rightarrow\infty$ and $\nu\rightarrow 0$ do not commute.
	
	Expanding the non-perturbative corrections as in \eqref{eq: f0 fnp flux decomp} and \eqref{eq: fnp_I N expansion}, we find that the primary coefficients are given by
	\begin{equation}
		\hat{f}^{\rm (np)}_{1/2, I} = - \frac{2\sqrt{2}\pi}{3} (1-\sqrt{\alpha}),\quad \hat{f}^{\rm (np)}_{{\rm log}, I} = \frac{3}{2}. 
	\end{equation}
	Correspondingly, there should be a supersymmetric configuration of M2-branes wrapping a 3-cycle in the internal space $S^7$ whose classical action is
	\begin{equation}
		S_{\rm cl} =   \frac{2\sqrt{2}\pi}{3} \sqrt{N} - \frac{2\sqrt{2}\pi}{3} \sqrt{\alpha N}.
	\end{equation}
	We present this expression in the hope that it is another clue to the problem of determining the bulk geometry that is dual to the interpolating solutions.
    
    In the special limit where $n \sim N^{1/2}$ in $\nu=\frac{n}{N-2n}$, the non-perturbative corrections are no-longer exponentially suppressed in $N$:
    \begin{equation}
        S_{\rm cl} \sim \sqrt{2}\pi \frac{n}{\sqrt{N}}.
    \end{equation}
    In the same limit, the non-perturbative corrections for the $n$-pair dressed solutions with non-isolated points at $\Delta_1 = \Delta_2$ in \eqref{eq: dressed 1/(n+1) instanton expansion} is no longer exponentially suppressed either. This may explain why the free energy does not reduce to the dressed one-cut free energy smoothly in the $\alpha \rightarrow 1$ limit discussed above. 
	
	%%%%%%%%%%%%%%%%%%%%%%%%
	\section{Superconformal index in the Cardy limit}
	\label{sec:SCI}
	%%%%%%%%%%%%%%%%%%%%%%%%

    A precise knowledge of the Bethe ansatz equations \eqref{eq:BAEexp} also allows us to compute the superconformal index in the Cardy limit, since the index in this limit reduces to an asymptotic integral in which the leading saddle point equation coincides with the Bethe ansatz equations \cite{Choi:2019zpz, Bobev:2022wem}. Using this fact, we will evaluate the superconformal index at the saddle points corresponding to the 1,3-cut solutions and interpolating solutions. Apart from the 1-cut or black hole solution, which was already considered in \cite{Bobev:2022wem}, the other results are new.
	
	%%%%%%%%%%%%%%%%%%%%%%%%
	\subsection{The superconformal index of ADHM and a review of the Cardy limit}
	\label{subsec:SCI:intro}
	%%%%%%%%%%%%%%%%%%%%%%%%
	
	The superconformal index \cite{Kim:2009wb, Imamura:2011su} of ADHM theory is defined as the following trace over the Hilbert space $\cH_{S^2}$ on $S^2$ 
    \begin{gather}
    \cI_{\rm sc}(q,\tilde{t}_1,\tilde{t}_2,\tilde{\zeta})\equiv \Tr_{\cH_{S^2}}\left[(-1)^Fe^{-\beta (\Delta - R_{\rm sc}-j_3)}q^{\Delta+j_3}\tilde{t}_1^{F_1}\tilde{t}_2^{F_2}\tilde{\zeta}^{F_C}\right]\notag\\
    =\Tr_{\rm BPS}\left[(-1)^F q^{R_{\rm sc}+2j_3}\tilde{t}_1^{F_1}\tilde{t}_2^{F_2}\tilde{\zeta}^{F_C}\right]\,,
    \end{gather}
    where $\Delta$ is the Hamiltonian under radial quantization, $R_{\rm sc}$ is the superconformal $R$-charge, and $j_3$ is the third component of the angular momentum rotating $S^2$. $F_{1,2}$ and $F_C$ are the same flavour symmetries that we have defined in Section \ref{subsec:TTI:ADHM}. Since
    \be
    \Delta - R_{\rm sc}-j_3 = \{\cQ^\dagger,\cQ\}\,,
    \ee
    where $\cQ$ is a specially chosen supercharge in the superconformal algebra, the index only receives contributions from BPS states that are annihilated by $\cQ$ and $\cQ^\dagger$. Hence the index is independent of $\beta$, and we can replace $q^{\Delta+j_3}$ with $q^{R_{\rm sc}+2j_3}$ in the trace because the two insertions are equal on BPS states. The index also admits additional flavour magnetic fluxes \cite{Kapustin:2011jm}, but we shall restrict to the case of zero flavour fluxes since that is the context for which the Cardy limit has been established in \cite{Bobev:2022wem}. The chemical potentials $\omega$, $\delta_{1,2}$, $\delta_C$ corresponding to $q$ and the flavour fugacities are defined as
    \be\label{eq: SCI flav chem pot def}
        q = e^{\pi i\omega}\,,\qquad \tilde{t}_{1,2} = e^{i\delta_{1,2}}\,,\qquad \tilde{\zeta} = e^{i\delta_C}\,.
    \ee
    For later convenience, we also define the auxiliary variables
    \be\label{eq: SCI orig flav aux}
        \tilde{t}_3 \equiv \tilde{t}_1^{-1}\tilde{t}_2^{-1}=e^{i\delta_3}\,,\quad \delta_3\equiv 2\pi - \delta_1-\delta_2\,. 
    \ee

	Similarly to the topologically twisted index, supersymmetric localization reduces the superconformal index to an $N$-dimensional contour integral over gauge fugacities $z_a=e^{i h_a}$ in each sector of gauge magnetic fluxes $\fm_a \in \bZ$ \cite{Imamura:2011su}\footnote{Under the identifications $\xi_F=\tilde{t}_1\tilde{t}_2$, $\xi_l=\tilde{t}_1\tilde{t}_2^{-1}$, and $\xi_T=\tilde{\zeta}$, one can check that the expression for the superconformal index in \cite{Bobev:2022wem} coincides with the above. }:
    \begin{gather}
    \frac{1}{N!}\sum_{\fm\in\bZ^N}\int \left(\prod_{a=1}^N\frac{dz_a}{2\pi i z_a}\tilde{\zeta}^{\fm_a}\right)\prod_{\substack{a,b=1 \\ a\neq b}}^Nq^{-\frac{1}{2}|\fm_{ab}|}(1-z_{ab}q^{|\fm_{ab}|})\prod_{a,b=1}^N\left[(q^{\frac{1}{2}}z_{ba}\tilde{t}_1^{-1})^{\frac{1}{2}|\fm_{ab}|}\vphantom{\frac{(z_{ba}\tilde{t}_1^{-1}q^{\frac{3}{2}+|m_{ab}|};q^2)_\infty}{(z_{ab}\tilde{t}_1q^{\frac{1}{2}+|m_{ab}|};q^2)_\infty}}\right.\notag \\
    \left. \frac{(z_{ba}\tilde{t}_1^{-1}q^{\frac{3}{2}+|\fm_{ab}|};q^2)_\infty}{(z_{ab}\tilde{t}_1q^{\frac{1}{2}+|\fm_{ab}|};q^2)_\infty}(q^{\frac{1}{2}}z_{ba}\tilde{t}_2^{-1})^{\frac{1}{2}|\fm_{ab}|}\frac{(z_{ba}\tilde{t}_2^{-1}q^{\frac{3}{2}+|\fm_{ab}|};q^2)_\infty}{(z_{ab}\tilde{t}_2q^{\frac{1}{2}+|\fm_{ab}|};q^2)_\infty}(z_{ba}\tilde{t}_3^{-1})^{\frac{1}{2}|\fm_{ab}|}\right.\notag\\
    \left.\frac{(z_{ba}\tilde{t}_3^{-1}q^{1+|\fm_{ab}|};q^2)_\infty}{(z_{ab}\tilde{t}_3q^{1+|\fm_{ab}|};q^2)_\infty}\right]\prod_{a=1}^N\left[(q^{\frac{1}{2}}z_a^{-1}\tilde{t}_1^{-\frac{1}{2}}\tilde{t}_2^{-\frac{1}{2}})^{\frac{1}{2}|\fm_a|}\frac{(z_a^{-1}\tilde{t}_1^{-\frac{1}{2}}\tilde{t}_2^{-\frac{1}{2}}q^{\frac{3}{2}+|\fm_a|};q^2)_\infty}{(z_a\tilde{t}_1^{\frac{1}{2}}\tilde{t}_2^{\frac{1}{2}}q^{\frac{1}{2}+|\fm_a|};q^2)_\infty}\right.\notag\\
    \left.(q^{\frac{1}{2}}z_a\tilde{t}_1^{-\frac{1}{2}}\tilde{t}_2^{-\frac{1}{2}})^{\frac{1}{2}|\fm_a|}\frac{(z_a\tilde{t}_1^{-\frac{1}{2}}\tilde{t}_2^{-\frac{1}{2}}q^{\frac{3}{2}+|\fm_a|};q^2)_\infty}{(z_a^{-1}\tilde{t}_1^{\frac{1}{2}}\tilde{t}_2^{\frac{1}{2}}q^{\frac{1}{2}+|\fm_a|};q^2)_\infty}\right]\,,\quad \fm_{ab}\equiv\fm_a-\fm_b\,,\quad z_{ab}=z_a/z_b\,,\label{eq: ADHM SCI}
    \end{gather} 
    where $(a;q)_\infty$ is the $q$-Pochhammer symbol defined as
    \be
    (a;q)_\infty\equiv \prod_{k=0}^\infty (1-aq^k)\,.
    \ee

    For the sake of clarity, we now review how the Cardy limit is taken following roughly the same steps as \cite{Bobev:2022wem,Bobev:2024mqw}. Due to differences in convention between our topologically twisted index and theirs, the way that we redefine the chemical potentials later will be different. In particular, we will not introduce the extra variables $\Delta_{q}$, $\Delta_{\tilde{q}}$, $\fn_q$, and $\fn_{\tilde{q}}$, which are redundant. This computation will also be useful for pinning down the identifications between the equivariant weights and chemical potentials when computing the superconformal index from the bulk using equivariant localization. Firstly, using the identity
    \begin{gather}
    (aq^{1-r})^{\frac{1}{2}|n|}\frac{(aq^{2-r+|n|};q^2)_\infty}{(a^{-1}q^{r+|n|};q^2)_\infty}=(-1)^{\frac{1}{2}(n+|n|)}(aq^{1-r})^{-\frac{1}{2}n}\frac{(aq^{2-r-n};q^2)_\infty}{(a^{-1}q^{r-n};q^2)_\infty}\notag\\
    =(-1)^{\frac{1}{2}(|n|-n)}(aq^{1-r})^{\frac{1}{2}n}\frac{(aq^{2-r+n};q^2)_\infty}{(a^{-1}q^{r+n};q^2)_\infty}\,,\quad r\in\bR\,,\quad n\in\bZ\,,
    \end{gather}
    we can remove the absolute values in \eqref{eq: ADHM SCI} in various ways. For the ratios of Pochhammer symbols containing $|\fm_{ab}|$, we could either substitute $|\fm_{ab}|\rightarrow \fm_{ab}$ or $|\fm_{ab}|\rightarrow -\fm_{ab}$, at the expense of a sign. Similarly, for the terms containing $|\fm_a|$, one could either choose $|\fm_a|\rightarrow \fm_a$ or $|\fm_a|\rightarrow -\fm_a$. Following \cite{Choi:2019zpz, Bobev:2022wem}, we choose
    \be
    |\fm_{ab}|\rightarrow\begin{cases}\fm_{ab}\,, & a<b \\ -\fm_{ab}\,, & a\geq b\end{cases}\qquad |\fm_a|\rightarrow -\fm_a\,.
    \ee
    Subsequently, we also change variables from the original set of generically complex gauge and flavour fugacities $(z_a,\tilde{t}_{1,2},\tilde{\zeta})$, to the new fugacities $(s_a,t_{1,2},\zeta)$, which are pure phases, supplemented by the real variables $\fn_{1,2}$, $\fn_C$. They are related via 
    \begin{gather}
        z_a=s_aq^{-\fm_a}\,,\quad \tilde{t}_{1,2}=t_{1,2}q^{-(\fn_{1,2}+\frac{1}{2})}\,,\quad \tilde{\zeta} = \zeta q^{\fn_C}\,,\quad |s_a|=|t_{1,2}|=|\zeta|=1\,,\notag\\
        s_a = e^{iU_a}\,,\quad t_{1,2}=e^{i\Delta_{1,2}}\,,\quad \zeta = e^{i\Delta_C}\,.\label{eq: SCI aux flux def}
    \end{gather}
    The new variables are given the same names as the flavour fugacities and fluxes of the topologically twisted index because, as we will see shortly, when expressed in terms of these variables, the superconformal index at next-to-leading order in the Cardy limit coincides with the topologically twisted index. From \eqref{eq: SCI aux flux def}, the ``fluxes'' $\fn_{1,2}=-\log|\tilde{t}_{1,2}|/\log q-\frac{1}{2}$ and $\fn_C=\log |\tilde{\zeta}|/\log q$ are determined by the modulus of the original chemical potentials, while $t_{1,2}$ and $\zeta$ carry their phases. Thus the ``fluxes'' are not genuine integer quantized magnetic fluxes, and should be understood purely in terms of the change of variables. $\Delta_{1,2}$ and $\Delta_C$ are the chemical potentials corresponding to the new flavour fugacities. For convenience, we also define the auxiliary variables
    \be\label{eq: SCI new aux chem pot}
        t_3\equiv (t_1t_2)^{-1}=e^{i\Delta_3}\,,\quad \Delta_3 \equiv 2\pi-\Delta_1-\Delta_2\,,\quad \fn_3\equiv -2-\fn_1-\fn_2\,.
    \ee  
    Following \cite{Bobev:2024mqw}, we also insert the trivial phase
    \be
    1=e^{-2\pi i\sum_{a=1}^N\fm_an_a}=\prod_a |s_a|^{-\frac{2n_a}{\omega}}\,,\quad n_a\in\bZ
    \ee
    into the integrand of the index. The integers $n_a$ are arbitrary, and will later correspond to the integer ambiguities in the Bethe potential. After performing these steps, the index can be written as
    \begin{gather}
    \frac{(-1)^{\frac{N(N-1)}{2}}}{N!}\sum_{\fm\in\bZ^N}\int\left[\prod_{a=1}^N\frac{ds_a}{2\pi is_a}\left(\zeta q^{\frac{\fn_3}{2}+\fn_C}t_3^{-\frac{1}{2}}\right)^{\fm_a}|s_a|^{N-2a+1-\frac{2n_a}{\omega}}\right]\prod_{\substack{a,b=1 \\ a\neq b}}^N|1-s_{ab}|\notag\\
    \prod_{I=1}^3\left[\frac{(t_I^{-1}q^{\fn_I+2};q^2)_\infty^N}{(t_Iq^{-\fn_I};q^2)_\infty^N}\prod_{\substack{a,b=1 \\ a>b}}^N\frac{(s_{ba}t_I^{-1}q^{\fn_I+2};q^2)_\infty}{(\overline{s}_{ba}t_Iq^{-\fn_I};q^2)_\infty}\frac{(\overline{s}_{ba}t_I^{-1}q^{\fn_I+2};q^2)_\infty}{(s_{ba}t_Iq^{-\fn_I};q^2)_\infty}\right]\notag\\
    \prod_{a=1}^N\left[\frac{(s_a^{-1}t_1^{-\frac{1}{2}}t_2^{-\frac{1}{2}}q^{1-\frac{\fn_3}{2}};q^2)_\infty}{(\overline{s}_a^{-1}t_1^{\frac{1}{2}}t_2^{\frac{1}{2}}q^{\frac{\fn_3}{2}+1};q^2)_\infty}\frac{(\overline{s}_a^{-1}t_1^{-\frac{1}{2}}t_2^{-\frac{1}{2}}q^{1-\frac{\fn_3}{2}};q^2)_\infty}{(s_a^{-1}t_1^{\frac{1}{2}}t_2^{\frac{1}{2}}q^{\frac{\fn_3}{2}+1};q^2)_\infty}\right]\,.
    \end{gather}
    Throughout the algbebraic manipulations, we have assumed that $\omega\in i\bR^+$, so that $q\in (0,1)$.  

    The Cardy limit that we are interested in is 
    \begin{equation}
    		|q|\to1^- \quad \Longleftrightarrow \quad \omega\to\ii0^+,
    		\label{eq:qomega}
    	\end{equation}
    In this limit, the sum over fluxes can be approximated by an $N$-dimensional integral using the Euler-Maclaurin formula, and one can show that \cite{Bobev:2022wem}
    \be
    \sum_{\fm\in\bZ^N}\int \prod_{a=1}^N\frac{ds_a}{2\pi is_a}(\cdots) = \int \prod_{a=1}^N\frac{dU_ad\overline{U}_a}{-4i\pi^2\omega}(\cdots)(1+\cO(\omega))\,.
    \ee 
    The integrand of the index can also be simplified using the identity
    \be
    \lim_{\omega\rightarrow i0^+}(aq^m;q^2)_\infty = \exp\left[-\frac{i}{2\pi\omega}\Li_2(a)+\frac{m-1}{2}\Li_1(a)\right](1+\cO(\omega))\,.
    \ee
    The result is
    \begin{gather}
    \cI_{\rm sc}= \frac{1}{N!} \int \prod_{a=1}^N\frac{dU_ad\overline{U}_a}{-4i\pi^2\omega}\exp[\frac{1}{\pi\omega}\im W+ \re W^{(1)}+\log (-1)^\frac{N(N-1)}{2}+\cO(\omega)]\,,\notag\\
    W\equiv\left(\frac{\Delta_3}{2}-\Delta_C\right)\sum_{a=1}^NU_a+N\sum_{I=1}^3\Li_2(t_I^{-1})+\sum_{I=1}^3\sum_{a>b}\left[\Li_2(s_{ba}t_I^{-1})-\Li_2(s_{ba}t_I)\right]\notag\\
    +\sum_{a=1}^N\left[\Li_2\left(s_a^{-1}t_1^{-\frac{1}{2}}t_2^{-\frac{1}{2}}\right)-\Li_2\left(s_a^{-1}t_1^{\frac{1}{2}}t_2^{\frac{1}{2}}\right)\right]+\sum_{a=1}^N 2\pi n_a U_a\,,\notag\\
    W^{(1)}\equiv i\sum_{a=1}^N\left(\frac{\fn_3}{2}+\fn_C+N-2a+1\right)U_a-\sum_{\substack{a,b=1 \\ a\neq b}}^N\Li_1(s_{ba})+N\sum_{I=1}^3(\fn_I+1)\Li_1(t_I^{-1})\notag\\
    +\sum_{I=1}^3(\fn_I+1)\sum_{\substack{a,b=1 \\ a>b}}^N\left[\Li_1(s_{ba}t_I^{-1})+\Li_1(s_{ba}t_I)\right]-\frac{\fn_3}{2}\sum_{a=1}^N\left[\Li_1\left(s_a^{-1}t_1^{-\frac{1}{2}}t_2^{-\frac{1}{2}}\right)+\Li_1\left(s_a^{-1}t_1^{\frac{1}{2}}t_2^{\frac{1}{2}}\right)\right]\label{eq: Isc Cardy asymp int} 
    \end{gather}
    In the Cardy limit, the integral above can be computed in the saddle point approximation, where the saddle point equations are $\partial_{U_a}W = 0$. One can check that $\partial_{U_a}W$ exactly coincides with $W_a$ in the exponent of the Bethe ansatz equations \eqref{eq:CZZBAE}, up to redefinitions of the integer ambiguities $n_a$. Thus $W$ is precisely the Bethe potential, or the effective twisted superpotential of the theory reduced on $S^1$. The saddle point equations coincide with the Bethe ansatz equations, whose solutions are the Bethe roots $\lambda$. As is standard in the saddle point approximation, the integral up to next-to-leading order is given by a sum over saddle points (i.e. Bethe roots $\lambda$), where the contribution of each saddle is the integrand evaluated at that saddle, together with the one-loop determinant $(\det\frac{\partial^2 W}{\partial U_a\partial U_b})^{-\frac{1}{2}}$. Direct computation shows that the combination $\left|(\det\frac{\partial^2 W}{\partial U_a\partial U_b})^{-\frac{1}{2}}e^{W^{(1)}}\right|$ coincides with $|\cI_{0,\lambda}|$ in \eqref{eq: index BA formula}, which is the absolute value of the contribution of the Bethe root $\lambda$ to the $\fg = 0$ topologically twisted index (TTI). We thus obtain    
    \be\label{eq:SCItoTTIroot}
    \cI_{\rm sc}=\sum_{\lambda\in\text{BAE}}\exp\left[\frac{1}{\pi\omega}\im W|_\lambda +\re\log\cI_{0,\lambda} + i\varphi + \cO(\omega) \right]\equiv\sum_{\lambda\in\text{BAE}}\cI_{{\rm sc},\lambda}\,,
    \ee
    where $W|_\lambda$ is the Bethe potential evaluated on the Bethe root $\lambda$, and $i \varphi = \log(-1)^{\frac{N(N-1)}{2}}-\frac{1}{2}\log(-1)^N$. In the following subsections, we will evaluate the contributing summands $\cI_{{\rm sc},\lambda}$ in \eqref{eq:SCItoTTIroot} corresponding to the 1,3-cut solutions and the interpolating solution. 
	
	%%%%%%%%%%%%%%%%%%%%%%%%
	\subsection{Contribution of the one-cut solution and equivariant localization}
	\label{subsec:SCI:onecut}
	%%%%%%%%%%%%%%%%%%%%%%%%
	In the contribution of the one-cut solution in \eqref{eq:SCItoTTIroot}, the topologically twisted index $\cI_{0,\lambda}$ was already computed in Section \ref{subsec:oneCut:numerics}, and we only have to evaluate the Bethe potential $W$ on the solution. As was found in \cite{Bobev:2022wem}, the result is
	\begin{gather} 
			- \log \cI_{{\rm sc},{\rm BH}} = \frac{2}{\omega} \left[ \frac{1}{\pi} \frac{\sqrt{2\hat{\Delta}_1 \hat{\Delta}_2 \hat{\Delta}_3  \hat{\Delta}_4  }}{3} \hat{N}^{\frac{3}{2}} + \hat{g}_0 + \hat{g}_{\rm np} \right] - \frac{\sqrt{2\hat{\Delta}_1 \hat{\Delta}_2 \hat{\Delta}_3  \hat{\Delta}_4  }}{3} \sum_{I=1}^4 \frac{\hat{\fn}_I}{\hat{\Delta}_I} \left( \hat{N}^{\frac{3}{2}}\right.\notag\\
            \left.+ \fc_I^{(\rm BH)} \hat{N}^{\frac{1}{2}} \right) + \frac{1}{2} \log \hat{N} + \hat{f}_0 + \hat{f}_{\rm np} + O(\omega)\,,\quad \hat{N}\equiv N + \delta^{(\rm BH)}\,,\label{eq:SCI-BH-loc}
	\end{gather}
	where $\hat{g}_0\sim\cO(1)$, $\hat{g}_{\rm np}\sim \cO(e^{-\#\sqrt{N}})$, and the chemical potentials $\hat{\Delta}_I$ are defined as in \eqref{eq:CZZnotation}. The coefficients $\fc_I^{(\rm BH)}$ and $\delta^{(\rm BH)}$ are identical to those defined for the TTI in \eqref{eq:coeExactBH}. Similarly, $\hat{f}_0$ and $\hat{f}_{\rm np}$ are exactly those of the TTI, which are studied in detail in Section \ref{subsec:oneCut:numerics}. In the following, we will explain how \eqref{eq:SCI-BH-loc} can be reproduced by the equivariant localization computation in \cite{BenettiGenolini:2026cyc}.

    Recall that the superconformal index was initially written in terms of the chemical potentials $\delta_{1,2,3}$, $\delta_C$ defined in \eqref{eq: SCI flav chem pot def} and \eqref{eq: SCI orig flav aux}. Similarly to \eqref{eq:CZZnotation}, it is more convenient to work in terms of the ABJM chemical potentials
    \be
    \hat{\delta}_I=\left(\delta_1\,,\delta_2\,,\frac{\delta_3}{2}+\delta_C\,,\frac{\delta_3}{2}-\delta_C\right)\,,
    \ee
    which are directly conjugate to the toric action of $\U(1)^4$ on $\bC^4\supset S^7$. The setup of the bulk computation of \cite{BenettiGenolini:2026cyc} is mostly analogous to Section \ref{subsec:equiv loc BH} for the TTI. In this case, we want to compute the partition function of M-theory on $S^7\hookrightarrow M \rightarrow M_4$, where $M_4$ is a rotating, electrically charged (Kerr-Newman) AdS$_4$ black hole, and $M$ is its 11-dimensional uplift. The Killing vector constructed from the Killing spinor bilinear again has two fixed points at the north ($p=1$) and south ($p=2$) poles of the horizon $S^2$. The tangent weights $b_{p,i=1,2}$, $u_{p,I}$, and the signs $\sigma_{p,i=1,2}$ specifying the spinor projections at the fixed points should be identified as
    \bea\label{eq: ref SCI loc iden orig}
        (b_{1,1}\,,b_{1,2}) = (1\,,\omega)\,,\quad u_{1,I}&=-\frac{\hat{\delta}_I}{2\pi}-\frac{\omega}{4}\,,\quad \sigma_{1,1} = \sigma_{1,2} = 1\,,\\
        (b_{2,1}\,,b_{2,2}) = (1\,,-\omega)\,,\quad u_{2,I}&=-\frac{\hat{\delta}_I}{2\pi}-\frac{\omega}{4}\,,\quad \sigma_{2,1} = -\sigma_{2,2} = 1\,.
    \eea
    Note that with this identification, the zero charge constraint $\sigma_{p,1}b_{p,1}+\sigma_{p,2}b_{p,2}+\sum_Iu_{p,I}=0$ in \cite{BenettiGenolini:2026cyc} is automatically satisfied. Now, using \eqref{eq: SCI aux flux def} and \eqref{eq: SCI new aux chem pot}, it is straightforward to show that the old chemical potentials $\hat{\delta}_I$ and new chemical potentials $\hat{\Delta}_I$ are related as
    \be
    \hat{\delta}_I=\hat{\Delta}_I-\left(\hat{\fn}_I+\frac{1}{2}\right)\pi\omega\,.
    \ee
    The identifications in \eqref{eq: ref SCI loc iden orig} can then be written as  
    \bea\label{eq: ref SCI loc iden new}
        (b_{1,1}\,,b_{1,2}) = (1\,,\omega)\,,\quad u_{1,I}&=-\frac{\hat{\Delta}_I}{2\pi}+\omega\frac{\hat{\fn}_I}{2}\,,\quad \sigma_{1,1} = \sigma_{1,2} = 1\,,\\
        (b_{2,1}\,,b_{2,2}) = (1\,,-\omega)\,,\quad u_{2,I}&=-\frac{\hat{\Delta}_I}{2\pi}+\omega\frac{\hat{\fn}_I}{2}\,,\quad \sigma_{2,1} = -\sigma_{2,2} = 1\,.
    \eea
    Recall that the M-theory partition function is given in terms of the coefficients $B_p$ and $C_p$ in \eqref{eq: loc comb def}. Plugging the identifications in \eqref{eq: ref SCI loc iden new} into \eqref{eq: loc comb def} and expanding in the Cardy limit $\omega\rightarrow i0^+$ gives
    \begin{gather}
        B_1=B_2=-\delta^{(\text{BH})} + \frac{\pi \omega}{3}\sum_{I=1}^4\frac{\fc_I^{(\rm BH)}\hat{\fn}_I}{\hat{\Delta}_I}+\cO(\omega^2)\,,\notag\\
        C_1^{-\frac{1}{2}}=C_2^{-\frac{1}{2}}=\frac{1}{\sqrt{2}\pi\omega}\sqrt{\prod_{I=1}^4\hat{\Delta}_I}\left(1-\frac{\pi\omega}{2}\sum_{I=1}^4\frac{\hat{\fn}_I}{\hat{\Delta}_I} +\cO(\omega^2)\right)\,.
    \end{gather}
    Plugging these coefficients into the formula \eqref{eq: M th Z} for the M-theory partition function, and simplifying in the Cardy limit using \eqref{eq: Ai asymp}, one can check that the result exactly coincides with \eqref{eq:SCI-BH-loc} up to the constant map and non-perturbative corrections. Recently, an analytic expression for $\hat{g}_0$ was obtained in \cite{BenettiGenolini:2026kjy}.
    
	%%%%%%%%%%%%%%%%%%%%%%%%
	\subsection{The three-cut and interpolating solutions}
	\label{subsec:SCI:threecut}
	%%%%%%%%%%%%%%%%%%%%%%%%
	
	For the 3-cut solution, the next-to-leading contribution $\cI_{0,\lambda}$ in \eqref{eq:SCItoTTIroot} was already computed in Section \ref{subsubsec:multi-cut:3cut:numerics}, and we only need to compute the Bethe potential in the leading contribution. By evaluating the Bethe potential on the numerical 3-cut solutions in region $\tilde{3}$ of Table \ref{tab: 3cut reg} for a range of large $N$, we observe that the numerical data closely adhere to the ansatz  
	\begin{gather}
		- \log \cI_{{\rm sc}, {\rm 3-cut}} = \frac{2}{(3\omega)} \left[ \frac{\sqrt{2 [3\hat{\Delta}_1] [3\hat{\Delta}_2]  [3\hat{\Delta}_3]  [3\hat{\Delta}_4]  }}{9\pi} \hat{N}^{\frac{3}{2}} + \hat{g}_0 + \hat{g}_{\rm np} \right]\notag\\
		- \frac{\sqrt{2 [3\hat{\Delta}_1] [3\hat{\Delta}_2]  [3\hat{\Delta}_3]  [3\hat{\Delta}_4]  }}{9} \sum_{I=1}^4 \frac{\hat{\fn}_I}{ [3\hat{\Delta}_I] } \left( \hat{N}^{\frac{3}{2}} + \fc_I^{(\rm 3-cut)}([3\hat{\Delta}]) \hat{N}^{\frac{1}{2}} \right) + \frac{1}{2} \log \hat{N} + \hat{f}_0 + \hat{f}_{\rm np} + O(\omega)\,, \notag\\
        \hat{N} = N + \delta^{(\rm 3-cut)}([3\hat{\Delta}])\,.\label{eq: SCI 3cut Cardy}
	\end{gather}
    This is consistent with the M-theory partition function of the $\bZ_3$ orbifold of the uplifted Kerr-Newman black hole. Since the $\bZ_3$ orbifold projection acts simultaneously on Euclidean time, the azimuthal angle on $S^2$, and the angles $\phi_I$ internal to $S^7$ \cite{BenettiGenolini:2023rkq}, by the same arguments as Section \ref{subsec:multi-cut:orb grav}, the functional form of \eqref{eq: SCI 3cut Cardy} is exactly the same as the one-cut contribution in \eqref{eq:SCI-BH-loc}, with $\hat{\Delta}_I$ and $\omega$ being replaced by $[3\hat{\Delta}_I]$ and $3\omega$. The only difference is an overall factor of $1/3$ in \eqref{eq: SCI 3cut Cardy}, which is expected from the $\bZ_3$ identification in the orbifold. 
    
    This is easily generalized to $m$-cut solutions, which are expected to be dual to the $\bZ_m$ orbifolds of AdS$_4$ Kerr-Newman black holes constructed in \cite{BenettiGenolini:2023rkq}.\footnote{The $1/m$ scaling applies to the local action and boundary terms inherited from a covering solution, integrated with the orbifold measure at fixed bulk couplings \cite[Section~5.2]{BenettiGenolini:2023rkq}. Its Euler term uses the Euler--Satake characteristic $\chi_{\rm ES}(X/\mathbb Z_m)=\chi(X)/m$. Contributions localized at fixed sets, if required, must be added separately. This prescription differs from the conical completion assumed for the logarithmic correction in Section~\ref{subsubsec:holography:log:4dpers}, which yields the ordinary topological Euler characteristic.} Its contribution $-\log\cI_{{\rm sc},{\rm m-cut}}$ should be $1/m$ times the one-cut contribution in \eqref{eq:SCI-BH-loc}, with $\omega$ and $\hat{\Delta}_I$ being replaced by $m\omega$ and $[m\hat{\Delta}_I]$ respectively:
    \begin{gather}
    	-\log\cI_{{\rm sc},{\rm m-cut}} = \frac{2}{m\omega} \left[ \frac{\sqrt{2 [m\hat{\Delta}_1] [m\hat{\Delta}_2]  [m\hat{\Delta}_3]  [m\hat{\Delta}_4]  }}{3m\pi} \hat{N}^{\frac{3}{2}} + \hat{g}_0 + \hat{g}_{\rm np} \right]\notag\\
    	- \frac{\sqrt{2 [m\hat{\Delta}_1] [m\hat{\Delta}_2]  [m\hat{\Delta}_3]  [m\hat{\Delta}_4]  }}{3m} \sum_{I=1}^4 \frac{\hat{\fn}_I}{ [m\hat{\Delta}_I] } \left( \hat{N}^{\frac{3}{2}} + \fc_I^{(\rm m-cut)}([m\hat{\Delta}]) \hat{N}^{\frac{1}{2}} \right) + \frac{1}{2} \log \hat{N} + \hat{f}_0 + \hat{f}_{\rm np} + O(\omega)\,,\notag\\
        \hat{N} = N + \delta^{(\rm m-cut)}([m\hat{\Delta}])\,.
	\end{gather}
	
	For the interpolating solutions with parameter $\alpha$ (see Section \ref{sec:multi-cut:mixing}) at the superconformal point of region $\tilde{3}$, we find that the numerical data are closely reproduced by the ansatz
	\begin{gather}
			- \log \cI_{\text{sc},\alpha} = \frac{2}{(3\omega)} \frac{\sqrt{2 [3\hat{\Delta}_1] [3\hat{\Delta}_2]  [3\hat{\Delta}_3]  [3\hat{\Delta}_4]  }}{9\pi} \left( \hat{N}^{\frac{3}{2}} - N_\alpha^{\frac{3}{2}} \right) + \frac{2}{\omega} \left[   \frac{\sqrt{2\hat{\Delta}_1 \hat{\Delta}_2 \hat{\Delta}_3  \hat{\Delta}_4  }}{3\pi} N_\alpha^{\frac{3}{2}}  + \hat{g}_0 + \hat{g}_{\rm np} \right]\notag\\
			 -\frac{\sqrt{2 [3\hat{\Delta}_1] [3\hat{\Delta}_2]  [3\hat{\Delta}_3]  [3\hat{\Delta}_4]  }}{9} \sum_{I=1}^4 \frac{\hat{\fn}_I}{ [3\hat{\Delta}_I] } \left[ \left( \hat{N}^{\frac{3}{2}} - N_\alpha^{\frac{3}{2}} \right) + \fc_I^{(\rm 3-cut)} \hat{N}^{\frac{1}{2}} - \tilde{\fc}_I^{(\rm 3-cut)} N_\alpha^{\frac{1}{2}} \right] \notag\\
			 -\frac{\sqrt{2\hat{\Delta}_1 \hat{\Delta}_2 \hat{\Delta}_3  \hat{\Delta}_4  }}{3} \sum_{I=1}^4 \frac{\hat{\fn}_I}{\hat{\Delta}_I} \left( N_\alpha^{\frac{3}{2}} + \tilde{\fc}_I^{(\rm BH)} N_\alpha^{\frac{1}{2}} \right) + \frac{1}{2} \log \hat{N} + \frac{1}{2} \log N_\alpha + \hat{f}_0 + \hat{f}_{\rm np} + O(\omega)\,,\notag\\
             \hat{N}=N + \delta^{(\rm 3-cut)}\,,\qquad N_\alpha = \alpha N + \delta'\,,
	\end{gather}
	where $\fc_I^{(\rm 3-cut)}$ and $\delta^{(\rm 3-cut)}$ are defined in \eqref{eq:coeExact3Cut}, and $\tilde{\fc}_I^{(\rm BH)}$, $\tilde{\fc}_I^{(\rm 3-cut)}$, $\delta'$ are specified by \eqref{eq: interp coeff}. Expanding the non-perturbative corrections $\hat{f}_{\rm np}$, $\hat{g}_{\rm np}$ according to \eqref{eq: f0 fnp flux decomp} and \eqref{eq: fnp_I N expansion}, we find that
	\begin{equation}
		\hat{g}^{\rm (np)}_{1/2, I} = \hat{f}^{\rm (np)}_{1/2, I} = - \frac{2\sqrt{2}\pi}{3} (1-\sqrt{\alpha}),\quad \hat{g}^{\rm (np)}_{{\rm log}, I} = \frac{1}{2}. 
	\end{equation}

	%%%%%%%%%%%%%%%%%%%%%%%%
	\section{Bethe roots without large $N$ scaling and their holographic duals}
	\label{sec:gravitons}
	%%%%%%%%%%%%%%%%%%%%%%%%

    Besides the Bethe roots analysed so far, which have $\im u_a\sim\cO(N^\frac{1}{2})$ and free energies $F\sim\cO(N^{3/2})$, there are many other solutions in which the imaginary parts of $u_a$ do not scale with $N$ and whose free energies are of $\cO(1)$. A common feature of the latter type of solutions is that the distribution of $\re u_a$ on the complex plane is seemingly random. As such, the phases of the logarithms $\log(1-s_{ab})$, $\log(1-s_{ab}t_I)$, $\log(1-s_a^{\pm 1}t_1^{1/2}t_2^{1/2})$ in the free energy \eqref{eq: log Z} will be random, and their contributions tend to cancel, leaving a residual of $\cO(1)$. In the following, we will speculate about the gravitational configurations which are dual to the Bethe roots with free energies of $\cO(1)$.  

    %%%%%%%%%%%%%%%%%%%%%%%%
	\subsection{A restriction on bulk topologies due to nonzero flux}
	\label{subsec:gravitons:TTI}
	%%%%%%%%%%%%%%%%%%%%%%%%

    Firstly, if at least one of the fluxes $\hat{\fn}_I$ are nonzero, then there is a constraint on the topology of any contributing bulk configuration in the absence of magnetic sources. Note that the premise of a nonzero flux is always true for $\fg\neq 1$ due to the topological twist condition $\sum_{I=1}^4\hat{\fn}_I=2(\fg-1)$. The only exception is the case $\fg = 1$ and $\hat{\fn}_I=0$, which corresponds to the original Witten index defined in \cite{Witten:1982df}. To describe the constraint, let $M_4$ be a smooth Euclidean bulk with asymptotic boundary $\partial M_4 = \Sigma_\fg\times S^1$, and let $t_0$ be a fixed reference point on $S^1$. Define the inclusion
    \be\label{eq: incl bdy slice def}
    i:\Sigma_\fg\times \{t_0\}\hookrightarrow M_4
    \ee 
    mapping the spatial slice at a fixed Euclidean time into the bulk. Suppose that $\hat{\fn}_I\neq 0$, and that there are no magnetic sources in the bulk. Then the bulk gauge field $A_I$ whose restriction to the boundary has flux $\hat{\fn}_I$ is the connection on a line bundle $L\rightarrow M_4$ with nontrivial first Chern class $c_1(L)\neq 0$, and we have 
    \be\label{eq: flux expr top pairing}
    \hat{\fn}_I = \frac{g}{2\pi}\int_{\Sigma_\fg}dA_I|_{\partial M_4} = g\left\langle i^*c_1(L),[\Sigma_\fg]\right\rangle=g\left\langle c_1(L),i_*[\Sigma_\fg]\right\rangle \neq 0\,,
    \ee
    where $i^*$ and $i_*$ denote the pullback and pushforward under the inclusion $i$, $[\Sigma_\fg]$ is the homology class of $\Sigma_\fg$, and $\langle \cdot\,,\cdot \rangle$ is the standard pairing between cohomology and homology classes. Since this pairing is nonzero, we must have
    \be
    i_*[\Sigma_\fg]\neq 0 \in H_2(M_4,\bZ)\,.
    \ee 
    This holds for any choice of reference point $t_0$ in \eqref{eq: incl bdy slice def}. That is, the boundary surface $\Sigma_\fg$ must remain a nontrivial 2-cycle in the bulk. In particular, this excludes geometries such as thermal AdS$_4$ with topology $B_3\times S^1$ when $\fg = 0$ \cite{deBoer:2004yu}, and topologies $H_\fg\times S^1$ for $\fg \neq 0$ more generally, where $H_\fg$ is a genus $\fg$ handlebody. In these cases, since $\partial H_\fg = \Sigma_\fg$, $\Sigma_\fg$ is contractible in the bulk and therefore $i_*[\Sigma_\fg] =0$. 

    Equivalently, if $\Sigma_\fg\times\{t_0\}=\partial C$ for a bulk three-chain $C$, the source-free Bianchi identity and Stokes' theorem imply
    \be
    \hat{\fn}_I=\frac{g}{2\pi}\int_{\partial C}F_I
    =\frac{g}{2\pi}\int_CdF_I=0\,.
    \ee
    This obstruction is independent of the metric. By contrast, in the black-saddle or bolt topology $D_2\times\Sigma_\fg$, the Euclidean time circle bounds $D_2$ while $\Sigma_\fg$ remains a nontrivial two-cycle and can support magnetic flux without sources \cite{Toldo:2017qsh,Bobev:2020pjk}.

    %%%%%%%%%%%%%%%%%%%%%%%%
	\subsection{Remaining possibilities}
	\label{subsec:gravitons:possibilities}
	%%%%%%%%%%%%%%%%%%%%%%%%

    One possibility for the duals of Bethe roots with no large $N$ scaling are bulk geometries with $i_*[\Sigma_\fg]\neq 0$, but where the Euclidean time circle is non-contractible and there is no horizon or bolt carrying an entropy contribution. This is unlike the black saddles of \cite{Bobev:2020pjk} with $D_2\times\Sigma_\fg$ topology. This would require the existence of a completely new smooth supersymmetric solution of 11d supergravity (or of the truncated STU model) with boundary $\Sigma_\fg\times S^1$ and the required magnetic flux.
    
    A more economical way of satisfying the requirements of $i_*[\Sigma_\fg]\neq 0$ and zero classical action is to consider a magnetic source of $\hat{\fn}_I$ units of flux dressing the thermal AdS$_4$-like geometries $M_4$ with topology $H_\fg\times S^1$. If the magnetic source has worldline $W=\{x_0\}\times S^1$ where $x_0$ is some point in $H_\fg$, the Bianchi identity for the curvature 2-form $F_I$ is modified to
    \be
    dF_I = 2\pi\hat{\fn}_I\delta_3(W)\,,
    \ee
    where $\delta_3(W)$ is the delta-form that is Poincar\'{e} dual to $W$. Thus the connection 1-form $A_I$ with $dA_I = F_I$ cannot be smoothly extended to the whole of $M_4$ including $W$. The proper formulation is to remove $W$ and regard $A_I$ as the connection on a line bundle $L\rightarrow M_4\setminus W$. Then \eqref{eq: flux expr top pairing} still holds, but the inclusion $i$ must be modified to
    \be
    i:\Sigma_\fg\times \{t_0\}\hookrightarrow M_4\setminus W\,.
    \ee
    Now 
    \be
    i_*[\Sigma_\fg]\neq 0 \in H_2(M_4\setminus W,\bZ)\,.
    \ee
    This evades the previous argument forbidding $H_\fg\times S^1$, since a 2-cycle that is trivial in $M_4$ might be nontrivial in $M_4\setminus W$, as is the case here. In addition, since empty thermal AdS$_4$ has zero renormalized on-shell action under the appropriate supersymmetric scheme \cite{BenettiGenolini:2016tsn,BenettiGenolini:2019jdz}, it is reasonable to expect that the same is true of the dressed AdS$_4$ solutions. Strictly speaking, one should solve for the combined system of bulk fields in 11d and the brane embedding describing the magnetic sources, and evaluate its on-shell action. We leave this computation for future study.
	
	%%%%%%%%%%%%%%%%%%%%%%%%
  	\section{Future directions}
	\label{sec:discussion}
	%%%%%%%%%%%%%%%%%%%%%%%%

    From the results presented above, and in the 4d context in \cite{Aharony:2021zkr}, there is considerable evidence that the Bethe roots at large $N$ are in one-to-one correspondence with solutions of semiclassical gravity. Thus, the Bethe ansatz formula reduces to the classical sum over saddle points in the large $N$ limit, and provides a natural quantum generalization of this sum at finite $N$. This might prove to be a useful way of organizing the quantum gravity path integral. Since the algorithm that we have developed is capable of finding all the $p(N)$ Bethe solutions, assuming that only physical solutions in the integrable system can contribute to the index, we can potentially study every contribution to the gravitational path integral exhaustively. This opens up a variety of future directions: 
    \begin{itemize}
        \item In the previous work \cite{Crew:2024fom}, it was conjectured that the 1-cut (black hole) solution corresponds to a triangular partition for any given $N$. However, this is in fact only true for a certain range of chemical potentials. This is straightforward to demonstrate: consider swapping $\Delta_1\leftrightarrow\Delta_2$. From \eqref{eq: classical BAE sols}, it is clear that the initial conditions at $\zeta =0$ corresponding to a partition $\lambda$ and its transpose $\lambda^T$ are swapped. Thus a Bethe root originally associated with $\lambda$ will now be associated with $\lambda^T$. This is also expected from the Stokes phenomena relating holomorphic blocks in different chambers of chemical potentials \cite{Beem:2012mb}. It would be desirable to delineate the precise chambers of chemical potentials within which the identifications between partitions and Bethe roots are fixed, and investigate whether there are any relationships between the shape of the partition and the characteristics of the corresponding solution, such as the number of cuts, number of outlying points dressing a cut, etc. More speculatively, the correspondence between integer partitions, Bethe roots, and contributions to the gravitational path integral is reminiscent of similar correspondences found in the LLM solutions \cite{Lin:2004nb} and the Nekrasov instanton partition function \cite{Nekrasov:2002qd}. It would be interesting to understand if there is any deeper reason why the sum over partitions is appearing in these different contexts, and whether some of the methods and findings in these other areas can inform our study of the topologically twisted index.   
        
        \item Since the ADHM theory with $N_f = 1$ is dual to the ABJM theory with $k=1$, one expects that the BAEs of the two theories should correspond to two equivalent integrable systems, as was found for dual pairs in \cite{Orlando:2010uu, Gaiotto:2013bwa}. It would be interesting to identify the integrable system corresponding to ABJM theory, and map the Bethe roots of ADHM to those of ABJM. In particular, there should also exist a one-to-one correspondence between the Bethe roots of ABJM and partitions of $N$. Thus one might expect there to be a similar limit like $\zeta\rightarrow 0$ in ADHM where the Bethe roots simplify into a form that is specified by the partitions.     
        
        \item There are many large $N$ BAE solutions which are still poorly understood. In particular, there is a discrete family of 3-cut solutions with unequal filling fractions which interpolate between the 1-cut and 2-cut solutions. In this family, there is very limited overlap between the values of $\im u_a$ in the central cut along the imaginary axis, and those in the other two symmetrically placed cuts, as seen in Figure \ref{fig: eg 1to2interp}. Thus there seems to be a ``repulsive force'' between the central cut and the other two cuts. Contrast this with the interpolating solution in Section \ref{sec:multi-cut:mixing}, whose distribution of imaginary parts is plotted in Figure \ref{fig: cont sol eg}. One can observe that there is a range of $\im u_a$ where the cuts completely overlap, and the density of holonomies is higher in the region of overlap. Thus there seems to be an ``attractive force'' in this case.  
        
        \begin{figure}[H]
            \centering        
            \includegraphics[width=0.5\textwidth]{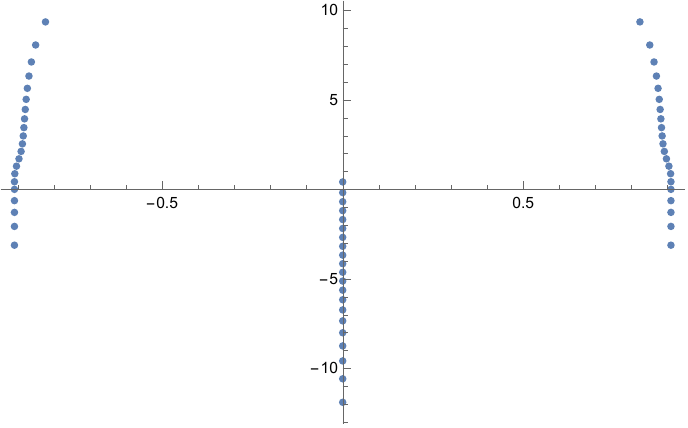}
            \caption{\small Plot of $u_a$ in the complex plane for a member of the family interpolating between the 1-cut and 2-cut solutions.}
            \label{fig: eg 1to2interp}
        \end{figure} 
        
        It would be interesting to analyse the formation of multi-cut solutions and interpolating solutions in terms of the matrix model framework of balancing attractive and repulsive forces \cite{Morita:2017oev}. Perhaps such a study might shed light on what other discrete families of solutions exist, besides those interpolating between the 1-cut and 3-cut solutions, and between the 1-cut and 2-cut solutions.  
        
        \item Recall that the conventional gravitational ansatz of \cite{Bobev:2022jte} fails to fit the numerical free energy of the multi-cut solutions when the corresponding orbifold action has a 7d fixed locus $\Sigma_\fg\times S^5$ on which a 7d SYM theory lives. If one can compute the free energy of the 7d SYM theory perturbatively in $1/N$, adding it to the gravitational ansatz would give an ansatz for the total free energy, which we can then compare with numerical data. A successful match would be a highly non-trivial precision test of the AdS/CFT correspondence.
          
        \item We have not explored the case of complex chemical potentials $\Delta_{1,2}$, $\Delta_C$. As the imaginary parts of $\Delta$ are turned on, the Bethe roots will cease to be organized into complex conjugate pairs or self-conjugate solutions, and there might be phase transitions along the lines of \cite{Copetti:2020dil, Chen:2022hbi} for the superconformal index. It would be interesting to study this further. 
        
        \item There is a direct generalization of the work here to ADHM theory with $N_f > 1$. The BAEs for general $N_f$ is 
        \begin{equation} 
        	B_a = (s_a - t_1^{\frac{1}{2}}t_2^{\frac{1}{2}} )^{N_f} \prod_{b\ne a} \prod_{I=1}^3 (s_b - t_I^{-1}s_a) +  \zeta t_1^{\frac{N_f}{2}}t_2^{\frac{N_f}{2}} (s_a - t_1^{-\frac{1}{2}}t_2^{-\frac{1}{2}} )^{N_f} \prod_{b\ne a} \prod_{I=1}^3 (s_b - t_I s_a) = 0\,,
        \end{equation}
        which coincides with the Bethe equations of quantum toroidal \(\fg\fl_1\) under a different module \(\cF(t_1^{\frac{1}{2}}t_2^{\frac{1}{2}})^{\otimes N_f}\), the \(N_f\)-fold tensor product of Fock modules \cite{Feigin:2015raa}. This would also enable us to consider the alternative large $N$ limit, $N,N_f\rightarrow \infty$ with $\lambda\equiv N/N_f$ fixed, where the theory is dual to type IIA string theory on AdS$_4\times \bC\bP^3$.
        
        \item Since the $V^{5,2}$ theory \cite{Martelli:2008si,Hosseini:2016ume} has the same field content as ADHM theory but a different superpotential, its BAEs are identical to those of ADHM that we have considered, with the only difference being a constraint on the chemical potentials
        \begin{equation}
        			\Delta_1 + \Delta_2 = \frac{4\pi}{3},\quad \Delta_3 = \frac{2\pi}{3}\,.
        \end{equation}
        It would be a straightforward exercise to repeat the computations in this paper for the $V^{5,2}$ theory and study the similarities and differences.
    \end{itemize}
    
	%%%%%%%%%%%%%%%%%%%%%%%%
	\bigskip
	
	\noindent\textbf{Acknowledgments}
	
	\medskip
	
	\noindent  We would like to thank Diandian Wang for the collaboration at an early stage of the project. We are grateful to Francesco Benini, Alberto Zaffaroni, Federico Faedo, Chiung Hwang, Jun Nian, Tomoki Nosaka, Alessia Segati, and Jiahua Tian for useful discussions. XZ would like to thank Nikolay Bobev, Junho Hong, and Valentin Reys for collaborations on related projects and useful discussions. ZZ would like to thank Daniel Zhang and Samuel Crew for collaborations on related projects. The work of CP, XZ, ZZ is supported by NSFC NO. 12675092, 12175237, the work of CP is also supported by NSFC NO. 12447108, and 12247103, the Fundamental Research Funds for the Central Universities. 
	%%%%%%%%%%%%%%%%%%%%%%%%
	
	%%%%%%%%%%%%%%%%%%%%%%%%
	\appendix
	\addtocontents{toc}{\protect\setcounter{tocdepth}{1}}
	%%%%%%%%%%%%%%%%%%%%%%%%
	
	%%%%%%%%%%%%%%%%%%%%%%%%
	\section{Comparison with the convention of \cite{Bobev:2022eus}}
	\label{Appsubsec:TTI:ConventionComparison}
	%%%%%%%%%%%%%%%%%%%%%%%%

	We now compare our notation, following CZZ \cite{Crew:2024fom} above, with the notation used in large $N$ studies of the ADHM topologically twisted index by Bobev, Hong, and Reys \cite{Bobev:2022eus, Bobev:2023lkx} and related works (denoted by ``BHR'' below).  Throughout this section, unadorned symbols such as $(\Delta_{1,2,3}$, $\Delta_C$, $\fn_{1,2,3}$, $\fn_C$, $t_{1,2,3}$, $\zeta$, $\hat{\Delta}_{I=1,\ldots,4}$, $\hat{\fn}_{I=1,\ldots,4})$ are in the CZZ notation by default. Variables under the convention of \cite{Bobev:2022eus} are denoted by the superscript ({\rm B}).
	
	\subsection{Fields}
	
	The field dictionary is
	\begin{equation}
		(A,B,C,I,J)
		\quad\longleftrightarrow\quad
		(\Psi_1,\Psi_2,\Psi_3,\psi,\widetilde\psi)_{\rm B}.
	\end{equation}
	The superpotential of \cite{Bobev:2022eus}
	\begin{equation}
		W_{\rm B}
		=
		\Tr\!\left(\widetilde\psi\,\Psi_3\psi+\Psi_3[\Psi_1,\Psi_2]\right)
	\end{equation}
	is the same as \eqref{eq:CZZsuperpotential}.
	
	\subsection{Chemical potentials}
	BHR uses the chemical potentials
	\begin{equation}
		y_{1,2,3}=e^{i\pi\Delta_{1,2,3}^{\rm B}},\qquad
		y_q=e^{i\pi\Delta_q^{\rm B}},\qquad
		y_{\tq}=e^{i\pi\Delta_{\tilde{q}}^{\rm B}},\qquad
		\xi=e^{i\pi\Delta_m^{\rm B}},
	\end{equation}
	with the superpotential constraints
	\begin{equation}
		\Delta_1^{\rm B}+\Delta_2^{\rm B}+\Delta_3^{\rm B}=2,\qquad
		\Delta_q^{\rm B}+\Delta_{\tq}^{\rm B}+\Delta_3^{\rm B}=2.
		\label{eq:BobevDeltaconstraint}
	\end{equation}
	For \(N_f=1\), the CZZ chemical potentials are
	\begin{equation}
		t_{1,2,3}=e^{i\Delta_{1,2,3}},\qquad
		\zeta=e^{i\Delta_C},\qquad
		\Delta_1+\Delta_2+\Delta_3=2\pi\,.
	\end{equation}
	  The dictionary is
	\begin{equation}
		\Delta_{1,2,3}=\pi\Delta_{1,2,3}^{\rm B},\qquad
		\Delta_C=-\pi\Delta_m^{\rm B}.
		\label{eq:Deltadictionary}
	\end{equation}
	Equivalently,
	\begin{equation}
		t_{1,2,3}=y_{1,2,3},\qquad
		\zeta\simeq \xi^{-1}.
	\end{equation}
	
	BHR has one extra free chemical potential, \(\Delta_q^{\rm B}\). Writing
	\begin{equation}
		\Delta_q^{\rm B}
		=
		\frac{\Delta_1^{\rm B}+\Delta_2^{\rm B}}{2}+\delta_q,\qquad
		\Delta_{\tq}^{\rm B}
		=
		\frac{\Delta_1^{\rm B}+\Delta_2^{\rm B}}{2}-\delta_q\,,
		\label{eq:deltaqdef}
	\end{equation}
	our convention here fixes
	\begin{equation}
		\delta_q=0,
		\qquad
		\Delta_q^{\rm B}=\Delta_{\tq}^{\rm B}
		=
		\frac{\Delta_1^{\rm B}+\Delta_2^{\rm B}}{2}
		=
		1-\frac{\Delta_3^{\rm B}}{2}.
		\label{eq:CZZdeltaqzero}
	\end{equation}
	In fugacity form this is
	\begin{equation}
		y_q=y_{\tq}=-\,t_3^{-1/2}
	\end{equation}
	up to the same phase convention. At the level of the Bethe equations, the extra parameter $\delta_q$ can be absorbed into a common phase shift of all the gauge fugacities together with an overall phase shift of the index \cite{Bobev:2023lkx}. This shows that $\Delta_q^{\rm B}$ is conjugate to the overall $\U(1)$ factor in the $\U(N)$ gauge group, and does not correspond to a genuine flavour symmetry.
	
	\subsection{Fluxes}
	
	BHR uses magnetic fluxes
	\begin{equation}
		\fn_{1,2,3}^{\rm B},\quad
		\fn_q^{\rm B},\quad \fn_{\tq}^{\rm B},\quad \mathfrak t^{\rm B},
	\end{equation}
	with constraints
	\begin{equation}
		\fn_1^{\rm B}+\fn_2^{\rm B}+\fn_3^{\rm B}=2,\qquad
		\fn_q^{\rm B}+\fn_{\tq}^{\rm B}+\fn_3^{\rm B}=2.
		\label{eq:BobevFluxconstraint}
	\end{equation}
	We will consider $\fg = 0$ here. The CZZ flux variables obey
	\begin{equation}
		\fn_1,\fn_2,\fn_C\in\bZ,\qquad
		\fn_1+\fn_2\in2\bZ,
	\end{equation}
	and it is convenient to define
	\begin{equation}
		\fn_3=-2-\fn_1-\fn_2.
	\end{equation}
	The dictionary between the two sets of flux variables is
	\begin{equation}
		\fn_{1,2,3}=-\fn_{1,2,3}^{\rm B},\qquad
		\fn_C=\mathfrak t^{\rm B}.
		\label{eq:Fluxdictionary}
	\end{equation}
	The remaining BHR flux freedom may be parametrized by $\eta_q$ in
	\begin{equation}
		\fn_q^{\rm B}
		=
		\frac{\fn_1^{\rm B}+\fn_2^{\rm B}}{2}+\eta_q,\qquad
		\fn_{\tq}^{\rm B}
		=
		\frac{\fn_1^{\rm B}+\fn_2^{\rm B}}{2}-\eta_q.
	\end{equation}
	CZZ fixes
	\begin{equation}
		\eta_q=0,\qquad
		\fn_q^{\rm B}=\fn_{\tq}^{\rm B}
		=	\frac{\fn_1^{\rm B}+\fn_2^{\rm B}}{2} =
		-\frac{\fn_1+\fn_2}{2}.
		\label{eq:CZZetaqzero}
	\end{equation}
	The condition \(\fn_1+\fn_2\in2\bZ\) guarantees that the
	right-hand side is integral. In the localization formula of the TTI, the redundant parameter
	\(\eta_q\) can be removed by a common shift of the gauge magnetic fluxes,
	\begin{equation}\label{eq: gauge flux shift redundant}
		\fm_a\longmapsto \fm_a-\eta_q.
	\end{equation}
	The adjoint contribution depends only on differences \(\fm_a-\fm_b\), and is
	therefore invariant under this shift. The only possible remnant is an overall contact term, which does not affect the Bethe equations or
	the \(K\)-theoretic interpretation emphasized in the CZZ convention. Again, \eqref{eq: gauge flux shift redundant} shows that $\eta_q$ is a flux corresponding to the overall $\U(1)$ in $\U(N)$, and is not a genuine flavour flux. 
	
	\subsection{\texorpdfstring{Four-variable large \(N\) parametrization}{Four-variable large N parametrization}}
	
	BHR often packages the variables as
	\begin{equation}
		\tDelta^{\rm B}
		=
		\left(\Delta_1^{\rm B},\Delta_2^{\rm B},
		\frac{\Delta_3^{\rm B}}{2}-\Delta_m^{\rm B},
		\frac{\Delta_3^{\rm B}}{2}+\Delta_m^{\rm B}\right),
	\end{equation}
	and
	\begin{equation}
		\tfn^{\rm B}
		=
		\left(\fn_1^{\rm B},\fn_2^{\rm B},
		\frac{\fn_3^{\rm B}}{2}+\mathfrak t^{\rm B},
		\frac{\fn_3^{\rm B}}{2}-\mathfrak t^{\rm B}\right),
	\end{equation}
	which sum to \(2\). The CZZ variables in \eqref{eq:CZZnotation} instead satisfy
	\begin{equation}
		\sum_{I=1}^4\hat{\Delta}_I=2\pi,\qquad
		\sum_{I=1}^4\hat{\fn}_I=-2.
	\end{equation}
	The compact relation between the two conventions is
	\begin{equation}
		\hat{\Delta}=\pi\,\tDelta^{\rm B},\qquad
		\hat{\fn}=-\,\tfn^{\rm B}.
	\end{equation}
	
	%%%%%%%%%%%%%%%%%%%%%%%%
	\section{The initial step for singular Jacobians}
	\label{App:tracking}
	%%%%%%%%%%%%%%%%%%%%%%%%

    In this appendix, we want to find an approximation of \(s_a(\epsilon)\), the solution the BAEs at \(v=\epsilon\ll 1\), when the Jacobian \(\frac{\partial B_a}{\partial s_b}\) is singular.

    Let \(s_a(\epsilon) = s_{\lambda,a} + \delta s_a\). The Bethe equation at \(v=\epsilon\) is \(B_a(s_\lambda +\delta s,\epsilon)=0\), and expanding it in \(\delta s\) and \(\epsilon\) gives
    \bea\label{eq: BAE expanded}
    &\sum_{n=1}^{3N-2}\sum_{b_1,\ldots,b_n}\frac{1}{n!}\partial_{b_1}\cdots\partial_{b_n}B_a(s_\lambda,0)\delta s_{b_1}\cdots\delta s_{b_n}\\
    &+ \sum_{n=0}^{3N-2}\sum_{c_1,\ldots,c_n}\frac{1}{n!}\partial_{c_1}\cdots\partial_{c_n}\partial_v B_a(s^*,0)\epsilon\, \delta s_{c_1}\cdots\delta s_{c_n} = 0\,.
    \eea
    The expansion terminates at order $3N-2$ because $B_a$ is a polynomial of degree $3N-2$. Let us separate $\delta s_a$ into pieces which scale with a fixed power of $\epsilon$, i.e.
    \be
    \delta s_a = \delta^{(D_a)}s_a + \delta^{(D_a+1)}s_a + \ldots\,,\qquad \delta^{(D)}s_a \sim \epsilon^D\,.
    \ee
    Recall that each label $a$ corresponds to a box in the Young diagram corresponding to the starting solution $s_\lambda$. We then assume that $D_a$ is a strictly monotonically decreasing function of $i_a$ and $j_a$. In other words, as the box $a$ ranges from top left to bottom right in the Young diagram, $D_a=\lim_{\epsilon\rightarrow 0}\frac{\log|s_a|}{\log\epsilon}$ strictly decreases. Given this assumption, it turns out that \eqref{eq: BAE expanded} uniquely determines the leading term $\delta^{(D_a)}s_a$ in each $\delta s_a$. This is achieved iteratively by first determining $\delta^{(D_a)}s_a$ for the boxes $a$ which are corners, that is, those boxes which can be hypothetically removed from the Young diagram to leave another valid Young diagram. Subsequently, we work towards the interior of the diagram, determining $\delta^{(D_a)}s_a$ for the neighbours of the corners, the neighbours of those neighbours, and so on.   
    
    Before delving into an analysis of \eqref{eq: BAE expanded}, we shall need to examine when the derivative components $\partial_{b_1}\cdots\partial_{b_n}B_a(s_\lambda,0)$ and $\partial_{b_1}\cdots\partial_{b_n}\partial_v B_a(s_\lambda,0)$ are nonzero. Given a box $a$ with coordinates $(i_a,j_a)\neq (1,1)$, for each of the neighbours $(i_a-1,j_a)$, $(i_a,j_a-1)$, $(i_a+1,j_a+1)$ which exist, one of the indices among $b_1,\ldots,b_n$ must either be $a$ or the index of the neighbour in question, in order for $\partial_{b_1}\cdots\partial_{b_n}B_a(s_\lambda,0)$ to be nonzero. This is because if there is no derivative acting on $s_a$ or the neighbour's variable $s_b$ in the factor $(s_b-t_I^{-1} s_a)$ in $B_a$, evaluating it on $s_\lambda$ will give zero. Otherwise, if $(i_a,j_a)=(1,1)$, one of the indices among $b_1,\ldots,b_n$ must be $a$ in order for $\partial_{b_1}\cdots\partial_{b_n}B_a(s_\lambda,0)$ to be nonzero. This is to prevent the factor $(s_a-t_1^{\frac{1}{2}}t_2^{\frac{1}{2}})$ from evaluating to zero on $s_\lambda$. In addition, if $(i_a+1,j_a+1)=(2,2)$ exists, its index must also be among $b_1,\ldots,b_n$ for $\partial_{b_1}\cdots\partial_{b_n}B_a(s_\lambda,0)$ to be nonzero. Similarly, for any box $a$, for each of the neighbours $(i_a+1,j_a)$, $(i_a,j_a+1)$, $(i_a-1,j_a-1)$ which exist, one of the indices among $c_1,\ldots,c_n$ must must either be $a$ or the index of the neighbour in order for $\partial_{c_1}\cdots\partial_{c_n}\partial_vB_a(s_\lambda,0)$ to be nonzero. 
    
    Let us now consider \eqref{eq: BAE expanded} for an exterior corner $a$, that is, a corner box that is in the top row or left-most column. In these cases, the Young diagram in the vicinity of $a$ looks like
    
    \begin{equation*}
    \begin{ytableau}
        \times & \times & \times \\
        b & a & \times \\
        ? & \times & \times
    \end{ytableau}\quad\quad \text{or}\quad\quad \begin{ytableau}
        \times & b & ? \\
        \times & a & \times \\
        \times & \times & \times
    \end{ytableau}
    \vspace{.5cm}
    \end{equation*}
    where $\times$ labels boxes which do not exist, and boxes with ``$?$'' could either exist or not exist, depending on the partition $\lambda$. Note that \eqref{eq: BAE expanded} must be satisfied at each order of $\epsilon$, while we shall only focus on the lowest order for which it is not trivially zero. This necessarily means focusing on the lowest number of derivatives in the two terms of \eqref{eq: BAE expanded}. Since the only existing neighbour among $(i_a-1,j_a)$, $(i_a,j_a-1)$, $(i_a+1,j_a+1)$ is either $(i_a,j_a-1)$ or $(i_a-1,j_a)$ labeled by $b$ above, The lowest order contribution to the first term of \eqref{eq: BAE expanded} is $\partial_aB_a(s_\lambda,0)\delta s_a$. The other non-vanishing first derivative term $\partial_bB_a(s_\lambda,0)\delta s_b$ has a higher order of $\epsilon$ due to our assumption that $D_a$ is strictly monotonically decreasing in $i_a$ and $j_a$. Considering the second term of \eqref{eq: BAE expanded}, since none of the neighbours $(i_a+1,j_a)$, $(i_a,j_a+1)$, $(i_a-1,j_a-1)$ exist, the lowest order contribution is $\partial_vB_a(s_\lambda,0)\epsilon$. Putting these together, \eqref{eq: BAE expanded} at the lowest order in $\epsilon$ is
    \be
    \partial_aB_a(s_\lambda,0)\delta s_a + \partial_vB_a(s_\lambda,0)\epsilon = 0\,,
    \ee
    which determines
    \be
    \delta^{(1)} s_a = -\frac{\partial_vB_a(s_\lambda,0)\epsilon}{\partial_aB_a(s_\lambda,0)}\,,\quad D_a = 1\,.
    \ee
    
    For the other (interior) corner boxes $a$ which are not exterior corners, the vicinity of $a$ looks like
    \begin{equation*}
    \begin{ytableau}
         c & b_1 & ? \\
        b_2 & a & \times \\
        ? & \times & \times
    \end{ytableau}
    \vspace{.5cm}
    \end{equation*}
    Of the possible neighbours $(i_a-1,j_a)$, $(i_a,j_a-1)$, $(i_a+1,j_a+1)$, two of them exist, labelled by $b_1=(i_a-1,j_a)$ and $b_2=(i_a,j_a-1)$ above. This means that the lowest order contribution to the first term of \eqref{eq: BAE expanded} is $\frac{1}{2}\partial_a^2B_a(s_\lambda,0)(\delta s_a)^2$. Considering the second term of \eqref{eq: BAE expanded}, since $c=(i_a-1,j_a-1)$ exists, the leading contribution is $\partial_a\partial_vB_a(s_\lambda,0)\epsilon\,\delta s_a$. In total, the Bethe equations set
    \be
    \delta s_a\left(\frac{1}{2}\partial_a^2B_a(s_\lambda,0)\delta s_a + \partial_a\partial_vB_a(s_\lambda,0)\epsilon\right)=0\,.
    \ee
    Assuming that $\delta s_a\neq 0$, this implies
    \be
    \delta^{(1)} s_a = -\frac{\partial_a\partial_vB_a(s_\lambda,0)\epsilon}{\frac{1}{2}\partial_a^2B_a(s_\lambda,0)}\,,\quad D_a = 1\,.
    \ee
    
    Having determined $\delta^{(D_a)} s_a$ for the corner boxes $a$, we want to move on to other boxes in the partition. As we shall see, it turns out that $\delta^{(D_a)} s_a$ is determined by \eqref{eq: BAE expanded} provided that we know $\delta^{(D_c)} s_c$ for all boxes $c$ in the Young diagram with $i_c\geq i_a$ and $j_c\geq j_a$ (these are all the boxes below and to the right of $a$). Thus we can proceed iteratively. The first step determines $\delta^{(D_a)} s_a$ for the corner boxes. Subsequently, in step number $n+1$, we determine $\delta^{(D_a)} s_a$ for which all boxes $c$ with $i_c\geq i_a$ and $j_c\geq j_a$ have already been determined in steps $1$ to $n$. For example, in the partition below, we have labeled each box by the step number in which it should be considered: 
    \begin{equation*}
    \begin{ytableau}
         5 & 4 & 3 & 2 & 1 \\
         4 & 3 & 2 & 1 \\
         2 & 1 \\
         1
        \end{ytableau}    
    \end{equation*}
    In this process, we have to distinguish between many cases for which the configuration of neighbours around the box $a$ are different. 
    
    Firstly, we consider the cases where $a$ is part of a vertical or horizontal edge and is either in the top row or left-most column. That is, the vicinity of $a$ looks like one of the four diagrams below:
    \begin{equation*}
    \begin{ytableau}
         \times & \times & \times \\
        b & a & c \\
        ? & \times & \times
    \end{ytableau} \qquad \begin{ytableau}
        \times & b & ? \\ \times & a & \times \\ \times & c & \times
    \end{ytableau} \qquad \begin{ytableau}
        \times & \times & \times \\ b & a & \times \\ {} & c & \times
    \end{ytableau} \qquad \begin{ytableau}
        \times & b & {} \\ \times & a & c \\ \times & \times & \times 
    \end{ytableau}
    \vspace{.5cm}
    \end{equation*}
    What they have in common is that one neighbour labelled by $b$ exists out of $(i_a-1,j_a)$, $(i_a,j_a-1)$, $(i_a+1,j_a+1)$, and one neighbour labelled by $c$ exists out of $(i_a+1,j_a)$, $(i_a,j_a+1)$, $(i_a-1,j_a-1)$. Following our previous analysis on when the derivative components are nonzero, we find that \eqref{eq: BAE expanded} at lowest order in $\epsilon$ is\footnote{Note that $\partial_a\partial_vB_a(s_\lambda,0) \epsilon\, \delta^{(D_a)} s_a$ is also nonzero but since $D_a > D_c$, this is higher order in $\epsilon$ than the terms that were kept.}
    \be
    \partial_aB_a(s_\lambda,0)\delta^{(D_a)}s_a + \partial_c\partial_vB_a(s_\lambda,0) \epsilon\, \delta^{(D_c)} s_c = 0\,.
    \ee
    This determines
    \be
    \delta^{(D_a)}s_a = -\frac{\partial_c\partial_vB_a(s_\lambda,0) \epsilon\, \delta^{(D_c)} s_c}{\partial_aB_a(s_\lambda,0)}\,,\quad D_a = D_c + 1\,.
    \ee
    
    Otherwise, if $a$ is part of a horizontal or vertical edge but where $i_a > 1$ and $j_a > 1$, as depicted below,
    \begin{equation*}
    \begin{ytableau}
        c_1 & b_1 & {} \\ b_2 & a & c_2 \\ ? & \times & \times 
    \end{ytableau} \qquad \begin{ytableau}
        c_1 & b_1 & ? \\ b_2 & a & \times \\ {} & c_2 & \times
    \end{ytableau}
    \end{equation*}
    there are two neighbours of each type in the positions labelled by $b_{1,2}$ and $c_{1,2}$. A similar analysis of the derivatives shows that \eqref{eq: BAE expanded} at lowest order in $\epsilon$ is
    \be
    \delta^{(D_a)}s_a\left(\frac{1}{2}\partial_a^2B_a(s_\lambda,0)\delta^{(D_a)}s_a + \partial_{c_2}\partial_a\partial_vB_a(s_\lambda,0)\epsilon\,\delta^{(D_{c_2})}s_{c_2}\right) = 0\,,
    \ee
    which determines
    \be
    \delta^{(D_a)}s_a = -\frac{\partial_{c_2}\partial_a\partial_vB_a(s_\lambda,0)\epsilon\,\delta^{(D_{c_2})}s_{c_2}}{\frac{1}{2}\partial_a^2B_a(s_\lambda,0)}\,,\quad D_a = D_{c_2} + 1\,.
    \ee
    
    In the following, since the analysis is repetitive and follows the exact same logic, we will just compile an exhaustive list of the remaining possible configurations of the neighbourhood of a chosen box \(a\), together with the corresponding formula for $\delta^{(D_a)}s_a$. For the cases
    \begin{equation*}
    \begin{ytableau}
        \times & b & {} \\ \times & a & c_2 \\ \times & c_1 & \times 
    \end{ytableau} \qquad \begin{ytableau}
        \times & \times & \times \\ b & a & c_2 \\ {} & c_1 & \times 
    \end{ytableau} \qquad \begin{ytableau}
        \times & \times & \times \\ \times & a & c_2 \\ \times & c_1 & \times
    \end{ytableau}
    \end{equation*}
    we have
    \be
    \delta^{(D_a)}s_a = -\frac{\partial_{c_2}\partial_{c_1}\partial_vB_a(s_\lambda,0)\epsilon\,\delta^{(D_{c_1})} s_{c_1}\delta^{(D_{c_2})} s_{c_2}}{\partial_aB_a(s_\lambda,0)}\,,\quad D_a = D_{c_1} + D_{c_2} + 1\,.
    \ee
    For
    \begin{equation*}
    \begin{ytableau}
        c_3 & b_1 & {} \\ b_2 & a & c_2 \\ {} & c_1 & \times
    \end{ytableau}
    \end{equation*}
    we have
    \be
    \delta^{(D_a)}s_a = -\frac{\partial_{c_2}\partial_{c_1}\partial_a\partial_vB_a(s_\lambda,0)\epsilon\,\delta^{(D_{c_1})} s_{c_1}\delta^{(D_{c_2})} s_{c_2}}{\frac{1}{2}\partial_a^2B_a(s_\lambda,0)}\,,\quad D_a = D_{c_1} + D_{c_2} + 1\,.
    \ee
    For the configurations 
    \begin{equation*}
    \begin{ytableau}
        \times & \times & \times \\ b_1 & a & c_2 \\ {} & c_1 & b_2 
    \end{ytableau}\qquad \begin{ytableau}
        \times & b_1 & {} \\ \times & a & c_2 \\ \times & c_1 & b_2
    \end{ytableau}\qquad \begin{ytableau}
    \times & \times & \times \\ \times & a & c_2 \\ \times & c_1 & b_2 
    \end{ytableau}
    \end{equation*}
    we have
    \be
    \delta^{(D_a)}s_a = -\frac{\partial_{c_2}\partial_{c_1}\partial_vB_a(s_\lambda,0)\epsilon\,\delta^{(D_{c_1})} s_{c_1}\delta^{(D_{c_2})} s_{c_2}}{\partial_{b_2}\partial_aB_a(s_\lambda,0)\delta^{(D_{b_2})} s_{b_2}}\,,\quad D_a = D_{c_1} + D_{c_2} - D_{b_2} + 1\,.
    \ee
    Lastly, for 
    \begin{equation*}
    \begin{ytableau}
        c_3 & b_1 & {} \\ b_2 & a & c_2 \\ {} & c_1 & b_3
    \end{ytableau}
    \end{equation*}
    we have 
    \be
    \delta^{(D_a)}s_a = -\frac{\partial_{c_2}\partial_{c_1}\partial_a\partial_vB_a(s_\lambda,0)\epsilon\,\delta^{(D_{c_1})} s_{c_1}\delta^{(D_{c_2})} s_{c_2}}{\frac{1}{2}\partial_{b_3}\partial_a^2B_a(s_\lambda,0)\delta^{(D_{b_3})} s_{b_3}}\,,\quad D_a = D_{c_1} + D_{c_2} - D_{b_3} + 1\,.
    \ee
    This exhausts all the possible cases one encounters when $a$ ranges over all the boxes of any partition $\lambda$. Since the corner boxes have $D_a=1$ and the orders of the other boxes are determined via the recursion relations above, one can show that the order of a general box $a$ is equal to the number of boxes with $i\geq i_a$ and $j\geq j_a$, with $a$ included. 
    
    Note that in the approximate solution $s=s_\lambda+\delta s$ that we have determined, there does not exist a group of $3$ gauge fugacities $s_{a,b,c}$ that satisfies the wheel condition \eqref{eq: wheel combi}. This is because for every such group existing in $s_\lambda$, $\delta s_{a,b,c}$ necessarily scale with different orders of $\epsilon$, spoiling the relation \eqref{eq: wheel combi}. In addition, since $s_{\lambda,a}\neq s_{\lambda,b}$ for all $a\neq b$ and $\delta s$ is a small correction, there are also no coincident fugacities in $s=s_\lambda + \delta s$. This ensures that the approximate solution that we have found is a physical solution.

    %%%%%%%%%%%%%%%%%%%%%%%%
    \section{Large $N$ limit of Bethe equations}
    \label{App: large N Bethe}
    %%%%%%%%%%%%%%%%%%%%%%%%

    In this appendix, we will derive the large $N$ limit of the Bethe equations under the various multi-cut ans\"atze considered in the paper. The one-cut ansatz has already been considered in \cite{Hosseini:2016tor,Hosseini:2016ume}, but we review the computation below to familiarize the reader with some key steps which will be common to the new 2-cut and 3-cut ans\"atze. 
    
    \subsection{1-cut ansatz}\label{App: large N Bethe 1cut}
    
    We start with the one-cut ansatz in \eqref{eq: BH ansatz}. Plugging it into the Bethe equations \eqref{eq: W'}, the first line can be simplified to
    \bea\label{eq: 1-cut fund fin}
    &-i\Delta_C-N^\frac{1}{2}t+if(t)+\log\left(1-e^{N^\frac{1}{2}t+i\left(\frac{\Delta_1+\Delta_2}{2}-f(t)\right)}\right)-\log\left(1-e^{-N^\frac{1}{2}t+i\left(\frac{\Delta_1+\Delta_2}{2}+f(t)\right)}\right)\\
    &=-i\Delta_C-i\sgn t\frac{\Delta_3}{2}-2\pi i\sgn t\left\lfloor\frac{\frac{\Delta_1+\Delta_2}{2}-\sgn(t) f(t)}{2\pi}\right\rfloor + \cO\left(e^{-N^\frac{1}{2}|t|}\right)\,.
    \eea
    In this step, we have used\footnote{We shall take $\log$ to mean to the principal value of the logarithm throughout.}
    \be\label{eq: log inv}
    \log(1-e^z)=z-2\pi i\left\lfloor\frac{\im z}{2\pi}\right\rfloor - i\pi+\log(1-e^{-z})
    \ee
    to put the logarithms into the form $\log(1-e^{-N^\frac{1}{2}|t|+\ldots})=\cO(e^{-N^\frac{1}{2}|t|})$, which is exponentially suppressed at large $N$. Similarly, in the second line of \eqref{eq: W'}, we can use \eqref{eq: log inv} and $\sum_{I=1}^3\Delta_I = 2\pi$ to write
    \bea\label{eq: 1cut aft int rem}
    &\sum_{I=1}^3\sum_{\substack{b=1 \\ b\neq a}}^N\left[i\pi-N^\frac{1}{2}(t_a-t_b)+i(f(t_a)-f(t_b))+\log\left(1-e^{-N^\frac{1}{2}(t_b-t_a)+i(\Delta_I+f(t_b)-f(t_a))}\right)\right.\\
    &\left.\hspace{7.8cm}-\log\left(1-e^{-N^\frac{1}{2}(t_a-t_b)+i(\Delta_I+f(t_a)-f(t_b))}\right)\right]\\
    &=2\pi i\sum_{\substack{b=1 \\ b\neq a}}^N\left[\Theta(t_b-t_a)\left(2+\sum_{I=1}^3\left\lfloor\frac{\Delta_I+f(t_a)-f(t_b)}{2\pi}\right\rfloor\right)\right. \\
    &\hspace{6.5cm}\left.+\Theta(t_a-t_b)\left(1-\sum_{I=1}^3\left\lfloor\frac{\Delta_I+f(t_b)-f(t_a)}{2\pi}\right\rfloor\right)\right]\\
    &+\sum_{I=1}^3\sum_{\substack{b=1 \\ t_b > t_a}}^N\left[\log\left(1-e^{-N^\frac{1}{2}(t_b-t_a)+i(\Delta_I+f(t_b)-f(t_a))}\right)-\log\left(1-e^{-N^\frac{1}{2}(t_b-t_a)-i(\Delta_I+f(t_a)-f(t_b))}\right)\right] \\
    &+\sum_{I=1}^3\sum_{\substack{b=1 \\ t_b < t_a}}^N\left[\log\left(1-e^{-N^\frac{1}{2}(t_a-t_b)-i(\Delta_I+f(t_b)-f(t_a))}\right)-\log\left(1-e^{-N^\frac{1}{2}(t_a-t_b)+i(\Delta_I+f(t_a)-f(t_b))}\right)\right]\,.
    \eea
    We choose to order the $u_a$ such that $t_a$ increases as $a$ increases, after which $t_b>t_a$ is equivalent to $b>a$ and $t_b<t_a$ is equivalent to $b<a$. We would then like to swap the last two sums in \eqref{eq: 1cut aft int rem} for integrals in $t$, but we have to be careful to keep track of $\cO(1)$ corrections using the Euler-Maclaurin formula because it turns out that the $\cO(1)$ part of the Bethe equations is what ultimately determines $\rho$. Since the following results will be used repeatedly for every subsequent ansatz as well, we shall keep the discussion more generic than what is necessary for simplifying \eqref{eq: 1cut aft int rem}.
    
    %%%%%%%%%%%%%%%%%%%%%%%%
    \subsection{Aside: Euler-Maclaurin formula}\label{App: E-M formula}
    %%%%%%%%%%%%%%%%%%%%%%%%
    
    We consider sums of the form
    \be\label{eq: EM form sum}
    \sum_{\substack{b=1 \\ \tilde{t}_b>t_a}}^M h(b) = \sum_{b=b_\text{min}}^M h(b)\,,\quad \sum_{\substack{b=1\\ \tilde{t}_b<t_a}}^M h(b) = \sum_{b=1}^{b_\text{max}}h(b)\,,
    \ee
    where $h$ is some smooth function and $M\in\bZ$ is a large integer. $t_a$ and $\tilde{t}_b$ are two a priori independent sequences of $M$ values. We order the sequences such that $t_a$ and $\tilde{t}_b$ are monotonically increasing with $a$ and $b$ respectively, so that $b_\text{min}$ is the minimum index $b$ such that $\tilde{t}_b>t_a$, and $b_\text{max}$ is the maximum index $b$ such that $\tilde{t}_b<t_a$. Suppose that both $t_a$ and $\tilde{t}_b$ are distributed according to the same density $\rho$, i.e. 
    \be
    \frac{1}{M}\frac{da}{dt} = \rho(t)\,,\quad \frac{1}{M}\frac{db}{d\tilde{t}} = \rho(\tilde{t})\,.
    \ee
    Then $a(t)$ and $b(t)$ must only differ by a constant, i.e. $a(t) = b(t)+C$. If $C\neq0$, this implies that $b_\text{min}=b(t_a)+r$ for $r\in(0,1)$ and $b_\text{max}=b(t_a)+r-1$. In this case, the sums can be approximated by integrals via a shifted Euler-Maclaurin formula \cite{Olver:1974as}
    \bea\label{eq: mod eulermac}
    \sum_{b=b_\text{min}}^M h(b) = &\int_{b(t_a)}^{M+1-r}db\, h(b) +\sum_{j=1}^p\frac{B_j(r)}{j!}\left[h^{(j-1)}(M+1-r)-h^{(j-1)}(b(t_a))\right]\\
    &-(-1)^p\int_{b(t_a)}^{M+1-r}db\frac{1}{p!}B_p(b-\lfloor b\rfloor)h^{(p)}(b)\,,\\
    \sum_{b=1}^{b_\text{max}}h(b) = &\int_{1-r}^{b(t_a)}db\,h(b) +\sum_{j=1}^p\frac{B_j(r)}{j!}\left[h^{(j-1)}(b(t_a))-h^{(j-1)}(1-r)\right] \\
    &-(-1)^p\int_{1-r}^{b(t_a)}db\frac{1}{p!}B_p(b-\lfloor b\rfloor)h^{(p)}(x)\,,
    \eea
    where $B_p$ are the degree $p$ Bernoulli polynomials. Otherwise, if $C = 0$, meaning that $\{\tilde{t}_b\}$ and $\{t_a\}$ are the same sequence of values, we have $b_\text{min}=a+1=b(t_a)+1$ and $b_\text{max}=a-1=b(t_a)-1$. Thus we should use the first line of \eqref{eq: mod eulermac} with $r=1$ and second line with $r=0$.
    
    For our purpose of finding the corrections to the integral approximation up to $\cO(1)$, it will be sufficient to consider the Euler-Maclaurin formula \eqref{eq: mod eulermac} at $p=1$, because the remainder term at $p=1$ is already of order $\cO(M^{-\frac{1}{2}})$. One can show this as follows. For the sum with $b=b_\text{min}$ as the lower limit, the functions $h$ that we will consider are sums of terms, each of the form
    \be\label{eq: eg h low lim}
    h(b) = \log\left(1-e^{-M^\frac{1}{2}(t_b-t_a)+iC(t_b)}\right)\approx \log\left(1-e^{-M^\frac{1}{2}(t_b-t_a)+iC}\right)\,,\quad C = \text{const.}\,,
    \ee
    where $C(t)$ is some real-valued function that is approximately constant at large $M$. The remainder term at $p=1$ is then
    \bea
    R_1&\equiv\int_{b(t_a)}^{M+1-r}db B_1(b-\lfloor b\rfloor)h'(b) \\
    &= M^{-\frac{1}{2}}\int_{b(t_a)}^{M+1-r}db \left(b-\lfloor b\rfloor-\frac{1}{2}\right)\frac{1}{\rho(t_b)}\frac{e^{-M^\frac{1}{2}(t_b-t_a)+iC}}{1-e^{-M^\frac{1}{2}(t_b-t_a)+iC}}\,,
    \eea
    where we have used $\frac{db}{dt}=M^{-1}\rho(t)^{-1}$. The integrand is the product of a sawtooth function with period $1$ and a decaying exponential with a much longer half-life $\sim M^\frac{1}{2}$. Splitting the domain of integration into unit periods and redefining the integration variable, we get
    \be\label{eq: R1 aft periodization}
    R_1 = M^{-\frac{1}{2}}\sum_{l=0}^{M-b_\text{min}}\int_0^1 db\left(b-r-\lfloor b-r\rfloor-\frac{1}{2}\right)\frac{1}{\rho(t_{b+b(t_a)+l})}\frac{e^{-M^\frac{1}{2}(t_{b+b(t_a)+l}-t_a)+iC}}{1-e^{-M^\frac{1}{2}(t_{b+b(t_a)+l}-t_a)+iC}}\,.
    \ee
    To keep the computation tractable, we consider a constant density $\rho(t) = 1$ and let $t_b = \frac{b}{M}$. Specifying $\rho$ does not change the $M$ dependence in \eqref{eq: R1 aft periodization}, so we expect that the large $M$ scaling that we obtain for $R_1$ when $\rho=1$ should also hold for generic $\rho$. After expanding the denominator in \eqref{eq: R1 aft periodization} and summing over $l$, this gives
    \bea\label{eq: remainder Euler-Mac}
    R_1 &= M^{-\frac{1}{2}}\sum_{k=1}^\infty\frac{e^{ikC}(1-e^{-(M-b_\text{min}+1)M^{-\frac{1}{2}}k})}{1-e^{-M^{-\frac{1}{2}}k}}\int_0^1db\left(b-r-\lfloor b-r\rfloor-\frac{1}{2}\right)e^{-M^{-\frac{1}{2}}kb}\\
    &=M^{-\frac{1}{2}}\sum_{k=1}^\infty\frac{e^{ikC}(1-e^{-(M-b_\text{min}+1)M^{-\frac{1}{2}}k})}{1-e^{-M^{-\frac{1}{2}}k}}\left[-\frac{1}{2}M^{-\frac{1}{2}}k\left(r^2-r+\frac{1}{6}\right)+\cO(M^{-1})\right]\\
    &=-\frac{1}{2}M^{-\frac{1}{2}}\left(r^2-r+\frac{1}{6}\right)\frac{e^{iC}}{1-e^{iC}}+\cO(M^{-1})\,.
    \eea
    Thus we have $R_1\sim\cO(M^{-\frac{1}{2}})$, as anticipated. For the sum in \eqref{eq: mod eulermac} with $b=b_\text{max}$ as the upper limit, the functions $h$ that we will consider have the form
    \be\label{eq: eg h upp lim}
    h(b) = \log\left(1-e^{-M^\frac{1}{2}(t_a-t_b)+iC}\right)\,.
    \ee
    A similar computation shows that the remainder term at $p=1$ for this sum is also suppressed as $\cO(M^{-\frac{1}{2}})$. In addition, for the function $h$ in \eqref{eq: eg h low lim}, $h(M+1-r)$ is exponentially suppressed at large $M$, while for the $h$ in \eqref{eq: eg h upp lim}, $h(1-r)$ is exponentially suppressed. Thus the shifted Euler-Maclaurin formula at $p=1$ simplifies to
    \bea\label{eq: Euler-Mac simp}
    \sum_{b=b_\text{min}}^M h(b) &= \int_{b(t_a)}^{M+1-r}db\,h(b)-\left(r-\frac{1}{2}\right)h(b(t_a))+\cO\left(M^{-\frac{1}{2}}\right)\,,\\
    \sum_{b=1}^{b_\text{max}}h(b) &= \int_{1-r}^{b(t_a)}db\,h(b)+\left(r-\frac{1}{2}\right)h(b(t_a))+\cO\left(M^{-\frac{1}{2}}\right)\,.
    \eea
    This is the version that we will be using to simplify the Bethe equations.
    
    \begin{figure}[H]
        \centering        
        \includegraphics[width=\textwidth]{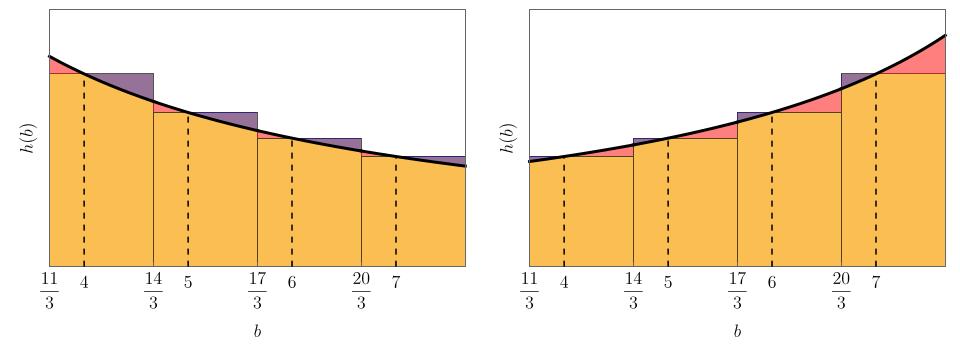}
        \caption{\small Plots illustrating the modified Euler-Maclaurin formula when $b(t_a)=b_\text{min}-\frac{1}{3}$. The left plot shows an example of a sum with lower limit $b_\text{min}=4$, while the right plot shows an example of a sum with upper limit $b_\text{max}=7$. The sum, given by the area of the orange rectangles, is equal to the area of under the black curve (the integral) plus the area of the blue regions minus the area of the red regions.}
        \label{fig: side eulermac}
    \end{figure}
    
    Visually, the $p=1$ corrections in the Euler-Maclaurin formula can be understood by looking at Figure \ref{fig: side eulermac}. The difference between the sums and its approximating integral is given by adding and subtracting the areas of the blue and red curvilinear triangles. For $p=1$, the curvilinear triangles are approximated by ordinary triangles with straight edges, with an error given by the remainder term. 
    
    It is important to note that the remainder term is only subleading, as we saw above, if $C\neq 0$ in \eqref{eq: eg h low lim} and \eqref{eq: eg h upp lim}. If $C=0$, as we will encounter for some cases when analysing the 3-cut solution in Section \ref{subsec:multi-cut:3cut}, this is no longer true because \eqref{eq: remainder Euler-Mac} diverges. Indeed, one can check numerically that the remainder is not suppressed at large $M$ in such cases, even if $p>1$ corrections are added. Fundamentally, this is because the integrands \eqref{eq: eg h low lim} and \eqref{eq: eg h upp lim} have logarithmic singularities at the endpoint $t_b = t_a$ when $C=0$. One might surmise that generalized Euler-Maclaurin formulas catered to such singularities, like the one in \cite{Sidi2012}, can be used instead. We leave such an exploration to future work. 

    %%%%%%%%%%%%%%%%%%%%%%%%
    \subsection{1-cut ansatz, continued}
    %%%%%%%%%%%%%%%%%%%%%%%%
    
    Having introduced how the Euler-Maclaurin formula can be used to find the $\cO(1)$ corrections incurred by approximating sums as integrals, we are ready to resume our analysis of the one-cut ansatz by applying it to the sums in \eqref{eq: 1cut aft int rem}. Since $\{t_b\}$ and $\{t_a\}$ are the same sequence of values in this case, we should take $r=1$ and $r=0$ in the first and second lines of \eqref{eq: Euler-Mac simp} respectively. This gives
    \bea\label{eq: 1cut sums aft EM}
    &\sum_{I=1}^3\sum_{\substack{b=1 \\ t_b > t_a}}^N\left[\log\left(1-e^{-N^\frac{1}{2}(t_b-t_a)+i(\Delta_I+f(t_b)-f(t_a))}\right)-\log\left(1-e^{-N^\frac{1}{2}(t_b-t_a)-i(\Delta_I+f(t_a)-f(t_b))}\right)\right] \\
    &+\sum_{I=1}^3\sum_{\substack{b=1 \\ t_b < t_a}}^N\left[\log\left(1-e^{-N^\frac{1}{2}(t_a-t_b)-i(\Delta_I+f(t_b)-f(t_a))}\right)-\log\left(1-e^{-N^\frac{1}{2}(t_a-t_b)+i(\Delta_I+f(t_a)-f(t_b))}\right)\right]\\
    &\approx\sum_{I=1}^3\int_{b(t_a)}^M db\left[\log\left(1-e^{-N^\frac{1}{2}(t_b-t_a)+i(\Delta_I+f(t_b)-f(t_a))}\right)-\log\left(1-e^{-N^\frac{1}{2}(t_b-t_a)-i(\Delta_I+f(t_a)-f(t_b))}\right)\right]\\
    &+\sum_{I=1}^3\int_1^{b(t_a)}db\left[\log\left(1-e^{-N^\frac{1}{2}(t_a-t_b)-i(\Delta_I+f(t_b)-f(t_a))}\right)-\log\left(1-e^{-N^\frac{1}{2}(t_a-t_b)+i(\Delta_I+f(t_a)-f(t_b))}\right)\right]\\
    &+\frac{1}{2}\sum_{I=1}^3\left[-\log(1-e^{i\Delta_I})+\log(1-e^{-i\Delta_I})\right]-\frac{1}{2}\sum_{I=1}^3\left[\log(1-e^{-i\Delta_I})-\log(1-e^{i\Delta_I})\right]\\
    &=N\sum_{I=1}^3\int_t^{t_+}dt'\rho(t')\left[\log\left(1-e^{-N^\frac{1}{2}(t'-t)+i(\Delta_I+f(t')-f(t))}\right)-\log\left(1-e^{-N^\frac{1}{2}(t'-t)-i(\Delta_I+f(t)-f(t'))}\right)\right]\\
    &+N\sum_{I=1}^3\int_{t_-}^t dt'\rho(t')\left[\log\left(1-e^{-N^\frac{1}{2}(t-t')-i(\Delta_I+f(t')-f(t))}\right)-\log\left(1-e^{-N^\frac{1}{2}(t-t')+i(\Delta_I+f(t)-f(t'))}\right)\right]\,,
    \eea
    where we have used $db = Ndt \rho(t)$, and $t_\pm$ are the maximal and minimal values in $\{t_a\}$. In the case of the one-cut solution, we see from the above that the $\cO(1)$ corrections coming from the Euler-Maclaurin formula exactly cancel between the two sums, and it would have been fine to na\"ively ignore the corrections from the beginning. However, this should not be taken for granted. As we shall see, these $\cO(1)$ corrections play an important role in the 2-cut and 3-cut solutions. Next, we use the expansion $\log(1-z)=-\sum_{k=1}^\infty \frac{z^k}{k}$ for $|z|<1$ and the subsequent taylor expansion of $\rho(t')e^{\pm ik(f(t')-f(t))}$ around $t'=t$ to write \eqref{eq: 1cut sums aft EM} as
    \bea\label{eq: 1cut sums aft taylor}
    &N\sum_{I=1}^3\sum_{k=1}^\infty\left(-\frac{e^{ik\Delta_I}}{k}+\frac{e^{-ik\Delta_I}}{k}\right)\sum_{m=0}^\infty \frac{\partial_{t'}^m\left[\rho(t')e^{ik(f(t')-f(t))}\right]_{t}}{m!}\int_t^{t_+}dt'(t'-t)^m e^{-kN^\frac{1}{2}(t'-t)}\\
    &+N\sum_{I=1}^3\sum_{k=1}^\infty\left(-\frac{e^{-ik\Delta_I}}{k}+\frac{e^{ik\Delta_I}}{k}\right)\sum_{m=0}^\infty \frac{(-1)^m\partial_{t'}^m\left[\rho(t')e^{-ik(f(t')-f(t))}\right]_{t}}{m!}\int_{t_-}^t dt'(t-t')^m e^{-kN^\frac{1}{2}(t-t')}\,.
    \eea
    By integrating by parts and neglecting exponentially suppressed terms at large $N$, it is straightforward to show that 
    \bea\label{eq: low upp ints t}
    &\int_t^{t_+}dt'(t'-t)^m e^{-kN^\frac{1}{2}(t'-t)}=\frac{m!}{(kN^\frac{1}{2})^{m+1}}+\cO\left(e^{-kN^\frac{1}{2}(t_+-t)}\right)\,,\\
    &\int_{t_-}^{t}dt'(t-t')^m e^{-kN^\frac{1}{2}(t-t')}=\frac{m!}{(kN^\frac{1}{2})^{m+1}}+\cO\left(e^{-kN^\frac{1}{2}(t-t_-)}\right)\,.
    \eea
    After substituting \eqref{eq: low upp ints t} and using the identities
    \bea\label{eq: Li identities}
    \sum_{k=1}^\infty\frac{e^{ik\Delta}}{k^n}=\Li_n(e^{i\Delta})&\,, \quad \Li_3(e^{i\Delta})-\Li_3(e^{-i\Delta})=ig_+([\Delta])\,, \quad \Li_2'(e^{i\Delta})+\Li_2'(e^{-i\Delta})=g_+'([\Delta])\,, \\
    &[\Delta]\equiv \Delta-2\pi\left\lfloor\frac{\Delta}{2\pi}\right\rfloor\,, \quad g_+(\Delta) \equiv \frac{1}{6}\Delta^3-\frac{\pi}{2}\Delta^2+\frac{\pi^2}{3}\Delta\,, 
    \eea
    \eqref{eq: 1cut sums aft taylor} can be simplified to
    \be\label{eq: 1cut adj sum fin}
    -2i\rho'(t)\sum_{I=1}^3g_+([\Delta_I])+\cO\left(N^{-\frac{1}{2}}\right)\,.
    \ee
    Setting the integer ambiguities $n_a$ in the Bethe equations \eqref{eq: W'} to
    \bea
    n_a = &\sum_{\substack{b=1 \\ b\neq a}}^N\left[\Theta(t_b-t_a)\left(2+\sum_{I=1}^3\left\lfloor\frac{\Delta_I+f(t_a)-f(t_b)}{2\pi}\right\rfloor\right)\right. \\
    &\hspace{5.5cm}\left.+\Theta(t_a-t_b)\left(1-\sum_{I=1}^3\left\lfloor\frac{\Delta_I+f(t_b)-f(t_a)}{2\pi}\right\rfloor\right)\right]\,,
    \eea
    the $2\pi i\bZ$ terms in \eqref{eq: 1cut aft int rem} are cancelled, and the remaining contributions from \eqref{eq: 1-cut fund fin} and \eqref{eq: 1cut adj sum fin} exactly give the large $N$ Bethe equation in \eqref{eq: large N W'}. Note that because the $2\pi i\bZ$ terms in \eqref{eq: 1cut aft int rem} are $\cO(N)$, while the remaining contributions in \eqref{eq: 1-cut fund fin} and \eqref{eq: 1cut adj sum fin} are $\cO(1)$, cancelling the $2\pi i\bZ$ terms with integer ambiguities is the only way to solve the Bethe equations. However, there is still the freedom to include an additional $\cO(1)$ integer in the ambiguities $n_a$. This would have modified the large $N$ Bethe equation in \eqref{eq: large N W'} with shifts by $2\pi\bZ$. We have fixed this freedom by demanding that \eqref{eq: large N W'} is invariant under the vacuum-preserving symmetries and joint symmetry 1 in Section \ref{subsec:TTI:FundamentalRegions}.

    %%%%%%%%%%%%%%%%%%%%%%%%
    \subsection{Reflection symmetric 2-cut ansatz}\label{App: large N Bethe 2cut}
    %%%%%%%%%%%%%%%%%%%%%%%%
    
    We now proceed to consider the 2-cut ansatz in \eqref{eq: large N 2 col ansatz}. Due to the symmetry of the ansatz under $f\mapsto -f$ and $u_a\leftrightarrow v_a$, it is enough to consider the $M$ Bethe equations corresponding to $u_a$. The remaining equations corresponding to $v_a$ can be obtained by taking $f\mapsto -f$. As in the analysis of the one-cut ansatz, we can use \eqref{eq: log inv} to simplify the first line of \eqref{eq: W'} as
    \bea\label{eq: 2cut fund}
    &-i\Delta_C-M^\frac{1}{2}t + if(t) + \log\left(1-e^{M^\frac{1}{2}t + i\left(\frac{\Delta_1+\Delta_2}{2}-f(t)\right)}\right) - \log\left(1-e^{-M^\frac{1}{2}t + i\left(\frac{\Delta_1+\Delta_2}{2}+f(t)\right)}\right)\\
    &=-i\Delta_C-i\sgn t\frac{\Delta_3}{2}-2\pi i\sgn(t)\left\lfloor\frac{\frac{\Delta_1+\Delta_2}{2}-\sgn(t)f(t)}{2\pi}\right\rfloor + \cO\left(e^{-M^\frac{1}{2}|t|}\right)\,.
    \eea
    In the second line of \eqref{eq: W'}, we separate the sum into two pieces:
    \bea\label{eq: adj sum split}
    &\sum_{I=1}^3\sum_{\substack{b=1 \\ b\neq a}}^M\left[i\pi+i(u_a-u_b)+\log\left(1-e^{i(u_b-u_a+\Delta_I)}\right)-\log\left(1-e^{i(u_a-u_b+\Delta_I)}\right)\right]\\
    &+\sum_{I=1}^3\sum_{b=1}^M\left[i\pi+i(u_a-v_b)+\log\left(1-e^{i(v_b-u_a+\Delta_I)}\right)-\log\left(1-e^{i(u_a-v_b+\Delta_I)}\right)\right]\,.
    \eea
    The first sum over $u_{b\neq a}$ in \eqref{eq: adj sum split} is identical to the sum \eqref{eq: 1cut aft int rem} we already encountered when analyzing the one-cut ansatz, if we take $M\rightarrow N$. Therefore the result \eqref{eq: 1cut adj sum fin} can be directly reused to give
    \bea\label{eq: 2cut u-u sum}
    &\sum_{I=1}^3\sum_{\substack{b=1 \\ b\neq a}}^M\left[i\pi+i(u_a-u_b)+\log\left(1-e^{i(u_b-u_a+\Delta_I)}\right)-\log\left(1-e^{i(u_a-u_b+\Delta_I)}\right)\right]\\
    &=2\pi i\sum_{\substack{b=1 \\ b\neq a}}^M\left[\Theta(t_b-t_a)\left(2+\sum_{I=1}^3\left\lfloor\frac{\Delta_I+f(t_a)-f(t_b)}{2\pi}\right\rfloor\right)\right. \\
    &\left.+\Theta(t_a-t_b)\left(1-\sum_{I=1}^3\left\lfloor\frac{\Delta_I+f(t_b)-f(t_a)}{2\pi}\right\rfloor\right)\right]-2i\rho'(t)\sum_{I=1}^3g_+([\Delta_I])\,.
    \eea
    For the second sum over $v_b$ in \eqref{eq: adj sum split}, we first isolate the term with $b=a$, for which $\im u_a = \im v_a$. Isolating its real and imaginary parts, this term can be written as
    \bea\label{eq: 2cut vb eq a}
    &\sum_{I=1}^3\left[i\pi+2if(t)+\log\left(1-e^{i(\Delta_I-2f(t))}\right)-\log\left(1-e^{i(\Delta_I+2f(t))}\right)\right]\\
    &\log\prod_{I=1}^3\frac{\sin\frac{1}{2}[\Delta_I-2f(t)]}{\sin\frac{1}{2}[\Delta_I+2f(t)]}+i\pi\sum_{I=1}^3\left(1+\left\lfloor\frac{\Delta_I+2f(t)}{2\pi}\right\rfloor-\left\lfloor\frac{\Delta_I-2f(t)}{2\pi}\right\rfloor\right)\,.
    \eea
    The remaining sum over $v_{b\neq a}$ can be analysed following the same steps that were performed on \eqref{eq: 1cut aft int rem}: we put the logarithms into the form $\log(1-z)$ for $|z|<1$, and use the Euler-Maclaurin formula for the case where $\{t_b\}$ and $\{t_a\}$ are identical sequences. This gives
    \bea\label{eq: 3cut aft int & EM}
    &\sum_{I=1}^3\sum_{\substack{b=1\\ b\neq a}}^M\left[i\pi+i(u_a-v_b)+\log\left(1-e^{i(v_b-u_a+\Delta_I)}\right)-\log\left(1-e^{i(u_a-v_b+\Delta_I)}\right)\right]\\
    &\approx 2\pi i\sum_{\substack{b=1 \\ b\neq a}}^M\left[\Theta(t_b-t_a)\left(2+\sum_{I=1}^3\left\lfloor\frac{\Delta_I+f(t_a)+f(t_b)}{2\pi}\right\rfloor\right)\right.\\
    &\left.+\Theta(t_a-t_b)\left(1-\sum_{I=1}^3\left\lfloor\frac{\Delta_I-f(t_a)-f(t_b)}{2\pi}\right\rfloor\right)\right]+\log\prod_{I=1}^3\frac{\sin\frac{1}{2}[\Delta_I+2f(t)]}{\sin\frac{1}{2}[\Delta_I-2f(t)]}\\
    &+M\sum_{I=1}^3\int_t^{t_+}dt'\rho(t')\left[\log\left(1-e^{-M^\frac{1}{2}(t'-t)+i(\Delta_I-f(t')-f(t))}\right)-\log\left(1-e^{-M^\frac{1}{2}(t'-t)-i(\Delta_I+f(t)+f(t'))}\right)\right]\\
    &+M\sum_{I=1}^3\int_{t_-}^t dt'\rho(t')\left[\log\left(1-e^{-M^\frac{1}{2}(t-t')-i(\Delta_I-f(t')-f(t))}\right)-\log\left(1-e^{-M^\frac{1}{2}(t-t')+i(\Delta_I+f(t)+f(t'))}\right)\right]\,,
    \eea
    where the term $\log\prod_{I=1}^3\frac{\sin\frac{1}{2}[\Delta_I+2f(t)]}{\sin\frac{1}{2}[\Delta_I-2f(t)]}$ comes from the $\cO(1)$ corrections in the Euler-Maclaurin formula. Unlike what we found in \eqref{eq: 1cut sums aft EM}, the corrections are nonzero, and exactly cancel the real part of \eqref{eq: 2cut vb eq a}. The simplification of the integrals in \eqref{eq: 3cut aft int & EM} then follows the same steps as in \eqref{eq: 1cut sums aft taylor}. We expand the logs, use the taylor expansion of the integrand around $t'=t$, and simplify using the identities \eqref{eq: Li identities}. The result is
    \bea
    &{M\sum_{I=1}^3\int_t^{t_+}dt'\rho(t')\left[\log\left(1-e^{-M^\frac{1}{2}(t'-t)+i(\Delta_I-f(t')-f(t))}\right)-\log\left(1-e^{-M^\frac{1}{2}(t'-t)-i(\Delta_I+f(t)+f(t'))}\right)\right]}\\
    &{+M\sum_{I=1}^3\int_{t_-}^t dt'\rho(t')\left[\log\left(1-e^{-M^\frac{1}{2}(t-t')-i(\Delta_I-f(t')-f(t))}\right)-\log\left(1-e^{-M^\frac{1}{2}(t-t')+i(\Delta_I+f(t)+f(t'))}\right)\right]}\\
    &{=(M^\frac{1}{2}-if'(t))\rho(t)\sum_{I=1}^3\left[g_+'([\Delta_I+2f(t)])-g_+'([\Delta_I-2f(t)])\right]}\\
    &{-i\rho'(t)\sum_{I=1}^3\left[g_+([\Delta_I+2f(t)])+g_+([\Delta_I-2f(t)])\right]+\cO\left(M^{-\frac{1}{2}}\right)}\,.
    \eea
    We choose the integer ambiguities $n_a$ in the Bethe equations \eqref{eq: W'} to be
    \bea\label{eq: 2cut int amb}
    &n_a = \sum_{\substack{b=1 \\ b\neq a}}^M\left[\Theta(t_b-t_a)\left(2+\sum_{I=1}^3\left\lfloor\frac{\Delta_I+f(t_a)-f(t_b)}{2\pi}\right\rfloor\right)\right. \\
    &\hspace{5.5cm}\left.+\Theta(t_a-t_b)\left(1-\sum_{I=1}^3\left\lfloor\frac{\Delta_I+f(t_b)-f(t_a)}{2\pi}\right\rfloor\right)\right]\\
    &+\sum_{\substack{b=1 \\ b\neq a}}^M\left[\Theta(t_b-t_a)\left(2+\sum_{I=1}^3\left\lfloor\frac{\Delta_I+f(t_a)+f(t_b)}{2\pi}\right\rfloor\right)\right.\\
    &\left.+\Theta(t_a-t_b)\left(1-\sum_{I=1}^3\left\lfloor\frac{\Delta_I-f(t_a)-f(t_b)}{2\pi}\right\rfloor\right)\right]+2+\sum_{I=1}^3\left\lfloor\frac{\Delta_I+2f(t)}{2\pi}\right\rfloor\,.
    \eea
    Subtracting $2\pi in_a$ from the total contributions in \eqref{eq: 2cut fund}, \eqref{eq: 2cut u-u sum}, \eqref{eq: 2cut vb eq a}, and \eqref{eq: 3cut aft int & EM}, we exactly obtain the large $N$ Bethe equation in \eqref{eq: large N W' 2 col}. As we mentioned above \eqref{eq: 2cut fund}, the remaining $M$ Bethe equations corresponding to $v_a$ can be obtained by taking $f\mapsto -f$. Under this transformation, the real part of \eqref{eq: large N W' 2 col} flips sign, while the imaginary part is invariant up to a $2\pi\bZ$ shift. Thus the remaining equations are redundant and do not give any extra information. As is the case in the one-cut ansatz, there is a freedom to add $\cO(1)$ terms to the integer ambiguities in \eqref{eq: 2cut int amb} without affecting the consistency of \eqref{eq: large N W' 2 col}. We have fixed this freedom by demanding that \eqref{eq: large N W' 2 col} should be invariant under the vacuum-preserving symmetries and joint symmetry 1 introduced in Section \ref{subsec:TTI:FundamentalRegions}, as these are symmetries of the original finite $N$ Bethe equations. 

    %%%%%%%%%%%%%%%%%%%%%%%%
    \subsection{2-cut ansatz with $\re u = 0, \pi$}\label{App: large N Bethe 2cut 0pi}
    %%%%%%%%%%%%%%%%%%%%%%%%

    We move on to the 2-cut ansatz in \eqref{eq: 2 col ansatz 0 pi} with one cut at $\re u =0$ and the other at $\re u =\pi$, and look at the $M$ Bethe equations corresponding to $u_a$ first. As is standard, we use \eqref{eq: log inv} to simplify the first line of \eqref{eq: W'} as
    \bea
    &{-i\Delta_C-M^\frac{1}{2}t_a+\log\left(1-e^{M^\frac{1}{2}t_a+i\frac{\Delta_1+\Delta_2}{2}}\right)-\log\left(1-e^{-M^\frac{1}{2}t_a+i\frac{\Delta_1+\Delta_2}{2}}\right)}\\
    &{=-i\Delta_C-i\sgn(t)\frac{\Delta_3}{2}-2\pi i\sgn(t)\left\lfloor\frac{\Delta_1+\Delta_2}{4\pi}\right\rfloor+\cO\left(e^{-M^\frac{1}{2}|t|}\right)\,.}
    \eea
    As before, the second line of \eqref{eq: W'} is split into the two sums in \eqref{eq: adj sum split}. The first sum is identical to \eqref{eq: adj sum split} if we set $f=0$. Therefore
    \bea
    &{\sum_{I=1}^3\sum_{\substack{b=1 \\ b\neq a}}^M\left[i\pi+i(u_a-u_b)+\log\left(1-e^{i(u_b-u_a+\Delta_I)}\right)-\log\left(1-e^{i(u_a-u_b+\Delta_I)}\right)\right]}\\
    &{=2\pi i\sum_{\substack{b=1 \\ b\neq a}}^M\left[\Theta(t_b-t_a)\left(2+\sum_{I=1}^3\left\lfloor\frac{\Delta_I}{2\pi}\right\rfloor\right)+\Theta(t_a-t_b)\left(1-\sum_{I=1}^3\left\lfloor\frac{\Delta_I}{2\pi}\right\rfloor\right)\right]-2i\rho'(t)\sum_{I=1}^3g_+([\Delta_I])\,.}
    \eea
    For the second sum in \eqref{eq: adj sum split}, since $t_a \neq \tilde{t}_b$ for all $a,b$, there will be no need to isolate the term with $a=b$. We can directly put the logarithms into the form $\log(1-z)$ with $|z|<1$, obtaining
    \bea\label{eq: 2cut 0pi logz decaying}
    &2\pi i\sum_{b=1}^M\left[\Theta(\tilde{t}_b-t_a)\left(2+\sum_{I=1}^3\left\lfloor\frac{\Delta_I-\pi}{2\pi}\right\rfloor\right)+\Theta(t_a-\tilde{t}_b)\left(1-\sum_{I=1}^3\left\lfloor\frac{\Delta_I+\pi}{2\pi}\right\rfloor\right)\right]\\
    &+\sum_{I=1}^3\sum_{\substack{b=1 \\ \tilde{t}_b>t_a}}^M\left[\log\left(1-e^{-M^\frac{1}{2}(\tilde{t}_b-t_a)+i(\Delta_I+\pi)}\right)-\log\left(1-e^{-M^\frac{1}{2}(\tilde{t}_b-t_a)-i(\Delta_I-\pi)}\right)\right]\\
    &+\sum_{I=1}^3\sum_{\substack{b=1 \\ \tilde{t}_b<t_a}}^M\left[\log\left(1-e^{-M^\frac{1}{2}(t_a-\tilde{t}_b)-i(\Delta_I+\pi)}\right)-\log\left(1-e^{-M^\frac{1}{2}(t_a-\tilde{t}_b)+i(\Delta_I-\pi)}\right)\right]\,.
    \eea
    Since $t_a$ and $\tilde{t}_b$ have the same distribution $\rho$ but $t_a \neq \tilde{t}_b$ for all $a,b$, we are exactly in the situation described in Appendix \ref{App: E-M formula}, where $\lceil b(t_a)\rceil-b(t_a) = r$ for some piecewise constant $r\in (0,1)$. Using the Euler-Maclaurin formula in \eqref{eq: Euler-Mac simp}, the $\cO(1)$ corrections from approximating the sums in \eqref{eq: 2cut 0pi logz decaying} by integrals is
    \bea
    -&\sum_{I=1}^3(2r-1)\left[\log\left(1-e^{i(\Delta_I+\pi)}\right)-\log\left(1-e^{-i(\Delta_I+\pi)}\right)\right]\\
    &=-2\pi i(2r-1)\left(1-\sum_{I=1}^3\left\lfloor\frac{\Delta_I+\pi}{2\pi}\right\rfloor\right)\,.
    \eea
    The remaining integrals can be simplified via the standard steps in \eqref{eq: 1cut sums aft taylor} to give
    \bea
    &{M\sum_{I=1}^3\int_t^{t_+} dt'\rho(t')\left[\log\left(1-e^{-M^\frac{1}{2}(t'-t)+i(\Delta_I+\pi)}\right)-\log\left(1-e^{-M^\frac{1}{2}(t'-t)-i(\Delta_I-\pi)}\right)\right]}\\
    +&{M\sum_{I=1}^3\int^t_{t_-} dt'\rho(t')\left[\log\left(1-e^{-M^\frac{1}{2}(t-t')-i(\Delta_I+\pi)}\right)-\log\left(1-e^{-M^\frac{1}{2}(t-t')+i(\Delta_I-\pi)}\right)\right]}\\
    &{=-2i\rho'(t)\sum_{I=1}^3 g_+([\Delta_I+\pi])+\cO\left(M^{-\frac{1}{2}}\right)\,.}
    \eea
    Choosing the integer ambiguities $n_a$ corresponding to $u_a$ to be
    \bea\label{eq: 2cut 0pi u eq amb}
    &n_a = \sum_{\substack{b=1 \\ b\neq a}}^M\left[\Theta(t_b-t_a)\left(2+\sum_{I=1}^3\left\lfloor\frac{\Delta_I}{2\pi}\right\rfloor\right)+\Theta(t_a-t_b)\left(1-\sum_{I=1}^3\left\lfloor\frac{\Delta_I}{2\pi}\right\rfloor\right)\right]\\
    &+\sum_{b=1}^M\left[\Theta(\tilde{t}_b-t_a)\left(2+\sum_{I=1}^3\left\lfloor\frac{\Delta_I-\pi}{2\pi}\right\rfloor\right)+\Theta(t_a-\tilde{t}_b)\left(1-\sum_{I=1}^3\left\lfloor\frac{\Delta_I+\pi}{2\pi}\right\rfloor\right)\right]-1\,,
    \eea
    the large $N$ limit of the Bethe equations corresponding to $u_a$ is
    \bea\label{eq: 2cut 0pi u largeN}
    &2\rho'(t)\sum_{I=1}^3\left[g_+([\Delta_I])+g_+([\Delta_I+\pi])\right]+\Delta_C+\sgn(t)\frac{\Delta_3}{2}+2\pi\sgn(t)\left\lfloor\frac{\Delta_1+\Delta_2}{4\pi}\right\rfloor\\
    &\hspace{3cm}+2\pi(2r-1)\left(1-\sum_{I=1}^3\left\lfloor\frac{\Delta_I+\pi}{2\pi}\right\rfloor\right)-2\pi=0\,.
    \eea
    We have fixed the freedom to add $\cO(1)$ integers to $n_a$ by matching \eqref{eq: 2cut 0pi u eq amb} with the integer ambiguities of a numerical solution. 
    
    The large $N$ limit of the remaining Bethe equations corresponding to $v_a$ can be computed in an identical fashion, so we will just quote the result. After choosing the integer ambiguities corresponding to $v_a$ to be
    \bea
    &n_a = \sum_{\substack{b=1 \\ b\neq a}}^M\left[\Theta(\tilde{t}_b-\tilde{t}_a)\left(2+\sum_{I=1}^3\left\lfloor\frac{\Delta_I}{2\pi}\right\rfloor\right)+\Theta(\tilde{t}_a-\tilde{t}_b)\left(1-\sum_{I=1}^3\left\lfloor\frac{\Delta_I}{2\pi}\right\rfloor\right)\right]\\
    &+\sum_{b=1}^M\left[\Theta(t_b-\tilde{t}_a)\left(2+\sum_{I=1}^3\left\lfloor\frac{\Delta_I+\pi}{2\pi}\right\rfloor\right)+\Theta(\tilde{t}_a-t_b)\left(1-\sum_{I=1}^3\left\lfloor\frac{\Delta_I-\pi}{2\pi}\right\rfloor\right)\right]+1\,,
    \eea
    the large $N$ limit of the equations is
    \bea\label{eq: 2cut 0pi v largeN}
    &2\rho'(t)\sum_{I=1}^3\left[g_+([\Delta_I])+g_+([\Delta_I+\pi])\right]+\Delta_C+\sgn(t)\frac{\Delta_3}{2}+2\pi\sgn(t)\left\lfloor\frac{\frac{\Delta_1+\Delta_2}{2}-\sgn(t)\pi}{2\pi}\right\rfloor\\
    &\hspace{3cm}-2\pi(2r-1)\left(1-\sum_{I=1}^3\left\lfloor\frac{\Delta_I+\pi}{2\pi}\right\rfloor\right)+2\pi=0\,.
    \eea
    Note that \eqref{eq: 2cut 0pi u largeN} is precisely the first equation in \eqref{eq: 2cut large N W' 0 pi}, while the difference of \eqref{eq: 2cut 0pi u largeN} and \eqref{eq: 2cut 0pi v largeN} is the second equation in \eqref{eq: 2cut large N W' 0 pi}.  

    %%%%%%%%%%%%%%%%%%%%%%%%
    \subsection{3-cut ansatz with one cut at $\re u =0$}\label{App: 3cut w 1cut 0}
    %%%%%%%%%%%%%%%%%%%%%%%%

    In this section, we compute the large $N$ limit of the Bethe equations \eqref{eq: W'} under the 3-cut ansatz \eqref{eq: large N 3 col ansatz}, where the central cut is along the imaginary axis. 
    
    We consider the Bethe equations corresponding to $u_a$ first. The first line of \eqref{eq: W'} gives
    \be\label{eq: 3cut u fund}
    -i\Delta_C-i\sgn(t)\frac{\Delta_3}{2}-2\pi i\sgn(t)\left\lfloor\frac{\Delta_1+\Delta_2}{4\pi}\right\rfloor+\cO(e^{-M^\frac{1}{2}|t|})\,,
    \ee
    where \eqref{eq: log inv} was used. The sum in the second line of \eqref{eq: W'} can be split into $3$ parts ranging over $u$, $v$, and $w$ respectively:
    \begin{align}\label{eq: u eq sum split}
    &\sum_{I=1}^3\sum_{\substack{b=1 \\ b\neq a}}^M\left[i\pi + i(u_a-u_b)+\log\left(1-e^{i(u_b-u_a+\Delta_I)}\right)-\log\left(1-e^{i(u_a-u_b+\Delta_I)}\right)\right]\notag\\
    &+\sum_{I=1}^3\sum_{b=1}^M\left[i\pi + i(u_a-v_b)+\log\left(1-e^{i(v_b-u_a+\Delta_I)}\right)-\log\left(1-e^{i(u_a-v_b+\Delta_I)}\right)\right]\notag\\
    &+\sum_{I=1}^3\sum_{b=1}^M\left[i\pi + i(u_a-w_b)+\log\left(1-e^{i(w_b-u_a+\Delta_I)}\right)-\log\left(1-e^{i(u_a-w_b+\Delta_I)}\right)\right]\,.
    \end{align}
    Let us first consider the sum over $u_b$. It is identical to the sum in \eqref{eq: 1cut aft int rem} under $N\mapsto M$ and $f\mapsto 0$. We can thus reuse the result in \eqref{eq: 1cut adj sum fin} to get
    \begin{align}\label{eq: 3cut uu sum}
    &\sum_{I=1}^3\sum_{\substack{b=1 \\ b\neq a}}^M\left[i\pi + i(u_a-u_b)+\log\left(1-e^{i(u_b-u_a+\Delta_I)}\right)-\log\left(1-e^{i(u_a-u_b+\Delta_I)}\right)\right]\notag\\
    &=2\pi i\sum_{\substack{b=1 \\ b\neq a}}^M\left[\Theta(t_{u,b}-t_{u,a})\left(2+\sum_{I=1}^3\left\lfloor\frac{\Delta_I}{2\pi}\right\rfloor\right)+\Theta(t_{u,a}-t_{u,b})\left(1-\sum_{I=1}^3\left\lfloor\frac{\Delta_I}{2\pi}\right\rfloor\right)\right]\notag\\
    &-2i\rho'(t)\sum_{I=1}^3g_+([\Delta_I])+\cO\left(M^{-\frac{1}{2}}\right)\,.
    \end{align}
    
    We now consider the sum over $v_b$ in the second line of \eqref{eq: u eq sum split}. As before, we split the sum into terms with $t_{v,b}>t_{u,a}$ and those with $t_{v,b}<t_{u,a}$, and put the logarithms into the form $\log(1-z)$ for $|z|<1$. Since $\lceil b(t_{u,a})\rceil - b(t_{u,a})=r_v$, as introduced in \eqref{eq: 3cut rv rw def}, we approximate the remaining sums by integrals using the Euler-Maclaurin formula \eqref{eq: Euler-Mac simp} with $r=r_v$. The integrals can then be simplified using the standard steps following \eqref{eq: 1cut sums aft taylor} to give
    \begin{align}\label{eq: 3cut uv sum}
    &\sum_{I=1}^3\sum_{b=1}^M\left[i\pi + i(u_a-v_b)+\log\left(1-e^{i(v_b-u_a+\Delta_I)}\right)-\log\left(1-e^{i(u_a-v_b+\Delta_I)}\right)\right]\notag\\
    &=2\pi i\sum_{b=1}^M\left[\Theta(t_{v,b}-t_{u,a})\left(2+\sum_{I=1}^3\left\lfloor\frac{\Delta_I-f(t_{v,b})}{2\pi}\right\rfloor\right)+\Theta(t_{u,a}-t_{v,b})\left(1-\sum_{I=1}^3\left\lfloor\frac{\Delta_I+f(t_{v,b})}{2\pi}\right\rfloor\right)\right]\notag\\
    &-(M^\frac{1}{2}+if'(t))\rho(t)\sum_{I=1}^3[g_+'([\Delta_I+f(t)])-g_+'([\Delta_I-f(t)])]-i\rho'(t)\sum_{I=1}^3[g_+([\Delta_I+f(t)])\notag\\
    &+g_+([\Delta_I-f(t)])]+2\pi i\left(r_v-\frac{1}{2}\right)\left(1+\sum_{I=1}^3\left\lfloor\frac{\Delta_I+f(t)}{2\pi}\right\rfloor+\sum_{I=1}^3\left\lfloor\frac{\Delta_I-f(t)}{2\pi}\right\rfloor\right)+\cO\left(M^{-\frac{1}{2}}\right)\,.
    \end{align}
    Since the ansatz for $v$ and $w$ in \eqref{eq: large N 3 col ansatz} are exchanged under $f\mapsto -f$, the sum over $w$ in the third line of \eqref{eq: u eq sum split} can be obtained from \eqref{eq: 3cut uv sum} by taking $f\mapsto -f$ and $r_v\mapsto r_w$:
    \begin{align}\label{eq: 3cut uw sum}
    &\sum_{I=1}^3\sum_{b=1}^M\left[i\pi + i(u_a-w_b)+\log\left(1-e^{i(w_b-u_a+\Delta_I)}\right)-\log\left(1-e^{i(u_a-w_b+\Delta_I)}\right)\right]\notag\\
    &=2\pi i\sum_{b=1}^M\left[\Theta(t_{w,b}-t_{u,a})\left(2+\sum_{I=1}^3\left\lfloor\frac{\Delta_I+f(t_{w,b})}{2\pi}\right\rfloor\right)+\Theta(t_{u,a}-t_{w,b})\left(1-\sum_{I=1}^3\left\lfloor\frac{\Delta_I-f(t_{w,b})}{2\pi}\right\rfloor\right)\right]\notag\\
    &+(M^\frac{1}{2}-if'(t))\rho(t)\sum_{I=1}^3[g_+'([\Delta_I+f(t)])-g_+'([\Delta_I-f(t)])]-i\rho'(t)\sum_{I=1}^3[g_+([\Delta_I+f(t)])\notag\\
    &+g_+([\Delta_I-f(t)])]+2\pi i\left(r_w-\frac{1}{2}\right)\left(1+\sum_{I=1}^3\left\lfloor\frac{\Delta_I+f(t)}{2\pi}\right\rfloor+\sum_{I=1}^3\left\lfloor\frac{\Delta_I-f(t)}{2\pi}\right\rfloor\right)+\cO\left(M^{-\frac{1}{2}}\right)\,.
    \end{align}
    Taking \eqref{eq: 3cut u fund}, \eqref{eq: 3cut uu sum}, \eqref{eq: 3cut uv sum}, \eqref{eq: 3cut uw sum} together, and subtracting $2\pi i$ times the integer ambiguities
    \begin{align}\label{eq: 3cut u int amb}
    &n_a = \sum_{\substack{b=1 \\ b\neq a}}^M\left[\Theta(t_{u,b}-t_{u,a})\left(2+\sum_{I=1}^3\left\lfloor\frac{\Delta_I}{2\pi}\right\rfloor\right)+\Theta(t_{u,a}-t_{u,b})\left(1-\sum_{I=1}^3\left\lfloor\frac{\Delta_I}{2\pi}\right\rfloor\right)\right]-\Theta\left(\Delta_C-\frac{\pi}{3}\right)\notag\\
    &+\sum_{b=1}^M\left[\Theta(t_{v,b}-t_{u,a})\left(2+\sum_{I=1}^3\left\lfloor\frac{\Delta_I-f(t_{v,b})}{2\pi}\right\rfloor\right)+\Theta(t_{u,a}-t_{v,b})\left(1-\sum_{I=1}^3\left\lfloor\frac{\Delta_I+f(t_{v,b})}{2\pi}\right\rfloor\right)\right]\notag\\
    &+\sum_{b=1}^M\left[\Theta(t_{w,b}-t_{u,a})\left(2+\sum_{I=1}^3\left\lfloor\frac{\Delta_I+f(t_{w,b})}{2\pi}\right\rfloor\right)+\Theta(t_{u,a}-t_{w,b})\left(1-\sum_{I=1}^3\left\lfloor\frac{\Delta_I+f(t_{w,b})}{2\pi}\right\rfloor\right)\right]\,,
    \end{align}
    we obtain that the large $N$ limit of the Bethe equations corresponding to $u$ is
    \begin{align}\label{eq: 3cut u eqn}
    &2\partial_t\left\{\rho(t)\sum_{I=1}^3\left[g_+([\Delta_I])+g_+([\Delta_I+f(t)])+g_+([\Delta_I-f(t)])\right]\right\}+\Delta_C+\sgn(t)\frac{\Delta_3}{2}\notag\\
    &+2\pi\sgn(t)\left\lfloor\frac{\Delta_1+\Delta_2}{4\pi}\right\rfloor+(1-r_v-r_w)\left(1+\sum_{I=1}^3\left\lfloor\frac{\Delta_I+f(t)}{2\pi}\right\rfloor+\sum_{I=1}^3\left\lfloor\frac{\Delta_I-f(t)}{2\pi}\right\rfloor\right)\notag\\
    &-2\pi\Theta\left(\Delta_C-\frac{\pi}{3}\right)=0\,.
    \end{align}
    The $\cO(1)$ part of \eqref{eq: 3cut u int amb} has been fixed by comparing with numerical solutions. Using the fact that $\left\lfloor\frac{\Delta_1+\Delta_2}{4\pi}\right\rfloor = 0$ in the fundamental domain \eqref{eq: fund dom}, \eqref{eq: 3cut u eqn} coincides with the third equation in \eqref{eq: 3col large N BAE LO and rho}.
    
    The analysis of the Bethe equations corresponding to $v$ follows the same steps. The first line of \eqref{eq: W'} simplifies to
    \be\label{eq: 3cut v eqn fund}
    -i\Delta_C-i\sgn(t)\frac{\Delta_3}{2}-2\pi i\sgn(t)\left\lfloor\frac{\frac{\Delta_1+\Delta_2}{2}-\sgn(t)f(t)}{2\pi}\right\rfloor+\cO\left(e^{-M^\frac{1}{2}|t|}\right)\,.
    \ee
    The second line of \eqref{eq: W'} can be split into sums over $v$, $u$, and $w$:
    \begin{align}\label{eq: v eq sum split}
    &\sum_{I=1}^3\sum_{\substack{b=1 \\ b\neq a}}^M\left[i\pi + i(v_a-v_b)+\log\left(1-e^{i(v_b-v_a+\Delta_I)}\right)-\log\left(1-e^{i(v_a-v_b+\Delta_I)}\right)\right]\notag\\
    &+\sum_{I=1}^3\sum_{b=1}^M\left[i\pi + i(v_a-u_b)+\log\left(1-e^{i(u_b-v_a+\Delta_I)}\right)-\log\left(1-e^{i(v_a-u_b+\Delta_I)}\right)\right]\notag\\
    &+\sum_{I=1}^3\sum_{b=1}^M\left[i\pi + i(v_a-w_b)+\log\left(1-e^{i(w_b-v_a+\Delta_I)}\right)-\log\left(1-e^{i(v_a-w_b+\Delta_I)}\right)\right]\,.
    \end{align}
    Like before, we can reuse \eqref{eq: 1cut adj sum fin} under $N\mapsto M$ to get the first sum over $v$:
    \begin{align}\label{eq: 3cut vv sum}
    &\sum_{I=1}^3\sum_{\substack{b=1 \\ b\neq a}}^M\left[i\pi + i(v_a-v_b)+\log\left(1-e^{i(v_b-v_a+\Delta_I)}\right)-\log\left(1-e^{i(v_a-v_b+\Delta_I)}\right)\right]\notag\\
    &=2\pi i\sum_{\substack{b=1\\ b\neq a}}^M\left[\Theta(t_{v,b}-t_{v,a})\left(2+\sum_{I=1}^3\left\lfloor\frac{\Delta_I+f(t_{v,a})-f(t_{v,b})}{2\pi}\right\rfloor\right)\right.\notag\\
    &\left.+\Theta(t_{v,a}-t_{v,b})\left(1-\sum_{I=1}^3\left\lfloor\frac{\Delta_I+f(t_{v,b})-f(t_{v,a})}{2\pi}\right\rfloor\right)\right]-2i\rho'(t)\sum_{I=1}^3g_+([\Delta_I])+\cO\left(M^{-\frac{1}{2}}\right)\,.
    \end{align}
    For the second sum over $u$, it is important to note that $\lceil b(t_{u,a})\rceil - b(t_{u,a}) = r_v$ implies $\lceil a(t_{v,b})\rceil -a(t_{v,b}) = 1-r_v$. Therefore we should use the Euler-Maclaurin formula \eqref{eq: Euler-Mac simp} with $r=1-r_v$ when approximating sums by integrals. The other steps are identical to \eqref{eq: 3cut uv sum}, and we get
    \begin{align}\label{eq: 3cut vu sum}
    &\sum_{I=1}^3\sum_{b=1}^M\left[i\pi + i(v_a-u_b)+\log\left(1-e^{i(u_b-v_a+\Delta_I)}\right)-\log\left(1-e^{i(v_a-u_b+\Delta_I)}\right)\right]\notag\\
    &=2\pi i\sum_{b=1}^M\left[\Theta(t_{u,b}-t_{v,a})\left(2+\sum_{I=1}^3\left\lfloor\frac{\Delta_I+f(t_{v,a})}{2\pi}\right\rfloor\right)+\Theta(t_{v,a}-t_{u,b})\left(1-\sum_{I=1}^3\left\lfloor\frac{\Delta_I-f(t_{v,a})}{2\pi}\right\rfloor\right)\right]\notag\\
    &+M^\frac{1}{2}\rho(t)\sum_{I=1}^3[g_+'([\Delta_I+f(t)])-g_+'([\Delta_I-f(t)])]-i\rho'(t)\sum_{I=1}^3[g_+([\Delta_I+f(t)])+g_+([\Delta_I-f(t)])]\notag\\
    &-2\pi i\left(r_v-\frac{1}{2}\right)\left(1+\sum_{I=1}^3\left\lfloor\frac{\Delta_I+f(t)}{2\pi}\right\rfloor+\sum_{I=1}^3\left\lfloor\frac{\Delta_I-f(t)}{2\pi}\right\rfloor\right)+\cO\left(M^{-\frac{1}{2}}\right)\,.
    \end{align}
    For the third sum over $w$, we need to differentiate between the ``symmetric'' ansatz where $\{t_{v,a}\}$ and $\{t_{w,b}\}$ are identical (i.e. $r_v = r_w$), and the ``deformed'' ansatz where $r_v\neq r_w$. In the former case, the sum includes a term with $t_{w,b}=t_{v,a}$, which must be treated separately so that the remaining logarithms can be put into the form $\log(1-z)$ for $|z|<1$. This gives
    \begin{align}\label{eq: 3cut vw int rem}
    &\sum_{I=1}^3\sum_{b=1}^M\left[i\pi + i(v_a-w_b)+\log\left(1-e^{i(w_b-v_a+\Delta_I)}\right)-\log\left(1-e^{i(v_a-w_b+\Delta_I)}\right)\right]\notag\\
    &=2\pi i\sum_{b=1}^M\left[\Theta(t_{w,b}-t_{v,a})\left(2+\sum_{I=1}^3\left\lfloor\frac{\Delta_I+f(t_{v,a})+f(t_{w,b})}{2\pi}\right\rfloor\right)\right.\notag\\
    &\left.+\Theta(t_{v,a}-t_{w,b})\left(1-\sum_{I=1}^3\left\lfloor\frac{\Delta_I-f(t_{w,b})-f(t_{v,a})}{2\pi}\right\rfloor\right)\right]+\delta_{r_v,r_w}\left[i\pi\sum_{I=1}^3\left(1+\left\lfloor\frac{\Delta_I+2f(t)}{2\pi}\right\rfloor\right.\right.\notag\\
    &\hspace{3.5cm}\left.\left.-\left\lfloor\frac{\Delta_I-2f(t)}{2\pi}\right\rfloor\right)+\log\prod_{I=1}^3\frac{\sin\frac{1}{2}[\Delta_I-2f(t)]}{\sin\frac{1}{2}[\Delta_I+2f(t)]}\right]\notag\\
    &+\sum_{I=1}^3\sum_{\substack{b=1\\ t_{w,b}>t_{v,a}}}^M\left[\log\left(1-e^{-M^\frac{1}{2}(t_{w,b}-t_{v,a})+i(\Delta_I-f(t_{v,a})-f(t_{w,b}))}\right)\right.\notag\\
    &\left.\hspace{5cm}-\log\left(1-e^{-M^\frac{1}{2}(t_{w,b}-t_{v,a})-i(\Delta_I+f(t_{v,a})+f(t_{w,b}))}\right)\right]\notag\\
    &+\sum_{I=1}^3\sum_{\substack{b=1\\ t_{w,b}<t_{v,a}}}^M\left[\log\left(1-e^{-M^\frac{1}{2}(t_{v,a}-t_{w,b})-i(\Delta_I-f(t_{v,a})-f(t_{w,b}))}\right)\right.\notag\\
    &\left.\hspace{5cm}-\log\left(1-e^{-M^\frac{1}{2}(t_{v,a}-t_{w,b})+i(\Delta_I+f(t_{v,a})+f(t_{w,b}))}\right)\right]\,.
    \end{align}
    When approximating the last two sums above by integrals, if $r_v = r_w$, we should use the Euler-Maclaurin formula \eqref{eq: Euler-Mac simp} with $r=1$ for the sum with lower limit $t_{w,b}>t_{v,a}$, and $r=0$ for the sum with upper limit $t_{w,b}<t_{v,a}$, as explained in Appendix \ref{App: E-M formula}. Otherwise, if $r_v\neq r_w$, \eqref{eq: 3cut rv rw def} implies
    \be\label{eq: 3cut vw sum deformed r}
    r=\lceil c(t_{v,b})\rceil -c(t_{v,b}) =\begin{cases}r_w -r_v\,, & r_w-r_v>0 \\ 1+r_w-r_v\,, & r_w-r_v <0\end{cases}
    \ee
    and we should use the Euler-Maclaurin formula \eqref{eq: Euler-Mac simp} with $r$ determined as \eqref{eq: 3cut vw sum deformed r}. The integrals are then simplified using the standard sequence of steps in \eqref{eq: 1cut sums aft taylor}. In summary, the last two sums in \eqref{eq: 3cut vw int rem} become
    \bea\label{eq: 3cut vw sum rem}
    &\begin{cases}-\log\prod_{I=1}^3\frac{\sin\frac{1}{2}[\Delta_I-2f(t)]}{\sin\frac{1}{2}[\Delta_I+2f(t)]}\,, & r_v=r_w \\
    2\pi i\sgn(r_w-r_v)\left(|r_w-r_v|-\frac{1}{2}\right)\left(1+\sum_{I=1}^3\left\lfloor\frac{\Delta_I+2f(t)}{2\pi}\right\rfloor+\sum_{I=1}^3\left\lfloor\frac{\Delta_I-2f(t)}{2\pi}\right\rfloor\right)\,, & r_v\neq r_w\end{cases}\\
    &\hspace{2.5cm}+(M^\frac{1}{2}-if'(t))\rho(t)\sum_{I=1}^3[g_+'([\Delta_I+2f(t)])-g_+'([\Delta_I-2f(t)])]\\
    &\hspace{3cm}-i\rho'(t)\sum_{I=1}^3[g_+([\Delta_I+2f(t)])+g_+([\Delta_I-2f(t)])]+\cO\left(M^{-\frac{1}{2}}\right)\,,
    \eea
    and we get
    \begin{align}\label{eq: 3cut vw sum}
    &\sum_{I=1}^3\sum_{b=1}^M\left[i\pi + i(v_a-w_b)+\log\left(1-e^{i(w_b-v_a+\Delta_I)}\right)-\log\left(1-e^{i(v_a-w_b+\Delta_I)}\right)\right]\notag\\
    &=2\pi i\sum_{b=1}^M\left[\Theta(t_{w,b}-t_{v,a})\left(2+\sum_{I=1}^3\left\lfloor\frac{\Delta_I+f(t_{v,a})+f(t_{w,b})}{2\pi}\right\rfloor\right)\right.\notag\\
    &\hspace{3cm}\left.+\Theta(t_{v,a}-t_{w,b})\left(1-\sum_{I=1}^3\left\lfloor\frac{\Delta_I-f(t_{w,b})-f(t_{v,a})}{2\pi}\right\rfloor\right)\right]\notag\\
    &\hspace{2cm}+(M^\frac{1}{2}-if'(t))\rho(t)\sum_{I=1}^3[g_+'([\Delta_I+2f(t)])-g_+'([\Delta_I-2f(t)])]\notag\\
    &\hspace{2cm}-i\rho'(t)\sum_{I=1}^3[g_+([\Delta_I+2f(t)])+g_+([\Delta_I-2f(t)])]\notag\\
    &+\begin{cases}i\pi\sum_{I=1}^3\left(1+\left\lfloor\frac{\Delta_I+2f(t)}{2\pi}\right\rfloor-\left\lfloor\frac{\Delta_I-2f(t)}{2\pi}\right\rfloor\right)\,, & r_v=r_w \\[.2cm]
    2\pi i\left(r_w-r_v-\frac{\sgn(r_w-r_v)}{2}\right)\left(1+\sum_{I=1}^3\left\lfloor\frac{\Delta_I+2f(t)}{2\pi}\right\rfloor+\sum_{I=1}^3\left\lfloor\frac{\Delta_I-2f(t)}{2\pi}\right\rfloor\right)\,, & r_v\neq r_w\end{cases}\notag\\
    &+\cO\left(M^{-\frac{1}{2}}\right)\,.
    \end{align}
    Summing the contributions from \eqref{eq: 3cut v eqn fund}, \eqref{eq: 3cut vv sum}, \eqref{eq: 3cut vu sum}, \eqref{eq: 3cut vw sum}, and subtracting $2\pi i$ times the integer ambiguities
    \begin{align}\label{eq: 3cut v eqn int amb}
    &n_a = \sum_{\substack{b=1\\ b\neq a}}^M\left[\Theta(t_{v,b}-t_{v,a})\left(2+\sum_{I=1}^3\left\lfloor\frac{\Delta_I+f(t_{v,a})-f(t_{v,b})}{2\pi}\right\rfloor\right)\right.\notag\\
    &\hspace{3cm}\left.+\Theta(t_{v,a}-t_{v,b})\left(1-\sum_{I=1}^3\left\lfloor\frac{\Delta_I+f(t_{v,b})-f(t_{v,a})}{2\pi}\right\rfloor\right)\right]\notag\\
    &+\sum_{b=1}^M\left[\Theta(t_{u,b}-t_{v,a})\left(2+\sum_{I=1}^3\left\lfloor\frac{\Delta_I+f(t_{v,a})}{2\pi}\right\rfloor\right)+\Theta(t_{v,a}-t_{u,b})\left(1-\sum_{I=1}^3\left\lfloor\frac{\Delta_I-f(t_{v,a})}{2\pi}\right\rfloor\right)\right]\notag\\
    &+\sum_{b=1}^M\left[\Theta(t_{w,b}-t_{v,a})\left(2+\sum_{I=1}^3\left\lfloor\frac{\Delta_I+f(t_{v,a})+f(t_{w,b})}{2\pi}\right\rfloor\right)\right.\notag\\
    &\hspace{3cm}\left.+\Theta(t_{v,a}-t_{w,b})\left(1-\sum_{I=1}^3\left\lfloor\frac{\Delta_I-f(t_{w,b})-f(t_{v,a})}{2\pi}\right\rfloor\right)\right]+4\delta_{r_v,r_w}\,,
    \end{align}
    we get that the large $N$ limit of the Bethe equations corresponding to $v$ is
    \begin{align}\label{eq: 3cut v eqn}
    &M^\frac{1}{2}\rho(t)\sum_{I=1}^3[g_+'([\Delta_I+f(t)])-g_+'([\Delta_I-f(t)])+g_+'([\Delta_I+2f(t)])-g_+'([\Delta_I-2f(t)])]\notag\\
    &-i\rho'(t)\sum_{I=1}^3[2g_+([\Delta_I])+g_+([\Delta_I+f(t)])+g_+([\Delta_I-f(t)])+g_+([\Delta_I+2f(t)])+g_+([\Delta_I-2f(t)])]\notag\\
    &-if'(t)\rho(t)\sum_{I=1}^3[g_+'([\Delta_I+2f(t)])-g_+'([\Delta_I-2f(t)])]-i\Delta_C-i\sgn(t)\frac{\Delta_3}{2}\notag\\
    &-2\pi i\sgn(t)\left\lfloor\frac{\frac{\Delta_1+\Delta_2}{2}-\sgn(t)f(t)}{2\pi}\right\rfloor-2\pi i\left(r_v-\frac{1}{2}\right)\left(1+\sum_{I=1}^3\left\lfloor\frac{\Delta_I+f(t)}{2\pi}\right\rfloor+\sum_{I=1}^3\left\lfloor\frac{\Delta_I-f(t)}{2\pi}\right\rfloor\right)\notag\\
    &+\begin{cases}i\pi\sum_{I=1}^3\left(1+\left\lfloor\frac{\Delta_I+2f(t)}{2\pi}\right\rfloor-\left\lfloor\frac{\Delta_I-2f(t)}{2\pi}\right\rfloor\right)-8\pi i\,, & r_v = r_w \\ 2\pi i\sgn(r_w-r_v)\left(|r_w-r_v|-\frac{1}{2}\right)\left(1+\sum_{I=1}^3\left\lfloor\frac{\Delta_I+2f(t)}{2\pi}\right\rfloor+\sum_{I=1}^3\left\lfloor\frac{\Delta_I-2f(t)}{2\pi}\right\rfloor\right)\,, & r_v\neq r_w\end{cases}\notag\\
    &=0\,.
    \end{align}
    Like before, the $\cO(1)$ part of \eqref{eq: 3cut v eqn int amb} was fixed by matching with numerical solutions.
    
    Due to the symmetry of the ansatz \eqref{eq: large N 3 col ansatz} under $f\mapsto -f$ and $v\leftrightarrow w$, the Bethe equation corresponding to $w$ can be obtained by taking \eqref{eq: 3cut v eqn fund}, \eqref{eq: 3cut vv sum}, \eqref{eq: 3cut vu sum}, \eqref{eq: 3cut vw int rem}, \eqref{eq: 3cut vw sum rem} with $f\mapsto -f$ and $r_v\leftrightarrow r_w$, and subtracting $2\pi i$ times the integer ambiguities
    \begin{align}
    &n_a=\sum_{\substack{b=1\\ b\neq a}}^M\left[\Theta(t_{w,b}-t_{w,a})\left(2+\sum_{I=1}^3\left\lfloor\frac{\Delta_I-f(t_{w,a})+f(t_{w,b})}{2\pi}\right\rfloor\right)\right.\notag\\
    &\hspace{3cm}\left.+\Theta(t_{w,a}-t_{w,b})\left(1-\sum_{I=1}^3\left\lfloor\frac{\Delta_I-f(t_{w,b})+f(t_{w,a})}{2\pi}\right\rfloor\right)\right]\notag\\
    &+\sum_{b=1}^M\left[\Theta(t_{u,b}-t_{w,a})\left(2+\sum_{I=1}^3\left\lfloor\frac{\Delta_I-f(t_{w,a})}{2\pi}\right\rfloor\right)+\Theta(t_{w,a}-t_{u,b})\left(1-\sum_{I=1}^3\left\lfloor\frac{\Delta_I+f(t_{w,a})}{2\pi}\right\rfloor\right)\right]\notag\\
    &+\sum_{b=1}^M\left[\Theta(t_{v,b}-t_{w,a})\left(2+\sum_{I=1}^3\left\lfloor\frac{\Delta_I-f(t_{w,a})-f(t_{v,b})}{2\pi}\right\rfloor\right)\right.\notag\\
    &\hspace{3cm}\left.+\Theta(t_{w,a}-t_{v,b})\left(1-\sum_{I=1}^3\left\lfloor\frac{\Delta_I+f(t_{v,b})+f(t_{w,a})}{2\pi}\right\rfloor\right)\right]-\delta_{r_v,r_w}\,.
    \end{align}
    The result is
    \begin{align}\label{eq: 3cut w eqn}
    &-M^\frac{1}{2}\rho(t)\sum_{I=1}^3[g_+'([\Delta_I+f(t)])-g_+'([\Delta_I-f(t)])+g_+'([\Delta_I+2f(t)])-g_+'([\Delta_I-2f(t)])]\notag\\
    &-i\rho'(t)\sum_{I=1}^3[2g_+([\Delta_I])+g_+([\Delta_I+f(t)])+g_+([\Delta_I-f(t)])+g_+([\Delta_I+2f(t)])+g_+([\Delta_I-2f(t)])]\notag\\
    &-if'(t)\rho(t)\sum_{I=1}^3[g_+'([\Delta_I+2f(t)])-g_+'([\Delta_I-2f(t)])]-i\Delta_C-i\sgn(t)\frac{\Delta_3}{2}\notag\\
    &-2\pi i\sgn(t)\left\lfloor\frac{\frac{\Delta_1+\Delta_2}{2}+\sgn(t)f(t)}{2\pi}\right\rfloor-2\pi i\left(r_w-\frac{1}{2}\right)\left(1+\sum_{I=1}^3\left\lfloor\frac{\Delta_I+f(t)}{2\pi}\right\rfloor+\sum_{I=1}^3\left\lfloor\frac{\Delta_I-f(t)}{2\pi}\right\rfloor\right)\notag\\
    &+\begin{cases}i\pi\sum_{I=1}^3\left(1+\left\lfloor\frac{\Delta_I-2f(t)}{2\pi}\right\rfloor+\left\lfloor\frac{\Delta_I+2f(t)}{2\pi}\right\rfloor\right)+2\pi i\,, & r_v = r_w \\ 2\pi i\sgn(r_v-r_w)\left(|r_v-r_w|-\frac{1}{2}\right)\left(1+\sum_{I=1}^3\left\lfloor\frac{\Delta_I+2f(t)}{2\pi}\right\rfloor+\sum_{I=1}^3\left\lfloor\frac{\Delta_I-2f(t)}{2\pi}\right\rfloor\right)\,, & r_v\neq r_w\end{cases}\notag\\
    &=0\,.
    \end{align}
    The $\cO(M^\frac{1}{2})$ parts of \eqref{eq: 3cut v eqn} and \eqref{eq: 3cut w eqn} give
    \be\label{eq: 3 col eq rho LO}
    \sum_{I=1}^3\left[g_+'([\Delta_I+f(\tilde{t})])-g_+'([\Delta_I-f(\tilde{t})])+g_+'([\Delta_I+2f(\tilde{t})])-g_+'([\Delta_I-2f(\tilde{t})])\right]=0\,.
    \ee
    This coincides with the first equation in \eqref{eq: 3col large N BAE LO and rho}. Equation \eqref{eq: 3 col eq rho LO} and its $t$ derivative then imply that \eqref{eq: 3cut v eqn} and \eqref{eq: 3cut w eqn} can be written as
    \begin{align}\label{eq: 3 col W' v rho eq rhot}
    &\partial_t\left\{\rho(t)\sum_{I=1}^3\left[2g_+([\Delta_I])+g_+([\Delta_I+f(t)])+g_+([\Delta_I-f(t)])+g_+([\Delta_I+2f(t)])\right.\right.\notag\\
    &\left.\left.+g_+([\Delta_I-2f(t)])\right]\vphantom{\sum_{I=1}^3}\right\}+\Delta_C+\sgn(t)\frac{\Delta_3}{2}+2\pi\sgn(t)\left\lfloor\frac{\frac{\Delta_1+\Delta_2}{2}-\sgn(t)f(t)}{2\pi}\right\rfloor\notag\\
    &+2\pi\left(r_v-\frac{1}{2}\right)\left(1+\sum_{I=1}^3\left\lfloor\frac{\Delta_I+f(t)}{2\pi}\right\rfloor+\sum_{I=1}^3\left\lfloor\frac{\Delta_I-f(t)}{2\pi}\right\rfloor\right)\notag\\
    &+\begin{cases}-\pi\sum_{I=1}^3\left(1+\left\lfloor\frac{\Delta_I+2f(t)}{2\pi}\right\rfloor-\left\lfloor\frac{\Delta_I-2f(t)}{2\pi}\right\rfloor\right)+8\pi\,, & r_v = r_w \\ -2\pi\sgn(r_w-r_v)\left(|r_w-r_v|-\frac{1}{2}\right)\left(1+\sum_{I=1}^3\left\lfloor\frac{\Delta_I+2f(t)}{2\pi}\right\rfloor+\sum_{I=1}^3\left\lfloor\frac{\Delta_I-2f(t)}{2\pi}\right\rfloor\right)\,, & r_v\neq r_w\end{cases}\notag\\
    &=0\,,
    \end{align}
    and 
    \begin{align}\label{eq: 3 col W' w rho eq rhot}
    &\partial_t\left\{\rho(t)\sum_{I=1}^3\left[2g_+([\Delta_I])+g_+([\Delta_I+f(t)])+g_+([\Delta_I-f(t)])+g_+([\Delta_I+2f(t)])\right.\right.\notag\\
    &\left.\left.+g_+([\Delta_I-2f(t)])\right]\vphantom{\sum_{I=1}^3}\right\}+\Delta_C+\sgn(t)\frac{\Delta_3}{2}+2\pi\sgn(t)\left\lfloor\frac{\frac{\Delta_1+\Delta_2}{2}+\sgn(t)f(t)}{2\pi}\right\rfloor\notag\\
    &+2\pi\left(r_w-\frac{1}{2}\right)\left(1+\sum_{I=1}^3\left\lfloor\frac{\Delta_I+f(t)}{2\pi}\right\rfloor+\sum_{I=1}^3\left\lfloor\frac{\Delta_I-f(t)}{2\pi}\right\rfloor\right)\notag\\
    &+\begin{cases}-\pi\sum_{I=1}^3\left(1+\left\lfloor\frac{\Delta_I-2f(t)}{2\pi}\right\rfloor-\left\lfloor\frac{\Delta_I+2f(t)}{2\pi}\right\rfloor\right)-2\pi\,, & r_v = r_w \\ -2\pi\sgn(r_v-r_w)\left(|r_v-r_w|-\frac{1}{2}\right)\left(1+\sum_{I=1}^3\left\lfloor\frac{\Delta_I+2f(t)}{2\pi}\right\rfloor+\sum_{I=1}^3\left\lfloor\frac{\Delta_I-2f(t)}{2\pi}\right\rfloor\right)\,, & r_v\neq r_w\end{cases}\notag\\
    &=0\,.
    \end{align}
    For \eqref{eq: 3 col W' v rho eq rhot} and \eqref{eq: 3 col W' w rho eq rhot} to be consistent with \eqref{eq: 3cut u eqn}, the coefficients multiplying $\rho(t)$ must be equal, and the remaining terms outside $\partial_t\{\ldots\}$ must also be equal. The first condition requires
    \be\label{eq: 3 col coeff consistency cond}
    \sum_{I=1}^3\left[g_+([\Delta_I+f(t)])+g_+([\Delta_I-f(t)])-g_+([\Delta_I+2f(t)])-g_+([\Delta_I-2f(t)])\right]=0\,,
    \ee
    which coincides with the second equation in \eqref{eq: 3col large N BAE LO and rho}, while the second condition gives \eqref{eq: 3col large N BAE phases}. Recall that the third equation in \eqref{eq: 3col large N BAE LO and rho} is already accounted for by \eqref{eq: 3cut u eqn}. 

    %%%%%%%%%%%%%%%%%%%%%%%%
    \subsection{3-cut ansatz with one cut at $\re u=\pi$}\label{App: 3cut w 1cut pi}
    %%%%%%%%%%%%%%%%%%%%%%%%

    We proceed to compute the large $N$ limit of the Bethe equations under the variant of the 3-cut ansatz \eqref{eq: large N 3 col ansatz pi} where one cut is along the vertical line $\re u =\pi$, and the other two cuts are reflection symmetric about the imaginary axis. The difference between \eqref{eq: large N 3 col ansatz pi} and the ansatz \eqref{eq: large N 3 col ansatz} considered in the previous section is simply that the real parts of $u_a$ have been shifted by $\pi$.
    
    In the Bethe equations corresponding to $u$, the first line of \eqref{eq: W'} gives
    \be\label{eq: 3cut pi u fund}
    -i\Delta_C-i\sgn(t)\frac{\Delta_3}{2}-2\pi i\left\lfloor\frac{\frac{\Delta_1+\Delta_2}{2}-\sgn(t)\pi}{2\pi}\right\rfloor+\cO\left(e^{-M^\frac{1}{2}|t|}\right)\,.
    \ee
    We split the second line of \eqref{eq: W'} into sums over $u$, $v$, and $w$ as in \eqref{eq: u eq sum split}. Since the constant $\re u_a=\pi$ cancels in $u_a-u_b$, the sum over $u$ is identical to \eqref{eq: 3cut uu sum}. In the sum over $v$, $u$ only appears in the combination $u_a-v_b$. Therefore, shifting $\re u$ by $\pi$ is the same as shifting $f\mapsto f-\pi$. Applying this shift to \eqref{eq: 3cut uv sum} gives
    \begin{align}\label{eq: 3cut pi uv sum}
    &\sum_{I=1}^3\sum_{b=1}^M\left[i\pi + i(u_a-v_b)+\log\left(1-e^{i(v_b-u_a+\Delta_I)}\right)-\log\left(1-e^{i(u_a-v_b+\Delta_I)}\right)\right]\notag\\
    &= 2\pi i\sum_{b=1}^M\left[\Theta(t_{v,b}-t_{u,a})\left(2+\sum_{I=1}^3\left\lfloor\frac{\Delta_I-f(t_{v,b})+\pi}{2\pi}\right\rfloor\right)+\Theta(t_{u,a}-t_{v,b})\left(\vphantom{\Bigg(}1\right.\right.\notag\\
    &\left.\left.-\sum_{I=1}^3\left\lfloor\frac{\Delta_I+f(t_{v,b})-\pi}{2\pi}\right\rfloor\right)\right]-(M^\frac{1}{2}+if'(t))\rho(t)\sum_{I=1}^3[g_+'([\Delta_I+f(t)-\pi])\notag\\
    &-g_+'([\Delta_I-f(t)+\pi])]-i\rho'(t)\sum_{I=1}^3[g_+([\Delta_I+f(t)-\pi])+g_+([\Delta_I-f(t)+\pi])]\notag\\
    &+2\pi i\left(r_v-\frac{1}{2}\right)\left(1+\sum_{I=1}^3\left\lfloor\frac{\Delta_I+f(t)-\pi}{2\pi}\right\rfloor+\sum_{I=1}^3\left\lfloor\frac{\Delta_I-f(t)+\pi}{2\pi}\right\rfloor\right)+\cO\left(M^{-\frac{1}{2}}\right)\,.
    \end{align}
    Similarly, since $u$ only appears in the sum over $w$ as $u_a-w_b$, shifting $\re u_a$ by $\pi$ is the same as shifting $f\mapsto f+\pi$. The sum over $w$ can thus be obtained from \eqref{eq: 3cut uw sum} via $f\mapsto f+\pi$: 
    \begin{align}\label{eq: 3cut pi uw sum}
    &\sum_{I=1}^3\sum_{b=1}^M\left[i\pi + i(u_a-w_b)+\log\left(1-e^{i(w_b-u_a+\Delta_I)}\right)-\log\left(1-e^{i(u_a-w_b+\Delta_I)}\right)\right]\notag\\
    &=2\pi i\sum_{b=1}^M\left[\Theta(t_{w,b}-t_{u,a})\left(2+\sum_{I=1}^3\left\lfloor\frac{\Delta_I+f(t_{w,b})+\pi}{2\pi}\right\rfloor\right)+\Theta(t_{u,a}-t_{w,b})\left(1\vphantom{\Bigg(}\right.\right.\notag\\
    &\left.\left.-\sum_{I=1}^3\left\lfloor\frac{\Delta_I-f(t_{w,b})-\pi}{2\pi}\right\rfloor\right)\right]+(M^\frac{1}{2}-if'(t))\rho(t)\sum_{I=1}^3[g_+'([\Delta_I+f(t)-\pi])\notag\\
    &-g_+'([\Delta_I-f(t)+\pi])]-i\rho'(t)\sum_{I=1}^3[g_+([\Delta_I+f(t)-\pi])+g_+([\Delta_I-f(t)+\pi])]\notag\\
    &+2\pi i\left(r_w-\frac{1}{2}\right)\left(1+\sum_{I=1}^3\left\lfloor\frac{\Delta_I+f(t)-\pi}{2\pi}\right\rfloor+\sum_{I=1}^3\left\lfloor\frac{\Delta_I-f(t)+\pi}{2\pi}\right\rfloor\right)+\cO\left(M^{-\frac{1}{2}}\right)\,,
    \end{align}
    where we have used $\left\lfloor\frac{\Delta_I+f(t)+\pi}{2\pi}\right\rfloor=\left\lfloor\frac{\Delta_I+f(t)-\pi}{2\pi}\right\rfloor+1$, $\left\lfloor\frac{\Delta_I-f(t)-\pi}{2\pi}\right\rfloor=\left\lfloor\frac{\Delta_I-f(t)+\pi}{2\pi}\right\rfloor-1$, $[\Delta_I+f(t)+\pi]=[\Delta_I+f(t)-\pi]$, and $[\Delta_I-f(t)-\pi]=[\Delta_I-f(t)+\pi]$. By matching with numerical solutions, we fix the integer ambiguities to be
    \bea
    &n_a = \sum_{\substack{b=1 \\ b\neq a}}^M\left[\Theta(t_{u,b}-t_{u,a})\left(2+\sum_{I=1}^3\left\lfloor\frac{\Delta_I}{2\pi}\right\rfloor\right)+\Theta(t_{u,a}-t_{u,b})\left(1-\sum_{I=1}^3\left\lfloor\frac{\Delta_I}{2\pi}\right\rfloor\right)\right]\notag\\
    &+\sum_{b=1}^M\left[\Theta(t_{v,b}-t_{u,a})\left(2+\sum_{I=1}^3\left\lfloor\frac{\Delta_I-f(t_{v,b})+\pi}{2\pi}\right\rfloor\right)+\Theta(t_{u,a}-t_{v,b})\left(\vphantom{\Bigg(}1\right.\right.\notag\\
    &\left.\left.-\sum_{I=1}^3\left\lfloor\frac{\Delta_I+f(t_{v,b})-\pi}{2\pi}\right\rfloor\right)\right]+\sum_{b=1}^M\left[\Theta(t_{w,b}-t_{u,a})\left(2+\sum_{I=1}^3\left\lfloor\frac{\Delta_I+f(t_{w,b})+\pi}{2\pi}\right\rfloor\right)\right.\notag\\
    &\left.+\Theta(t_{u,a}-t_{w,b})\left(1-\sum_{I=1}^3\left\lfloor\frac{\Delta_I-f(t_{w,b})-\pi}{2\pi}\right\rfloor\right)\right]+\Theta\left(\frac{2\pi}{3}-\Delta_C\right)\,.
    \eea
    Subtracting $2\pi in_a$ from the combined contributions of \eqref{eq: 3cut pi u fund}, \eqref{eq: 3cut uu sum}, \eqref{eq: 3cut pi uv sum}, and \eqref{eq: 3cut pi uw sum} then gives
    \begin{align}
    &2\partial_t\left\{\rho(t)\sum_{I=1}^3\left[g_+([\Delta_I])+g_+([\Delta_I+f(t)-\pi])+g_+([\Delta_I-f(t)+\pi])\right]\right\}+\Delta_C+\sgn(t)\frac{\Delta_3}{2}\notag\\
    &+2\pi\sgn(t)\left\lfloor\frac{\frac{\Delta_1+\Delta_2}{2}-\sgn(t)\pi}{2\pi}\right\rfloor+2\pi(1-r_v-r_w)\left(1+\sum_{I=1}^3\left\lfloor\frac{\Delta_I+f(t)-\pi}{2\pi}\right\rfloor\right.\notag\\
    &\hspace{3cm}\left.+\sum_{I=1}^3\left\lfloor\frac{\Delta_I-f(t)+\pi}{2\pi}\right\rfloor\right)+2\pi\Theta\left(\frac{2\pi}{3}-\Delta_C\right)=0\label{eq: 3cut pi u eqn}\,.
    \end{align}
    This is the third equation in \eqref{eq: bae 3cut pi LO rho' consis}.
    
    For the Bethe equations corresponding to $v$, since the ansatz for $v$ is identical in \eqref{eq: large N 3 col ansatz pi} and \eqref{eq: large N 3 col ansatz}, the first line of \eqref{eq: W'} is still given by \eqref{eq: 3cut v eqn fund}, while the second line is split into sums over $v$, $u$, and $w$ as in \eqref{eq: v eq sum split}. Once again, because the ansatz for $v$ is unchanged, the sum over $v$ is still simplified to \eqref{eq: 3cut vv sum}. In the sum over $u$, $u$ only appears via the combination $v_a-u_b$, and shifting $\re u$ by $\pi$ is equivalent to shifting $f\mapsto f-\pi$. The sum over $u$ is thus given by \eqref{eq: 3cut vu sum} under $f\mapsto f-\pi$, which is
    \begin{align}\label{eq: 3cut pi vu sum}
    &\sum_{I=1}^3\sum_{b=1}^M\left[i\pi + i(v_a-u_b)+\log\left(1-e^{i(u_b-v_a+\Delta_I)}\right)-\log\left(1-e^{i(v_a-u_b+\Delta_I)}\right)\right]\notag\\
    &=2\pi i\sum_{b=1}^M\left[\Theta(t_{u,b}-t_{v,a})\left(2+\sum_{I=1}^3\left\lfloor\frac{\Delta_I+f(t_{v,a})-\pi}{2\pi}\right\rfloor\right)+\Theta(t_{v,a}-t_{u,b})\left(1\vphantom{\Bigg(}\right.\right.\notag\\
    &\left.\left.-\sum_{I=1}^3\left\lfloor\frac{\Delta_I-f(t_{v,a})+\pi}{2\pi}\right\rfloor\right)\right]+M^\frac{1}{2}\rho(t)\sum_{I=1}^3[g_+'([\Delta_I+f(t)-\pi])-g_+'([\Delta_I-f(t)+\pi])]\notag\\
    &\hspace{2.6cm}-i\rho'(t)\sum_{I=1}^3[g_+([\Delta_I+f(t)-\pi])+g_+([\Delta_I-f(t)+\pi])]\notag\\
    &-2\pi i\left(r_v-\frac{1}{2}\right)\left(1+\sum_{I=1}^3\left\lfloor\frac{\Delta_I+f(t)-\pi}{2\pi}\right\rfloor+\sum_{I=1}^3\left\lfloor\frac{\Delta_I-f(t)+\pi}{2\pi}\right\rfloor\right)+\cO\left(M^{-\frac{1}{2}}\right)\,.
    \end{align}
    Since the ansatz for $v$ and $w$ are identical in \eqref{eq: large N 3 col ansatz pi} and \eqref{eq: large N 3 col ansatz}, the sum over $w$ is still given by \eqref{eq: 3cut vw sum}. By matching with numerical solutions, the integer ambiguities are fixed to
    \begin{align}
    &n_a = \sum_{\substack{b=1\\ b\neq a}}^M\left[\Theta(t_{v,b}-t_{v,a})\left(2+\sum_{I=1}^3\left\lfloor\frac{\Delta_I+f(t_{v,a})-f(t_{v,b})}{2\pi}\right\rfloor\right)+\Theta(t_{v,a}-t_{v,b})\left(1\vphantom{\Bigg(}\right.\right.\notag\\
    &\left.\left.-\sum_{I=1}^3\left\lfloor\frac{\Delta_I+f(t_{v,b})-f(t_{v,a})}{2\pi}\right\rfloor\right)\right]+\sum_{b=1}^M\left[\Theta(t_{u,b}-t_{v,a})\left(2+\sum_{I=1}^3\left\lfloor\frac{\Delta_I+f(t_{v,a})-\pi}{2\pi}\right\rfloor\right)\right.\notag\\
    &\left.+\Theta(t_{v,a}-t_{u,b})\left(1-\sum_{I=1}^3\left\lfloor\frac{\Delta_I-f(t_{v,a})+\pi}{2\pi}\right\rfloor\right)\right]+\sum_{b=1}^M\left[\Theta(t_{w,b}-t_{v,a})\left(2\vphantom{\Bigg(}\right.\right.\notag\\
    &\left.\left.+\sum_{I=1}^3\left\lfloor\frac{\Delta_I+f(t_{v,a})+f(t_{w,b})}{2\pi}\right\rfloor\right)+\Theta(t_{v,a}-t_{w,b})\left(1-\sum_{I=1}^3\left\lfloor\frac{\Delta_I-f(t_{w,b})-f(t_{v,a})}{2\pi}\right\rfloor\right)\right]\notag\\
    &+\begin{cases}2\,, & r_v=r_w \\ -\Theta\left(\Delta_C-\frac{2\pi}{3}\right)\,, & r_v\neq r_w\end{cases}\,.
    \end{align}
    Summing the contributions from \eqref{eq: 3cut v eqn fund}, \eqref{eq: 3cut vv sum}, \eqref{eq: 3cut pi vu sum}, \eqref{eq: 3cut vw sum}, and subtracting $2\pi in_a$, the large $N$ limit of the Bethe equations corresponding to $v$ is
    \begin{gather}
    iM^\frac{1}{2}\rho(t)\sum_{I=1}^3\left[g_+'([\Delta_I+f(t)-\pi])-g_+'([\Delta_I-f(t)+\pi])+g_+'([\Delta_I+2f(t)])\right.\notag\\
    \left.-g_+'([\Delta_I-2f(t)])\right]+\rho'(t)\sum_{I=1}^3[2g_+([\Delta_I])+g_+([\Delta_I+f(t)-\pi])+g_+([\Delta_I-f(t)+\pi])\notag\\
    +g_+([\Delta_I+2f(t)])+g_+([\Delta_I-2f(t)])]+f'(t)\rho(t)\sum_{I=1}^3[g_+'([\Delta_I+2f(t)])\notag\\
    -g_+'([\Delta_I-2f(t)])]+\Delta_C+\sgn(t)\frac{\Delta_3}{2}+2\pi\sgn(t)\left\lfloor\frac{\frac{\Delta_1+\Delta_2}{2}-\sgn(t)f(t)}{2\pi}\right\rfloor\notag\\
    +2\pi\left(r_v-\frac{1}{2}\right)\left(1+\sum_{I=1}^3\left\lfloor\frac{\Delta_I+f(t)-\pi}{2\pi}\right\rfloor+\sum_{I=1}^3\left\lfloor\frac{\Delta_I-f(t)+\pi}{2\pi}\right\rfloor\right)\notag\\
    -\begin{cases}\pi\sum_{I=1}^3\left(1+\left\lfloor\frac{\Delta_I+2f(t)}{2\pi}\right\rfloor-\left\lfloor\frac{\Delta_I-2f(t)}{2\pi}\right\rfloor\right)-4\pi\,, & r_v=r_w \\[.2cm]
    \scalebox{0.78}{$\begin{aligned}2\pi\left(r_w-r_v-\frac{\sgn(r_w-r_v)}{2}\right)\left(1+\sum_{I=1}^3\left\lfloor\frac{\Delta_I+2f(t)}{2\pi}\right\rfloor\right.\\
    \left.+\sum_{I=1}^3\left\lfloor\frac{\Delta_I-2f(t)}{2\pi}\right\rfloor\right)+2\pi\Theta\left(\Delta_C-\frac{2\pi}{3}\right)\,,\end{aligned}$} & r_v\neq r_w\end{cases}=0\label{eq: 3cut pi v eqn}\,.
    \end{gather}
    
    Since swapping $v\leftrightarrow w$ in the ansatz \eqref{eq: large N 3 col ansatz pi} is equivalent to taking $f\mapsto -f$, the Bethe equation corresponding to $w$ can be obtained by summing \eqref{eq: 3cut v eqn fund}, \eqref{eq: 3cut vv sum}, \eqref{eq: 3cut pi vu sum}, \eqref{eq: 3cut vw sum} with $v\leftrightarrow w$ and $f\mapsto -f$, and subtracting $2\pi in_a$ for a distinct set of integer ambiguities $n_a$. By matching with numerical solutions, we fix
    \begin{align}
    &n_a = \sum_{\substack{b=1\\ b\neq a}}^M\left[\Theta(t_{w,b}-t_{w,a})\left(2+\sum_{I=1}^3\left\lfloor\frac{\Delta_I-f(t_{w,a})+f(t_{w,b})}{2\pi}\right\rfloor\right)+\Theta(t_{w,a}-t_{w,b})\left(1\vphantom{\Bigg(}\right.\right.\notag\\
    &\left.\left.-\sum_{I=1}^3\left\lfloor\frac{\Delta_I-f(t_{w,b})+f(t_{w,a})}{2\pi}\right\rfloor\right)\right]+\sum_{b=1}^M\left[\Theta(t_{u,b}-t_{w,a})\left(2+\sum_{I=1}^3\left\lfloor\frac{\Delta_I-f(t_{w,a})-\pi}{2\pi}\right\rfloor\right)\right.\notag\\
    &\left.+\Theta(t_{w,a}-t_{u,b})\left(1-\sum_{I=1}^3\left\lfloor\frac{\Delta_I+f(t_{w,a})+\pi}{2\pi}\right\rfloor\right)\right]+\sum_{b=1}^M\left[\Theta(t_{v,b}-t_{w,a})\left(2\vphantom{\Bigg(}\right.\right.\notag\\
    &\left.\left.+\sum_{I=1}^3\left\lfloor\frac{\Delta_I-f(t_{w,a})-f(t_{v,b})}{2\pi}\right\rfloor\right)+\Theta(t_{w,a}-t_{v,b})\left(1-\sum_{I=1}^3\left\lfloor\frac{\Delta_I+f(t_{v,b})+f(t_{w,a})}{2\pi}\right\rfloor\right)\right]\,.
    \end{align}
    The resulting large $N$ Bethe equation is
    \begin{gather}
    -iM^\frac{1}{2}\sum_{I=1}^3[g_+'([\Delta_I+f(t)-\pi])-g_+'([\Delta_I-f(t)+\pi])+g_+'([\Delta_I+2f(t)])\notag\\
    -g_+'([\Delta_I-2f(t)])]+\rho'(t)\sum_{I=1}^3[2g_+([\Delta_I])+g_+([\Delta_I+f(t)-\pi])+g_+([\Delta_I-f(t)+\pi])\notag\\
    +g_+([\Delta_I+2f(t)])+g_+([\Delta_I-2f(t)])]+f'(t)\rho(t)\sum_{I=1}^3[g_+'([\Delta_I+2f(t)])\notag\\
    -g_+'([\Delta_I-2f(t)])]+\Delta_C+\sgn(t)\frac{\Delta_3}{2}+2\pi\sgn(t)\left\lfloor\frac{\frac{\Delta_1+\Delta_2}{2}+\sgn(t)f(t)}{2\pi}\right\rfloor\notag\\
    +2\pi\left(r_w-\frac{1}{2}\right)\left(1+\sum_{I=1}^3\left\lfloor\frac{\Delta_I+f(t)-\pi}{2\pi}\right\rfloor+\sum_{I=1}^3\left\lfloor\frac{\Delta_I-f(t)+\pi}{2\pi}\right\rfloor\right)\notag\\
    -\begin{cases}\pi\sum_{I=1}^3\left(1+\left\lfloor\frac{\Delta_I-2f(t)}{2\pi}\right\rfloor-\left\lfloor\frac{\Delta_I+2f(t)}{2\pi}\right\rfloor\right)\,, & r_v=r_w \\ 2\pi\left(r_v-r_w-\frac{\sgn(r_v-r_w)}{2}\right)\left(1+\sum_{I=1}^3\left\lfloor\frac{\Delta_I+2f(t)}{2\pi}\right\rfloor+\sum_{I=1}^3\left\lfloor\frac{\Delta_I-2f(t)}{2\pi}\right\rfloor\right)\,, & r_v\neq r_w \end{cases}=0\label{eq: 3cut pi w eqn}\,.
    \end{gather}
    
    The leading $\cO(M^\frac{1}{2})$ part of the $v$ and $w$ equations \eqref{eq: 3cut pi v eqn} and \eqref{eq: 3cut pi w eqn} both set
    \be\label{eq: 3cut pi v w eqn LO}
    \sum_{I=1}^3[g_+'([\Delta_I+f(t)-\pi])-g_+'([\Delta_I-f(t)+\pi])+g_+'([\Delta_I+2f(t)])-g_+'([\Delta_I-2f(t)])]=0\,.
    \ee
    This is the first equation in \eqref{eq: bae 3cut pi LO rho' consis}. Using \eqref{eq: 3cut pi v w eqn LO}, what remains in \eqref{eq: 3cut pi v eqn} and \eqref{eq: 3cut pi w eqn} can be written as 
    \begin{gather}
    \partial_t\left\{\rho(t)\sum_{I=1}^3[2g_+([\Delta_I])+g_+([\Delta_I+f(t)-\pi])+g_+([\Delta_I-f(t)+\pi])+g_+([\Delta_I+2f(t)])\right.\notag\\
    \left.\vphantom{\sum_{I=1}^3}+g_+([\Delta_I-2f(t)])]\right\}+\Delta_C+\sgn(t)\frac{\Delta_3}{2}+2\pi\sgn(t)\left\lfloor\frac{\frac{\Delta_1+\Delta_2}{2}-\sgn(t)f(t)}{2\pi}\right\rfloor\notag\\
    +2\pi\left(r_v-\frac{1}{2}\right)\left(1+\sum_{I=1}^3\left\lfloor\frac{\Delta_I+f(t)-\pi}{2\pi}\right\rfloor+\sum_{I=1}^3\left\lfloor\frac{\Delta_I-f(t)+\pi}{2\pi}\right\rfloor\right)\notag\\
    -\begin{cases}\pi\sum_{I=1}^3\left(1+\left\lfloor\frac{\Delta_I+2f(t)}{2\pi}\right\rfloor-\left\lfloor\frac{\Delta_I-2f(t)}{2\pi}\right\rfloor\right)-4\pi\,, & r_v=r_w \\[.2cm]
    \scalebox{0.78}{$\begin{aligned}2\pi\left(r_w-r_v-\frac{\sgn(r_w-r_v)}{2}\right)\left(1+\sum_{I=1}^3\left\lfloor\frac{\Delta_I+2f(t)}{2\pi}\right\rfloor\right.\\
    \left.+\sum_{I=1}^3\left\lfloor\frac{\Delta_I-2f(t)}{2\pi}\right\rfloor\right)+2\pi\Theta\left(\Delta_C-\frac{2\pi}{3}\right)\,,\end{aligned}$} & r_v\neq r_w\end{cases}=0\label{eq: 3cut pi v eqn wo f'}\,,
    \end{gather}
    and 
    \begin{gather}
    \partial_t\left\{\rho(t)\sum_{I=1}^3[2g_+([\Delta_I])+g_+([\Delta_I+f(t)-\pi])+g_+([\Delta_I-f(t)+\pi])+g_+([\Delta_I+2f(t)])\right.\notag\\
    \left.\vphantom{\sum_{I=1}^3}+g_+([\Delta_I-2f(t)])]\right\}+\Delta_C+\sgn(t)\frac{\Delta_3}{2}+2\pi\sgn(t)\left\lfloor\frac{\frac{\Delta_1+\Delta_2}{2}+\sgn(t)f(t)}{2\pi}\right\rfloor\notag\\
    +2\pi\left(r_w-\frac{1}{2}\right)\left(1+\sum_{I=1}^3\left\lfloor\frac{\Delta_I+f(t)-\pi}{2\pi}\right\rfloor+\sum_{I=1}^3\left\lfloor\frac{\Delta_I-f(t)+\pi}{2\pi}\right\rfloor\right)\notag\\
    -\begin{cases}\pi\sum_{I=1}^3\left(1+\left\lfloor\frac{\Delta_I-2f(t)}{2\pi}\right\rfloor-\left\lfloor\frac{\Delta_I+2f(t)}{2\pi}\right\rfloor\right)\,, & r_v=r_w \\ 2\pi\left(r_v-r_w-\frac{\sgn(r_v-r_w)}{2}\right)\left(1+\sum_{I=1}^3\left\lfloor\frac{\Delta_I+2f(t)}{2\pi}\right\rfloor+\sum_{I=1}^3\left\lfloor\frac{\Delta_I-2f(t)}{2\pi}\right\rfloor\right)\,, & r_v\neq r_w \end{cases}=0\label{eq: 3cut pi w eqn wo f'}
    \end{gather}
    respectively. The $u$ equation \eqref{eq: 3cut pi u eqn} determines $\rho$ to be
    \begin{gather}
    \rho(t) = \frac{\mu-\left(\sgn(t)\frac{\Delta_3}{2}+\Delta_C+2\pi q\right)t}{2G}\,,\notag\\
    G\equiv \sum_{I=1}^3[g_+([\Delta_I])+g_+([\Delta_I+f(t)-\pi])+g_+([\Delta_I-f(t)+\pi])]\,,\notag\\
    q \equiv\sgn(t)\left\lfloor\frac{\frac{\Delta_1+\Delta_2}{2}-\sgn(t)\pi}{2\pi}\right\rfloor+(1-r_v-r_w)\left(1+\sum_{I=1}^3\left\lfloor\frac{\Delta_I+f(t)-\pi}{2\pi}\right\rfloor\right.\notag\\
    \left.+\sum_{I=1}^3\left\lfloor\frac{\Delta_I-f(t)+\pi}{2\pi}\right\rfloor\right)+\Theta\left(\frac{2\pi}{3}-\Delta_C\right)\,,\label{eq: 3cut pi u rho}
    \end{gather}
    while \eqref{eq: 3cut pi v eqn wo f'} and \eqref{eq: 3cut pi w eqn wo f'} determine $\rho$ to be
    \begin{gather}
    \rho(t) = \frac{\tilde{\mu}-\left(\sgn(t)\frac{\Delta_3}{2}+\Delta_C+2\pi \tilde{q}\right)t}{2\wt G}\,,\notag\\
    \wt G\equiv\frac{1}{2}\sum_{I=1}^3[2g_+([\Delta_I])+g_+([\Delta_I+f(t)-\pi])+g_+([\Delta_I-f(t)+\pi])\notag\\
    +g_+([\Delta_I+2f(t)])+g_+([\Delta_I-2f(t)])]\notag\\
    \tilde{q}\equiv \sgn(t)\left\lfloor\frac{\frac{\Delta_1+\Delta_2}{2}-\sgn(t)f(t)}{2\pi}\right\rfloor+\left(r_v-\frac{1}{2}\right)\left(1+\sum_{I=1}^3\left\lfloor\frac{\Delta_I+f(t)-\pi}{2\pi}\right\rfloor+\sum_{I=1}^3\left\lfloor\frac{\Delta_I-f(t)+\pi}{2\pi}\right\rfloor\right)\notag\\
    -\begin{cases}\frac{1}{2}\sum_{I=1}^3\left(1+\left\lfloor\frac{\Delta_I+2f(t)}{2\pi}\right\rfloor-\left\lfloor\frac{\Delta_I-2f(t)}{2\pi}\right\rfloor\right)-2\,, & r_v=r_w \\[.2cm]
    \scalebox{0.78}{$\begin{aligned}\left(r_w-r_v-\frac{\sgn(r_w-r_v)}{2}\right)\left(1+\sum_{I=1}^3\left\lfloor\frac{\Delta_I+2f(t)}{2\pi}\right\rfloor\right.\\
    \left.+\sum_{I=1}^3\left\lfloor\frac{\Delta_I-2f(t)}{2\pi}\right\rfloor\right)+\Theta\left(\Delta_C-\frac{2\pi}{3}\right)\,,\end{aligned}$} & r_v\neq r_w\end{cases}\notag\\
    =\sgn(t)\left\lfloor\frac{\frac{\Delta_1+\Delta_2}{2}+\sgn(t)f(t)}{2\pi}\right\rfloor
    +\left(r_w-\frac{1}{2}\right)\left(1+\sum_{I=1}^3\left\lfloor\frac{\Delta_I+f(t)-\pi}{2\pi}\right\rfloor+\sum_{I=1}^3\left\lfloor\frac{\Delta_I-f(t)+\pi}{2\pi}\right\rfloor\right)\notag\\
    -\begin{cases}\frac{1}{2}\sum_{I=1}^3\left(1+\left\lfloor\frac{\Delta_I-2f(t)}{2\pi}\right\rfloor-\left\lfloor\frac{\Delta_I+2f(t)}{2\pi}\right\rfloor\right)\,, & r_v=r_w \\ \left(r_v-r_w-\frac{\sgn(r_v-r_w)}{2}\right)\left(1+\sum_{I=1}^3\left\lfloor\frac{\Delta_I+2f(t)}{2\pi}\right\rfloor+\sum_{I=1}^3\left\lfloor\frac{\Delta_I-2f(t)}{2\pi}\right\rfloor\right)\,, & r_v\neq r_w \end{cases}\label{eq: 3cut pi v/w rho}
    \end{gather}
    The constraint on $\tilde{q}$ above comes from the difference of equations \eqref{eq: 3cut pi v eqn wo f'} and \eqref{eq: 3cut pi w eqn wo f'}, and it is the second equality in \eqref{eq: bae 3cut pi rho' numer}. For \eqref{eq: 3cut pi u rho} and $\eqref{eq: 3cut pi v/w rho}$ to be consistent with each other, we need 
    \be\label{eq: rho u v consis}
    \frac{q}{G} = \frac{\tilde{q}}{\wt G}\,.
    \ee
    For simplicity, we will assume the stronger conditions
    \be\label{eq: rho u v consis num denom}
    G = \wt G\,,\quad q = \tilde{q}\,,
    \ee
    which are the second equation in \eqref{eq: bae 3cut pi LO rho' consis} and the first equality in \eqref{eq: bae 3cut pi rho' numer}. It turns out that almost all solutions satisfy \eqref{eq: rho u v consis num denom}, except for the solutions in regions $\tilde{2}$ and $\tilde{2}'$ introduced in Section \ref{subsec:multi-cut:3cut}, which satisfy the weaker condition \eqref{eq: rho u v consis} with modified numerators. 

    %%%%%%%%%%%%%%%%%%%%%%%%
    \section{Supersymmetry of M2-branes}
    \label{App:M2 susy}
	%%%%%%%%%%%%%%%%%%%%%%%%

    Our explanation for the non-perturbative corrections as wrapped M2-branes only makes sense if the wrappings considered in \eqref{eq: wrapped S3 def} are supersymmetric, so that these configurations can contribute to the index. In the following, we shall show this explicitly.

    The uplifted black hole solution in \eqref{eq: 11d uplifted g G} preserves two real supercharges in Lorentzian signature, i.e. it is $\frac{1}{16}$-BPS in 11d supergravity. Following equation (4.1) of \cite{Gauntlett:2001qs}, we choose the 11d coframe $e^{\hat{A}}$ as
    \begin{gather}
    e^{\hat{0}} = \Delta^{\frac{1}{3}}e^Udt\,,\quad e^{\hat{x}} = \Delta^{\frac{1}{3}}e^{\psi-U+h}dx\,,\quad e^{\hat{y}} = \Delta^{\frac{1}{3}}e^{\psi-U+h}dy\,,\notag\\
    e^{\hat{\rho}_I} = \Delta^{-\frac{1}{6}}\left(\mu_I\sqrt{L_I}e^{-U}dr+\frac{2}{\sqrt{L_I}}d\mu_I\right)\,,\quad e^{\hat{\phi}_I} = 2\Delta^{-\frac{1}{6}}\frac{\mu_I}{\sqrt{L_I}}\left(d\varphi_I+\frac{1}{\sqrt{2}}A_I\right)\,, 
    \end{gather}
    where the gauge coupling has been set to $g=\frac{1}{\sqrt{2}}$ and $h(x,y)$ is defined by $d\Omega_{\Sigma_\fg}^2=e^{2h}(dx^2+dy^2)$. The gamma matrices associated with $e^{\hat{A}}$ are denoted by $\Gamma_{\hat{A}}$, and they satisfy the 11d Clifford algebra $\{\Gamma_{\hat{A}},\Gamma_{\hat{B}}\}=2\delta_{\hat{A}\hat{B}}$. The Killing spinor in this frame is then given by
    \be\label{eq: 11d Killing spinor sing frame}
    \chi = \Delta^{\frac{1}{6}}e^{\frac{U}{2}}e^{-\frac{1}{2}\sum_I\varphi_I\Gamma_{\hat{x}\hat{y}}}\chi_0\,,
    \ee
    where $\Gamma_{\hat{A}_1\cdots\hat{A}_p}\equiv\Gamma_{[\hat{A}_1}\Gamma_{\hat{A}_2}\cdots\Gamma_{\hat{A}_p]}$, and the constant spinor $\chi_0$ satisfies the projections
    \be\label{eq: const 11d spinor proj}
    \Gamma_{\hat{\rho}_I\hat{\phi}_I}\chi_0 = -\Gamma_{\hat{x}\hat{y}}\chi_0\,.
    \ee
    These 4 projections commute, and thus the 32 components of $\chi_0$ only has 2 free components. With straightforward but lengthy algebra, one can check that \eqref{eq: 11d Killing spinor sing frame} satisfies the 11d Killing spinor equation \cite{Gauntlett:2002fz}
    \be
    \nabla_{\hat{A}}\chi + \frac{1}{288}\left[\Gamma_{\hat{A}}{}^{\hat{B}\hat{C}\hat{D}\hat{E}}-8\delta_{\hat{A}}^{\hat{B}}\Gamma^{\hat{C}\hat{D}\hat{E}}\right](G_4)_{\hat{B}\hat{C}\hat{D}\hat{E}}\chi=0\,,
    \ee
    where $\nabla_{\hat{A}}$ contains the spin connection.
    
    A probe Euclidean M2-brane configuration (the worldvolume does not wrap time) preserves the same supersymmetry as the ambient geometry if the Killing spinor $\chi$ satisfies the $\kappa$-symmetry condition \cite{Boido:2023ojv,Becker:1995kb}
    \be\label{eq: kappa symm cond}
    \bP \chi = \pm \chi\,,\quad \bP \equiv \frac{i}{3!\sqrt{h}}\varepsilon^{ijk}\partial_ix^M\partial_jx^N\partial_kx^P\Gamma_{MNP}\,,\quad \Gamma_M = \Gamma_{\hat{A}}e^{\hat{A}}_M
    \ee
    on the worldvolume of the brane. In the following, flat frame indices will always be denoted with a hat, while ordinary coordinate indices will not have hats. Also note that the Euclidean projector $\bP$ above is only compatible with complex Dirac spinors $\chi$, and not with Majorana spinors of the Lorentzian 11d supergravity. 
    
    To test the condition \eqref{eq: kappa symm cond} at the location \eqref{eq: wrapped S3 def} of the brane, we must first change frames. This is because $dt$ in $e^{\hat{0}}$ is not well-defined at $r=0$, and the coefficient of $dr$ in $e^{\hat{\rho}_I}$ diverges as $1/r$. Relatedly, the Killing spinor \eqref{eq: 11d Killing spinor sing frame} vanishes at $r=0$ since $e^{\frac{U(0)}{2}}=0$. To fix this, we transform $(e^{\hat{0}},e^{\hat{\rho}_I})\mapsto (f^{\hat{0}},f^{\hat{\rho}_I})$, where
    \begin{gather}
    f^{\hat{0}}=\cosh Ue^{\hat{0}} - \sinh U\sum_I\nu_I e^{\hat{\rho}_I}=\Delta^{\frac{1}{3}}\left(\frac{e^{2U}+1}{2}dv-dr\right)\,,\notag \\
    f^{\hat{\rho}_I}=-\sinh U\nu_Ie^{\hat{0}}+\sum_J\left[\delta_{IJ}+(\cosh U-1)\nu_I\nu_J\right]e^{\hat{\rho}_J}\notag\\
    =\nu_I\Delta^{\frac{1}{3}}\left(\frac{1-e^{2U}}{2}dv+dr\right)+\frac{2\Delta^{-\frac{1}{6}}}{\sqrt{L_I}}d\mu_I\,,\notag\\
    \nu_I\equiv \frac{\mu_I\sqrt{L_I}}{\sqrt{\Delta}}\,,\quad \sum_I\nu_I^2 = 1\,.
    \end{gather}
    One can check that this is a Lorentz transformation, since $-(e^{\hat{0}})^2+\sum_I(e^{\hat{\rho}_I})^2 = -(f^{\hat{0}})^2+\sum_I(f^{\hat{\rho}_I})^2$. Note that $f^{\hat{0}}$ and $f^{\hat{\rho}_I}$ are perfectly regular at $r=0$, once they are written in terms of the Eddington-Finkelstein coordinate $v$ defined in \eqref{eq: EF coord def}. In matrix form, the transformation can be written as
    \be
    e^{\hat{A}}\mapsto\Lambda^{\hat{A}}_{\hat{B}}e^{\hat{B}}\,,\quad \Lambda = \begin{pmatrix}\cosh U & -\sinh U\nu^T \\ -\sinh U\nu & \delta + (\cosh U-1)\nu\nu^T \end{pmatrix}\oplus\mathbbm{1}_{6\times 6}\,.
    \ee
    Its spinor representation is
    \begin{gather}
    S(\Lambda)=e^{-\frac{1}{2}U\Gamma_{\hat{0}}\sum_I\nu_I\Gamma_{\hat{\rho}_I}}=\cosh\frac{U}{2}-\sinh\frac{U}{2}\Gamma_{\hat{0}}\sum_I\nu_I\Gamma_{\hat{\rho}_I}=e^{-\frac{U}{2}}\Pi_+ + e^{\frac{U}{2}}\Pi_-\,,\notag\\
    \Pi_\pm \equiv \frac{1}{2}\left(1\pm \Gamma_{\hat{0}}\sum_I\nu_I\Gamma_{\hat{\rho}_I}\right)\,.
    \end{gather}
    Thus the Killing spinor in the transformed frame is 
    \be
    \tilde{\chi}=S(\Lambda)\chi=\Delta^{\frac{1}{6}}(\Pi_+ + e^U\Pi_-)e^{-\frac{1}{2}\sum_I\varphi_I\Gamma_{\hat{x}\hat{y}}}\chi_0\,.
    \ee
    Its value at $r=0$ is
    \be
    \tilde{\chi}|_{r=0}=\Delta^{\frac{1}{6}}\Pi_+e^{-\frac{1}{2}\sum_I\varphi_I\Gamma_{\hat{x}\hat{y}}}\chi_0\,,
    \ee
    which is nonzero.
    
    We are now ready to test \eqref{eq: kappa symm cond}. The projector $\bP$ is computed as
    \begin{gather}
    \Gamma_{\theta}=\Delta^{-\frac{1}{6}}\left(-\frac{\sin\frac{\theta}{2}}{\sqrt{L_1}}\Gamma_{\hat{\rho}_1}+\frac{\cos\frac{\theta}{2}}{\sqrt{L_2}}\Gamma_{\hat{\rho}_2}\right)\,,\quad \Gamma_{\varphi_1}=2\Delta^{-\frac{1}{6}}\frac{\cos\frac{\theta}{2}}{\sqrt{L_1}}\Gamma_{\hat{\varphi}_1}\,,\quad \Gamma_{\varphi_2}=2\Delta^{-\frac{1}{6}}\frac{\sin\frac{\theta}{2}}{\sqrt{L_2}}\Gamma_{\hat{\varphi}_2}\,,\notag\\
    \bP=\frac{i}{\sqrt{h}}\Gamma_{\theta}\Gamma_{\varphi_1}\Gamma_{\varphi_2}=i(-\nu_2\Gamma_{\hat{\rho}_1}+\nu_1\Gamma_{\hat{\rho}_2})\Gamma_{\hat{\varphi}_1}\Gamma_{\hat{\varphi}_2}\,.
    \end{gather}
    Using $[\bP,\Pi_\pm]=0$, the projections satisfied by $\chi_0$ in \eqref{eq: const 11d spinor proj}, and evaluating at the position \eqref{eq: wrapped S3 def} of the brane, one can show that
    \be
    \bP\tilde{\chi}|_{r=\mu_3=\mu_4=0}=i\Delta^{\frac{1}{6}}\Pi_+e^{-\frac{1}{2}\sum_I\varphi_I\Gamma_{\hat{x}\hat{y}}}\Gamma_{\hat{x}\hat{y}}\chi_0\,.
    \ee
    Since $i\Gamma_{\hat{x}\hat{y}}$ commutes with $\Gamma_{\hat{\rho}_I\hat{\phi}_I}$, and $(i\Gamma_{\hat{x}\hat{y}})^2=1$, we can further impose the projection
    \be\label{eq: complex proj}
    i\Gamma_{\hat{x}\hat{y}}\chi_0=\pm\chi_0\,,
    \ee
    in addition to \eqref{eq: const 11d spinor proj}. Then
    \be
    \bP\tilde{\chi}|_{r=\mu_3=\mu_4=0}=\pm\Delta^{\frac{1}{6}}\Pi_+e^{-\frac{1}{2}\sum_I\varphi_I\Gamma_{\hat{x}\hat{y}}}\chi_0 =\pm \tilde{\chi}|_{r=\mu_3=\mu_4=0}\,,
    \ee
    and the $\kappa$-symmetry condition \eqref{eq: kappa symm cond} is satisfied. Note that spinors satisfying \eqref{eq: kappa symm cond} are necessarily complex, not Majorana. The combined projections \eqref{eq: complex proj} and \eqref{eq: const 11d spinor proj} reduce $\chi_0$ to one complex component, which is consistent with the amount of supersymmetry preserved by the black hole and the TTI. Generically, one can check that M2-branes wrapping the $S^3$'s defined by $Z_I=Z_J=0$ for $I\neq J$ are supersymmetric, following the identical steps above. 

    \subsection*{M2-branes wrapping $\Sigma_\fg\times(S^1\subset S^7)$}

    We can also verify the supersymmetry of the M2-brane configurations considered in Section \ref{subsec:dressed:DressVSinst}, where the brane is located at $v=v_0$, $r=0$, and wraps a $S^1\in S^7$ given by $\mu_I=1$, $\mu_{J\neq I}=0$. The projector $\bP$ in this case is
    \be
    \bP = i\Gamma_{\hat{x}\hat{y}\hat{\varphi}_I}\,.
    \ee
    Using the projections \eqref{eq: const 11d spinor proj} and \eqref{eq: complex proj}, one can directly show that
    \be
    \bP\tilde{\chi}|_{r=0,\mu_I=1} = \pm\tilde{\chi}|_{r=0,\mu_I=1}\,. 
    \ee
    Thus the $\kappa$-symmetry condition is satisfied for these configurations and they are supersymmetric.
    
	%%%%%%%%%%%%%%%%%%%%%%%%
	\section{Regions of existence of multi-cut solutions}
	\label{App:multicut}
	%%%%%%%%%%%%%%%%%%%%%%%%
	
	In this section, we provide more details about the regions of existence of the $\bZ_m$-orbifold solutions as dictated by the supersymmetry condition, which is only briefly summarized in section \ref{subsubsec:multi-cut:mcut:distribution} of the main text. In section \ref{Appsubsec:multicut:Carry}, we use a more convenient normalization for the chemical potentials and translate the task of finding the regions of existence into the more transparent form of a four-weight coloring problem. In Section \ref{Appsubsec:multicut:generic-distribution}, we present the generic regions of existence of $\bZ_m$ orbifolds at fixed $m$. The exceptional points where no $\bZ_m$ orbifolds exist for any $m$ are discussed in Section \ref{Appsubsec:multicut:exceptional-points}. We conclude with a proof in Section \ref{Appsubsec:multicut:large-m} that away from the exceptional points, an infinite number of $\bZ_m$ orbifolds exist. 
	
	\subsection{The associated colouring problem}
	\label{Appsubsec:multicut:Carry}
	
	We first translate the problem of finding the regions of existence of the $m$-cut solutions into a problem of colouring the $(\Delta_1,\Delta_2)$ plane based on the value of a integer-valued function $T_m$. We shall denote the order $m$ of the $\bZ_m$ orbifold as the ``scale'' of the colouring since, as we shall see shortly, it determines the size of the unit cell of the periodic colouring. For later convenience, we define normalized versions of the chemical potentials
	\begin{gather}
    (x,y)\equiv\frac{1}{2\pi}(\Delta_1,\Delta_2)\,,\quad z\equiv\frac{\Delta_C}{\pi}\,,\notag\\
    w_I\equiv\frac{\hat{\Delta}_I}{2\pi}=\left(
    		x,\,
    		y,\,
    		\frac{1-x-y+z}{2},\,
    		\frac{1-x-y-z}{2}
    		\right)\,.\label{eq:wI}
    \end{gather}
	The weights $w_I$ satisfy $\sum_Iw_I=1$. For a positive integer \(m\), define
	the reduced quantities
	\be\label{eq: app bracket def}
	[m w_I]\equiv m w_I-s_I,\qquad I=1,2,3,4,
	\ee
	where the integral shifts $s_I$ are determined by requiring
	$
	[m w_I]\in[0,1).
	$\footnote{$[m w_I]$ here is the same as $\frac{1}{2\pi}[m\hat{\Delta}_I]$ in the main text. The two definitions of $[\cdot]$ should not be confused. We will stick to \eqref{eq: app bracket def} in this appendix.}
	This uniquely fixes
	\[
	s_I=\floor{m w_I},
	\qquad
	[m w_I]=\fracpart{m w_I}.
	\]
	The scale-\(m\) color at a point $(x,y,z)$ will be determined by the number
	\begin{equation}
		\label{eq:carry}
		T_m(x,y,z)
		:=
		\sum_{I=1}^4\fracpart{mw_I}
		=
		m-\sum_{I=1}^4\floor{mw_I}.
	\end{equation}
	The last expression shows that \(T_m\) is an integer.
	
	\begin{definition}
		At scale \(m\), a point is called black if \(T_m=1\), gray if \(T_m=3\), and white (uncolored) otherwise (i.e., $T_m = 0, 2$). The value \(T_m=4\) cannot occur because each reduced phase lies in \([0,1)\). In the following, black and gray are both regarded as colored. For fixed \(z\), a point \((x,y)\) is called exceptional if it is white at every positive scale \(m\).
	\end{definition}
    Comparing this definition with the existence condition in \eqref{eq: orbifold susy cond}, we see that a point is colored black if an orbifold in the $+$ branch exists, gray if an orbifold in the $-$ branch exists, or white otherwise. 
	
	\subsection{The generic distribution of $\bZ_m$ orbifolds}
	\label{Appsubsec:multicut:generic-distribution}
	
	This section concerns the geometry of the coloured set at a fixed scale $m$. We analyse the periodicity of the colouring in Section \ref{Appsubsubsec:multicut:generic-distribution:lattice} and prove the universal coloring pattern within a single cell in Section \ref{Appsubsubsec:multicut:generic-distribution:microcell}.
	
	\subsubsection{The lattice at fixed scale $m$}
	\label{Appsubsubsec:multicut:generic-distribution:lattice}
	
	For fixed \(m\), the coloring is periodic in the \((x,y)\)-plane.
	
	\begin{proposition}[Period lattice]
		\label{prop:period-lattice}
		The function \(T_m(x,y,z)\) is invariant under the lattice generated by
		\[
		\left(\frac{2}{m},0\right),
		\qquad
		\left(0,\frac{2}{m}\right),
		\qquad
		\left(\frac{1}{m},\frac{1}{m}\right).
		\]
		A fundamental period has area \(2/m^2\).
	\end{proposition}
	
	\begin{proof}
		Under \(x\mapsto x+2/m\), the four quantities \(mw_I\) change by
		\[
		(2,0,-1,-1),
		\]
		so their fractional parts are unchanged.  The same argument applies to \(y\mapsto y+2/m\).  Under \((x,y)\mapsto(x+1/m,y+1/m)\), they change by
		\[
		(1,1,-1,-1),
		\]
		which again leaves all fractional parts unchanged.  The first and third
		displayed translations generate a lattice of covolume \(2/m^2\).
	\end{proof}
	
	We now refine this period into a standard square grid. For
	\(p,q\in\Z\), consider the half-open square
	\[
	Q_{p,q}^{(m)}
	=
	\left[\frac{p}{m},\frac{p+1}{m}\right)
	\times
	\left[\frac{q}{m},\frac{q+1}{m}\right),
	\]
	in the $x-y$ plane, which we will refer to as a cell. For \((x,y)\in Q_{p,q}^{(m)}\), write
	\begin{equation}
		\label{eq:local-coordinates}
		mx=p+a,\qquad my=q+b,\qquad
		0\le a,b<1,
	\end{equation}
	and set
	\begin{equation}
		\label{eq:local-data}
		s=a+b,\qquad
		h\equiv m-p-q\pmod2,\quad h\in\{0,1\}.
	\end{equation}
	Thus \(h\) alternates between neighboring cells in the $x-y$ plane, and we will sometimes refer to it as the parity of the cell.
	
	For convenience, we also define the quantities
	\begin{equation}
		\label{eq:delta}
		\delta_m(z)
		:=
		\dist(mz,2\Z)
		\in[0,1]\,,
	\end{equation}
	and
	\begin{equation}
		\label{eq:lambda}
		\lambda_h(\delta)
		:=
		\begin{cases}
			\delta, & h=0,\\
			1-\delta, & h=1.
		\end{cases}
	\end{equation}
    We will see shortly that $\lambda_h(\delta_m(z))$ is the width of the coloured regions at fixed $z$.
    
	\subsubsection{The universal colouring law}
	\label{Appsubsubsec:multicut:generic-distribution:microcell}
	
	\begin{theorem}[Three-band decomposition]
		\label{thm:microcell}
		Fix \(m,z\), let \((x,y)\in Q_{p,q}^{(m)}\), and define
		\(a,b,s,h\) by \eqref{eq:local-coordinates}--\eqref{eq:local-data}.  Put
		\[
		\lambda=\lambda_h(\delta_m(z)).
		\]
		When \(0<\lambda\le1\), we have the exact rule
		\begin{equation}
			\label{eq:microcell-law}
			T_m=
			\begin{cases}
				1, & 0\le s\le\lambda,\\
				2, & \lambda<s\le2-\lambda,\\
				3, & 2-\lambda<s<2.
			\end{cases}
		\end{equation}
		If \(\lambda=0\), the interior of the square cell $Q_{p,q}^{(m)}$ is white and its lower-left corner has \(T_m=0\). In particular, every cell at $\lambda\neq 0$ consists of a black lower-left triangle, a white diagonal band, and a gray upper-right
		triangle.
	\end{theorem}
	
	\begin{proof}
		From \eqref{eq:wI} and \eqref{eq:local-coordinates},
		\[
		mw_3
		=
		\frac{m-p-q+mz-s}{2},
		\qquad
		mw_4
		=
		\frac{m-p-q-mz-s}{2}.
		\]
		Removing the integer part of \((m-p-q)/2\) gives
		\begin{equation}
			\label{eq:local-carry}
			T_m
			=
			s
			+
			\fracpart{\frac{h+mz-s}{2}}
			+
			\fracpart{\frac{h-mz-s}{2}}.
		\end{equation}
		The last two terms are invariant under \(mz\mapsto mz+2k\) and under \(mz\mapsto-mz\).  We may therefore replace \(mz\) by
		\(\delta=\delta_m(z)\in[0,1]\) defined in \eqref{eq:delta}.
		
		For \(h=0\), direct evaluation of the two fractional parts in
		\eqref{eq:local-carry} gives
		\[
		T_m=
		\begin{cases}
			1,&0\le s\le\delta,\\
			2,&\delta<s\le2-\delta,\\
			3,&2-\delta<s<2.
		\end{cases}
		\]
		For \(h=1\), the same calculation gives the transition values
		\[
		s=1-\delta,\qquad s=1+\delta,
		\]
		where $T_m$ changes from $1$ to $2$, and from $2$ to $3$ respectively. This is \eqref{eq:microcell-law} with \(\lambda=1-\delta\).  At
		\(\lambda=0\), the two coloured triangles collapse, and the stated corner value follows directly from \eqref{eq:local-carry}.
	\end{proof}
	
\begin{figure}[htbp]
		\centering
		\caption{The universal colouring pattern of two neighbouring cells at \(\delta_m=0.32\).  Black denotes \(T_m=1\), gray denotes \(T_m=3\), and white denotes \(T_m=2\). The coloured regions in each cell make up a square with width \(\lambda_0=\delta_m\) or \(\lambda_1=1-\delta_m\) for $h=0$ or $h=1$ respectively.}
		\vspace{1em}
		\includegraphics[width=\textwidth]{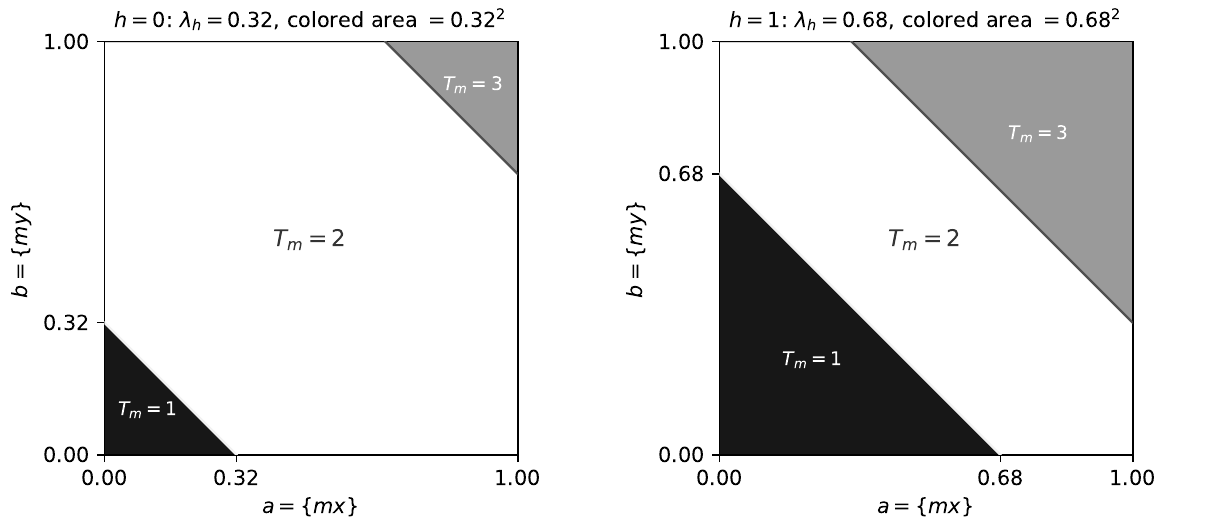}
		\label{fig:microcell}
	\end{figure}

	To illustrate the what the colouring is like within a cell, Figure~\ref{fig:microcell} displays two cells of different parities $h=0$ or $1$. The patterns are qualitatively the same; the parity only changes the widths $\delta_m$ and $1-\delta_m$ of the coloured regions.
	
\begin{figure}[htbp]
		\centering
		\caption{Coloring of the unit square in the $x-y$ plane for the irrational value \(z=1/(2\sqrt5)\) and various $m$. Black, gray, and white denote \(T_m=1,3,2\), respectively. The panel label \(A_m^{\rm sq}\) is the exact colored area in \([0,1)^2\).}
		\includegraphics[width=\textwidth]{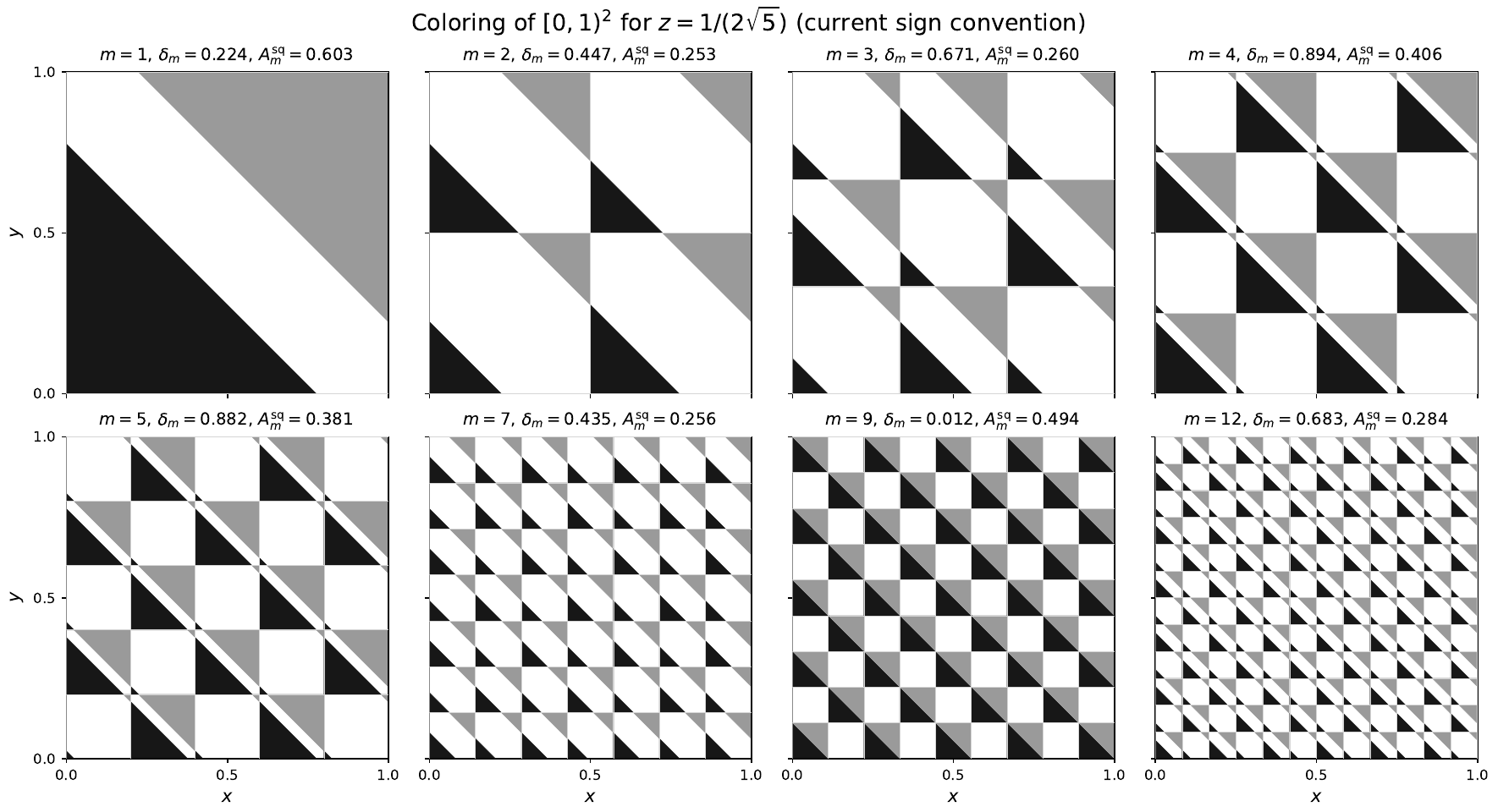}
		\label{fig:irrational-regions}
	\end{figure}

	More broadly, within the unit square $[0,1)^2$ in the $x-y$ plane, the coloured cells are repeated in a checkerboard-like pattern, since the parity $h$ alternates between neighbouring cells. We use
	\begin{equation}
		\label{eq:irrational-example}
		z=\frac{1}{2\sqrt5}
		\approx0.22360679775
	\end{equation}
	as a representative irrational value. Figure~\ref{fig:irrational-regions} then shows the colouring of \([0,1)^2\) at eight different values of $m$.  As $m$ increases, the dimensions of the cells are reduced as \(\sim 1/m\).  
    
	In summary, for fixed \(m\), the apparently complicated four-weight coloring reduces to a universal two-cell checkerboard controlled by \(\delta_m(z)\). The scale \(1/m\) determines the resolution of the grid of cells, while \(\delta_m\) determines the relative sizes of the black, gray, and white regions within each cell.
	
	\subsection{The exceptional points}
	\label{Appsubsec:multicut:exceptional-points}
	
	In this subsection, we study the values of $(x,y,z)$ that are always uncolored for any $m$, which we call the exceptional points. For \(z=0\), we completely classify the points that remain uncolored at every $m$: they are precisely the diagonal lines \(x-y\in 2\bZ+1\). For \(0<z<1\), the exceptional points lie along the anti-diagonal lines $x+y=n$, and diagonal lines $y-x=1\pm z+2n$, where $n\in\bZ$. If $z$ is irrational, all of these lines are exceptional, except for countably many points that we will specify. The closure of the exceptional points are the lines themselves. On the other hand, if $z$ is rational, every irrational point ($x\notin \bQ$, $y\notin\bQ$) on the diagonal lines are exceptional, while a rational point ($x\in \bQ$, $y\in\bQ$) on a diagonal line is only exceptional if it satisfies a specific condition. On the anti-diagonal lines, there are countably many exceptional points when $z$ is rational. Exact computations at finite $m$ support, but do not prove, the exhaustiveness of these exceptional points.
	
	\subsubsection{Complete Classification at \texorpdfstring{\(z=0\)}{z=0}}
	
	We first specialize to \(z=0\). Then
	\[
	w_3=w_4=\frac{1-x-y}{2}.
	\]
	Hence
	\[
	T_m(x,y,0)
	=
	\fracpart{mx}
	+
	\fracpart{my}
	+
	2\fracpart{\frac{m(1-x-y)}{2}}.
	\]
	Using
	\[
	T_m(x,y,0)
	=
	m-\floor{mx}-\floor{my}
	-2\floor{\frac{m(1-x-y)}{2}},
	\]
	we obtain the identity
	\[
	T_m(x,y,0)
	\equiv
	\floor{mx}+\floor{my}-m
	\pmod 2.
	\]
	Since \(T_m\) is an integer, the condition \(T_m\in\{1,3\}\) is equivalent to
	\[
	\floor{mx}+\floor{my}-m\equiv 1\pmod 2.
	\]
	Equivalently,
	\[
	\floor{mx}+\floor{m(y-1)}\equiv 1\pmod 2.
	\]
	Thus the problem is reduced to determining when the parity constraint above is violated for every $m$, so that $T_m\in\{0,2\}$ for every $m$ and the point is exceptional.
	
	\begin{lemma}[Parity separation]
		\label{lem:parity-separation}
		Let \(a,b\in\R\).  Then
		\[
		\floor{ma}+\floor{mb}\equiv 0\pmod 2
		\qquad\text{for every }m\ge 1
		\]
		if and only if $a-b\in 2\Z.$
	\end{lemma}
	
	\begin{proof}
		If \(a-b=2r\) with \(r\in\Z\), then
		\[
		\floor{ma}=\floor{mb}+2mr,
		\]
		so \(\floor{ma}+\floor{mb}\) is even for every \(m\).
		
		Conversely, assume that \(a-b\notin2\Z\). Choose representatives $t$, $t'$ of the two distinct classes \(a,b\in\R/2\Z\) such that \(t<t'\le t+1\). It suffices to show that the half-open interval $J=(t,t']$ contains an odd number of points of the lattice \(q^{-1}\Z\) for some positive integer \(q\), because then
		\[
		\floor{qt'}-\floor{qt}
		=
		\#\bigl(J\cap q^{-1}\Z\bigr)
		\]
		is odd and the parities of \(\floor{qa}\) and \(\floor{qb}\) are different.
		
		Choose a rational point \(p/q\in J\), written in lowest terms with \(q>0\), whose denominator \(q\) is minimal among all rational points of \(J\).  We claim that \(p/q\) is the unique point of \(J\cap q^{-1}\Z\). Indeed, if two distinct points of \(q^{-1}\Z\) lay in \(J\), then \(q>1\) because \(J\) has length at most \(1\) and is open at its left endpoint. Two adjacent such points, say \(r/q\) and \((r+1)/q\), would also lie in \(J\).  If either endpoint has reduced denominator \(<q\), this contradicts the choice of \(q\). Otherwise, after translating by an integer, \(1\le r\le q-2\), and the rational number $\frac{r}{q-1}$ lies strictly between \(r/q\) and \((r+1)/q\), again contradicting the minimality of \(q\). Thus \(\#(J\cap q^{-1}\Z)=1\), giving the required odd parity difference. Therefore, equality of the parities of $\floor{ma}$ and $\floor{mb}$ for every \(m\) is possible only when \(a\equiv b\pmod 2\), i.e. \(a-b\in2\Z\).
	\end{proof}
	
	\begin{theorem}
		\label{thm:z0-exceptional}
		For the colouring above at \(z=0\), a point \((x,y)\in\R^2\) is exceptional if and only if
		\[
		x-y\in 2\Z+1.
		\]
	\end{theorem}
	
	\begin{proof}
		Recall that by the arguments above, \((x,y)\) is colored at scale \(m\) if and only if
		\[
		\floor{mx}+\floor{m(y-1)}
		\equiv 1\pmod 2.
		\]
		Thus \((x,y)\) is exceptional if and only if
		\[
		\floor{mx}+\floor{m(y-1)}
		\equiv 0\pmod 2
		\qquad\text{for every }m\ge 1.
		\]
		Applying Lemma~\ref{lem:parity-separation} with
		\[
		a=x,\qquad b=y-1,
		\]
		this holds if and only if
		\[
		x-(y-1)\in 2\Z.
		\]
		Equivalently,
		\[
		x-y\in 2\Z-1=2\Z+1.
		\]
		This proves the claim.
	\end{proof}
	
	\begin{corollary}
		In the window \(-1<x,y<1\), the exceptional set consists of the two open line segments
		\[
		y=x+1,\qquad -1<x<0,
		\]
		and
		\[
		y=x-1,\qquad 0<x<1.
		\]
	\end{corollary}
	
	\subsubsection{Sufficient conditions for exceptionality at $z\neq 0$}
	
	For $z\neq 0$, we consider points where the four weights $w_I$ can be grouped into two pairs such that the sum of each pair of weights is an integer. As we will show shortly, if the points additionally satisfy a simple arithmetic condition, this is sufficient to show that they are exceptional. Whether this is also a necessary condition is addressed computationally in Section~\ref{sec:computations}.
	
	\begin{definition}
		For \(t\in\R\), we define the denominator ideal of $t$ in $\bZ_{>0}$
		\[
		H(t):=\{m\in\Z_{>0}: mt\in\Z\}.
		\]
		If \(t\notin\Q\), then \(H(t)=\varnothing\).  If \(t\in\Q\) and let
		\(\den(t)\) denote its reduced positive denominator, then $H(t)=\den(t)\Z_{>0}.$ This motivates its name; $H(t)$ is the set of possible positive denominators in a fractional representation of $t\in\bQ$. This set is closed under integer addition and multiplication in $\bZ_{>0}$, making it an ideal. 
	\end{definition}
	
	It follows immediately that
	\[
	H(u)=H(v)
	\]
	if and only if either \(u\) and \(v\) are both irrational, or \(u\) and \(v\) are both rational with the same reduced denominator.
	
	\begin{lemma}[Integer-pair contribution]
		\label{lem:pair-contribution}
		If \(u+v\in\Z\), then for every \(m\ge1\),
		\[
		\fracpart{mu}+\fracpart{mv}
		=
		\begin{cases}
			0, & m\in H(u),\\
			1, & m\notin H(u).
		\end{cases}
		\]
	\end{lemma}
	
	\begin{proof}
		Since \(m(u+v)\in\Z\), we have \(mv\equiv -mu\pmod\Z\).  If \(mu\in\Z\), both fractional parts vanish.  Otherwise the two fractional parts are complementary and sum to \(1\).
	\end{proof}
	
	If the four weights $w_I$ split into two pairs, say $(w_1,w_2)$ and $(w_3,w_4)$, such that $w_1+w_2\in\bZ$, $w_3+w_4\in\bZ$, and $H(w_1)=H(w_3)$, Lemma~\ref{lem:pair-contribution} makes the contributions of the pairs to $T_m$ equal at every $m$. Hence \(T_m\in\{0,2\}\) for all \(m\), and the point is exceptional. The three possible pairings yield the following families.

	\paragraph{The anti-diagonal family.}
	The pairing \((w_1,w_2)\), \((w_3,w_4)\) gives
	\[
	x+y=n,\qquad n\in\Z,
	\]
	and the pointwise condition
	\[
	H(x)=H\left(\frac{1-n+z}{2}\right).
	\]
	In the square \(0\le x,y<2\) and for \(0<z<1\), the relevant line segments are
	\[
	x+y=1,\qquad x+y=2,\qquad x+y=3.
	\]
	
	\paragraph{The first diagonal family.}
	The pairing \((w_1,w_3)\), \((w_2,w_4)\) gives
	\[
	y-x=1+z+2k,\qquad k\in\Z,
	\]
	and the pointwise condition
	\[
	H(x)=H(x+z).
	\]
	In \(0\le x,y<2\), \(0<z<1\), this produces the two segments
	\[
	y-x=1+z,\qquad y-x=-1+z.
	\]
	
	\paragraph{The second diagonal family.}
	The pairing \((w_1,w_4)\), \((w_2,w_3)\) gives
	\[
	y-x=1-z+2k,\qquad k\in\Z,
	\]
	and the pointwise condition
	\[
	H(x)=H(x-z).
	\]
	In \(0\le x,y<2\), \(0<z<1\), this produces the two segments
	\[
	y-x=1-z,\qquad y-x=-1-z.
	\]
	
	Thus the pairing construction produces seven line segments in the square: the 3 anti-diagonals
	\[
	x+y=1,\quad x+y=2,\quad x+y=3,
	\]
	together with the 4 diagonals
	\[
	 y-x=1+z,\quad y-x=-1+z;\quad y-x=1-z,\quad y-x=-1-z,
	\]
	with the pointwise conditions stated above, where the denominator ideals of the pairs of weights must be equal.
	
	\subsubsection*{Rational and Irrational $z$}
	
	The conditions on the denominator ideals determine which points on these segments are actually retained. If \(z\notin\Q\), then \((1-n+z)/2\) is irrational.  Hence, within the anti-diagonal line \(x+y=n\), the retained points are precisely those with
	\[
	x\notin\Q.
	\]
	Within the diagonal lines \(y-x=1\pm z+2k\), the retained points are
	those with
	\[
	x\notin \Q\cup(\mp z+\Q).
	\]
	Thus each of the seven segments is filled by exceptional points up to a countable set that is excluded; its closure is the whole segment.
	
	If \(z=a/b\in\Q\) is reduced, the diagonal segments contain every irrational point on the segment as exceptional points. A rational point is exceptional only when the relevant reduced denominators agree:
	\[
	\den(x)=\den(x\pm z)
	\]
	for \(y-x=1\pm z+2k\). On the other hand, the number of exceptional points on the anti-diagonal segments is finite. They all have $x,y\in\bQ$, and their count is given as follows. 
	\begin{lemma}[anti-diagonal count]
		\label{lem:off-slope-count}
		Let \(z=a/b\in(0,1)\cap\Q\) be reduced.  If the four diagonal
		segments are removed, the remaining number of exceptional points in the anti-diagonal segments inside \([0,2)^2\) depends only on \(b\), and is
		\begin{equation} \label{eq:slopeNegativeCount}
			N(a/b)=
			\begin{cases}
				4\varphi(b)-8, & b \text{ odd},\\[2mm]
				8\varphi(b)-8, & b \text{ even}.
			\end{cases}
		\end{equation}
		Here \(\varphi\) is Euler's totient function.
	\end{lemma}
	
	\begin{proof}
		On the anti-diagonal \(x+y=n\), \(n=1,2,3\), the condition on the denominator ideals can be written as
		\[
		\den(x)=\den\left(\frac{1-n+a/b}{2}\right).
		\]
		For \(n=1\) and \(n=3\), this denominator is
		\[
		q_1=q_3=\frac{2b}{\gcd(a,2)}.
		\]
		For \(n=2\), it is
		\[
		q_2=\frac{2b}{\gcd(a-b,2)}.
		\]
		If \(b\) is even, then \(a\) is odd, and hence
		\[
		q_1=q_2=q_3=2b.
		\]
		Since \(\varphi(2b)=2\varphi(b)\) for even \(b\), the three anti-diagonals
		contribute
		\[
		\varphi(2b)+2\varphi(2b)+\varphi(2b)=4\varphi(2b)=8\varphi(b)
		\]
		points before removing intersections with the diagonal segments. If \(b\) is odd, then one of \(q_1,q_2\) is \(b\) and the other is \(2b\), while \(\varphi(2b)=\varphi(b)\). The same count inside $[0,2)^2$ gives
		\[
		2\varphi(q_1)+2\varphi(q_2)=4\varphi(b)
		\]
		points before removing intersections with diagonal segments.
		
		It remains to subtract the overlaps with the four diagonal segments. Put
		\[
		c\in\{1-z,1+z,-1+z,-1-z\}.
		\]
		The intersections with \(x+y=n\) are given by
		\[
		x=\frac{n-c}{2},\qquad y=\frac{n+c}{2}.
		\]
		For \(c=1-z\in(0,1)\), the values \(n=1,2,3\) all lie in \([0,2)^2\). For \(c=1+z\in(1,2)\), only \(n=2\) lies in \([0,2)^2\). For \(c=-1+z\in(-1,0)\), again \(n=1,2,3\) all lie in \([0,2)^2\).  Finally, for \(c=-1-z\in(-2,-1)\), only \(n=2\) lies in \([0,2)^2\). Hence there are
		\[
		3+1+3+1=8
		\]
		distinct intersection points: the pair \((n,c)\) determines \((x,y)=((n-c)/2,(n+c)/2)\). It remains only to check that all eight were indeed included in the anti-diagonal count, and are exceptional points. On \(x+y=n\), the exceptionality requirement is $H(x)=H(t_n)$ with
		\[
		t_n=\frac{1-n+z}{2}.
		\]
		For the four possible values of \(c\), one has
		\[
		\begin{array}{ccl}
			c=1-z   &:& x=t_n+n-1,\\
			c=1+z   &:& x=-t_n,\\
			c=-1+z  &:& x=1-t_n,\\
			c=-1-z  &:& x=t_n+n.
		\end{array}
		\]
		Thus \(x\) differs from \(\pm t_n\) by an integer in every case, so \(H(x)=H(t_n)\). Equivalently, the reduced denominators agree when the point is rational. Each of the eight diagonal intersections has therefore been counted exactly once in the anti-diagonal total. Removing them gives the stated formula.
	\end{proof}
	
	\subsubsection{Numerical evidence for necessity}
	\label{sec:computations}
	
	Here, we present numerical evidence that the subsets of the segments presented above are the only exceptional points. For rational \(x,y,z\), the sequence \(T_m(x,y,z)\) in $m$ is periodic, with period dividing the least common multiple of the denominators of the four weights. Exceptionality is therefore decidable by checking one period in $m$, and removing all points with $T_m\in\{1,3\}$ for some $m$ in the period. All computations in the following use exact rational arithmetic.
	
	\subsubsection*{Checks at rational $(x,y,z)$}
	
\begin{figure}[htbp]
		\centering
		\includegraphics[width=\textwidth]{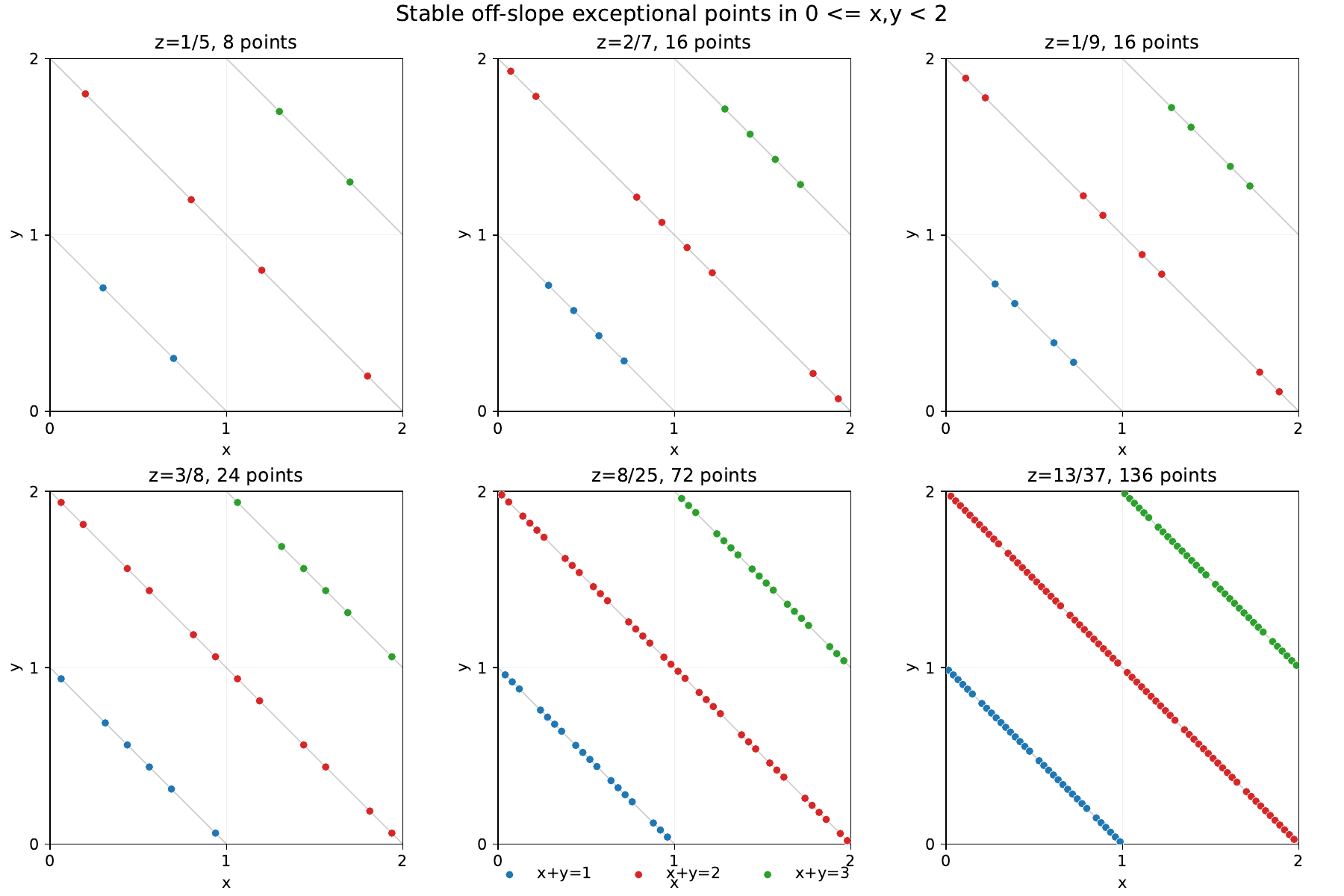}
		\caption{Exceptional points in \(0\le x,y<2\), excluding those on the diagonal segments, for several rational values of \(z\). The colors correspond to \(x+y=1,2,3\).}
		\label{fig:rational-examples}
	\end{figure}

	After removing the 4 diagonal segments, we find that the remaining exceptional points all lie on the three anti-diagonals \(x+y=1,2,3\). Figure~\ref{fig:rational-examples} shows the locations of the exceptional points for several choices of rational $z$. The point counts agree precisely with the formula above \eqref{eq:slopeNegativeCount}:
	\renewcommand{\arraystretch}{1}
	\[
	\begin{array}{c|c|c|c}
		z & b & N_{\mathrm{count}}(z) & N_{\mathrm{formula}}(z)\\
		\hline
		1/5 & 5 & 8 & 8\\
		2/7 & 7 & 16 & 16\\
		1/9 & 9 & 16 & 16\\
		3/8 & 8 & 24 & 24\\
		8/25 & 25 & 72 & 72\\
		13/37 & 37 & 136 & 136
	\end{array}
	\]
	
	In addition, the same count was evaluated for every reduced \(0<z<1\) of denominator \(b\leq 60\). In every case, the sweep of $[0,2)^2$ was performed along a grid of size \(1/2b\). For all reduced \(z\) with denominators \(b\leq 24\), a broader scan with grid size \(1/4b\) found no exceptional points outside the diagonal and anti-diagonal segments. These checks support the exhaustiveness of the exceptional points that we have found along the diagonal and anti-diagonal segments, at least in the tested range of $z$. They do not prove it in general.
	
	\subsubsection*{Approximating an irrational $z$}
	
	It is interesting to consider the limit where $z$ goes to an irrational number. Consider the continued-fraction convergents of
	\[
	\frac1{\sqrt2}=[0;1,2,2,\ldots]\,,
	\]
	where the RHS lists the sequence of partial quotients. For each convergent \(p/q\), Lemma~\ref{lem:off-slope-count} applies. As \(q\) grows, the finite sets of exceptional points become dense on the three anti-diagonal segments. Table~\ref{tab:sqrt2} records their sizes and largest gaps in $x$. At the exact value \(z=1/\sqrt2\), rational \(x\)-values on the anti-diagonal segments are no longer exceptional. This countable deletion does not change its closure, which is the union of \(x+y=1,2,3\). The diagonal segments have analogous countable deletions that we have already described.
	
	\renewcommand{\arraystretch}{1}
	\begin{table}[htbp]
		\centering
		\begin{tabular}{rrrrr}
			\toprule
			\(z=p/q\) & \(q\) & \(N(z)\) & formula & max gap\\
			\midrule
			\(5/7\) & \(7\) & \(16\) & \(16\) & \(4/7\)\\
			\(12/17\) & \(17\) & \(56\) & \(56\) & \(2/17\)\\
			\(70/99\) & \(99\) & \(232\) & \(232\) & \(1/33\)\\
			\(169/239\) & \(239\) & \(944\) & \(944\) & \(2/239\)\\
			\(408/577\) & \(577\) & \(2296\) & \(2296\) & \(2/577\)\\
			\(985/1393\) & \(1393\) & \(4744\) & \(4744\) & \(3/1393\)\\
			\(5741/8119\) & \(8119\) & \(30968\) & \(30968\) & \(3/8119\)\\
			\(13860/19601\) & \(19601\) & \(73720\) & \(73720\) & \(3/19601\)\\
			\(33461/47321\) & \(47321\) & \(186568\) & \(186568\) & \(3/47321\)\\
			\bottomrule
		\end{tabular}
		\caption{Exact rational checks along the convergents to \(1/\sqrt2\).}
		\label{tab:sqrt2}
	\end{table}
	
	\subsection{The infinite count of $\bZ_m$ orbifolds at non-exceptional points}
	\label{Appsubsec:multicut:large-m}
	
	In this section, we prove that if a point $(x,y,z)$ is coloured at some $m\in\bZ_{>0}$, it is coloured at infinitely many such values of $m$. This implies that at a non-exceptional point where a $\bZ_m$ orbifold exists for some $m$, there are in fact infinitely many other orbifolds existing at the same point.
	
	\begin{lemma}[Simultaneous recurrence on a torus]
		\label{lem:simultaneous-recurrence}
		For every \(r\in\R^d\), there exists a strictly increasing sequence of positive
		integers \(q_j\) such that
		\[
		q_jr\longrightarrow 0\pmod{\Z^d}.
		\]
	\end{lemma}
	
	\begin{proof}
		Write
		\[
		\|u\|_{\mathbb T^d}
		:=
		\max_{1\le i\le d}\min_{n\in\Z}|u_i-n|
		\]
		for the distance to the origin on \(\mathbb T^d=\R^d/\Z^d\).  Fix
		\(N\ge1\), partition \([0,1)^d\) into \(N^d\) half-open cubes of side
		\(1/N\), and consider the \(N^d+1\) points
		\[
		0,r,2r,\ldots,N^dr\pmod{\Z^d}.
		\]
		Two of them lie in the same cube.  Subtracting their indices produces an
		integer \(q_N\) with
		\[
		1\le q_N\le N^d,
		\qquad
		\|q_Nr\|_{\mathbb T^d}<\frac1N.
		\]
		
		If \(r\) has finite order on \(\mathbb T^d\), say
		\(Qr=0\pmod{\Z^d}\), one may simply take \(q_j=jQ\).  Otherwise the
		integers \(q_N\) obtained above must be unbounded.  Indeed, if they were
		bounded, some fixed positive integer \(q\) would occur for infinitely many
		values of \(N\); along that subsequence,
		\[
		\|qr\|_{\mathbb T^d}<\frac1N\longrightarrow0,
		\]
		which would imply \(qr=0\pmod{\Z^d}\), contradicting the assumption that
		\(r\) has infinite order.  Extracting a strictly increasing subsequence of
		the \(q_N\) proves the claim.
	\end{proof}
	
	This is the standard pigeonhole proof of the simultaneous form of
	Dirichlet's approximation theorem.
	
	\begin{theorem}[Recurrence of colored $m$]
		\label{thm:colored-recurrence}
		For every \((x,y,z)\in\R^3\), the set
		\[
		\mathcal M(x,y,z)
		:=\{m\in\Z_{>0}:T_m(x,y,z)\in\{1,3\}\}
		\]
		is either empty or infinite.
	\end{theorem}
	
	\begin{proof}
		Suppose that \(m_0\in\mathcal M(x,y,z)\), and set
		\[
		r_I=\fracpart{m_0w_I},\qquad
		R=\sum_{I=1}^4r_I=T_{m_0}\in\{1,3\}.
		\]
		For every positive integer \(k\),
		\[
		T_{km_0}=\sum_{I=1}^4\fracpart{k r_I}.
		\]
		If all \(r_I\) are rational, let \(Q\) be a common denominator.  Then
		\[
		\fracpart{(1+jQ)r_I}=r_I
		\]
		for every \(j\ge0\), and hence
		\[
		T_{(1+jQ)m_0}=T_{m_0}.
		\]
		
		Now suppose that at least one \(r_I\) is irrational. Simultaneous recurrence	on \(\R^4/\Z^4\) gives an unbounded sequence \(q_j\) such that
		\[
		q_jr\longrightarrow0\pmod{\Z^4},
		\qquad r=(r_1,r_2,r_3,r_4).
		\]
		Put \(k_j=q_j+1\).  Then \(k_jr\to r\) on the torus.  If \(r_I=0\), one has \(\fracpart{k_jr_I}=0\) identically.  For every nonzero \(r_I\), the point lies strictly inside \((0,1)\), so for sufficiently large \(j\) the torus convergence is ordinary convergence of the representatives in \([0,1)\).
		Consequently,
		\[
		T_{k_jm_0}=\sum_I\fracpart{k_jr_I}\longrightarrow\sum_Ir_I=R.
		\]
		Both sides are integers.  They are therefore equal for all sufficiently large \(j\), which supplies infinitely many colored multiples of \(m_0\).
	\end{proof}
	
	Thus a point cannot be colored at exactly one, or at any finite nonzero number of integers $m$. 
    
    	%%%%%%%%%%%%%%%%%%%%%%%%
	\section{Logarithmic corrections from the ADE field theory}
	\label{App:Log}
	%%%%%%%%%%%%%%%%%%%%%%%%
	
	In section \ref{subsubsec:holography:log} of the main text, we discuss the extra logarithmic corrections to the free energy from the orbifold fixed point $\Sigma_\fg \times S^5$ in the special case $s_I = (m,0,0,1)$ and $(m+1,0,0,0)$. Now we consider the most general $s_I$ satisfying $\sum_I s_I = m\pm 1$. 
	
	\subsection{Two vanishing components of $s_I$}
	
	We start with a simple example where $s_I$ has two vanishing components, where $s_I = (m+1,0,0,-2)$, where the generator acts as
	\begin{equation}
		(w, z_1, z_2, z_3, z_4)  \mapsto (\zeta w, \zeta z_1, z_2, z_3, \zeta^{-2} z_4).
	\end{equation}
	The fixed locus is five-dimensional $F_5 = \Sigma_{\fg} \times S^3$ with $w = z_1 = z_4 = 0$, and the normal slice is $\mathbb{C}^3/\mathbb{Z}_m(1,1,-2)$.\footnote{The fixed-point is identical to $s_I = (m+2,0,0,-1)$, where the generator acts as 
	\begin{equation}
			(w, z_1, z_2, z_3, z_4)  \mapsto (\zeta w, \zeta^2 z_1, z_2, z_3, \zeta^{-1} z_4).
	\end{equation} Using the complex coordinates $(w,\bar{z}_1, \bar{z}_4)$ on the normal six-plane, the weights are again equivalent, after relabeling, to $(1,1,m-2)$.
	} The theory defined by M-theory on a local singularity $\mathbb{C}_3/\Gamma, \Gamma\subset SU(3)$, is a strongly coupled five-dimensional $\cN=1$ SCFT rather than a fixed-plane Yang–Mills theory. \cite{Intriligator:1997pq, Tian:2021cif} This is the key difference from the $A$-type surface singularities living on the seven-dimensional fixed point. The latter give genuine seven dimensional ADE gauge fields on a codimension-four fixed plane, whereas a $\mathbb{C}_3/\Gamma$ singularity gives a strongly coupled five-dimensional fixed point. Gauge-theory descriptions, when present, are Coulomb-branch or mass-deformed phases of that SCFT, not intrinsic localized gauge fields at the orbifold point. Thus we expect it to give zero extra contributions to the logarithmic corrections for odd $m$. 
	
	However, when $m$ is even, there is an additional larger fixed stratum associated with the element $g^k = g^{m/2}$. For $s_I = (m+1,0,0,-2)$, the stratum is 
	\begin{equation}
		F_7 = \Sigma_\fg \times S^5, \quad w = z_1 = 0.
	\end{equation}
	Away from the smaller locus $F_5$, the stabilizer is only $\mathbb{Z}_2$, and the normal singularity is $\bC^2/\bZ_2$. This stratum supports a genuine seven-dimensional $SU(2)$ Yang–Mills theory described in section \ref{subsubsec:holography:orbifold:ADE}. Recall that the ADE gauge field with gauge group $SU(m)$ contributes to the logarithmic corrections \eqref{eq:ADElog}
	\begin{equation}
		\Delta F_{SU(m)-{\rm gauge\ zm}} = \frac{1}{2} (1-\fg)(m^2-1) \log N,
	\end{equation}
	thus the total ${\rm log}\, N$ term with $m = 2$ should be
	\begin{equation}
		c_{\rm log, total} = \left( \frac{1}{2}(1-\fg) + \frac{3}{2} (1-\fg) \right) \log N = 2 (1 - \fg) \log N.
	\end{equation}
	The most general $s_I$ with two vanishing components can be written as $s_I = (m+p+1,0,0,-p)$, or equivalently $s_I = (m+p, 0,0,-p-1)$. Both choices have the same local Calabi–Yau threefold quotient $\bC^3/\bZ_m(1,p,m-p-1)$. Let 
	\begin{equation}
		d_p = {\rm gcd}(m, p),\quad d_{p+1} = {\rm gcd}(m,p+1).
	\end{equation}
	If $d_p > 1$, then the coordinate with weight $p$ has a nontrivial maximal stabilizer $\bZ_{d_p}$. The	corresponding seven-dimensional stratum has normal surface singularity $\bC^2/\bZ_{d_p}$ and therefore supports a seven-dimensional $\cN = 1\ SU(d_p)$ Yang–Mills theory. Similarly for $d_{p+1} > 1$. The most general formula for the logarithmic coefficient is
	\begin{equation}
		c_{\rm log, total} = \frac{1}{2}(1-\fg) (d_p^2 + d_{p+1}^2 -1)\log N.
	\end{equation}
	In the special case $p = 0$ or $m-1$, it reduces to the $A_{m-1}$ seven-dimensional fixed-plane result of section \ref{subsubsec:holography:orbifold:ADE}. Let us take another example of $p = 3$ for demonstration:
	the maximal-stratum rule gives
	\renewcommand{\arraystretch}{1.3}
	\[
	\begin{array}{c|c|c}
		\text{condition on }m & \text{seven-dimensional strata}
		& \Delta c_{\rm log, total} \\ \hline
		3\nmid m,\ 2\nmid m & \text{none}
		& \frac{1}{2} (1-\fg)\log N \\
		2\mid m,\ 4\nmid m,\ 3\nmid m & SU(2)
		& 2 (1-\fg)\log N \\
		4\mid m,\ 3\nmid m & SU(4)
		& 8 (1-\fg)\log N \\
		3\mid m,\ 2\nmid m & SU(3)
		& \frac{9}{2} (1-\fg)\log N \\
		6\mid m,\ 4\nmid m & SU(2)\times SU(3)
		& 6 (1-\fg)\log N \\
		12\mid m & SU(4)\times SU(3)
		& 12 (1-\fg)\log N
	\end{array}
	\]
	
	\subsection{General four-weight orbifolds}
	Now consider the most general internal weights $s_I$ with $\sum_I s_I = m\pm 1$. The fixed strata are most cleanly organized by their generic stabilizer. Let $A \subset \{1,2,3,4\}$ be the subset of internal coordinates that are allowed to be nonzero, and set $d_A = {\rm gcd}(m, \{s_I | I \in A\})$. If $d_A = 1$, the corresponding locus has no nontrivial generic stabilizer and is not a singular stratum. If $d_A > 1$, the stratum is $F_A = \Sigma_\fg \times S^{2|A|-1}$ up to a finite residual quotient, and its generic stabilizer is $\bZ_{d_A}$. The normal slice is $\bC^{5-|A|}/\bZ_{d_A}$. Since $d_A$ divides all $s_I$ with $I \in A$, the normal weights obey the Calabi–Yau condition modulo $d_A$.
	
	There is one further point which is invisible if one only records the generic stabilizer. Let 
	\begin{equation}
		H_A = \langle g^{m/d_A} \rangle \simeq \bZ_{d_A},\quad R_A = \bZ_m / H_A \simeq \bZ_{m/d_A}.
	\end{equation}
	The $H_A$-cover gives the local ADE singularity, but the full orbifold is the residual quotient of this cover. Hence the localized twisted complex on $F_A$ must be projected to $R_A$-invariant states. We denote by $D_A^{\rm eff}$ the resulting effective multiplicity of the seven-dimensional ADE vector complex. In general four-weight orbifolds, $R_A$ can act as a discrete holonomy on the parent $A_{d_A - 1}$ twisted sector, and $D_A^{\rm eff}$ is the dimension of the invariant sector rather than the full $ d_A^2 - 1$. This is the correction to the na\"ive formula which simply sums over complete $SU(d_A)$ algebras.
	
	There is no $|A| = 4$ singular stratum. Indeed, if a common divisor $d > 1$ divided $m$ and all four $s_I$, then it would also divide $\sum_I s_I = m\pm 1$, which is impossible. Thus the possible singular strata are:
	\[
	\begin{array}{c|c|c}
		|A| & F_A & \text{localized fixed-point theory} \\ \hline
		3 & \Sigma_\fg \times S^5
		& 7d\ \mathcal N=1\ \text{ADE vector complex, }R_A\text{-projected} \\
		2 & \Sigma_\fg \times S^3
		& 5d\ \mathcal N=1\ \text{SCFT from }\bC^3/\bZ_{d_A} \\
		1 & \Sigma_\fg \times S^1
		& 3d\ \mathcal N=2\ \text{SCFT from }\bC^4/\bZ_{d_A}
	\end{array}
	\]
	The first line is the standard M-theory ADE fixed-plane construction. The second line is the Calabi–Yau threefold singularity construction of five-dimensional SCFTs \cite{Intriligator:1997pq, Tian:2021cif}. The third line is the familiar M2-brane/Calabi–Yau fourfold mechanism for three-dimensional superconformal theories; ABJM theory for $\bC^4/\bZ_k$ is the basic example, and quiver Chern–Simons constructions realize many toric $\bC^4$ singularities \cite{Aharony:2008ug,Hanany:2008gx}.
	
	The 3d and 5d fixed-point sectors are conformal and live in odd dimensions, for which there is no $\log L$ coefficients because of the lack of Weyl anomalies. Moreover, at the Coulomb-branch origin these SCFTs are not intrinsic localized gauge theories and can run in the loop. Possible exact moduli, special twists producing fermion zero modes, or deliberate moves to a gauge-theory phase would have to be treated as extra input; they are not included in the universal orbifold-saddle result. Thus the only localized contributions to the universal zero-mode logarithm come from the $|A| = 3$ ADE strata. With the same assumptions as above for the seven-dimensional vector complexes, but with the residual projection included, the total logarithmic coefficient should be
	\[
		c_{\rm log, total}
		=
		\frac12(1-\fg)
		\left[
		1+
		\sum_{\substack{|A|=3\\ d_A>1}}
		D_A^{\mathrm{eff}}
		\right]\log N. 
	\]
	Let us conclude the discussions with an explicit $m = 6$ example with $s_I = (3,3,4,3)$. The unique seven-dimensional maximal stratum is $A = \{1,2,4\}$ with $d_A = {\rm gcd}(6,3,3,3) = 3$, so the $H_A = \bZ_3$ cover has an $A_2$ singularity over $\Sigma_\fg \times S^5$. The remaining quotient is $R_A = \bZ_6 / H_A \simeq \bZ_2$. The naive full $SU(3)$ count would use $\dim SU(3) = 8$. However the residual $\bZ_2$ projects the $A_2$ twisted complex to an invariant subcomplex with $D_A^{\rm eff} = 4$. There is also a lower-dimensional stratum with $A = \{3\}$ and $d_A = 2$, but this is a three-dimensional $\cN = 2$ SCFT on $\Sigma_\fg \times S^1$, and gives no universal zero-mode logarithm in the present counting. Therefore
	\begin{equation}
		c_{\rm log, total} = \frac{1}{2} (1-\fg) (1 + 4) \log N = \frac{5}{2} (1-\fg) \log N.
	\end{equation}
	All of the examples we discuss here are exactly in accordance with the precise numerical results on the field theory side up to $m = 9$. This is a highly non-trivial precision test of holography.
	
	\newpage
	
	%%%%%%%%%%%%%%%%%%%%%%%%
	\bibliography{citationsTTI}
	\bibliographystyle{JHEP}
	%%%%%%%%%%%%%%%%%%%%%%%%
	
\end{document}